\documentclass[review,10pt]{elsarticle}

\usepackage{lineno}
\modulolinenumbers[5]

\usepackage{hyperref} 
\hypersetup{ 
  colorlinks   = true, 
  urlcolor     = blue, 
  linkcolor    = blue, 
  citecolor    = blue, 
} 
\usepackage[fleqn]{amsmath}
\usepackage{amsfonts}
\usepackage{amssymb}
\usepackage{amsthm}

\usepackage{pifont}
\usepackage{natbib}
\usepackage{geometry}
\usepackage{graphicx}
\usepackage{txfonts}

\usepackage{setspace}
\usepackage[utf8]{inputenc}
\usepackage{caption}
\usepackage{subcaption}
\usepackage[usenames, dvipsnames]{color}
\usepackage{color,soul}
\setstcolor{red}

\biboptions{sort&compress}

\begin{document}

\begin{frontmatter}

\title{Temporal atomization of a transcritical liquid \textit{n}-decane jet into oxygen}

\author{Jordi Poblador-Ibanez\fnref{myfootnote1}\corref{mycorrespondingauthor}}
\ead{poblador@uci.edu}
\author{William A. Sirignano\fnref{myfootnote3}}
\address{University of California, Irvine, CA 92697-3975, United States}
\fntext[myfootnote1]{Graduate Student Researcher, Department of Mechanical and Aerospace Engineering.}
\fntext[myfootnote3]{Professor, Department of Mechanical and Aerospace Engineering.}


\cortext[mycorrespondingauthor]{Corresponding author}


\begin{abstract}
The injection of liquid fuel at supercritical pressures is a relevant but overlooked topic in combustion. Typically, the role of two-phase dynamics is neglected under the assumption that the liquid will rapidly transition to a supercritical state. However, a transcritical domain exists where a sharp phase interface remains. This scenario is common in the early times of liquid fuel injection under real-engine conditions involving hydrocarbon fuels. Under such conditions, the dissolution of the oxidizer species into the liquid phase is accelerated due to local thermodynamic phase equilibrium (LTE) and vaporization or condensation can occur at multiple locations along the interface simultaneously. Fluid properties vary strongly under species and thermal mixing, with liquid and gas mixtures becoming more similar near the interface. As a result of the combination of low, varying surface-tension force and gas-like liquid viscosities, small surface instabilities develop early. \par

The mixing process, interface thermodynamics, and early deformation of a cool liquid \textit{n}-decane jet surrounded by a hotter moving gas initially composed of pure oxygen are analyzed at various ambient pressures and gas velocities. For this purpose, a two-phase, low-Mach-number flow solver for variable-density fluids is used. The interface is captured using a split Volume-of-Fluid method, generalized for the case where the liquid velocity is not divergence-free and both phases exchange mass across the interface. The importance of transcritical mixing effects over time for increasing pressures is shown. Initially, local deformation features appear that differ considerably from previous incompressible works. Then, the minimal surface-tension force is responsible for generating overlapping liquid layers in favor of the classical atomization into droplets. Thus, surface-area growth at transcritical conditions is mainly a consequence of gas-like deformations under shear rather than spray formation. Moreover, the interface can be easily perturbed in hotter regions submerged in the faster oxidizer stream under trigger events such as droplet or ligament impacts. The net mass exchange at high pressures limits the liquid-phase vaporization to small liquid structures.
\end{abstract}


\end{frontmatter}


\setlength\abovedisplayshortskip{0pt}
\setlength\belowdisplayshortskip{-5pt}
\setlength\abovedisplayskip{-5pt}

\section{Introduction}
\label{sec:intro}

Optimization of combustion efficiency and energy conversion per unit mass of fuel leads to the development of high-pressure combustion chambers. Diesel and gas turbines may operate in the range of 25 bar to 40 bar, while rocket engines can reach much higher pressures between 70 bar and 200 bar. Typical fuels used in these applications are liquid hydrocarbon mixtures (e.g., diesel, Jet-A, RP-1) that have to be injected, atomized and vaporized before the combustion chemical reaction occurs. Understanding the injection process and mixing with the surrounding oxidizer is necessary to design the combustion chamber correctly (e.g., chamber's size, injectors' distribution and shape). Many experimental and numerical studies have been conducted at subcritical pressures to address this issue. In this low-pressure regime, liquid and gas are easily distinguished. \par

However, typical components of hydrocarbon-based liquid fuels are \textit{n}-octane, \textit{n}-decane and \textit{n}-dodecane, which have critical pressures around 20 bar. Therefore, operating conditions of the discussed applications occur at near-critical or supercritical pressures for the fuel. Experiments carried out at these extreme pressures reveal the existence of a thermodynamic transition where it becomes difficult to distinguish between liquid and gas anymore. Both fluids present similar properties near the liquid-gas interface, which is immersed in a variable-density layer and rapidly affected by turbulence~\cite{mayer1996propellant,h1998atomization,mayer2000injection,oschwald2006injection,chehroudi2012recent,falgout2016gas,crua2017transcritical}. Despite being often described in the past as a sudden transition from a liquid to a gas-like supercritical state~\cite{spalding1959theory,rosner1967liquid}, the requirement that liquid and gas be in LTE at the interface provides evidence that a two-phase behavior exists within a certain region of the mixture thermodynamic space~\cite{hsieh1991droplet,delplanque1993numerical,yang1994vaporization,delplanque1995transcritical,delplanque1996transcritical,poblador2018transient}. For pressures above the fuel critical pressure, LTE enhances the dissolution of lighter gas species into the liquid fuel, which causes a local change in mixture critical properties. Moreover, mixing layers with large variations in the fluid properties develop in each phase~\cite{poblador2018transient}. Such domain of two-phase coexistence is described as ``transcritical" in the literature, with pressures above the critical pressure of the liquid but temperatures below the mixture critical temperature.  \par 

The phase-equilibrium assumption is not only bounded by the mixture critical point, beyond which the two-phase interface transitions to supercritical diffuse mixing. Other limitations apply to the validity of LTE and have been discussed in various works. The phase-equilibrium model breaks down in scenarios where the interface presents a large thermal resistivity, for which the temperature jump across the phase transition layer cannot be neglected~\cite{stierle2020selection}. Also, Dahms and Oefelein~\cite{dahms2013transition,dahms2015liquid,dahms2015non} and Dahms~\cite{dahms2016understanding} discuss and quantify the interface internal structure transition to the continuum domain at transcritical conditions. It is shown that the interface phase transition zone is only a few nanometers thick at supercritical pressures for the injected fuel but still subcritical based on the resulting mixture critical temperature. As the interface temperature approaches the critical temperature, the thickness of the phase transition region increases. The examples provided for hydrocarbon-nitrogen mixtures do not show interface thicknesses larger than 8 nm. The Knudsen number criteria \(Kn<0.1\), which is the ratio between the molecular mean free path and the interface thickness, is used to define whether the phase transition region enters continuum. Despite higher temperatures increasing the molecular mean free path, the high-pressure environment dominates and significantly decreases it. The suitable model for the interface, once continuum is established, is a diffuse region with sharp gradients, similar to the interface supercritical transition. That is, two phases cannot be identified anymore and classical LTE breaks down. For interface temperatures sufficiently below the mixture critical temperature at a given pressure, LTE is well established. Mixing regions quickly grow to the micron scale~\cite{poblador2018transient,davis2019development,poblador2021selfsimilar}; thus, the thickness of the interface becomes negligible and it can be considered as a discontinuity with a jump in fluid properties given by the proper interface modeling. For practical purposes, the non-equilibrium layer of compressive shocks in compressible flows is also treated as a discontinuity, despite its thickness being at least an order of magnitude greater than the phase transition region. \par

Crua et al.~\cite{crua2017transcritical} present experiments that further support the transcritical behavior of fuel injection in diesel engines. A wide range of high pressures and temperatures are analyzed, showing the presence of droplets that are strongly affected by the surrounding mixing and eventually show signs of a diffuse interface where liquid and gas cannot be identified. That is, heating of the liquid droplet increases the local temperature to near- or supercritical conditions for the interface mixture. Other experimental works also show a transcritical behavior where the liquid-gas interface appears and disappears~\cite{h1998atomization}. This transcritical situation is related to phase separation caused by mixture stability conditions~\cite{jofre2021transcritical}. As pressure, temperature and composition change within the physical domain, it may be possible for certain supercritical regions of the fluid mixture to become diffusionally unstable. That is, diffusion processes happen from low to high concentrations. This unstable situation drives phase separation and the reemergence of a liquid-gas interface. \par 

The temperature range over which two phases coexist decreases significantly with pressure, either because the critical temperature of the mixture is lower than typical injection temperatures or because the dense fluid has a molecular mean free path that is orders of magnitude shorter than the thickness of the phase transition layer. As a result, some studies, such as those by Zhang et al.~\cite{zhang2018supercritical} and Wang et al.~\cite{wang2019three}, look into liquid fuel injection at supercritical pressures and high temperatures without establishing a phase interface. The selected chamber pressure of 253 bar is significantly over any of the analyzed fluids' critical pressure, and two phases cannot coexist within the specified temperature range above 490 K. \par 

This transcritical behavior can explain why experimental observations fail to capture a two-phase environment with the presence of liquid structures. LTE at high pressures causes the liquid and gas phases to be more alike near the interface. That is, the composition, density and viscosity of both fluids are more similar (e.g., liquid viscosity drops to gas-like values)~\cite{yang2000modeling,poblador2018transient,davis2019development,poblador2021selfsimilar,poblador2021liquidjet}. As a result, the surface-tension coefficient drops substantially and becomes negligible near the mixture critical point. Therefore, the interface may experience a fast growth of small surface perturbations, resulting in the early breakup of small droplets and mixing enhancement. Despite some progress toward developing new experimental techniques for capturing two-phase behavior in supercritical pressure environments~\cite{minniti2018ultrashort,minniti2019femtosecond,traxinger2019experimental,klima2020quantification}, traditional visual techniques (e.g., shadowgraphy or ballistic imaging) may suffer from scattering and refraction issues under the presence of a cloud of small droplets submerged in a variable-density fluid. \par 

An analysis of such fast surface deformation processes under transcritical conditions is crucial to understanding the early mixing process and atomization in real-engine configurations prior to an eventual heating of the liquid phase and a transition to a mixture supercritical state. Such studies have been performed in the limit of incompressible two-phase flows without interface thermodynamics nor real-fluid considerations by Jarrahbashi and Sirignano~\cite{jarrahbashi2014vorticity}, Jarrahbashi et al.~\cite{jarrahbashi2016early} and Zandian et al.~\cite{zandian2017planar,zandian2018understanding,zandian2019length,zandian2019vorticity} and have provided valuable insights on atomization sub-domains that show common deformation patterns. The cases analyzed in the latter works are classified in a gas Weber number, \(We_G\), vs. liquid Reynolds number, \(Re_L\), diagram, which parametrizes the configurations based on the fluid properties of each phase, the surface-tension coefficient and the density ratio between liquid and gas. Three sub-domains are identified (see Figure~\ref{fig:Weg_vs_Rel_overview}): (a) the Lobe-Ligament-Droplet (LoLiD) sub-domain is characterized by low \(Re_L\) and \(We_G\). Surface-tension forces remain significant when compared to inertia forces. The formation and stretching of lobes, which eventually break up into large droplets due to capillary instabilities, precedes the breakup cascade process; (b) the Lobe-Hole-Bridge-Ligament-Droplet (LoHBrLiD) sub-domain is characterized by higher inertia effects compared to surface-tension forces (i.e., higher gas density or lower surface-tension coefficient). Lobes are easily perforated by the gas phase, which expand and form bridges that eventually break up into ligaments and droplets; and (c) the Lobe-Corrugation-Ligament-Droplet (LoCLiD) is a similar mechanism as LoLiD, but where higher inertia effects over viscous dissipation generate lobes that develop corrugations near the edge. Thus, ligaments form and stretch before capillary instabilities break them up into smaller droplets. The novelty in the analysis introduced by these authors is the inclusion of vorticity dynamics to explain the generation of liquid structures due to interactions of hairpin vortices with Kelvin-Helmholtz vortices. \par 

These incompressible results imply that a very fast liquid fuel atomization dominates real-engine configurations. Nonetheless, high-pressure effects must be accounted for in detail to fully comprehend the injection process in engines operating at high pressures. Poblador-Ibanez and Sirignano~\cite{pobladoribanez2021volumeoffluid} have proposed a physical and numerical model to analyze transcritical flows in the domain of two-phase coexistence. Real-fluid effects, as well as mass and thermal mixing, are considered. A detailed interface thermodynamic model is used based on the LTE assumption that can describe the local state of each phase at the interface, capture mass exchange and the variations in the surface-tension coefficient along the liquid surface. This model has been used to analyze a two-dimensional transcritical planar jet in detail~\cite{poblador2021liquidjet}. Similar to the present work, a binary configuration is considered where \textit{n}-decane at 450 K is injected into oxygen at 550 K. The gas phase moves relative to the liquid at 30 m/s and the thermodynamic pressure is 150 bar, well above the critical pressure of \textit{n}-decane. Characteristics of transcritical jets are presented (e.g., fast deformation, the importance of mixing effects). Notably, a detailed picture of the variation of the interface equilibrium state along the interface is provided. The surface-tension coefficient varies considerably and a reversal in net mass exchange at high pressures occurs, where certain interface regions can show net condensation even with a hotter gas. Moreover, some preliminary insight into three-dimensional configurations and how compressible vorticity dynamics explain the surface deformation is provided. \par

In this work, a detailed parametric, three-dimensional study is performed for the same binary mixture but with varying relative velocities and thermodynamic pressures. Section~\ref{sec:governing} presents the governing equations and interface matching relations that define the behavior of transcritical flows. The physical modeling has been simplified for binary mixtures under a low-Mach-number formulation. Then, Section~\ref{sec:thermo} describes the thermodynamic model used to define the fluid properties and the interface equilibrium state and Section~\ref{sec:numerics} presents a summary of the important details of the numerical methodology used to solve transcritical two-phase flows. Section~\ref{sec:results} presents the results and discussion of the paper, where the importance of mixing effects is emphasized, various deformation mechanisms intrinsic to transcritical atomization are shown and a discussion about the ease with which the liquid surface can be perturbed is presented. Moreover, details on the atomization process are presented, such as ligament and droplet formation, surface-area growth and the vaporization of the liquid. Lastly, Section~\ref{sec:summary_and_conclusions} concludes and summarizes the paper. To complement the results and discussion presented in this paper, the reader is strongly encouraged to access the Supplemental Material of this publication. \par 


\section{Governing equations}
\label{sec:governing}

This section presents the physical and thermodynamic modeling used to analyze non-reactive, transcritical liquid injection. That is, injection into a gaseous environment with a thermodynamic pressure above the critical pressure of the injected liquid but with a temperature still below the mixture critical temperature. \par 

The governing equations are modeled under a low-Mach-number assumption. Both fluids are compressible because of species and thermal mixing at elevated pressures, but the effect of pressure variations in the thermodynamic model is neglected. That is, the thermodynamic pressure is assumed constant and equal to the ambient pressure. Pressure variations are still important when determining the velocity field. Moreover, this work simplifies the problem by considering a binary configuration. Initially, the liquid phase is composed of a pure fuel species such a hydrocarbon (i.e., \(Y_2=Y_F=1\)) and the gas phase is composed of an oxidizer species (i.e., \(Y_1=Y_O=1\)). The mass fractions of both species are related as \(\sum_{i=1}^{N=2} Y_i=Y_O+Y_F = 1\). Despite this simplification, the methodology can easily be extended to multi-component configurations (i.e., \(N>2\)). Under these simplifications, the governing equations are the continuity equation, Eq. (\ref{eqn:cont}), the momentum equation, Eq. (\ref{eqn:mom}), the species continuity equation, Eq. (\ref{eqn:spcont}), and the energy equation, Eq. (\ref{eqn:energy}).

\begin{equation}
\label{eqn:cont}
\frac{\partial \rho}{\partial t} + \nabla \cdot (\rho\vec{u})=0
\end{equation}
\begin{equation}
\label{eqn:mom}
\frac{\partial}{\partial t}(\rho \vec{u})+\nabla \cdot (\rho \vec{u}\vec{u}) = -\nabla p + \nabla \cdot \bar{\bar{\tau}}
\end{equation}
\begin{equation}
\label{eqn:spcont}
\frac{\partial}{\partial t}(\rho Y_O) + \nabla\cdot(\rho Y_O \vec{u}) = \nabla \cdot (\rho D_m \nabla Y_O)
\end{equation}
\begin{equation}
\label{eqn:energy}
\frac{\partial}{\partial t}(\rho h) + \nabla\cdot(\rho h \vec{u}) = \nabla \cdot \bigg(\frac{\lambda}{c_p}\nabla h \bigg) + \sum_{i=1}^{N=2} \nabla \cdot \Bigg(\bigg[\rho D_m - \frac{\lambda}{c_p}\bigg]h_i \nabla Y_i\Bigg)
\end{equation}

In the momentum equation, \(p\) represents the dynamic pressure and \(\bar{\bar{\tau}}=\mu [\nabla\vec{u}+\nabla\vec{u}^\text{T}-\frac{2}{3}(\nabla\cdot\vec{u})\bar{\bar{I}}]\) is the viscous stress dyad, where \(\mu\) represents the dynamic viscosity of the fluid and \(\bar{\bar{I}}\) represents the identity dyad. The fluid density and velocity are represented by \(\rho\) and \(\vec{u}\), respectively. Notice that a Newtonian fluid under the Stokes' hypothesis has been considered. This hypothesis simplifies the momentum equation, but the high-pressure environment introduces non-idealities in the fluid behavior. Therefore, it might be necessary to consider the bulk viscosity or second coefficient of viscosity in future works for a more general treatment of the viscous term. As an example, estimates for the this coefficient could be obtained from Jaeger et al.~\cite{jaeger2018bulkvisc}. This work presents a methodology to obtain the bulk viscosity of molecular fluids and provides some results for \textit{n}-decane, which is used in the present work. \par 

Only the transport of the oxidizer mass fraction, \(Y_O\), is considered here due to the binary nature of the analyzed configurations. Fickian diffusion is assumed, but where the mass diffusion coefficient, \(D_m\), is obtained from a high-pressure, non-ideal correlation. In future works, more complex models may be considered to estimate mass diffusion, such as a generalized Maxwell-Stefan formulation for multi-component mixtures. Also, other diffusion mechanisms such as thermo-diffusion (i.e., Soret effect) may be considered if larger temperature differences are analyzed. \par 

The energy equation is represented by a transport equation for the mixture enthalpy, \(h\), where the temperature gradient has been replaced by \(\lambda \nabla T = (\lambda/c_p)\nabla h - \sum_{i=1}^{N=2}(\lambda/c_p)h_i\nabla Y_i\). The proper formulation for mixture enthalpy based on a real-fluid equation of state is used for the convection and conduction terms in Eq.~(\ref{eqn:energy}), while the partial derivative of mixture enthalpy with respect to mass fraction, \(h_i\equiv\partial h/\partial Y_i\), is used in the term for energy transport via mass diffusion. This definition of \(h_i\) is not exactly equal to the standard definition of partial enthalpy for ideal mixtures as the species' enthalpy at the same temperature and pressure as the mixture. Both approaches are equivalent only in the ideal case. \(\lambda\) and \(c_p\) are the thermal conductivity and the specific heat at constant pressure, respectively, and Fickian diffusion is considered in the term for energy transport via mass diffusion. Under the low-Mach-number assumption, viscous dissipation and pressure variation terms have been neglected. \par 

Note that turbulence models are not considered here. Although liquid atomization is a problem involving a transition from laminar to turbulent flow, the early times can be modeled with reasonable accuracy following a direct numerical approach. The scale of the domain is described in Subsection~\ref{subsec:description}, and the mesh size used in this work, \(\Delta x \sim \mathcal{O}(10^{-8}\text{m})\), is sufficiently small to consider the effect of under-resolved dissipation scales negligible, if not fully resolved, during the time frame analyzed here. Characteristic Reynolds numbers for the configurations analyzed in this work are shown in Table~\ref{tab:cases_2}. \par 

For two-phase flows, the liquid-gas interface acts as a moving boundary connecting both phases. The interface normal and tangential unit vectors are defined, respectively, as \(\vec{n}\) and \(\vec{t}\). At the interface, liquid and gas properties are identified with the subscripts \(l\) and \(g\), respectively. \par 

The governing equations for each phase are related across the interface via matching relations or jump conditions. Conservation jump conditions define a relation for the normal and tangential components of the velocity field as 

\begin{equation}
\label{eqn:veljump1}
(\vec{u}_g-\vec{u}_l) \cdot \vec{n} = \bigg(\frac{1}{\rho_g}-\frac{1}{\rho_l}\bigg)\dot{m}' \quad ; \quad \vec{u}_g \cdot \vec{t} = \vec{u}_l \cdot \vec{t}
\end{equation}

\noindent
where a velocity jump exists perpendicular to the interface in the presence of phase change and a no-slip condition defines a continuous tangential component of the velocity field. The mass flux per unit area across the interface is represented by \(\dot{m}'\) and is defined positive for net vaporization and negative for net condensation. \par

A pressure jump occurs across the interface caused by surface tension, a discontinuity in the normal viscous stresses and mass exchange given by

\begin{equation}
\label{eqn:momjump1}
p_l - p_g = \sigma \kappa + (\bar{\bar{\tau}}_l \cdot \vec{n}) \cdot \vec{n} - (\bar{\bar{\tau}}_g \cdot \vec{n} ) \cdot \vec{n} +\bigg(\frac{1}{\rho_g}-\frac{1}{\rho_l}\bigg)(\dot{m}')^2
\end{equation}

\noindent 
and the tangential stress is discontinuous because interface properties vary along the surface; thus, the surface-tension coefficient is not constant. The relation is given by

\begin{equation}
\label{eqn:momjump2}
(\bar{\bar{\tau}}_g \cdot \vec{n})\cdot \vec{t}-(\bar{\bar{\tau}}_l \cdot \vec{n}) \cdot \vec{t} = \nabla_s \sigma \cdot \vec{t}
\end{equation}

\(\sigma\) represents the surface-tension coefficient, \(\kappa=\nabla\cdot\vec{n}\) is the interface curvature, defined positive with a convex liquid shape, and \(\nabla_s=\nabla-\vec{n}(\vec{n}\cdot\nabla)\) represents the surface gradient. The surface-tension term in Eq. (\ref{eqn:momjump1}) tends to minimize the surface area per unit volume, which acts as a smoothing force for two-dimensional structures and some three-dimensional structures as well. However, this term is responsible for ligament thinning, neck formation and liquid breakup in three dimensions. On the other hand, the effect of the surface-tension coefficient gradient in Eq. (\ref{eqn:momjump2}), \(d\sigma/ds\) where \(s\) is the distance along the interface, drives the flow towards regions of higher surface-tension coefficient along the interface. In three dimensions, two tangential directions \(s_1\) and \(s_2\) are considered. \par

Lastly, Eq. (\ref{eqn:spcontmatch}) presents the jump condition for the species continuity equation and Eq. (\ref{eqn:energymatch}) presents the jump condition for the enthalpy transport equation, which has been further simplified for a binary mixture. \par 

\begin{equation}
\label{eqn:spcontmatch}
\dot{m}'(Y_{O,g}-Y_{O,l}) = (\rho D_m \nabla Y_O)_g \cdot \vec{n} - (\rho D_m \nabla Y_O)_l \cdot \vec{n}
\end{equation}

\begin{equation}
\label{eqn:energymatch}
\begin{split}
\dot{m}'(h_g-h_l) = \bigg(\frac{\lambda}{c_p}\nabla h\bigg)_g \cdot \vec{n} - \bigg(\frac{\lambda}{c_p}\nabla h\bigg)_l \cdot \vec{n} &+ \Bigg[\bigg(\rho D_m - \frac{\lambda}{c_p}\bigg)(h_O-h_F) \nabla Y_O\Bigg]_g \cdot \vec{n} \\ &- \Bigg[\bigg(\rho D_m - \frac{\lambda}{c_p}\bigg)(h_O-h_F) \nabla Y_O\Bigg]_l \cdot \vec{n}
\end{split}
\end{equation}

A closure for the interface matching is given by LTE. The equality in chemical potential for each species on each phase is expressed in terms of an equality in the fugacity of each species, \(f_{li}(T_l,p_l,X_{li}) = f_{gi}(T_g,p_g,X_{gi})\)~\cite{soave1972equilibrium,poling2001properties}. Further, this relation can be rewritten in terms of the fugacity coefficient, \(\Phi_i=f_i/pX_i\). Although fugacity depends on the pressure on each side of the interface, here the pressure jump across the interface is neglected for thermodynamic equilibrium purposes and pressure is assumed constant and equal to the ambient thermodynamic pressure under the low-Mach-number assumption. Thus, the equilibrium condition is expressed as \(X_{li}\Phi_{li}=X_{gi}\Phi_{gi}\), where \(X_{li}\) is the mole fraction of species \(i\) in the liquid phase and \(X_{gi}\) is the mole fraction of species \(i\) in the gas phase. Note that the equilibrium mass fraction of oxygen in each phase (i.e., \(Y_{O,g}\) and \(Y_{O,l}\)) is used in Eqs. (\ref{eqn:spcontmatch}) and (\ref{eqn:energymatch}). Moreover, temperature is assumed to be continuous across the sharp interface (i.e., \(T_g=T_l\)), although temperature gradients are different on each side. This assumption simplifies the LTE solution and a mixture equilibrium composition for each phase becomes readily available. \par 

The modeling of the interface as a contact discontinuity with negligible thickness and a sudden and sharp jump in fluid properties is a reasonable hypothesis for the specific part of the transcritical injection problem we discuss in this work. The various interface thermodynamic states as deformation and mixing occur are expected to be well below the mixture critical point as shown in Poblador-Ibanez and Sirignano~\cite{poblador2018transient,pobladoribanez2021volumeoffluid}. The interface has a negligible thickness of the order of a few nanometers~\cite{dahms2013transition,dahms2015liquid,dahms2016understanding} compared to the fast growth of mass and thermal mixing regions on either side of the interface to the order of micrometers~\cite{poblador2018transient,davis2019development,poblador2021selfsimilar,poblador2021liquidjet}. Only near the mixture critical point at very high pressures, the interface phase transition region might enter the continuum regime~\cite{dahms2015non}, thus invalidating the phase-equilibrium assumption. Another consideration that might invalidate the interface modeling relates to the interface thermal resistivity or heat transfer efficiency. For a large thermal resistivity, a substantial temperature jump exists across the interface, even across a distance of a few nanometers. Thus, the non-equilibrium interface state must be modeled~\cite{stierle2020selection}. However, as shown in Davis et al.~\cite{davis2019development}, the temperature jump across the interface is negligible when compared to the interface equilibrium temperature despite the small thermal conductivities involved. Further details regarding the validity of our approach are provided in Poblador-Ibanez and Sirignano~\cite{pobladoribanez2021volumeoffluid}. \par

\section{Thermodynamic model}
\label{sec:thermo}

The governing equations, matching relations and interface modeling are coupled to a thermodynamic model based on a volume-corrected Soave-Redlich-Kwong (SRK) cubic equation of state~\cite{lin2006volumetric}, described in~\ref{apn:srk}. Compared to the original SRK equation of state~\cite{soave1972equilibrium}, which is known to present density errors for dense fluids of up to 20\% when compared to experimental measurements~\cite{yang2000modeling,prausnitz2004thermodynamics}, the volume-corrected equation can accurately represent non-ideal fluid states for both the dense gas and the liquid phase. This correction is necessary for the accuracy of the dynamical behavior of the fluid. Moreover, it also affects the prediction of other transport properties such as viscosity and thermal conductivity, which are obtained from correlations that require the fluid density as an input parameter. Since the correction is implemented as a volume translation, thermodynamic variables obtained from the equation of state are equivalent between both the improved and the original SRK equation of state. \par

The direct solution of the equation of state provides the density of the mixture, \(\rho\). For the non-ideal fluid, \(h\), \(h_i\), \(c_p\) and \(\Phi_i\) are obtained from thermodynamic relations based on the equation of state and departure functions from the ideal state~\cite{poling2001properties}. These relations are available in Davis et al.~\cite{davis2019development} for the volume-corrected SRK equation of state. Viscosity and thermal conductivity are obtained from the generalized multiparameter correlation from Chung et al.~\cite{chung1988generalized}. The unified model for diffusion in non-ideal fluids presented in Leahy-Dios and Firoozabadi~\cite{leahy2007unified} is simplified for binary mixtures and used to determine \(D_m\). At the same time, the surface-tension coefficient is estimated as a function of the interface properties and composition from the Macleod-Sugden correlation as suggested in Poling et al.~\cite{poling2001properties}. This model estimates the surface-tension coefficient as \(\sigma = ([P_l]\rho_l - [P_g]\rho_g)^n\), where \([P_l]\) and \([P_g]\) are the parachors of the liquid and gas mixtures at the interface, \(\rho_l\) and \(\rho_g\) are the liquid and gas densities at the interface, and \(n\) is a coefficient usually set to 4. This approach is preferred over other models as it provides the correct limit where \(\sigma \rightarrow 0\) at the mixture critical point~\cite{poling2001properties}. Note that the thermodynamic pressure is assumed constant and equal to the ambient pressure throughout the thermodynamic model. \par 

Although this complex thermodynamic model might introduce some uncertainty in evaluating fluid and transport properties, the proposed equation of state and the other correlations are widely used in the literature and are accepted approaches to model high-pressure fluids. With the increased availability of experimental data in future years, we expect the accuracy of similar models will improve and reduce the uncertainty. \par

\section{Numerical method}
\label{sec:numerics}

This section provides relevant details about the numerical method used in this work. For a detailed description of the numerical approach used to solve the governing equations and capture the liquid-gas interface, the reader is referred to Poblador-Ibanez and Sirignano~\cite{pobladoribanez2021volumeoffluid}. The methodology is validated in the limit of incompressible, two-phase flows, while the extension to low-Mach-number, compressible, two-phase flows at high pressures is verified with various benchmark tests. Focus is given to the physical and numerical robustness of the results (e.g., grid independence). Different modeling blocks, such as the thermodynamic model, are validated independently. \par

The numerical approach is based on an interface-capturing Volume of Fluid (VOF) method adapted to compressible liquids with phase change. It is an extension of the incompressible VOF method by Baraldi et al.~\cite{baraldi2014mass}, which has been used to solve isotropic turbulence in droplet-laden flows~\cite{dodd2014fast}. More recently, it has been extended to simulate evaporating droplets and includes gas-phase compressibility~\cite{dodd2021analysis}. The base VOF method has been chosen due to its mass-conserving properties in the incompressible limit (i.e., volume-preserving scheme) and computational efficiency compared to other higher-order VOF methods. The Piecewise Linear Interface Construction (PLIC) method~\cite{youngs1982time} is used to maintain a sharp interface. The volume fraction distribution of the liquid phase, \(C\), is used to obtain geometrical information about the interface topology. The interface normal unit vector is evaluated from a Mixed-Youngs-Centered method~\cite{aulisa2007interface} and the curvature is estimated from an improved Height Function method~\cite{lopez2009improved}. \par 

The interface equilibrium state is solved at all interface cells based on the matching relations and LTE. Because of the low-Mach-number nature of the problem, the equilibrium solution can be solely determined from the species mass balance, energy balance and phase equilibrium at the interface. A normal-probe technique is implemented, whereby a probe is extended into each phase from the centroid of each local interface plane following the normal unit vector. Then, the mass fraction and the enthalpy values are linearly interpolated onto two different points on the probe in each phase. This way, the normal gradients of mass fraction and enthalpy needed in Eqs.~(\ref{eqn:spcontmatch}) and~(\ref{eqn:energymatch}) can be evaluated at the interface with a one-sided, second-order scheme. The system of equations that results can be solved with an iterative solver to determine the local interface state~\cite{poblador2018transient} (i.e., equilibrium composition and temperature, mass flux per unit area). \par 

The non-conservative forms of the species and energy transport equations are solved for each phase separately. That is, the interface acts as a moving phase boundary. A first-order explicit time integration is used, while the spatial discretization is based on a one-sided hybrid first- and second-order upwinding scheme for the convective terms and second-order central differences for the diffusive terms. The upwinding scheme ensures that the integration of the equations maintains numerical stability and boundedness (i.e., \(0\leq Y_O\leq 1\)). Then, the interface solution is directly embedded in the numerical stencils. \par 

The continuity and momentum equations are solved in conservative form using a one-fluid approach. Fluid properties (e.g., density) are volume-averaged at interface cells using the volume fraction of the liquid as \(\phi=\phi_g+(\phi_l-\phi_g)C\), where \(\phi\) is any fluid property. Convective terms are discretized using the SMART algorithm~\cite{gaskell1988curvature} and viscous terms are discretized using central differences. The Continuum Surface Force (CSF)~\cite{brackbill1992continuum} adapted to flows with variable surface-tension coefficient~\cite{kothe1996volume,seric2018direct} is implemented to satisfy the momentum jump conditions. A predictor-projection method with a first-order, explicit temporal integration is used to address the pressure-velocity coupling for low-Mach-number, two-phase flows. The method is outlined in~\ref{apn:pressuremethod} and relies on the split pressure-gradient technique for a fast and computationally efficient pressure update using an FFT or DFT solver~\cite{dodd2014fast,pobladoribanez2021volumeoffluid}. Moreover,~\ref{apn:pressuremethod} presents the necessary extrapolations of fluid compressibilities and phase-wise velocities into a narrow band of cells across the interface. Such phase-wise variables are needed for the physical and numerical consistency of the solution of the governing equations~\cite{pobladoribanez2021volumeoffluid}. \par

Poblador-Ibanez and Sirignano~\cite{pobladoribanez2021volumeoffluid} also offer a discussion on numerical issues that might appear with the proposed methodology and some solutions to mitigate them (e.g., treatment of under-resolved interface regions). A concern for the proposed approach is the influence of spurious currents around the interface. Due to the limited smoothness of VOF methods, spurious currents appear naturally because of the sharp interface treatment and a lack of an exact interfacial pressure balance. Furthermore, the addition of phase change as a localized source term and the extrapolation of certain variables may generate further spurious currents. Also, the split pressure-gradient method might become unstable for large pressure jumps across the interface caused by a combination of high density ratio, surface-tension coefficient and curvature. Nevertheless, the proposed method has been successfully used in a previous work that showed, primarily, a preliminary study on the characteristics and instability growth of a two-dimensional planar liquid jet~\cite{poblador2021liquidjet}. In any case, these numerical issues must be kept in mind for future improvements of the method. \par 

Lastly, some specific details about the numerical code have to be provided. For parallel computing, the domain is divided in a pencil-like decomposition with the contiguous memory in the transverse direction of the jet. This way, the computational resources are better distributed to capture the interface surface area given our domain configuration (see Figure~\ref{subfig:initial_domain}). The numerical code is written in Fortran 90 and uses the message-passing interface (MPI) and OpenMP. Necessary external open-source libraries are 2DECOMP\&FFT~\cite{li20102decomp} to perform the domain decomposition and FFTW3~\cite{frigo2005design} to solve the pressure field via Eq.~(\ref{eqn:ppe2}) using DFT. The simulations were performed in the local HPC3 cluster at the University of California Irvine and in various supercomputers from the XSEDE network~\cite{towns2014xsede}. \par

\section{Results and discussion}
\label{sec:results}

\subsection{Problem configuration}
\label{subsec:description}

This work focuses on analyzing temporal planar real-liquid jets at elevated pressures and transcritical conditions. The numerical cost of solving such two-phase flows is very high due to the required level of resolution to obtain a sufficiently smooth interface solution, as well as the additional coupling with a complete thermodynamic model. Smaller liquid structures are continuously generated during the breakup process and the interface surface area grows with time. Moreover, the necessary phase-wise extrapolations add extra computational cost. These issues become more problematic in atomization simulations where mass and thermal mixing is considered. Therefore, the added physics scale up during the simulation, continuously increasing the numerical cost compared to simpler incompressible atomization simulations. \par 

With these considerations in mind, the numerical domain has been reduced to analyze temporal planar liquid jets with a symmetry boundary condition in the centerline of the jet, periodic boundary conditions in the streamwise and spanwise directions and outflow boundary conditions in the top gaseous boundary away from the liquid-gas interface. The jet thickness is \(H=20\) \(\mu\)m (i.e., a half-thickness of \(H'=10\) \(\mu\)m under symmetry conditions) and two initial sinusoidal perturbations are superimposed in the streamwise and spanwise directions. In the streamwise direction, defined along the \(x\) axis, the wavelength of the perturbation is 30 \(\mu\)m and the amplitude is 0.5 \(\mu\)m. The spanwise direction follows the \(z\) axis and has a perturbation wavelength of 20 \(\mu\)m with an amplitude of 0.3 \(\mu\)m. The initial perturbation amplitude is sufficiently small to let unstable waves develop naturally while the superimposed perturbation in the spanwise direction accelerates the generation of three-dimensional structures. The choice of wavelengths is made following previous estimates obtained in Poblador-Ibanez and Sirignano~\cite{poblador2019axisymmetric} for an axisymmetric jet under a similar configuration. Also, the reduced surface tension at high pressures triggers unstable waves with a shorter wavelength than low-pressure cases analyzed in other works, where the most unstable wavelengths were around 100 \(\mu\)m~\cite{jarrahbashi2014vorticity,jarrahbashi2016early,zandian2017planar,zandian2018understanding,zandian2019length,zandian2019vorticity}. \par

Only one perturbation wave is considered in each surface direction. Enough mesh resolution can be achieved with this domain configuration, which yields meaningful results, by using a uniform mesh of 450 x 450 x 300 nodes~\cite{pobladoribanez2021volumeoffluid}. The domain size is \(L_x=L_y=30\) \(\mu\)m and \(L_z=20\) \(\mu\)m, which corresponds to a uniform spatial resolution of \(\Delta=0.0667\) \(\mu\)m. \(\Delta t \sim \mathcal{O}\)(10\(^{-11}\) s) for all the simulations. Notice the time step is not constant and it varies slightly during the computation according to the CFL conditions~\cite{pobladoribanez2021volumeoffluid}. \par

\begin{figure}[h!]
\centering
\begin{subfigure}{0.5\textwidth}
  \centering
  \includegraphics[width=0.8\linewidth]{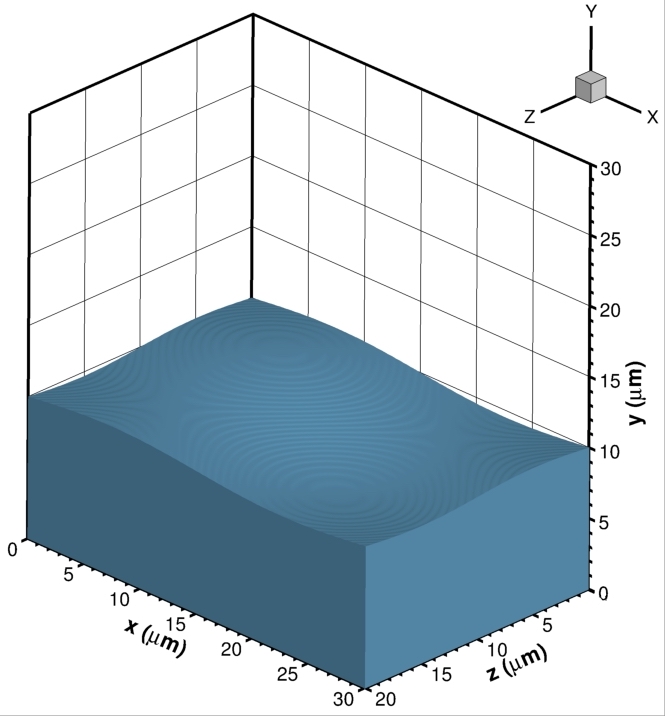}
  \caption{} 
  \label{subfig:initial_domain}
\end{subfigure}%
\begin{subfigure}{0.5\textwidth}
  \centering
  \includegraphics[width=0.8\linewidth]{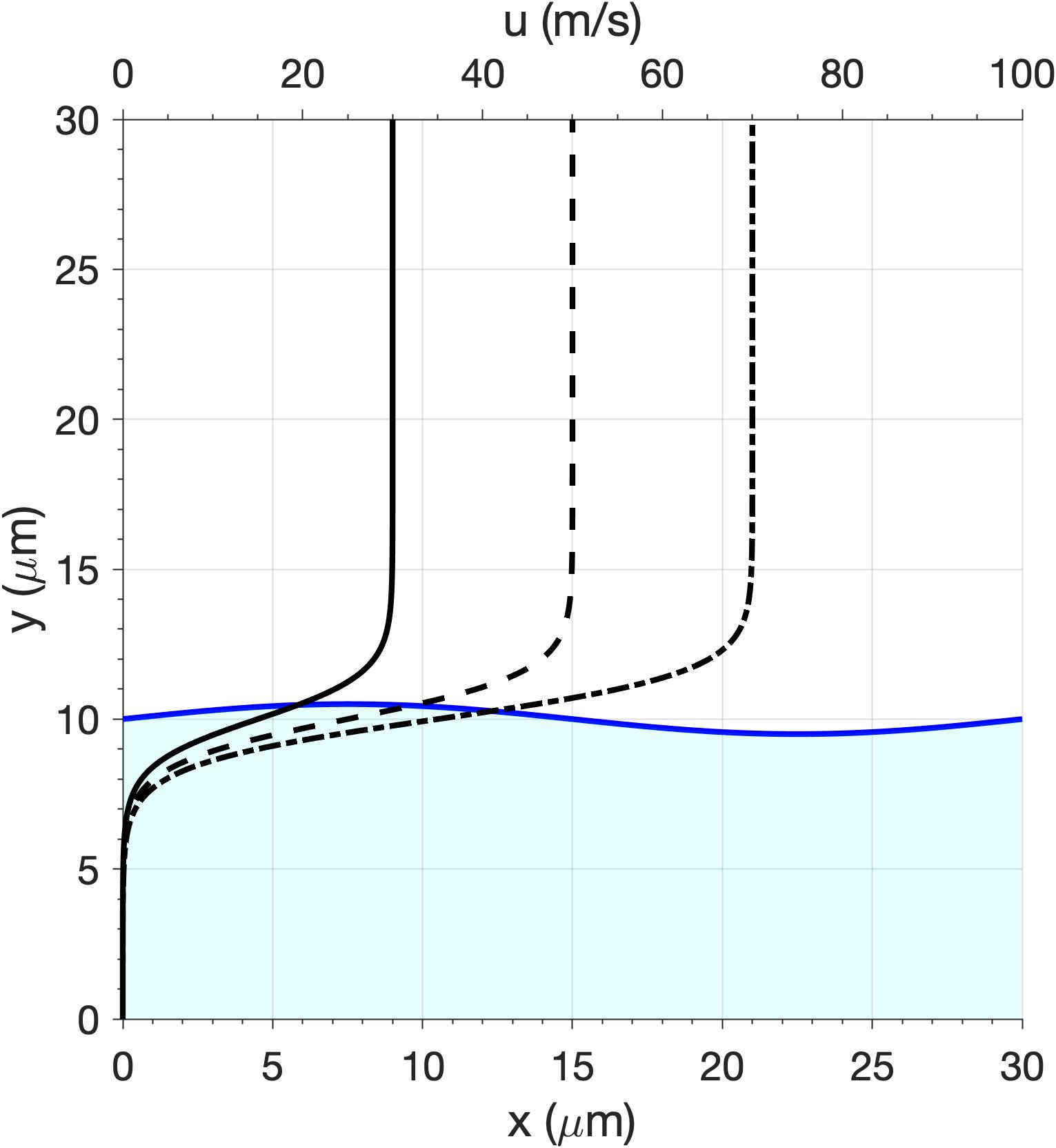}
  \caption{}
  \label{subfig:initial_vel}
\end{subfigure}%
\caption{Initial problem configuration. (a) size of the numerical domain and initial shape of the perturbed interface; (b) side view in the \(x-y\) plane showing the interface initial perturbation in the streamwise direction and the velocity profiles imposed for each gas freestream velocity, \(u_G\).}
\label{fig:initial_config}
\end{figure}

In this temporal study, the relative velocity between two streams rather than an absolute velocity is key. For all cases analyzed in this work, the higher-density central ``jet" starts at rest and as pure \textit{n}-decane at \(T_L=450\) K, while the lower-density ambient phase is initially composed of pure oxygen at \(T_G=550\) K and moving at a freestream velocity \(u_G\). The velocity field is initialized using a hyperbolic tangent function, which distributes the streamwise velocity from 0 m/s in the liquid phase to \(u_G\) within a region around the interface of about 6 \(\mu\)m following \(u(y)=(u_G/2)\big(\tanh{\big[6.5\times 10^{5}(y-H')\big]}+1\big)\). Thus, the shear layer thickness is well captured with 90 computational cells. The other velocity components are initialized with zero value (i.e., \(v=w=0\) m/s). Figure~\ref{fig:initial_config} shows the domain size and the initial velocity distribution for different \(u_G\) values. In some of the results, the domain has been enlarged using the periodic boundaries for a better visualization of certain liquid structures and features. \par  

The high-density, compressible fluid with decane as the dominant component will be described here as liquid, whether it is subcritical or supercritical locally. Similarly, the lower density fluid with oxygen as the dominant component will be described as gas over the transcritical domain. Note that a sharp initial discontinuity exists across the liquid-gas interface. Both species are representative of engines that operate with hydrocarbon fuels injected into enriched air or pure oxygen at high pressures (e.g., diesel engines, gas turbines or rocket engines). A summary of the molecular weight and critical properties of these two species (i.e., critical pressure, \(p_c\); critical temperature, \(T_c\); and critical density, \(\rho_c\)) is shown in Table~\ref{tab:crit_prop}. \par

\begin{table}[h!]
\begin{center}
\begin{tabular}{|r|r|r|r|r|} 
\multicolumn{1}{c|}{Species} & \multicolumn{1}{c|}{\(MW\) (kg/mol)} & \multicolumn{1}{c|}{\(p_c\) (bar)} & \multicolumn{1}{c|}{\(T_c\) (K)} &
\multicolumn{1}{c}{\(\rho_c\) (kg/m\(^3\))}\\ 
\hline
\hline
\multicolumn{1}{c|}{\(C_{10}H_{22}\)} & \multicolumn{1}{c|}{0.142280} & \multicolumn{1}{c|}{21.03} & \multicolumn{1}{c|}{617.70} &
\multicolumn{1}{c}{233.34}\\
\multicolumn{1}{c|}{\(O_2\)} & \multicolumn{1}{c|}{0.031999} & \multicolumn{1}{c|}{50.43} & \multicolumn{1}{c|}{154.58} &
\multicolumn{1}{c}{436.14}
\end{tabular}
\end{center}
\caption{Molecular weight (\(MW\)) and critical properties of \textit{n}-decane and oxygen. Source: NIST.}
\label{tab:crit_prop}
\end{table}

\begin{figure}[h!]
\centering
\includegraphics[width=0.5\linewidth]{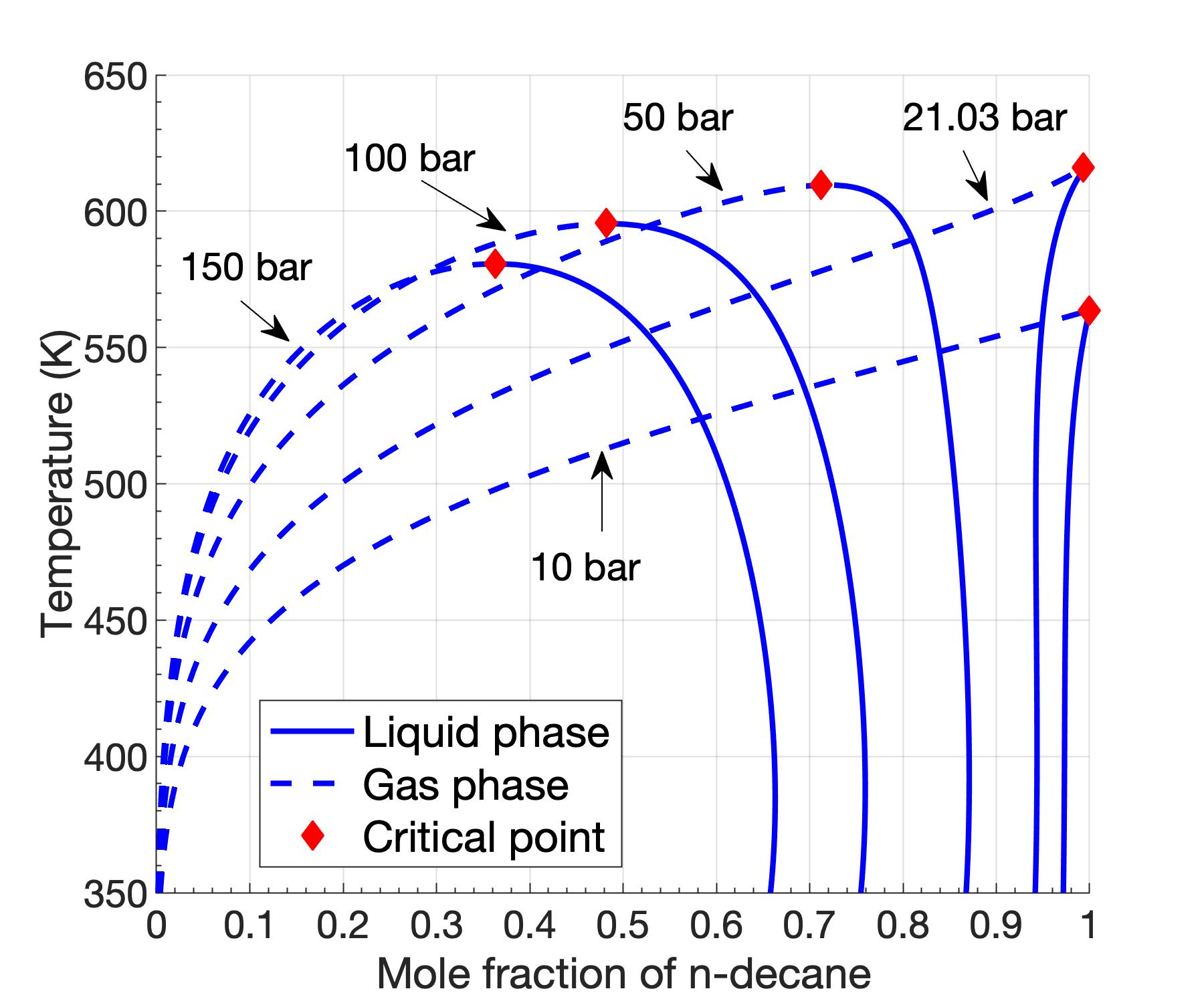}
\caption{Phase-equilibrium diagrams for the binary mixture of n-decane/oxygen obtained with the SRK equation of state as a function of interface temperature and pressure. The mole fraction of \textit{n}-decane and the mixture critical points at each pressure level are shown.}
\label{fig:pheq_diagram}
\end{figure}

A suitable choice of temperatures has been made, where the oxidizer stream is hotter than the injected decane to provide enough energy to vaporize the fuel. However, temperature is low enough in order to justify a two-phase modeling under LTE at these high pressures. The reduced temperature, \(T_r=T/T_{c,fuel}\), ranges between 0.73 and 0.89, and is slightly larger for the mixture. A thorough discussion about the validity of this physical modeling for the analyzed mixture and temperature range is provided in Poblador-Ibanez and Sirignano~\cite{pobladoribanez2021volumeoffluid}, and the two-phase domain of coexistence is emphasized by the phase-equilibrium diagrams for the considered mixture shown in Figure~\ref{fig:pheq_diagram}. \par

The thermodynamic model is used to calculate the speed of sound and verify the low-Mach-number domain in the limit of the dense gas. The speed of sound in the gas phase for the analyzed pressures is roughly 450 m/s. Therefore, the gas freestream velocity range analyzed in this work results in \(M \approx 0.067 - 0.16\). If a compressible, wave-like pressure equation is developed, compressible terms defining temporal pressure variations and linking density with pressure scale with \(M^2\), which becomes \(M^2\sim \mathcal{O}(10^{-3}-10^{-2})\). Thus, they can be safely neglected for this analysis. For faster gas freestream velocities of 100 m/s or more, the low-Mach-number model should be revised. This approach might be an important limitation when analyzing the transient injection of liquid fuels in some specific configurations. For instance, diesel is usually injected at velocities closer to or above the speed of sound of the oxidizer. However, applications with chamber pressures above 100 bar (e.g., rocket engines) may present much lower injection velocities.  \par

\begin{table}[h!]
\begin{center}
\begin{tabular}{|r|r|r|r|r|r|r|r|} 
\multicolumn{1}{c|}{Case} & \multicolumn{1}{c|}{\(p\) (bar)} & \multicolumn{1}{c|}{\(u_G\) (m/s)} & \multicolumn{1}{c|}{\(\rho_G\) (kg/m\(^3\))} &
\multicolumn{1}{c|}{\(\rho_L\) (kg/m\(^3\))} &
\multicolumn{1}{c|}{\(\mu_G\) (\(\mu\)Pa\(\cdot\)s)} &
\multicolumn{1}{c|}{\(\mu_L\) (\(\mu\)Pa\(\cdot\)s)} &
\multicolumn{1}{c}{\(\sigma\) (mN/m)} \\    
\hline
\hline
\multicolumn{1}{c|}{A1} & \multicolumn{1}{c|}{50} & 
\multicolumn{1}{c|}{50} & 
\multicolumn{1}{c|}{34.47} &
\multicolumn{1}{c|}{615.18} &
\multicolumn{1}{c|}{32.77} &
\multicolumn{1}{c|}{228.01} &
\multicolumn{1}{c}{7.10} \\
\multicolumn{1}{c|}{A2} & \multicolumn{1}{c|}{50} & 
\multicolumn{1}{c|}{70} & 
\multicolumn{1}{c|}{34.47} &
\multicolumn{1}{c|}{615.18} &
\multicolumn{1}{c|}{32.77} &
\multicolumn{1}{c|}{228.01} &
\multicolumn{1}{c}{7.10} \\
\multicolumn{1}{c|}{B1} & \multicolumn{1}{c|}{100} & 
\multicolumn{1}{c|}{50} & 
\multicolumn{1}{c|}{67.86} &
\multicolumn{1}{c|}{632.59} &
\multicolumn{1}{c|}{33.32} &
\multicolumn{1}{c|}{265.14} &
\multicolumn{1}{c}{5.00} \\
\multicolumn{1}{c|}{B2} & \multicolumn{1}{c|}{100} & 
\multicolumn{1}{c|}{70} & 
\multicolumn{1}{c|}{67.86} &
\multicolumn{1}{c|}{632.59} &
\multicolumn{1}{c|}{33.32} &
\multicolumn{1}{c|}{265.14} &
\multicolumn{1}{c}{5.00} \\
\multicolumn{1}{c|}{C1} & \multicolumn{1}{c|}{150} & 
\multicolumn{1}{c|}{30} & 
\multicolumn{1}{c|}{100.14} &
\multicolumn{1}{c|}{646.67} &
\multicolumn{1}{c|}{34.02} &
\multicolumn{1}{c|}{302.43} &
\multicolumn{1}{c}{3.40} \\
 \multicolumn{1}{c|}{C2} & \multicolumn{1}{c|}{150} & 
\multicolumn{1}{c|}{50} & 
\multicolumn{1}{c|}{100.14} &
\multicolumn{1}{c|}{646.67} &
\multicolumn{1}{c|}{34.02} &
\multicolumn{1}{c|}{302.43} &
\multicolumn{1}{c}{3.40} \\
\multicolumn{1}{c|}{C3} & \multicolumn{1}{c|}{150} & 
\multicolumn{1}{c|}{70} & 
\multicolumn{1}{c|}{100.14} &
\multicolumn{1}{c|}{646.67} &
\multicolumn{1}{c|}{34.02} &
\multicolumn{1}{c|}{302.43} &
\multicolumn{1}{c}{3.40}
\end{tabular}
\end{center}
\caption{List of analyzed cases using liquid \textit{n}-decane at 450 K and gaseous oxygen at 550 K. The subscripts \(G\) and \(L\) refer to values obtained using freestream conditions for the gas and the liquid phases, respectively.}
\label{tab:cases}
\end{table}

\begin{table}[h!]
\begin{center}
\begin{tabular}{|r|r|r|r|} 
\multicolumn{1}{c|}{Case} & 
\multicolumn{1}{c|}{\(Re_L\)} &
\multicolumn{1}{c|}{\(Re_G\)} &
\multicolumn{1}{c}{\(We_G\)}\\    
\hline
\hline
\multicolumn{1}{c|}{A1} &
\multicolumn{1}{c|}{2698} &
\multicolumn{1}{c|}{1052} &
\multicolumn{1}{c}{243} \\
\multicolumn{1}{c|}{A2} &
\multicolumn{1}{c|}{3777} &
\multicolumn{1}{c|}{1473} &
\multicolumn{1}{c}{476} \\
\multicolumn{1}{c|}{B1} &
\multicolumn{1}{c|}{2387} &
\multicolumn{1}{c|}{2037} &
\multicolumn{1}{c}{679} \\
\multicolumn{1}{c|}{B2} &
\multicolumn{1}{c|}{3342} &
\multicolumn{1}{c|}{2852} &
\multicolumn{1}{c}{1330} \\
\multicolumn{1}{c|}{C1} &
\multicolumn{1}{c|}{1285} &
\multicolumn{1}{c|}{1766} &
\multicolumn{1}{c}{530} \\
 \multicolumn{1}{c|}{C2} &
\multicolumn{1}{c|}{2141} &
\multicolumn{1}{c|}{2943} &
\multicolumn{1}{c}{1473} \\
\multicolumn{1}{c|}{C3} &
\multicolumn{1}{c|}{2998} &
\multicolumn{1}{c|}{4121} &
\multicolumn{1}{c}{2886}
\end{tabular}
\end{center}
\caption{List of analyzed three-dimensional cases using liquid \textit{n}-decane at 450 K and gaseous oxygen at 550 K and their characteristic Reynolds and Weber numbers. The subscripts \(G\) and \(L\) refer to the values obtained using freestream conditions for the gas and the liquid phases, respectively.}
\label{tab:cases_2}
\end{table}

The analyzed cases are summarized in Table~\ref{tab:cases}, which shows the ambient pressure, the freestream velocity of the gas phase, \(u_G\), and some freestream properties of each fluid (i.e., \(\rho_G\), \(\rho_L\), \(\mu_G\) and \(\mu_L\)). Moreover, a representative value of the surface-tension coefficient is provided based on the average value at the beginning of the simulation before substantial interface deformation occurs. The ambient pressure and the gas freestream velocity vary among the different cases. The analyzed pressures are 50 bar, 100 bar and 150 bar, and the analyzed gas freestream velocities are 30 m/s, 50 m/s and 70 m/s. All analyzed pressures are supercritical for the pure \textit{n}-decane with a reduced pressure, \(p_r=p/p_{c,fuel}\), between 2.38 and 7.15. Numerical stability limitations imposed by the FFT pressure solver~\cite{dodd2014fast} are to blame. For subcritical pressures, the density ratio \(\rho_G/\rho_L\) becomes too high for the pressure solver to handle efficiently (e.g., \(\rho_G/\rho_L=86\) at 10 bar). Considerable reduction of the time step delays the onset of numerical instabilities but becomes an unrealistic option. In future works, improvements to the methodology, such as those from Cifani~\cite{cifani2019analysis} and Turnquist and Owkes~\cite{turnquist2021fast}, can be sought to simulate lower pressures and obtain better comparisons between subcritical and transcritical atomization. At this point, two incompressible cases have been simulated (i.e., A2i and C1i) where phase change is neglected and thermal and species mixing are not considered. Thus, mixing effects on the atomization process at high pressures can be assessed. \par 

Using the variables presented in Table~\ref{tab:cases}, each case can be characterized in terms of a liquid Reynolds number, \(Re_L\), and a gas Weber number, \(We_G\), as done in Zandian et al.~\cite{zandian2017planar,zandian2018understanding,zandian2019length}. The use of the gas Weber number embeds the density ratio \(\rho_G/\rho_L\) between both fluids in the characterization. The Reynolds number is defined as \(Re_L=\rho_L u_G H/\mu_L\) and is a measure of the relative importance of inertia forces against viscous forces. The Weber number is the ratio of inertia and surface-tension forces and is defined as \(We_G=\rho_G u_G^2 H/\sigma\). In both cases, \(H\) represents the jet thickness. Table~\ref{tab:cases_2} shows \(Re_L\), \(Re_G\) and \(We_G\) for the studied configurations, highlighting the mild Reynolds number in both phases. Thus, the early liquid injection problem may be analyzed without turbulence models. A discussion about the characterization of high-pressure atomization problems is provided in Subsection~\ref{subsec:mixing_class}. \par 

The chosen configurations have some limitations. From the domain perspective, the analysis of a perturbed interface with just one wavelength per direction with periodic boundary conditions might prevent longer wavelengths from appearing naturally. This issue does not seem important a priori since all cases presented in Table~\ref{tab:cases} show a clear cascade deformation process with smaller perturbations and liquid structures forming. Also, the symmetric boundary in the centerline of the jet prevents antisymmetric modes from developing, which becomes important in some cases~\cite{zandian2018understanding}. Nevertheless, testing of non-symmetric configurations did not show any significant growth of antisymmetric modes during the time frame analyzed in this work. Moreover, as the liquid surface expands along the transverse direction (i.e., \(y\) axis), the top open boundary might influence the results once the surface gets very close. This issue limits the maximum physical time for each analyzed case. \par

Other limitations are linked to the numerical approach, the low-Mach-number assumption, and the two-phase modeling at high pressures. The VOF approach generates spurious currents around the interface due to the various reasons described in Section~\ref{sec:numerics}. These oscillations, which can be more detrimental around under-resolved regions, must be considered when evaluating whether the growth of surface instabilities has a physical origin or not, and are an important issue to be addressed in future works dealing with two-phase modeling of real liquids. Moreover, a broader thermodynamic modeling must be implemented to capture the transition from a two-phase behavior to a supercritical diffuse mixing between fluids. This topic is an ongoing research area, but some authors have already proposed methodologies to capture this transition~\cite{aggarwal2002transcritical}. \par

\subsection{Mixing effects and the classification of real liquid jets}
\label{subsec:mixing_class}

\begin{figure}[h!]
\centering
\includegraphics[width=0.5\linewidth]{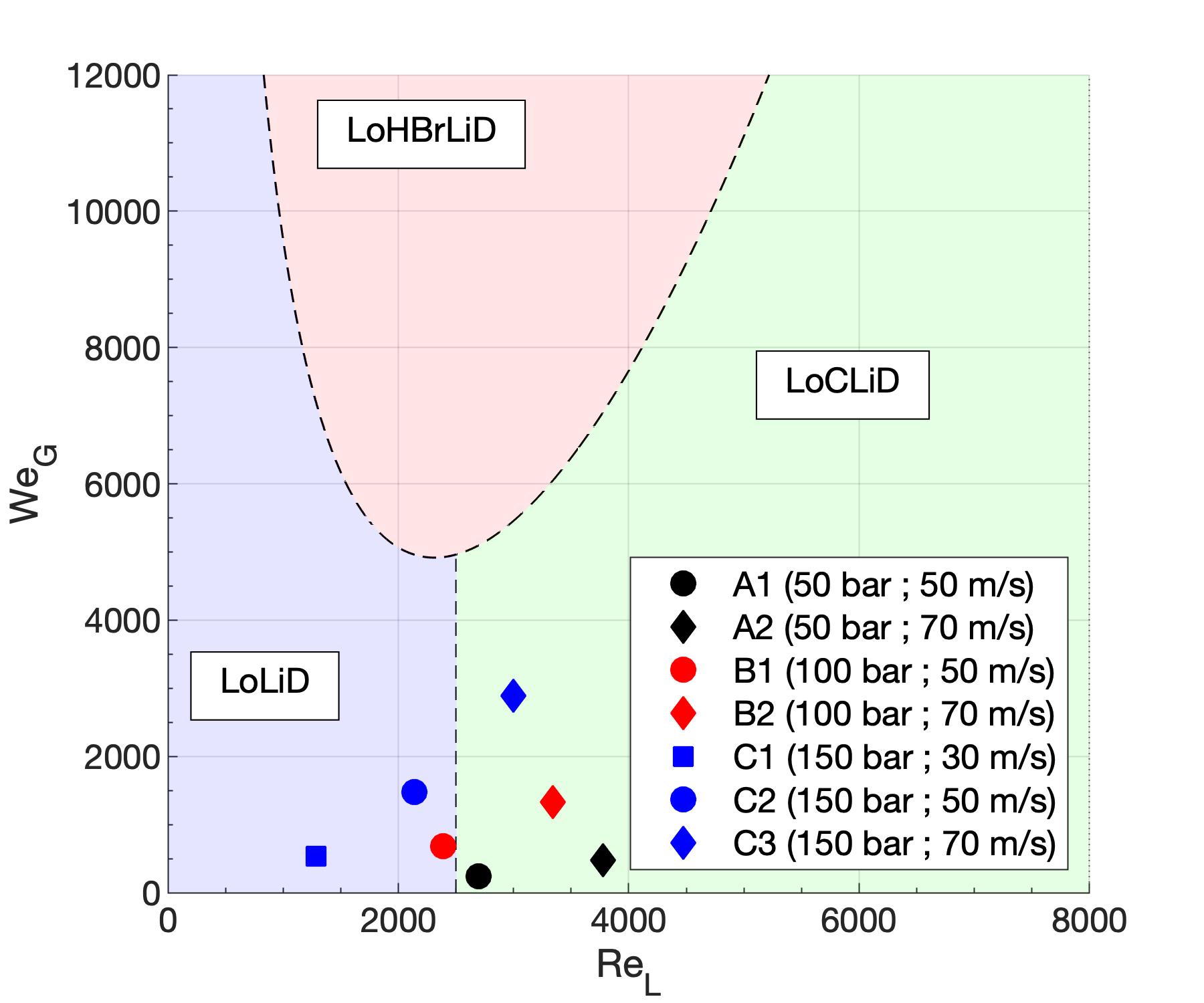}
\caption{Classification of the analyzed cases based on \(We_G\) and \(Re_L\) using freestream properties. The dashed curves separate the atomization sub-domains identified in the incompressible work by Zandian et al.~\cite{zandian2017planar}. The Lobe-Ligament-Droplet (LoLiD) sub-domain is shaded in blue, the Lobe-Corrugation-Ligament-Droplet (LoCLiD) sub-domain is shaded in green and the Lobe-Hole-Bridge-Ligament-Droplet (LoHBrLiD) sub-domain is shaded in red.}
\label{fig:Weg_vs_Rel_overview}
\end{figure}

The classification of the cases presented in Table~\ref{tab:cases_2} based on the gas Weber number and the liquid Reynolds number using freestream conditions (i.e., \(We_G\) and \(Re_L\)) follows the approach presented in Zandian et al.~\cite{zandian2017planar} and used in subsequent works~\cite{zandian2018understanding,zandian2019length}. Figure~\ref{fig:Weg_vs_Rel_overview} displays the analyzed cases in a Weber-Reynolds diagram that also identifies the three atomization sub-domains described by Zandian et al.~\cite{zandian2017planar} (i.e., LoLiD, LoHBrLiD and LoCLiD sub-domains). Despite using a reasonable high-pressure configuration, it is noticeable that all cases present a rather low gas Weber number. That is, the chosen jet size and gas velocity are representative of the lower end of real configurations. Moreover, a thermodynamic model is used to evaluate thermophysical properties and the surface-tension coefficient is small due to the high-pressure environment. Instead, the work by Zandian et al.~\cite{zandian2017planar,zandian2018understanding,zandian2019length} is performed in the incompressible limit without mass exchange or mixing. Only the liquid density, domain size and gas freestream velocity are chosen with realistic values, while all other fluid properties are obtained by fixing \(Re_L\), \(We_G\), density ratio and viscosity ratio; thus, making it possible to cover a wide range of Weber and Reynolds numbers. \par 

For comparison purposes, it would be interesting to analyze high-pressure cases that fall within the three atomization sub-domains. The proposed cases only cover the LoLiD and LoCLiD sub-domains according to Zandian et al.~\cite{zandian2017planar}, but higher gas Weber numbers are needed to cover the LoHBrLiD sub-domain. However, reaching the LoHBrLiD region would require various changes that might result in an unrealistic configuration, such as changing the working species, increasing the initial temperatures (i.e., reducing \(\sigma\) by having a higher interface temperature, which relates to more similar liquid and gas phases under phase equilibrium as discussed later) or modifying the jet size or the gas freestream velocity. A change in the working species might have a negligible effect on \(We_G\) if the focus of the work is to remain on the injection of liquid hydrocarbon fuels into oxygen or air. For instance, replacing \textit{n}-decane with \textit{n}-dodecane and oxygen with nitrogen might not make much of a difference. Also, increasing the fluid temperatures could cause the interface to fall into a high-temperature domain closer or above the mixture critical temperature, invalidating the two-phase assumption~\cite{pobladoribanez2021volumeoffluid}. Therefore, increasing the jet size or the gas freestream velocity seems the only way to proceed. \par

Looking at the closest case to the LoHBrLiD sub-domain, case C3, an increase in the gas freestream velocity from 70 m/s to about 120 m/s is needed, with all other variables fixed, to at least triple the gas Weber number such that \(We_G\approx 8650>4900\), which is the lower boundary of the LoHBrLiD sub-domain~\cite{zandian2017planar}. Nevertheless, the new case would still fall inside the LoCLiD sub-domain as \(Re_L\approx 5150\). Such high velocities, albeit not unrealistic, could invalidate the low-Mach-number assumption and might also be numerically problematic since the spurious currents generated with the implemented numerical model would be much higher and could affect the interface.  \par

The jet thickness needs to triple to achieve a similar increase in \(We_G\). However, both \(We_G\) and \(Re_L\) evolve linearly with \(H\) and the new case would still not fall inside the LoHBrLiD region. Moreover, an argument could be made to justify using the perturbation wavelength as characteristic length instead of the jet thickness. Even so, the cases presented in Zandian et al.~\cite{zandian2017planar,zandian2018understanding,zandian2019length} use a similar wavelength-to-thickness ratio as in the present work. Thus, a change in characteristic length does not justify the real configurations being so far from the LoHBrLiD sub-domain. \par 

Despite the analyzed cases being substantially away from the LoHBrLiD sub-domain identified by the incompressible theory, clear formation of holes and bridges is observed in some cases, as shown in Subsection~\ref{subsec:deformation}. The hole formation mechanism occurs more often at higher pressures and higher velocities, which is expected as \(\sigma\) drops and the Weber number increases. Moreover, the formation of holes can be related to two different sources: liquid perforation by the gas phase and liquid sheet tearing. \par

A key feature of high-pressure liquid injection is the enhanced mixing in both phases and the variations of fluid properties along the interface and across mixing regions. This feature is not captured in the definition of Weber and Reynolds numbers based on freestream properties and, therefore, the inertial effects may be underestimated. The enhanced mixing has been discussed in previous works dealing with simpler configurations (i.e., one-dimensional transient flow around a liquid-gas interface~\cite{poblador2018transient} or two-dimensional laminar mixing layer~\cite{davis2019development,poblador2021selfsimilar}) and in Poblador-Ibanez and Sirignano~\cite{poblador2021liquidjet}, where the effects of high-pressure mixing on the fluid properties are shown in detail for two-dimensional planar jets. Also, limited three-dimensional results are presented. \par

\begin{figure}[h!]
\centering
\begin{subfigure}{0.5\textwidth}
  \centering
  \includegraphics[width=0.9\linewidth]{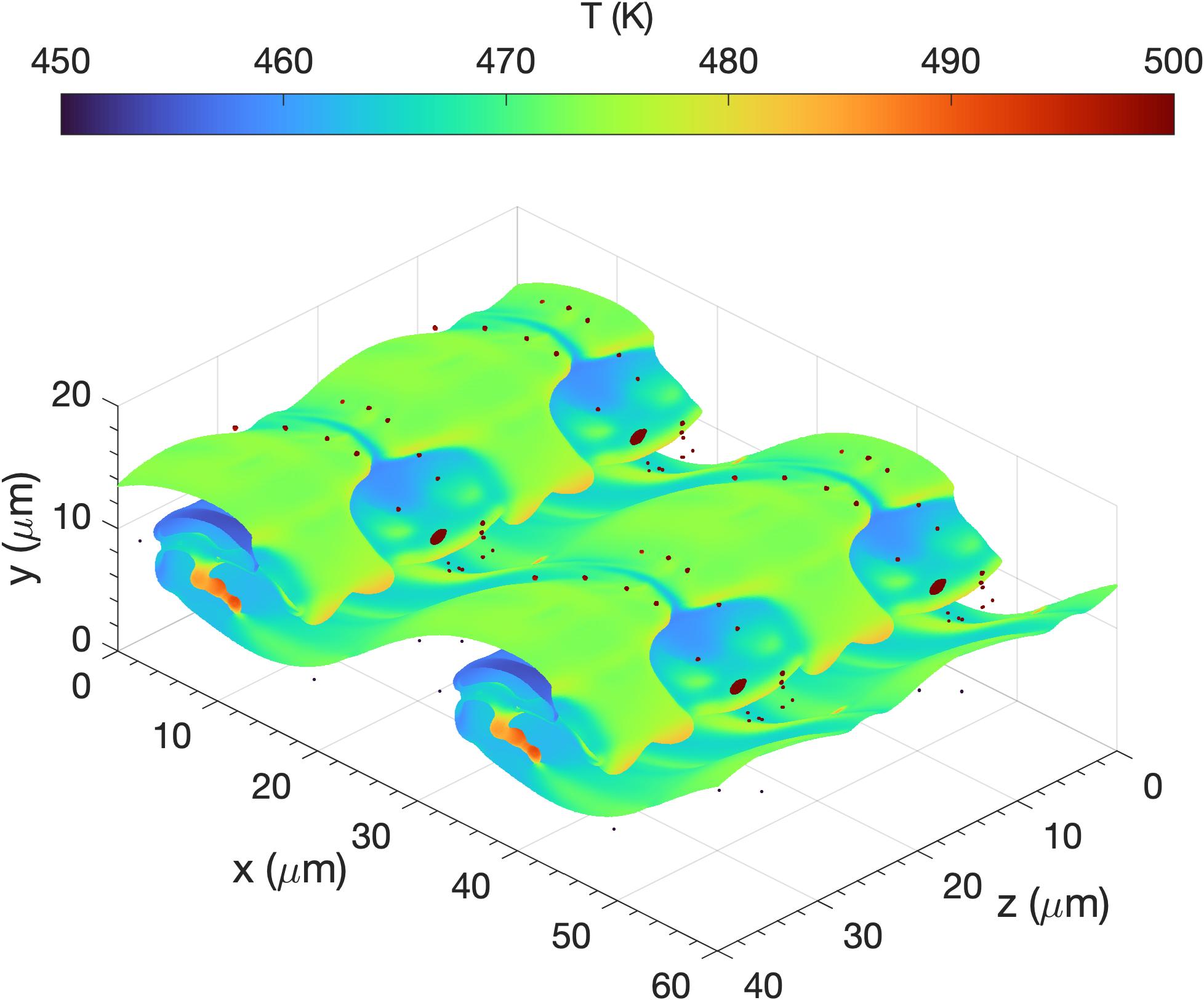}
  \caption{} 
  \label{subfig:150_30A_int_T_5mus}
\end{subfigure}%
\begin{subfigure}{0.5\textwidth}
  \centering
  \includegraphics[width=0.9\linewidth]{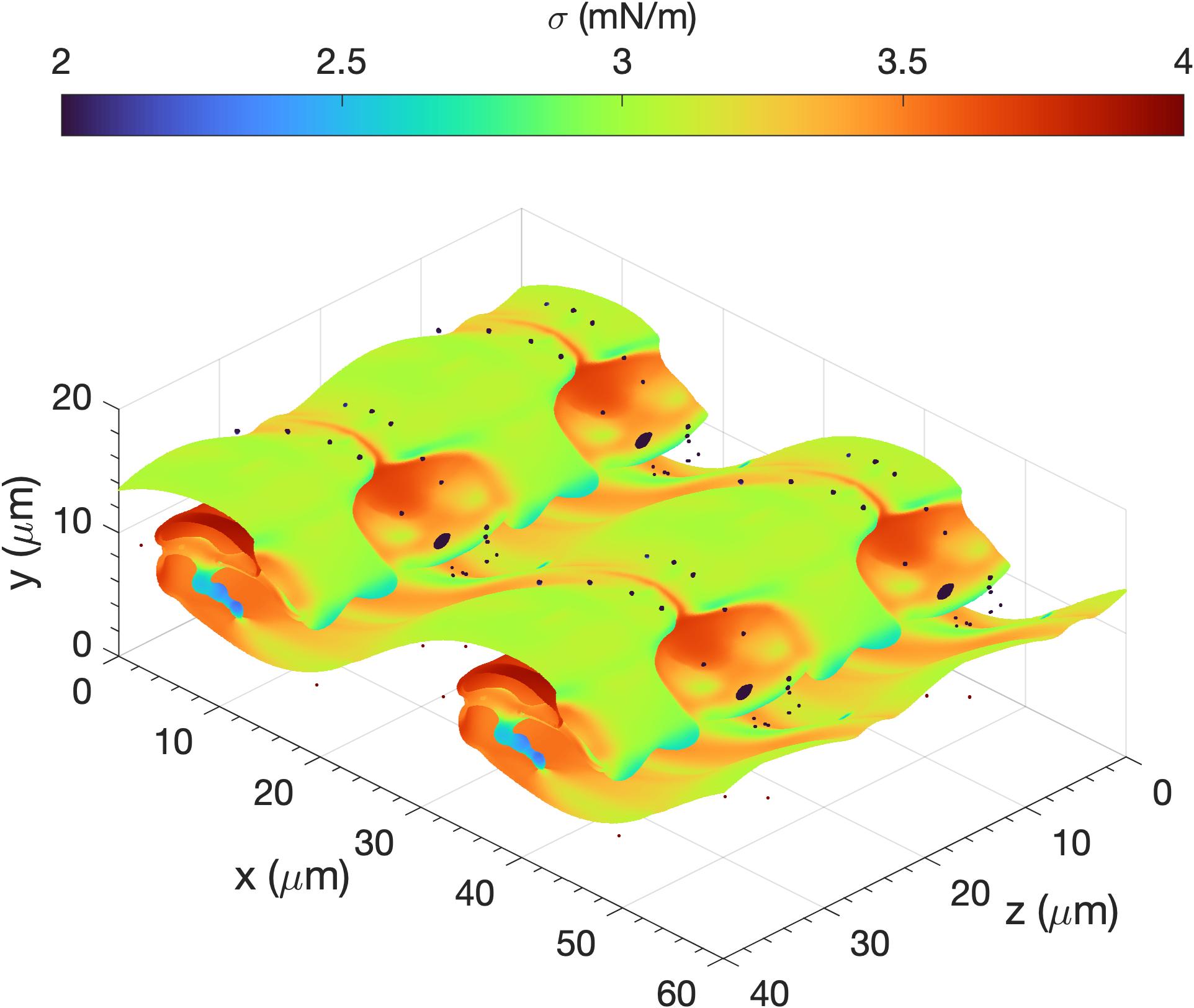}
  \caption{}
  \label{subfig:150_30A_int_sigma_5mus}
\end{subfigure}%
\caption{Interface temperature and surface-tension coefficient for case C1 at \(t=5\) \(\mu\)s with an ambient pressure of 150 bar and a gas freestream velocity of 30 m/s. Some small droplets are seen flowing above the main surface with a larger temperature and lower surface-tension coefficient than each respective color scale. (a) interface temperature; (b) surface-tension coefficient.}
\label{fig:150_30A_inter_5mus}
\end{figure}

As pressure increases, LTE enhances the dissolution of oxygen into the liquid phase. For reference, the reader is referred to Figure~\ref{fig:pheq_diagram} where phase-equilibrium diagrams for the binary mixture of \textit{n}-decane and oxygen at different pressures (e.g., 10, 50, 100 and 150 bar) have been presented. These diagrams are obtained with the SRK equation of state and show the mixture equilibrium composition of each phase and a temperature range of two-phase coexistence up to the mixture critical temperature at a given pressure. In the temperature range analyzed in this work (i.e., between 450 K and 550 K), liquid and gas become more similar as the interface temperature increases. That is, equilibrium compositions become more similar, liquid density drops and gas density increases, which reduces the surface-tension coefficient considerably. Therefore, the surface-tension coefficient can be substantially lower in hotter interface locations compared to colder regions as seen in Figure~\ref{fig:150_30A_inter_5mus}, where the interface local temperature and surface-tension coefficient are shown for case C1 at 5 \(\mu\)s. \par 

\begin{figure}[h!]
\centering
\begin{subfigure}{0.5\textwidth}
  \centering
  \includegraphics[width=0.95\linewidth]{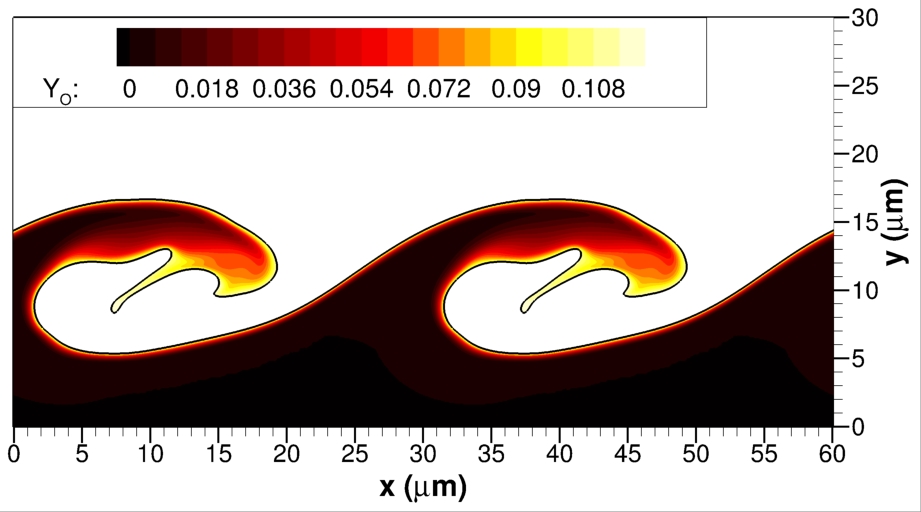}
  \caption{} 
  \label{subfig:150_30A_5mus_z15_YO}
\end{subfigure}%
\begin{subfigure}{0.5\textwidth}
  \centering
  \includegraphics[width=0.95\linewidth]{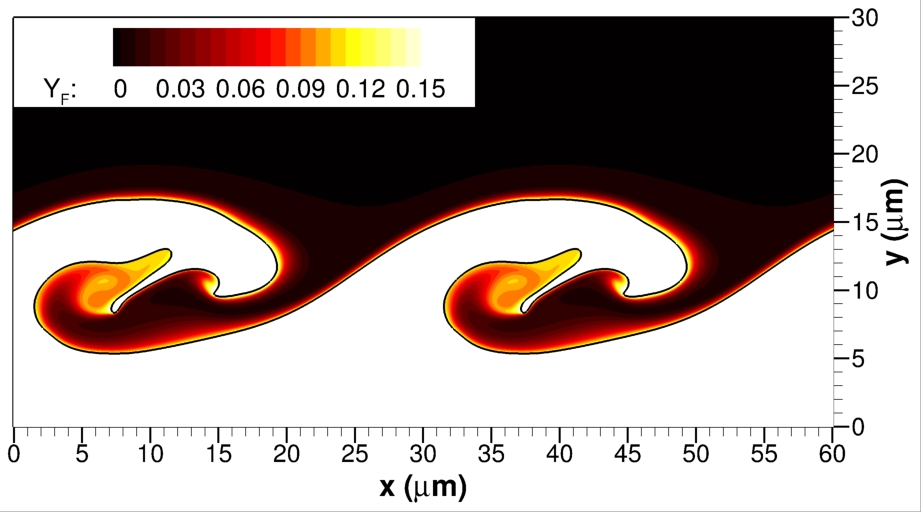}
  \caption{}
  \label{subfig:150_30A_5mus_z15_YF}
\end{subfigure}%
\\
\begin{subfigure}{0.5\textwidth}
  \centering
  \includegraphics[width=0.95\linewidth]{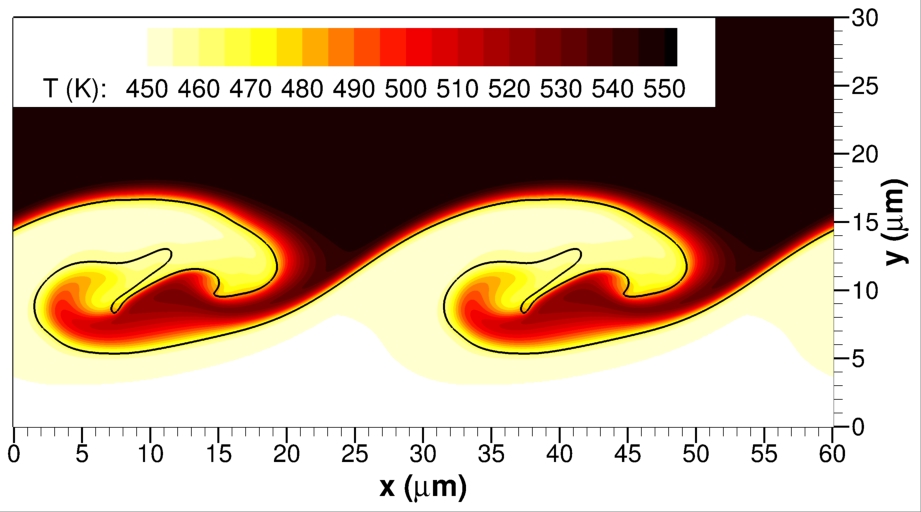}
  \caption{} 
  \label{subfig:150_30A_5mus_z15_T}
\end{subfigure}%
\begin{subfigure}{0.5\textwidth}
  \centering
  \includegraphics[width=0.95\linewidth]{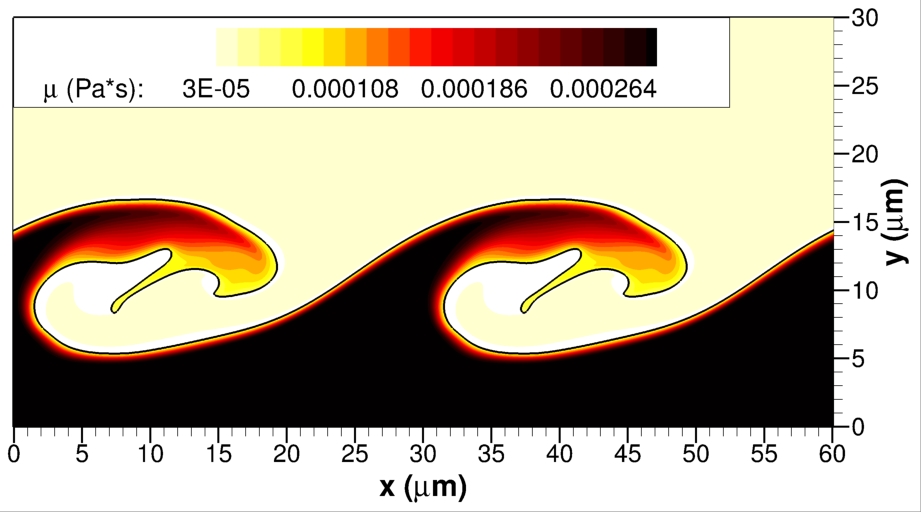}
  \caption{}
  \label{subfig:150_30A_5mus_z15_VIS}
\end{subfigure}%
\\
\begin{subfigure}{0.5\textwidth}
  \centering
  \includegraphics[width=0.95\linewidth]{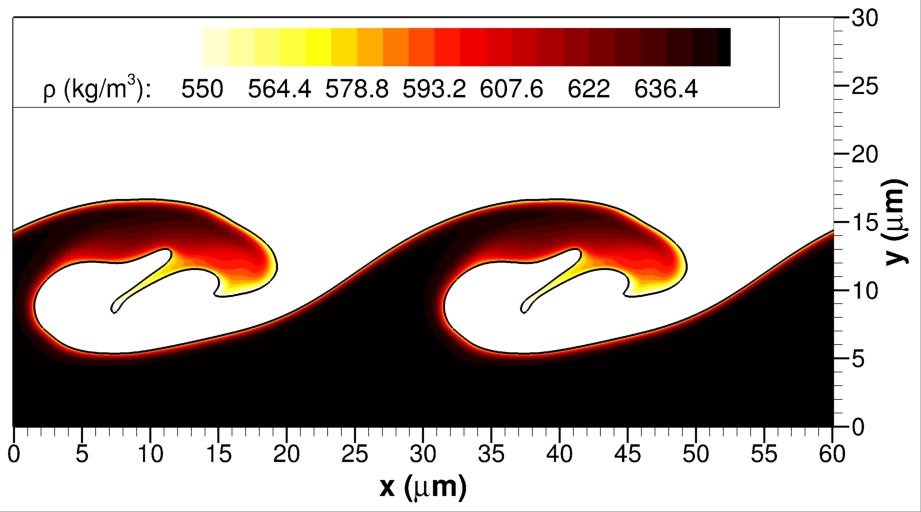}
  \caption{} 
  \label{subfig:150_30A_5mus_z15_DENL}
\end{subfigure}%
\begin{subfigure}{0.5\textwidth}
  \centering
  \includegraphics[width=0.95\linewidth]{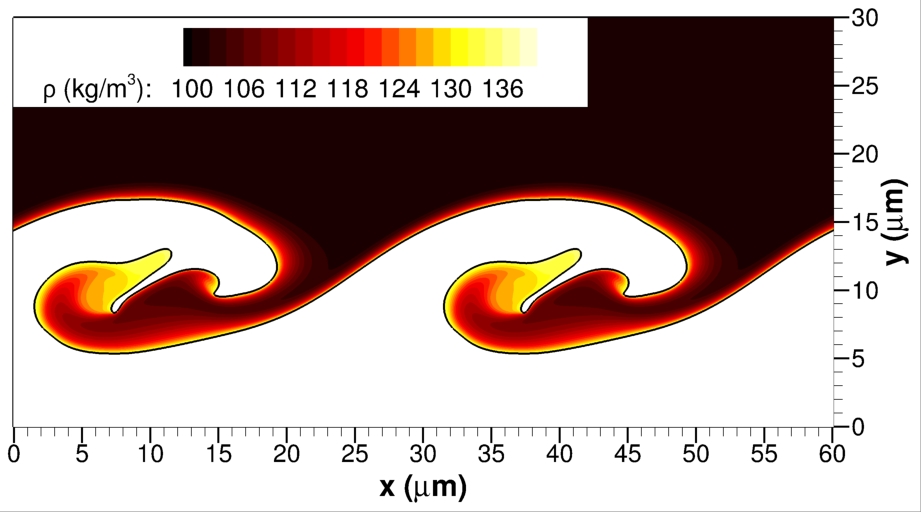}
  \caption{}
  \label{subfig:150_30A_5mus_z15_DENG}
\end{subfigure}%
\caption{Slice through the three-dimensional domain (\(xy\) plane with \(z=15\) \(\mu\)m) showing the variations of mixture composition, temperature, viscosity and density across the mixing region for case C1 at \(t=5\) \(\mu\)s with an ambient pressure of 150 bar and a gas freestream velocity of 30 m/s. (a) oxidizer or oxygen mass fraction in the liquid phase; (b) fuel or \textit{n}-decane mass fraction in the gas phase; (c) temperature; (d) viscosity of the two-phase mixture; (e) liquid density; (f) gas density.}
\label{fig:150_30A_slices_5mus}
\end{figure}

The variation of fluid properties across mixing regions is presented in Figure~\ref{fig:150_30A_slices_5mus} for case C1 at 5 \(\mu\)s as well. Here, a slice through the three-dimensional domain is shown for an \(xy\) plane located at \(z=15\) \(\mu\)m. Various features are observed. The vaporization of \textit{n}-decane, together with a large decrease in the gas temperature, causes a rise in gas density of up to 40\% from 100.14 kg/m\(^3\) in the freestream to around 140 kg/m\(^3\) near the interface. At the same time, the gas viscosity drops slightly as the increase in density cannot compensate for the sharp decrease in temperature (i.e., higher densities and hotter fluids relate to higher fluid viscosity). On the other hand, oxygen dissolves into the liquid phase, which also heats in the vicinity of the interface. Therefore, liquid density drops substantially from 646.67 kg/m\(^3\) in the freestream to around 550 kg/m\(^3\) near the interface. This drop represents about a 15\% decrease, but the absolute variation is more than twice the density variation in the gas phase. More importantly, the liquid viscosity drops almost an order of magnitude to gas-like values, creating liquid regions with fluid properties resembling more closely those of a gas. Moreover, the size of a particular liquid structure (e.g., droplet, ligament, lobe) influences how fast the liquid phase heats. Therefore, small or thin liquid structures immersed in the hotter gas present higher interface temperatures and stronger mixing effects. \par

\begin{figure}[h!]
\centering
\begin{subfigure}{0.5\textwidth}
  \centering
  \includegraphics[width=1.0\linewidth]{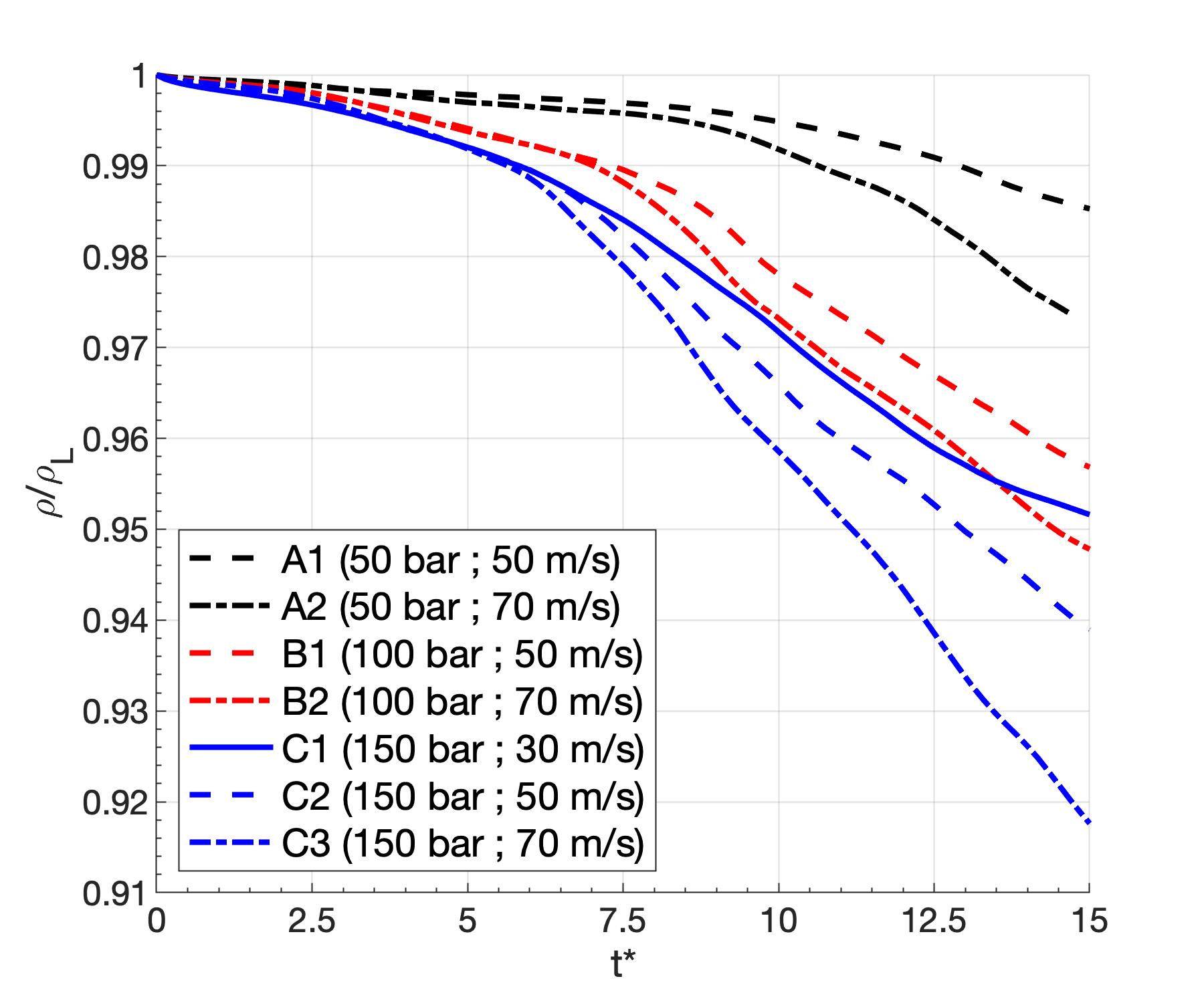}
  \caption{} 
  \label{subfig:avg_den}
\end{subfigure}%
\begin{subfigure}{0.5\textwidth}
  \centering
  \includegraphics[width=1.0\linewidth]{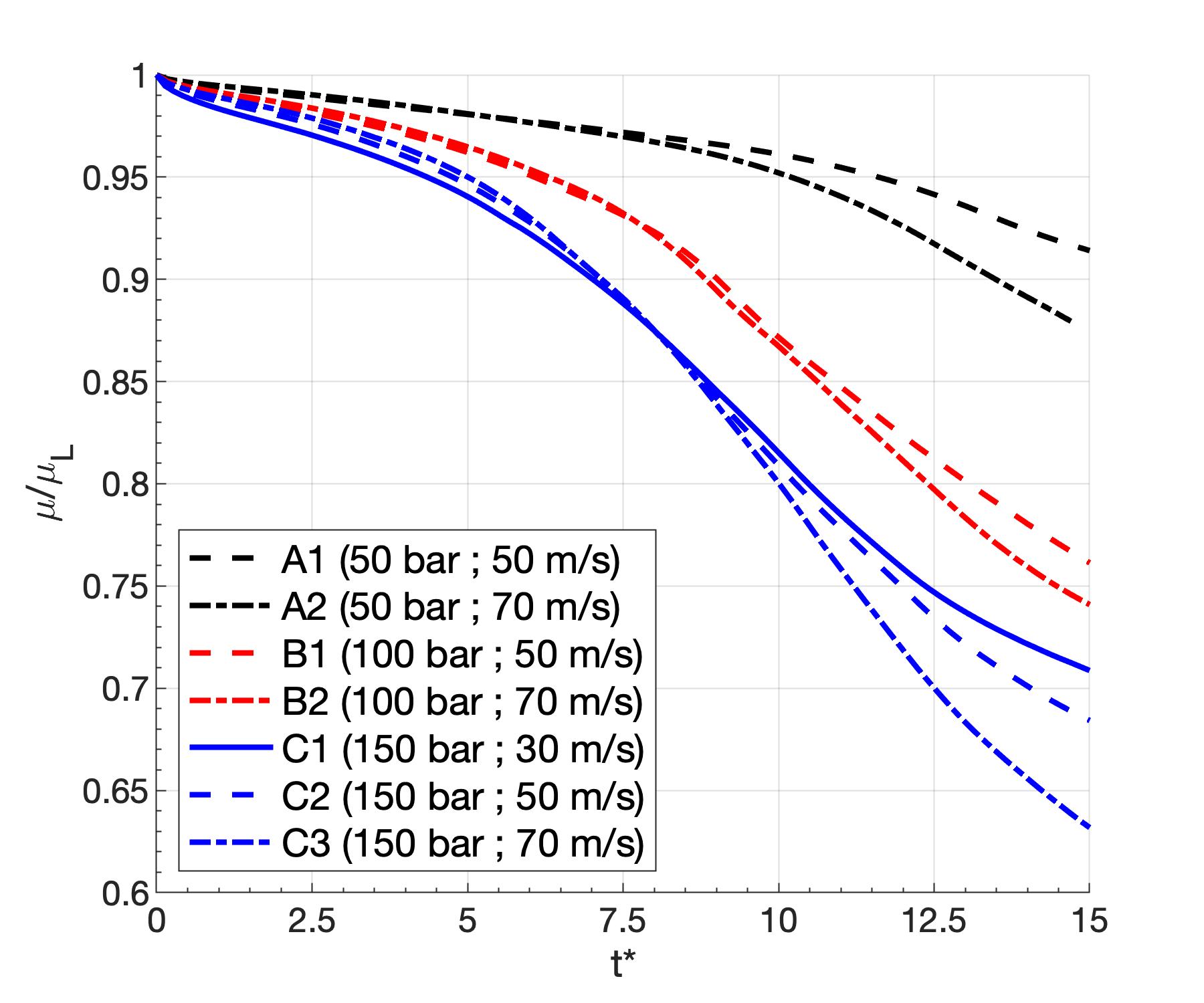}
  \caption{}
  \label{subfig:avg_vis}
\end{subfigure}%
\caption{Average liquid density and liquid viscosity over the liquid volume normalized by the freestream properties, \(\rho_L\) and \(\mu_L\), presented in Table~\ref{tab:cases}. The non-dimensional time is obtained as \(t^*=t/t_c=t\frac{u_G}{H}\). (a) average liquid density; and (b) average liquid viscosity.}
\label{fig:avg_den_vis}
\end{figure}

These observations provide valuable insights. The consequences of mass and thermal mixing are more extreme at 150 bar than at lower pressures (see Poblador-Ibanez et al.~\cite{poblador2021selfsimilar} or Poblador-Ibanez and Sirignano~\cite{poblador2021liquidjet}) and affect how the liquid deforms locally. That is, different liquid regions might show different behaviors depending not only on the local velocity and length scales, but also on the local interface state and fluid properties. Moreover, as mixing continuously occurs over time, the liquid may present a substantially different behavior at later times. Figure~\ref{fig:avg_den_vis} shows the volume-averaged liquid density and viscosity for the analyzed configurations. Over time, the average liquid properties change considerably, especially at very high pressures. \par 

\begin{figure}[h!]
\centering
\begin{subfigure}{0.33\textwidth}
  \centering
  \includegraphics[width=1.05\linewidth]{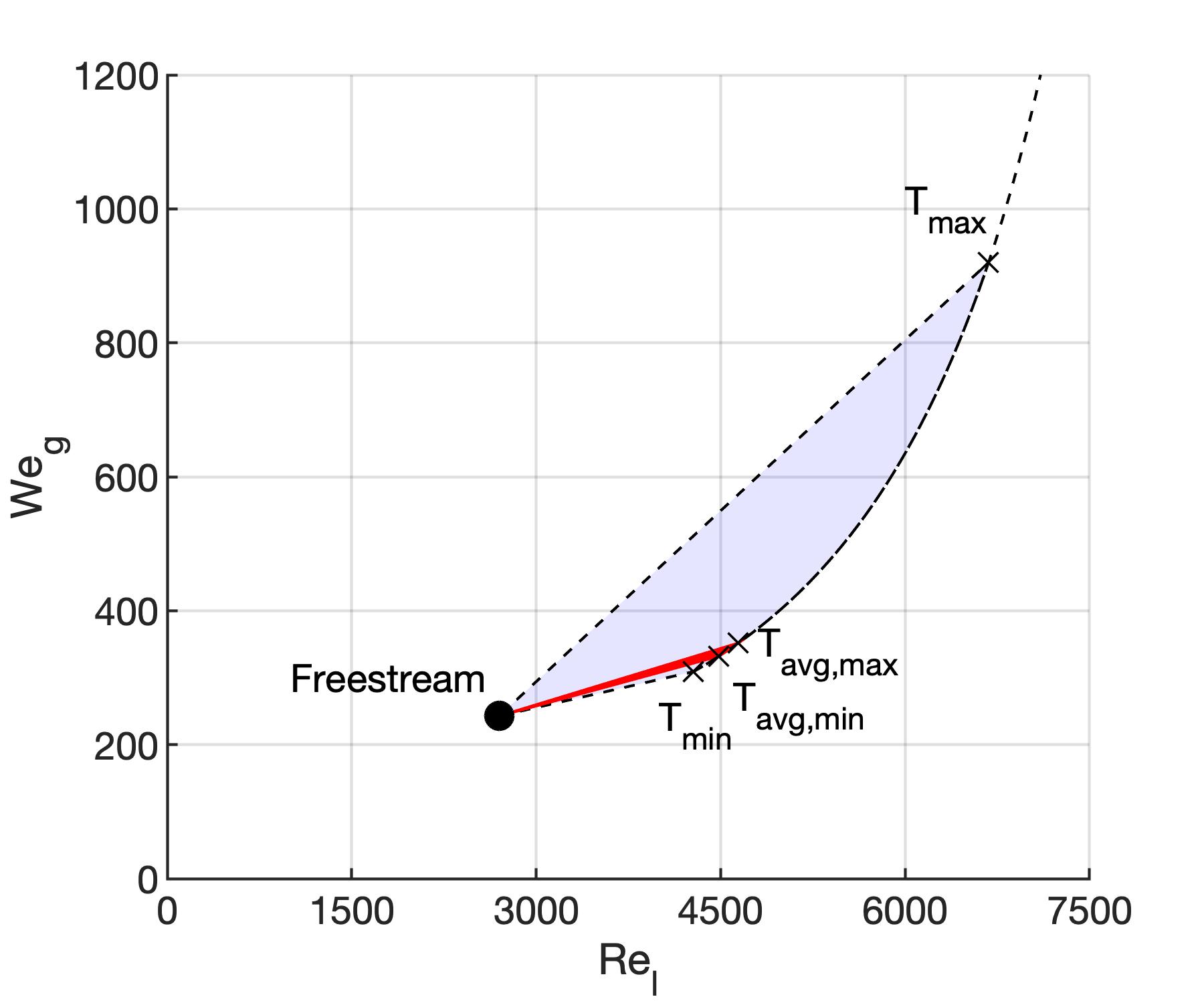}
  \caption{50 bar and \(u_G=50\) m/s (A1)}
  \label{subfig:Weg_vs_Rel_50_50A}
\end{subfigure}%
\begin{subfigure}{0.33\textwidth}
  \centering
  \includegraphics[width=1.05\linewidth]{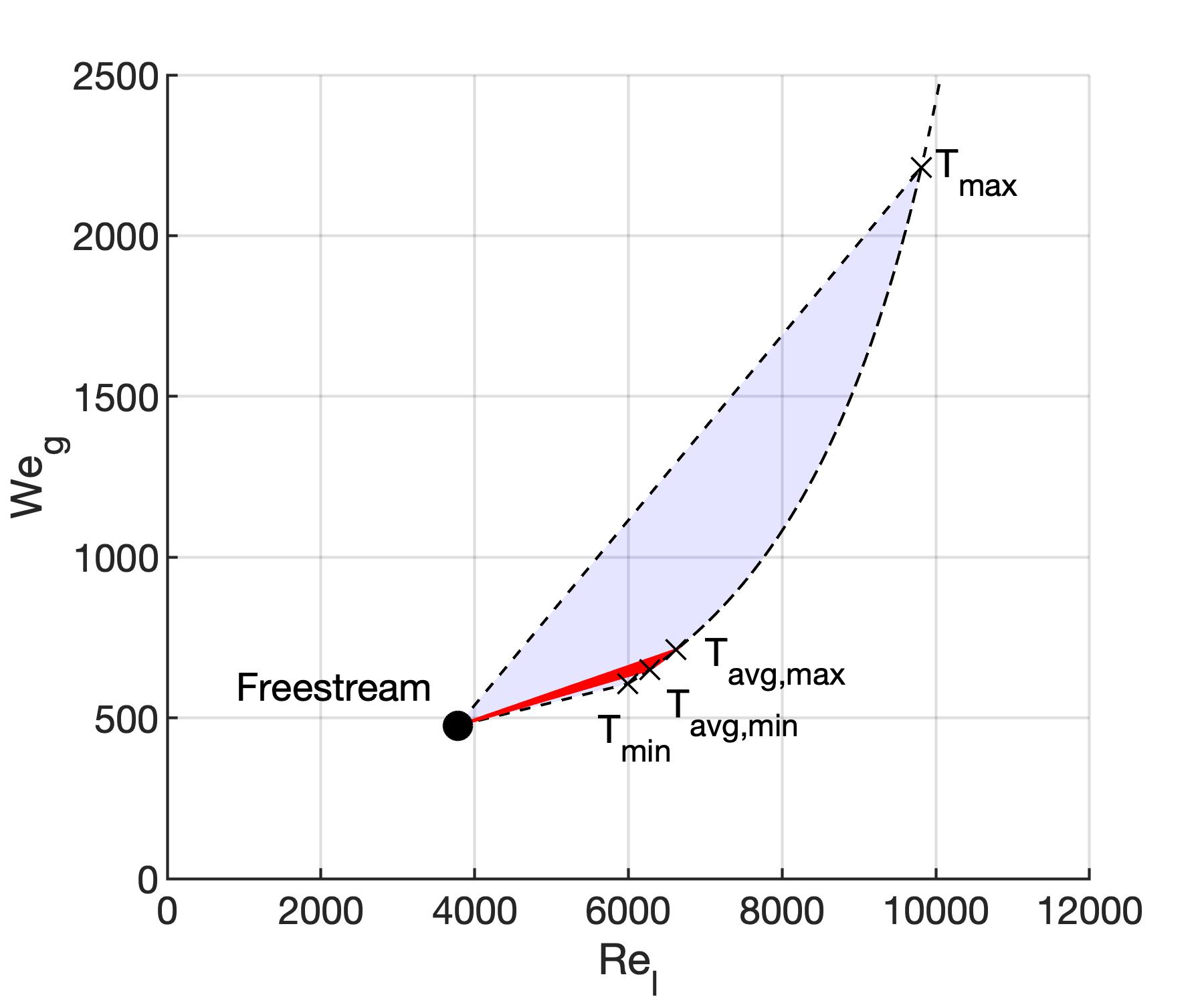}
  \caption{50 bar and \(u_G=70\) m/s (A2)}
  \label{subfig:Weg_vs_Rel_50_70A}
\end{subfigure}%
\\
\begin{subfigure}{0.33\textwidth}
  \centering
  \includegraphics[width=1.05\linewidth]{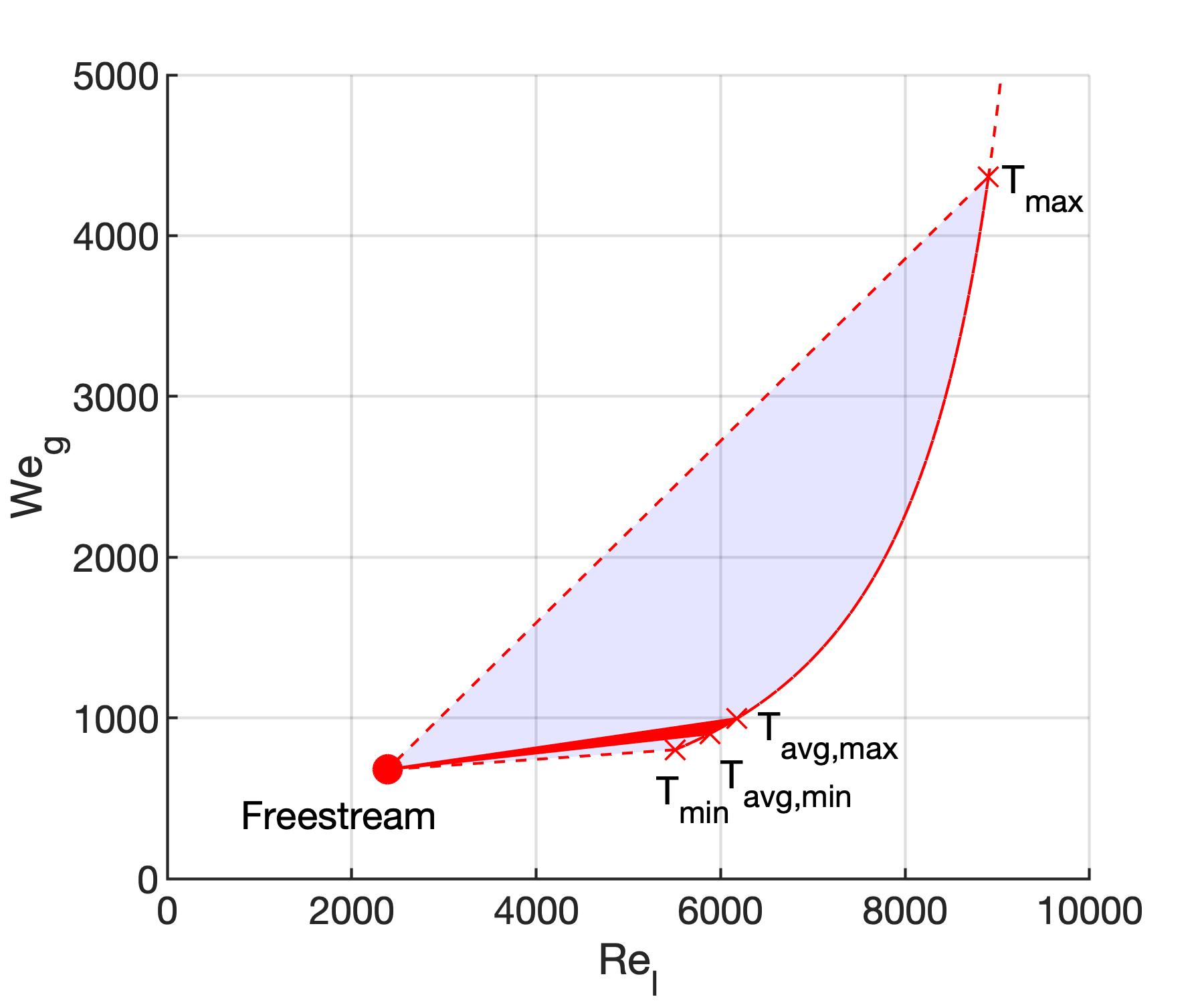}
  \caption{100 bar and \(u_G=50\) m/s (B1)}
  \label{subfig:Weg_vs_Rel_100_50A}
\end{subfigure}%
\begin{subfigure}{0.33\textwidth}
  \centering
  \includegraphics[width=1.05\linewidth]{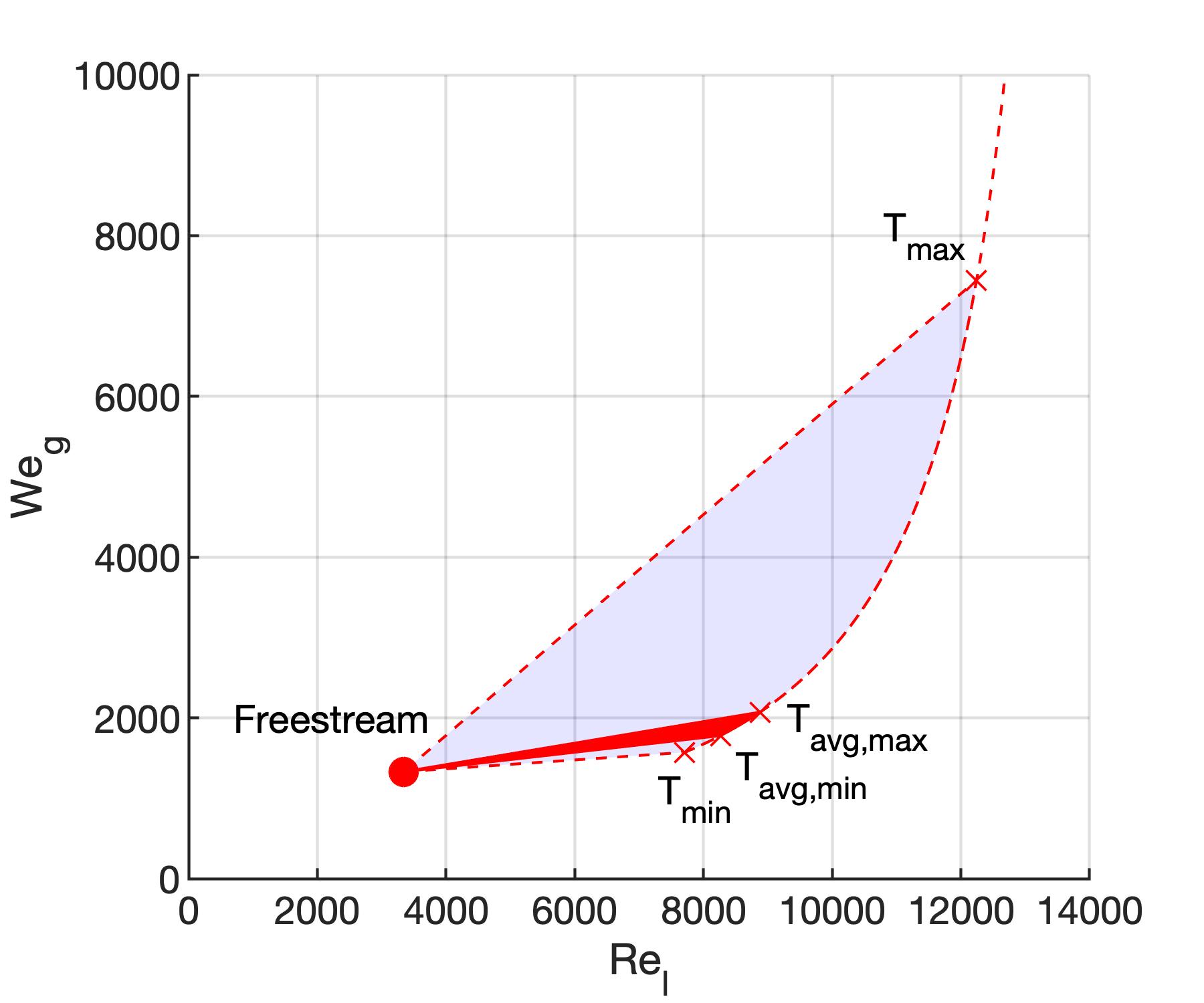}
  \caption{100 bar and \(u_G=70\) m/s (B2)}
  \label{subfig:Weg_vs_Rel_100_70A}
\end{subfigure}%
\\
\begin{subfigure}{0.33\textwidth}
  \centering
  \includegraphics[width=1.05\linewidth]{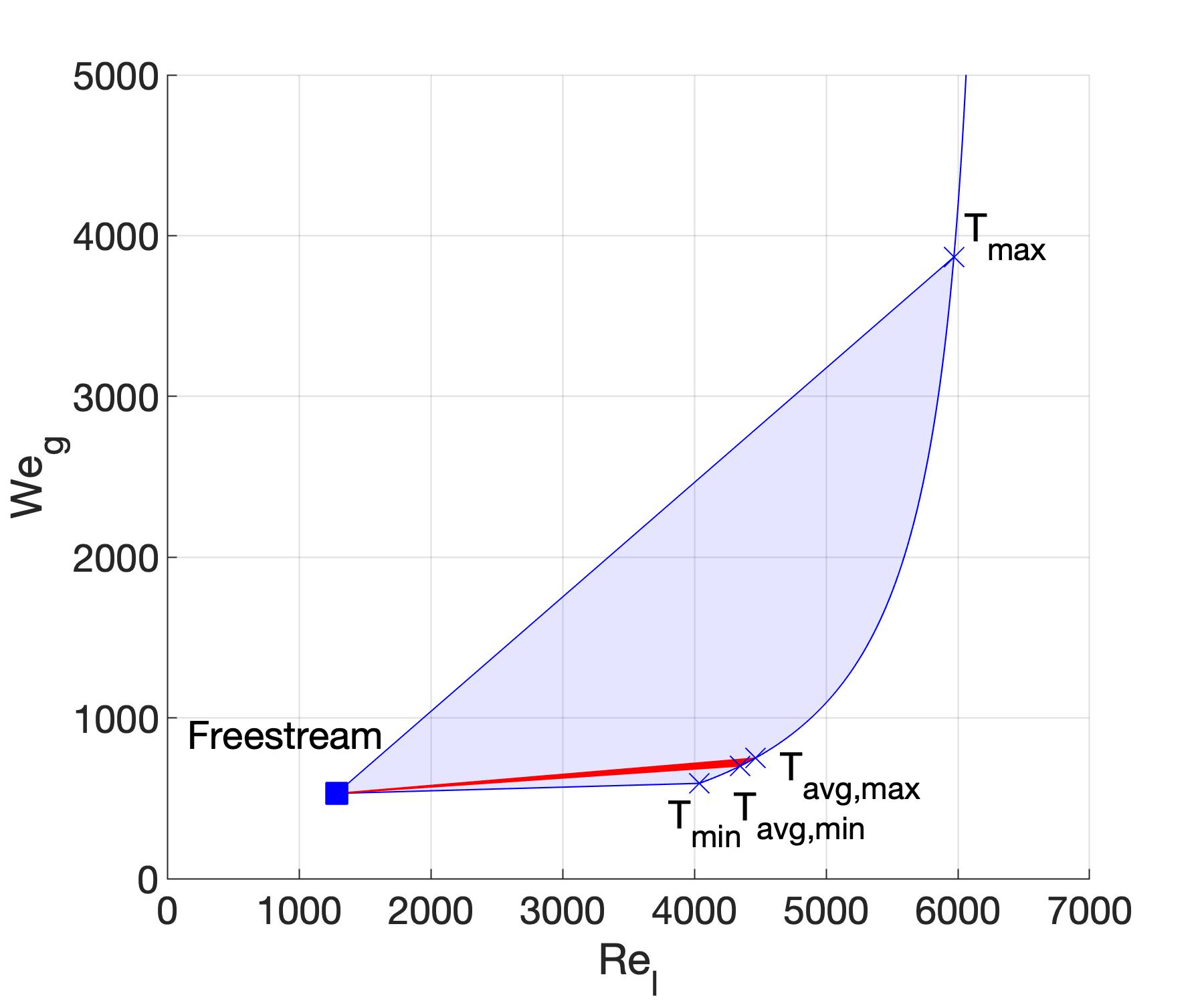}
  \caption{150 bar and \(u_G=30\) m/s (C1)} 
  \label{subfig:Weg_vs_Rel_150_30A}
\end{subfigure}%
\begin{subfigure}{0.33\textwidth}
  \centering
  \includegraphics[width=1.05\linewidth]{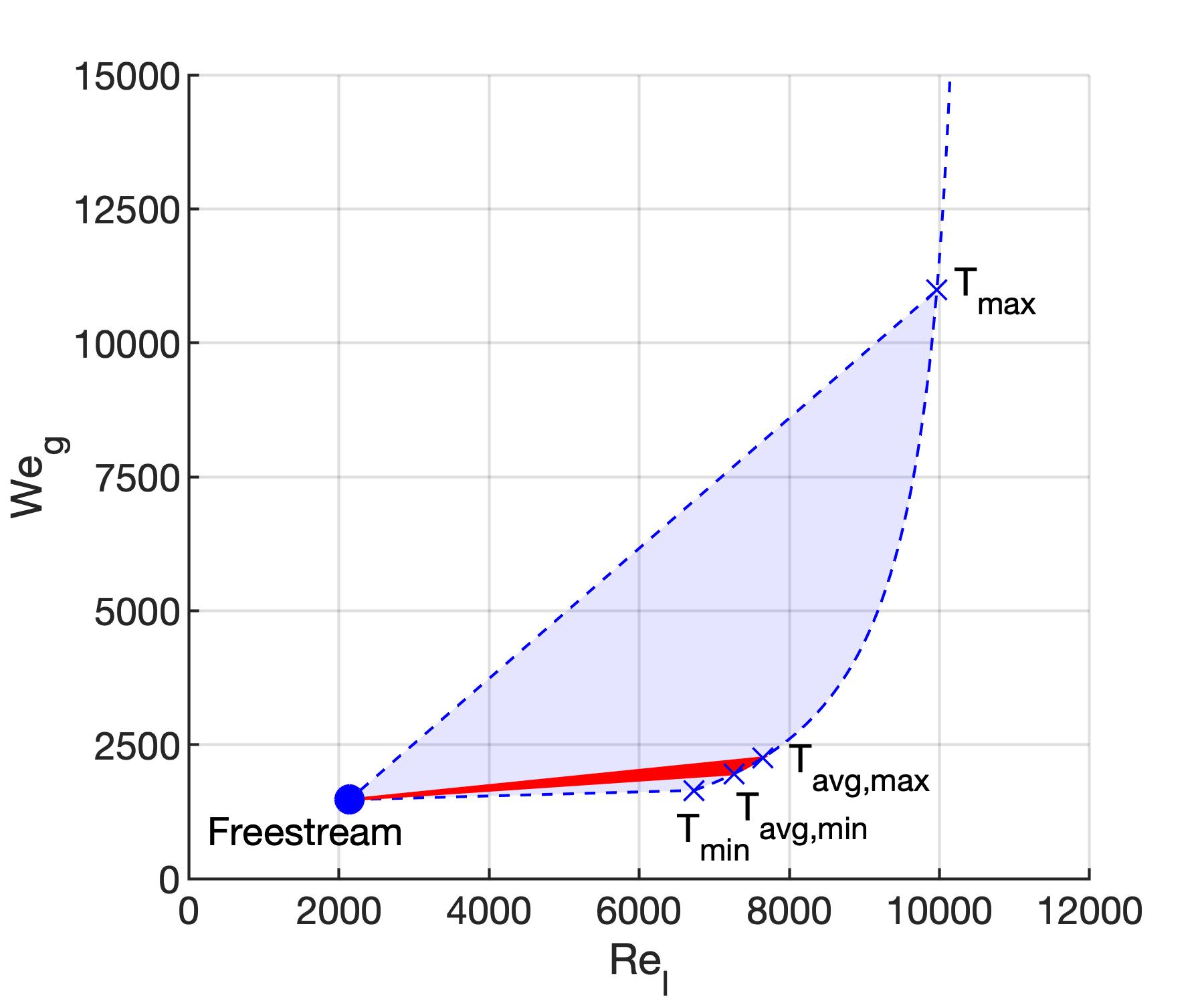}
  \caption{150 bar and \(u_G=50\) m/s (C2)}
  \label{subfig:Weg_vs_Rel_150_50A}
\end{subfigure}%
\begin{subfigure}{0.33\textwidth}
  \centering
  \includegraphics[width=1.05\linewidth]{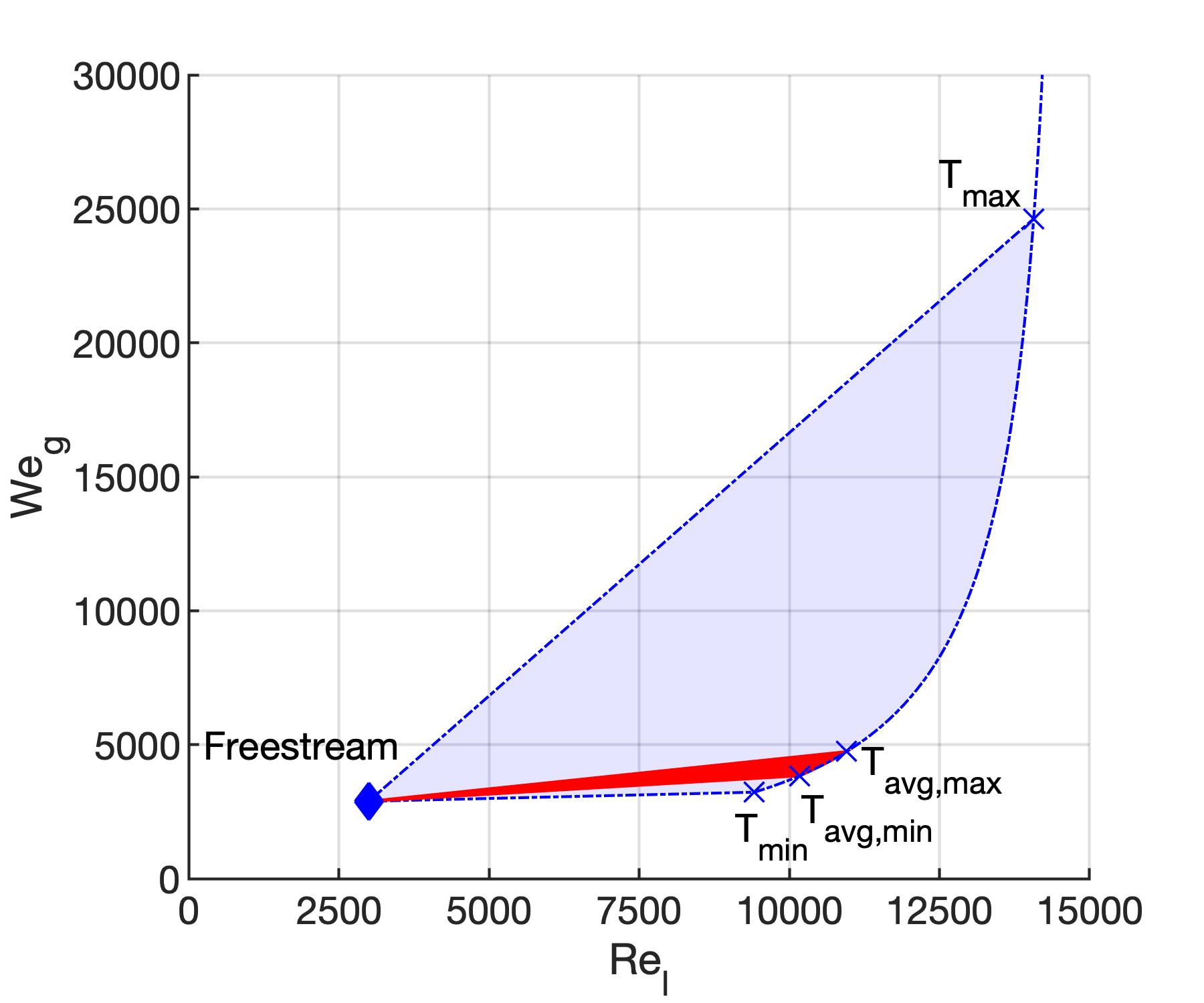}
  \caption{150 bar and \(u_G=70\) m/s (C3)}
  \label{subfig:Weg_vs_Rel_150_70A}
\end{subfigure}%
\caption{Effect of thermal and species mixing on the classification of each case in a \(We_g\) vs. \(Re_l\) diagram. The effective \(We_g\) and \(Re_l\) describing each configuration are located somewhere in the shaded area, most probably inside the region defined by the freestream condition and the minimum and maximum interface average temperatures. The values of \(We_g\) and \(Re_l\) change over time as both phases undergo mixing. (a) case A1; (b) case A2; (c) case B1; (d) case B2; (e) case C1; (f) case C2; and (g) case C3.}
\label{fig:variable_Re_We}
\end{figure}

Given this fluid behavior, it becomes clear that using freestream properties to characterize each case might be misleading. Determining an effective Weber number and Reynolds number is also complicated. As explained previously, different fluid regions will have different fluid properties and can behave substantially differently. However, a region in the \(We_g\) vs. \(Re_l\) diagram can be identified based on lower and upper physical boundaries where the effective parameters may be located (see Figure~\ref{fig:variable_Re_We}). \par

A lower boundary for \(We_g\) and \(Re_l\) is obtained by using the definition based on the freestream fluid properties and the average value of the surface-tension coefficient, \(We_G\) and \(Re_L\). For this analysis, the characteristic length and the characteristic velocity are still represented by the jet thickness, \(H\), and the gas freestream velocity, \(u_G\). The focus is still on the full jet characteristics while taking into account the variations in the fluid properties. Locally, the fluid is influenced by the local length scale (e.g., the radius of curvature or lobe thickness) and the local velocity. \par 

On the other hand, the upper boundary for \(We_g\) and \(Re_l\) is defined by the possible interface equilibrium states. LTE provides the fluid properties of each phase, the composition, and the surface-tension coefficient as a function of interface temperature. If the fluid properties at the interface are used to evaluate \(We_g\) and \(Re_l\), a curve is obtained following a locus of points that increases toward higher values of \(We_g\) and \(Re_l\) with the increase of the interface temperature (see any of the curves in Figure~\ref{fig:variable_Re_We}). That is, as temperature increases and more oxygen dissolves into the liquid phase, the liquid viscosity drops more than the liquid density causing an increase in \(Re_l\). Meanwhile, the gas density increases and the surface-tension coefficient drops substantially as both phases look more alike, causing a sharp increase in \(We_g\). \par 

The effective \(We_g\) and \(Re_l\) numbers defining the cases detailed in Table~\ref{tab:cases} are somewhere inside the two shaded areas shown in each respective sub-figure from Figure~\ref{fig:variable_Re_We}. The first shaded area covers the most probable region where the effective \(We_g\) and \(Re_l\) might be located. It is represented by the freestream condition and the interface solution between the minimum average interface temperature, \(T_\text{avg,min}\), and the maximum average interface temperature, \(T_\text{avg,max}\), observed during the computations. The second shaded area represents a much less probable region for the effective global \(We_g\) and \(Re_l\), but that might represent the local behavior of the jet. This region is enclosed by the freestream condition, the absolute minimum interface temperature throughout the simulation, \(T_\text{min}\), and the absolute maximum interface temperature observed during the simulation, \(T_\text{max}\). \par 

Early in the simulation, the jet classification is better represented by the freestream condition. Also, the two incompressible cases A2i and C1i are represented by each freestream condition. The effective \(We_g\) and \(Re_l\) increase as mixing occurs in both phases and the interface deforms. These changes are more pronounced at higher pressures as observed in Figure~\ref{fig:variable_Re_We} because of the enhanced mixing effects. At 150 bar with \(u_G=50\) m/s (i.e., case C2), \(We_g\) and \(Re_l\) could be as high as to 53\% and 257\% larger, respectively, at average interface conditions compared to the freestream condition. At lower pressures, such as 50 bar and \(u_G=50\) m/s (i.e., case A1), both parameters could increase up to 44\% and 70\%, respectively. Nevertheless, the reclassification of the jet still falls in the LoCLiD atomization sub-domain, although hole formation is apparent. \par

\subsection{Early-deformation characteristics}
\label{subsec:deformation}

This subsection describes and classifies some of the deformation patterns observed during the computations. Some features can be identified in various configurations, but emphasis is made on the main differences caused by the pressure increase. Similarities are observed for cases with similar \(We_G\) or \(Re_L\) based on freestream conditions, but the high-pressure mixing effects are responsible for variations of each feature. \par 

The features identified from Subsection~\ref{subsubsec:lobe_ext} to Subsection~\ref{subsubsec:layering} are: (a) lobe extension, bending and perforation; (b) lobe and crest corrugation; (c) ligament stretching and shredding; and (d) stretching, folding and layering of liquid sheets and liquid sheet tearing. A summary is provided in Table~\ref{tab:features}, which identifies which feature appears in each analyzed case. Also, the various deformation mechanisms are identified in a \(We_G\) vs. \(Re_L\) diagram in Figure~\ref{fig:Weg_Rel_deformation} (see Section~\ref{sec:summary_and_conclusions}). For brevity, we only present the necessary figures to support our discussion. Thus, the reader is referred to the Supplemental Material where slides have been provided showing the jet deformation for each case. Also, a non-dimensional time, \(t^*\), has been defined to compare cases properly. Based on the jet thickness as characteristic length and the gas freestream velocity as characteristic velocity, a characteristic time is defined as \(t_c=H/u_G\) such that \(t^*=t/t_c\). For reference, the physical times analyzed in this work are below 10 \(\mu\)s due to the fast interface distortion at high pressures and limited domain size. \par 

\begin{table}[h!]
\begin{center}
\begin{tabular}{|r|r|r|r|r|r|r|r|r|r|} 
\multicolumn{1}{c|}{} & \multicolumn{1}{c|}{A1} & \multicolumn{1}{c|}{A2} & \multicolumn{1}{c|}{B1} &
\multicolumn{1}{c|}{B2} &
\multicolumn{1}{c|}{C1} &
\multicolumn{1}{c|}{C2} &
\multicolumn{1}{c|}{C3} &
\multicolumn{1}{c|}{A2i} &
\multicolumn{1}{c}{C1i}\\ 
\hline
\hline
\multicolumn{1}{c|}{Lobe extension (a)} & 
\multicolumn{1}{c|}{\textcolor{red}{No}} & \multicolumn{1}{c|}{\textcolor{OliveGreen}{Yes}} & \multicolumn{1}{c|}{\textcolor{OliveGreen}{Yes}} &
\multicolumn{1}{c|}{\textcolor{red}{No}} &
\multicolumn{1}{c|}{\textcolor{OliveGreen}{Yes}} &
\multicolumn{1}{c|}{\textcolor{red}{No}} &
\multicolumn{1}{c|}{\textcolor{red}{No}} &
\multicolumn{1}{c|}{\textcolor{OliveGreen}{Yes}} &
\multicolumn{1}{c}{\textcolor{OliveGreen}{Yes}}\\ 
\multicolumn{1}{c|}{Lobe bending (a)} & 
\multicolumn{1}{c|}{\textcolor{red}{No}} & \multicolumn{1}{c|}{\textcolor{OliveGreen}{Yes}} & \multicolumn{1}{c|}{\textcolor{OliveGreen}{Yes}} &
\multicolumn{1}{c|}{\textcolor{red}{No}} &
\multicolumn{1}{c|}{\textcolor{OliveGreen}{Yes}} &
\multicolumn{1}{c|}{\textcolor{red}{No}} &
\multicolumn{1}{c|}{\textcolor{red}{No}} &
\multicolumn{1}{c|}{\textcolor{red}{No}} &
\multicolumn{1}{c}{\textcolor{red}{No}}\\ 
\multicolumn{1}{c|}{Lobe perforation (a)} & 
\multicolumn{1}{c|}{\textcolor{red}{No}} & \multicolumn{1}{c|}{\textcolor{red}{No}} & \multicolumn{1}{c|}{\textcolor{OliveGreen}{Yes}} &
\multicolumn{1}{c|}{\textcolor{red}{No}} &
\multicolumn{1}{c|}{\textcolor{OliveGreen}{Yes}} &
\multicolumn{1}{c|}{\textcolor{red}{No}} &
\multicolumn{1}{c|}{\textcolor{red}{No}} &
\multicolumn{1}{c|}{\textcolor{red}{No}} &
\multicolumn{1}{c}{\textcolor{red}{No}}\\ 
\multicolumn{1}{c|}{Lobe corrugation (b)} & 
\multicolumn{1}{c|}{\textcolor{red}{No}} & \multicolumn{1}{c|}{\textcolor{red}{No}} & \multicolumn{1}{c|}{\textcolor{red}{No}} &
\multicolumn{1}{c|}{\textcolor{OliveGreen}{Yes}} &
\multicolumn{1}{c|}{\textcolor{red}{No}} &
\multicolumn{1}{c|}{\textcolor{OliveGreen}{Yes}} &
\multicolumn{1}{c|}{\textcolor{OliveGreen}{Yes}} &
\multicolumn{1}{c|}{\textcolor{red}{No}} &
\multicolumn{1}{c}{\textcolor{red}{No}}\\ 
\multicolumn{1}{c|}{Crest corrugation (b)} & 
\multicolumn{1}{c|}{\textcolor{red}{No}} & \multicolumn{1}{c|}{\textcolor{red}{No}} & \multicolumn{1}{c|}{\textcolor{OliveGreen}{Yes}} &
\multicolumn{1}{c|}{\textcolor{OliveGreen}{Yes}} &
\multicolumn{1}{c|}{\textcolor{red}{No}} &
\multicolumn{1}{c|}{\textcolor{OliveGreen}{Yes}} &
\multicolumn{1}{c|}{\textcolor{OliveGreen}{Yes}} &
\multicolumn{1}{c|}{\textcolor{red}{No}} &
\multicolumn{1}{c}{\textcolor{red}{No}}\\ 
\multicolumn{1}{c|}{Ligament stretching (c)} & 
\multicolumn{1}{c|}{\textcolor{OliveGreen}{Yes}} & \multicolumn{1}{c|}{\textcolor{OliveGreen}{Yes}} & \multicolumn{1}{c|}{\textcolor{OliveGreen}{Yes}} &
\multicolumn{1}{c|}{\textcolor{OliveGreen}{Yes}} &
\multicolumn{1}{c|}{\textcolor{OliveGreen}{Yes}} &
\multicolumn{1}{c|}{\textcolor{OliveGreen}{Yes}} &
\multicolumn{1}{c|}{\textcolor{OliveGreen}{Yes}} &
\multicolumn{1}{c|}{\textcolor{OliveGreen}{Yes}} &
\multicolumn{1}{c}{\textcolor{OliveGreen}{Yes}}\\ 
\multicolumn{1}{c|}{Ligament shredding (c)} & 
\multicolumn{1}{c|}{\textcolor{red}{No}} & \multicolumn{1}{c|}{\textcolor{red}{No}} & \multicolumn{1}{c|}{\textcolor{red}{No}} &
\multicolumn{1}{c|}{\textcolor{OliveGreen}{Yes}} &
\multicolumn{1}{c|}{\textcolor{red}{No}} &
\multicolumn{1}{c|}{\textcolor{OliveGreen}{Yes}} &
\multicolumn{1}{c|}{\textcolor{OliveGreen}{Yes}} &
\multicolumn{1}{c|}{\textcolor{red}{No}} &
\multicolumn{1}{c}{\textcolor{red}{No}}\\
\multicolumn{1}{c|}{Layering and liquid sheet tearing (d)} & 
\multicolumn{1}{c|}{\textcolor{red}{No}} & \multicolumn{1}{c|}{\textcolor{red}{No}} & \multicolumn{1}{c|}{\textcolor{OliveGreen}{Yes}} &
\multicolumn{1}{c|}{\textcolor{OliveGreen}{Yes}} &
\multicolumn{1}{c|}{\textcolor{OliveGreen}{Yes}} &
\multicolumn{1}{c|}{\textcolor{OliveGreen}{Yes}} &
\multicolumn{1}{c|}{\textcolor{OliveGreen}{Yes}} &
\multicolumn{1}{c|}{\textcolor{red}{No}} &
\multicolumn{1}{c}{\textcolor{OliveGreen}{Yes}}\\ 
\end{tabular}
\end{center}
\caption{Summary of identified deformation features and in which analyzed configurations they appear.}
\label{tab:features}
\end{table}

\subsubsection{Lobe extension, bending and perforation}
\label{subsubsec:lobe_ext}

\begin{figure}[h!]
\centering
\begin{subfigure}{0.25\textwidth}
  \centering
  \includegraphics[width=1.0\linewidth]{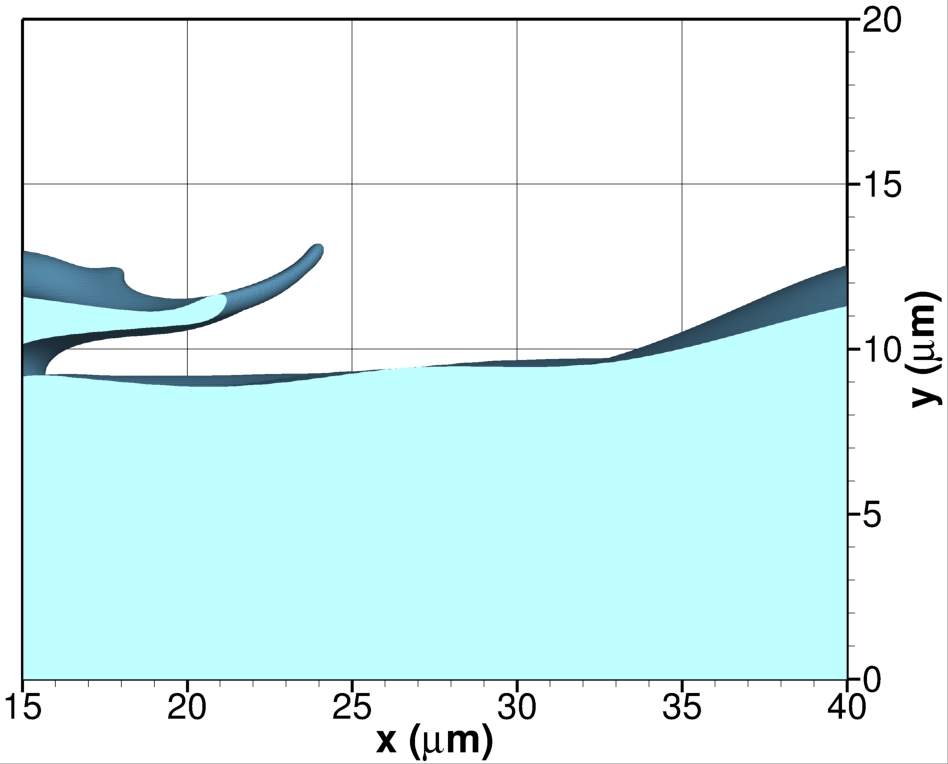}
  \label{subfig:150_30A_2p5mus_lobe_B}
\end{subfigure}%
\begin{subfigure}{0.25\textwidth}
  \centering
  \includegraphics[width=1.0\linewidth]{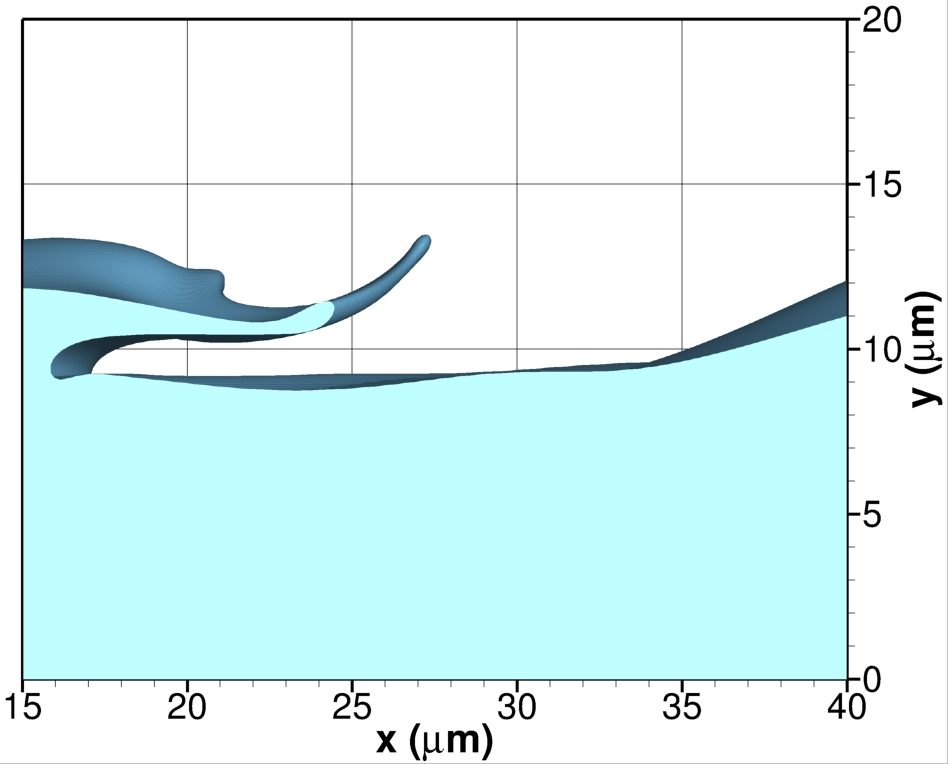}
  \label{subfig:150_30A_2p7mus_lobe_B}
\end{subfigure}%
\begin{subfigure}{0.25\textwidth}
  \centering
  \includegraphics[width=1.0\linewidth]{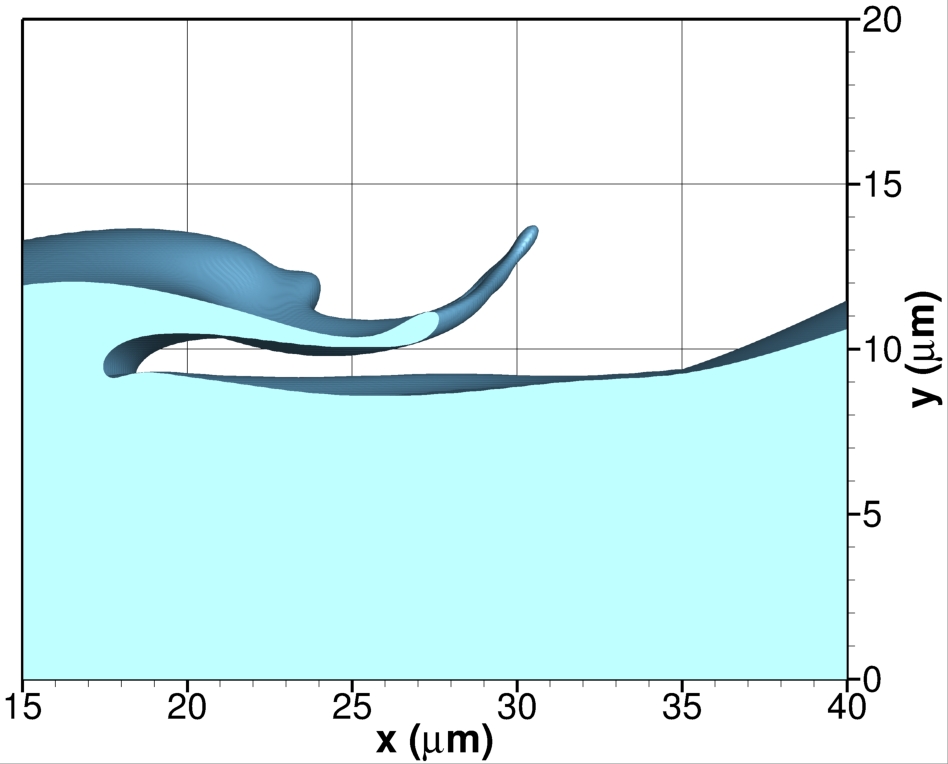}
  \label{subfig:150_30A_2p9mus_lobe_B}
\end{subfigure}%
\begin{subfigure}{0.25\textwidth}
  \centering
  \includegraphics[width=1.0\linewidth]{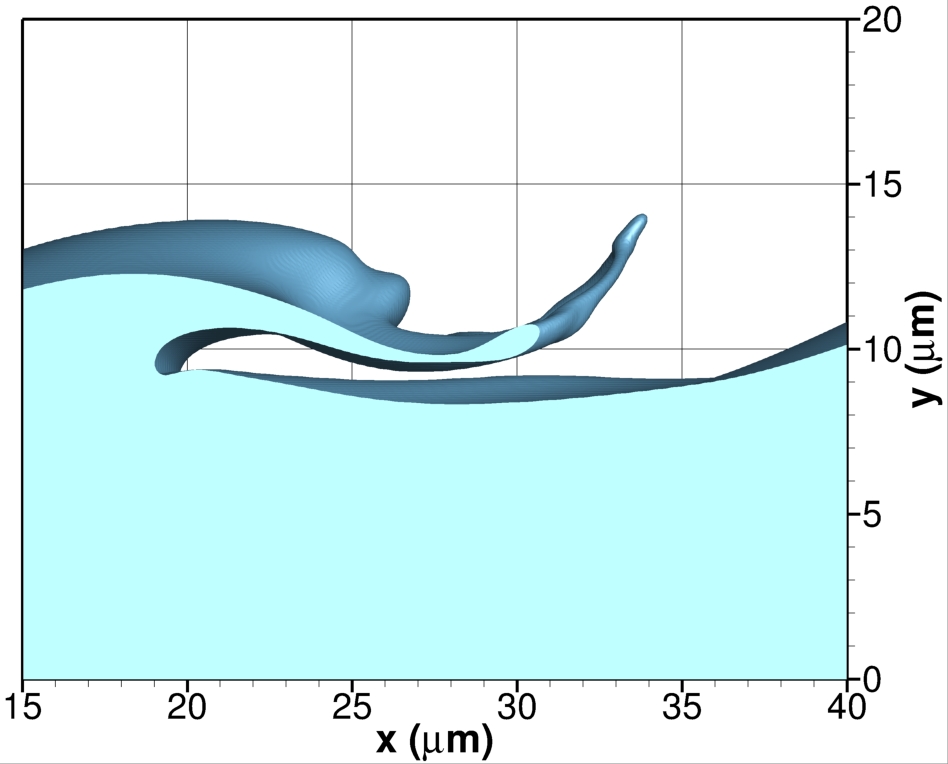}
  \label{subfig:150_30A_3p1mus_lobe_B}
\end{subfigure}%
\\[-3ex]
\begin{subfigure}{0.25\textwidth}
  \centering
  \includegraphics[width=1.0\linewidth]{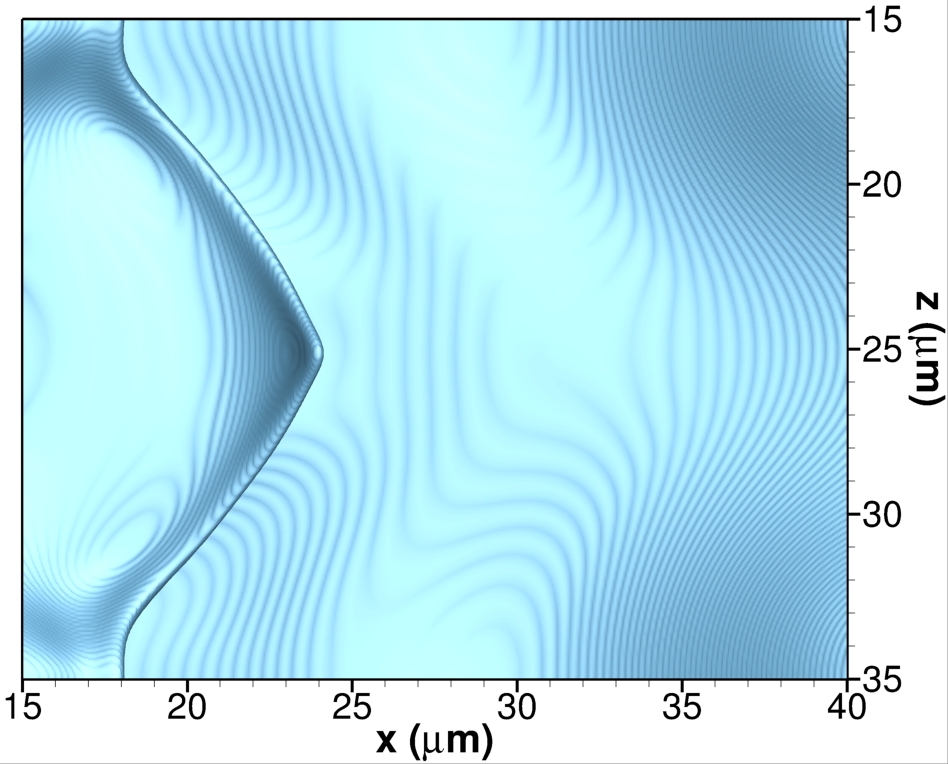}
  \caption{\(t^*=3.75\)} 
  \label{subfig:150_30A_2p5mus_lobe_D}
\end{subfigure}%
\begin{subfigure}{0.25\textwidth}
  \centering
  \includegraphics[width=1.0\linewidth]{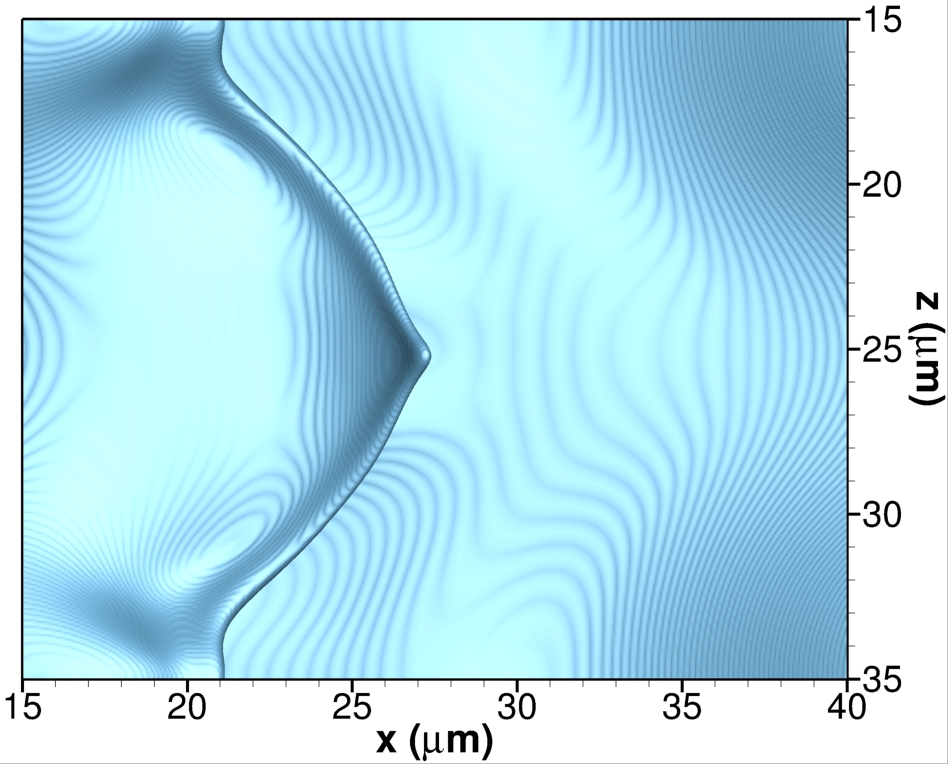}
  \caption{\(t^*=4.05\)}
  \label{subfig:150_30A_2p7mus_lobe_D}
\end{subfigure}%
\begin{subfigure}{0.25\textwidth}
  \centering
  \includegraphics[width=1.0\linewidth]{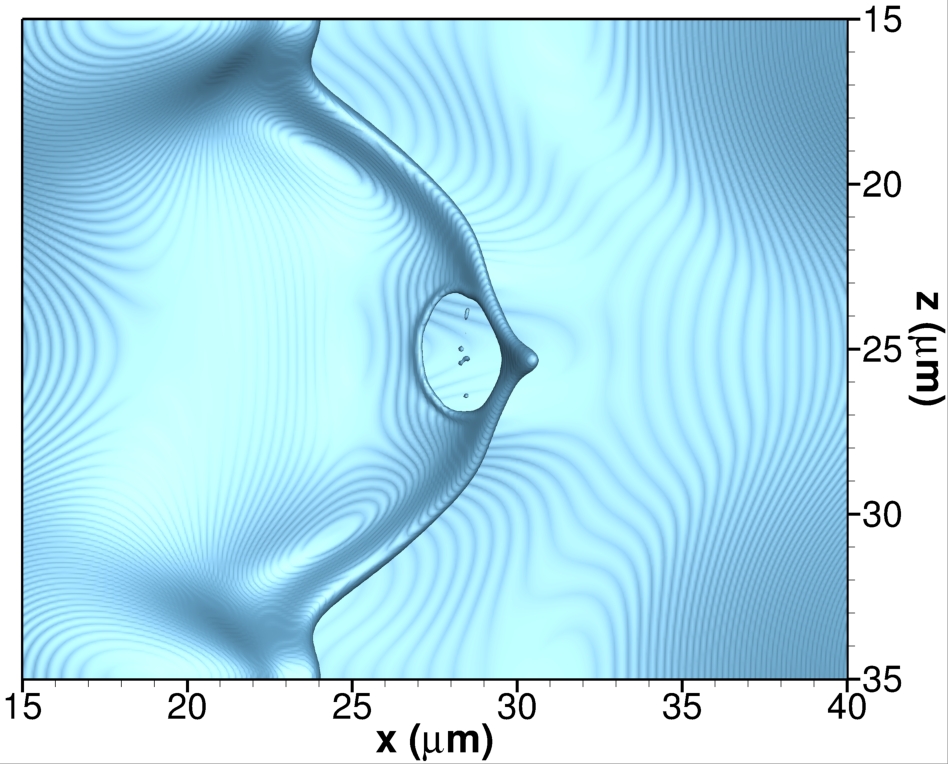}
  \caption{\(t^*=4.35\)}
  \label{subfig:150_30A_2p9mus_lobe_D}
\end{subfigure}%
\begin{subfigure}{0.25\textwidth}
  \centering
  \includegraphics[width=1.0\linewidth]{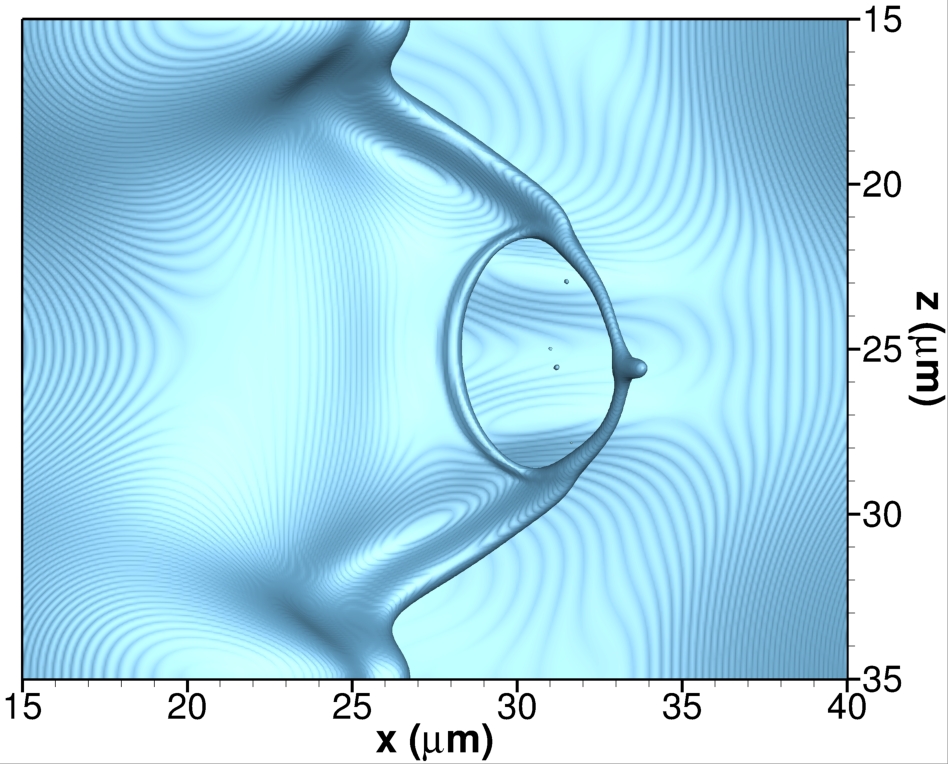}
  \caption{\(t^*=4.65\)}
  \label{subfig:150_30A_3p1mus_lobe_D}
\end{subfigure}%
\caption{Lobe extension, bending and perforation at 150 bar with gas freestream velocity of \(u_G=30\) m/s (i.e., case C1). The top figures show the side view from an \(xy\) plane located at \(z=40\) \(\mu\)m and the bottom figures show the top view from an \(xz\) plane located above the liquid surface. The interface location is identified as the isosurface with \(C=0.5\). A non-dimensional time is obtained as \(t^*=t/t_c=t\frac{u_G}{H}\). (a) \(t^*=3.75\); (b) \(t^*=4.05\); (c) \(t^*=4.35\); and (d) \(t^*=4.65\).}
\label{fig:lobe_ext_time}
\end{figure}

\begin{figure}[h!]
\centering
\begin{subfigure}{0.45\textwidth}
  \centering
  \includegraphics[width=0.9\linewidth]{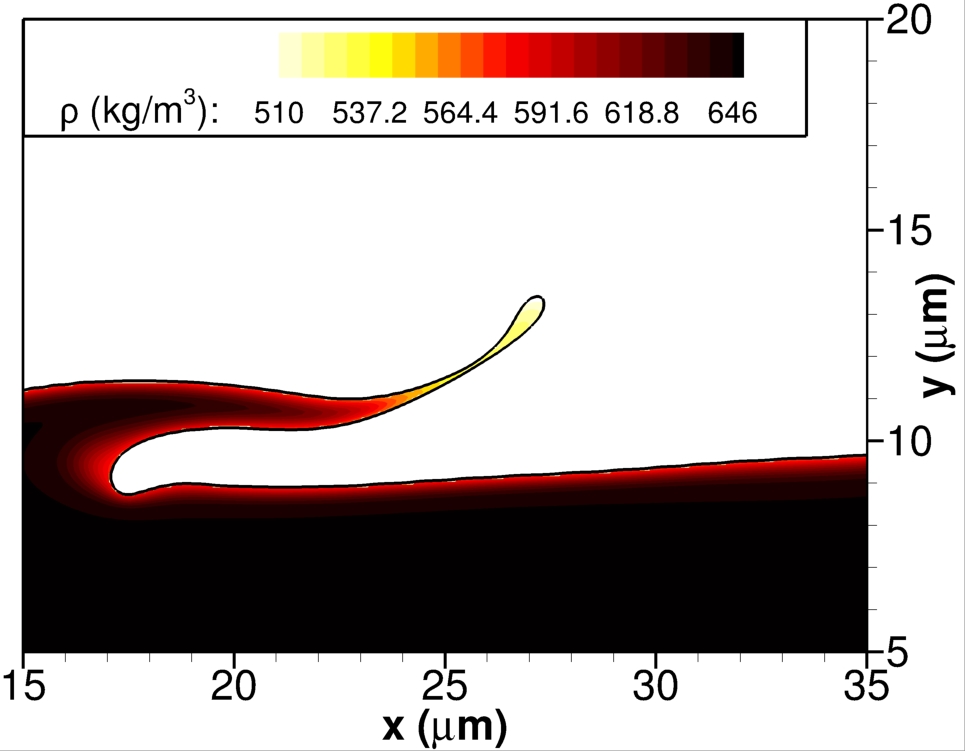}
  \caption{} 
  \label{subfig:150_30A_2p7mus_DENL_lobe_B_z25mum}
\end{subfigure}%
\begin{subfigure}{0.45\textwidth}
  \centering
  \includegraphics[width=0.9\linewidth]{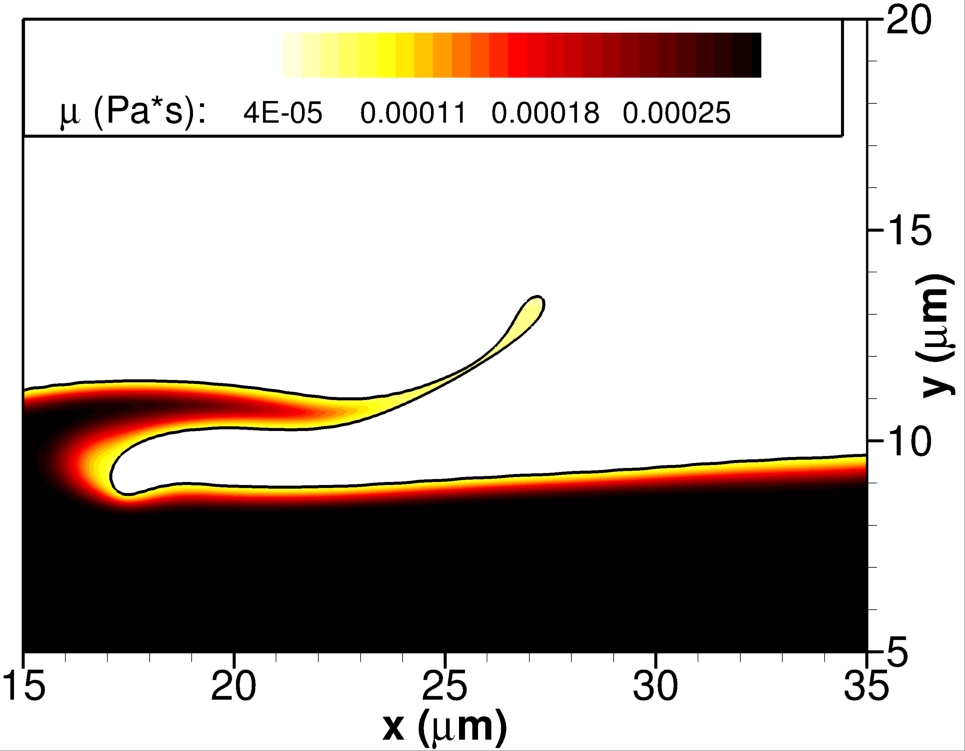}
  \caption{}
  \label{subfig:150_30A_2p7mus_VIS_lobe_B_z25mum}
\end{subfigure}%
\caption{Lobe extension, bending and perforation at 150 bar with gas freestream velocity of \(u_G=30\) m/s (i.e., case C1) at the non-dimensional time \(t^*=t/t_c=t\frac{u_G}{H}=4.05\). A slice through the three-dimensional domain is obtained from an \(xy\) plane located at \(z=25\) \(\mu\)m. The interface location is identified as the isosurface with \(C=0.5\). (a) liquid density; and (b) viscosity of the two-phase mixture.}
\label{fig:lobe_bending_den_vis}
\end{figure}

\begin{figure}[h!]
\centering
\begin{subfigure}{0.33\textwidth}
  \centering
  \includegraphics[width=1.0\linewidth]{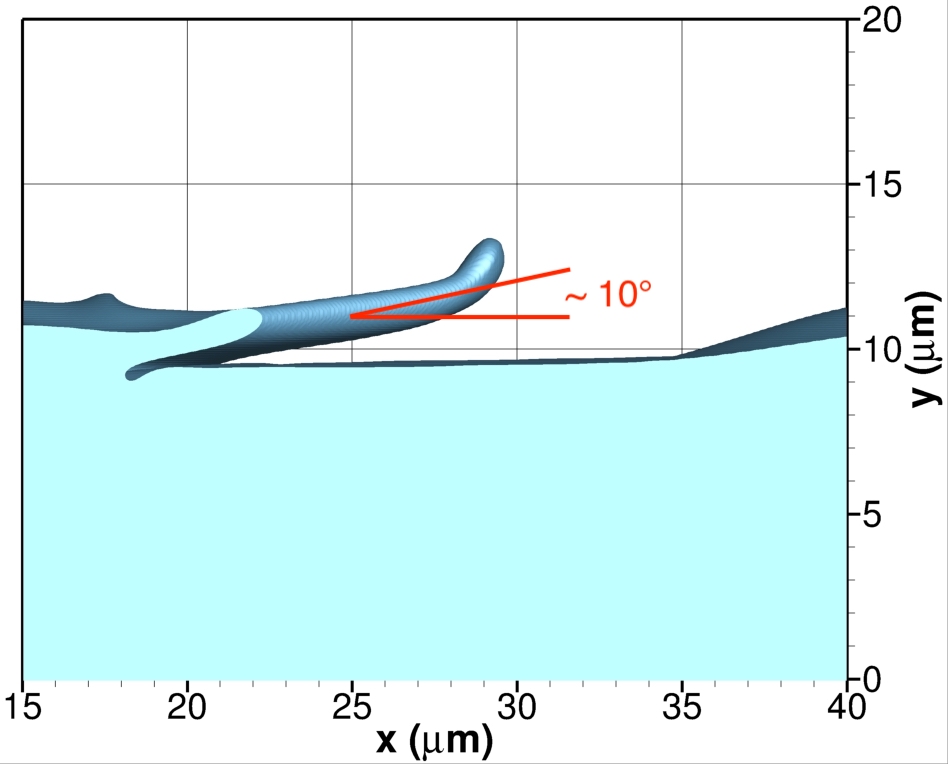}
  \label{subfig:50_70A_1p1mus_lobe_B}
\end{subfigure}%
\begin{subfigure}{0.33\textwidth}
  \centering
  \includegraphics[width=1.0\linewidth]{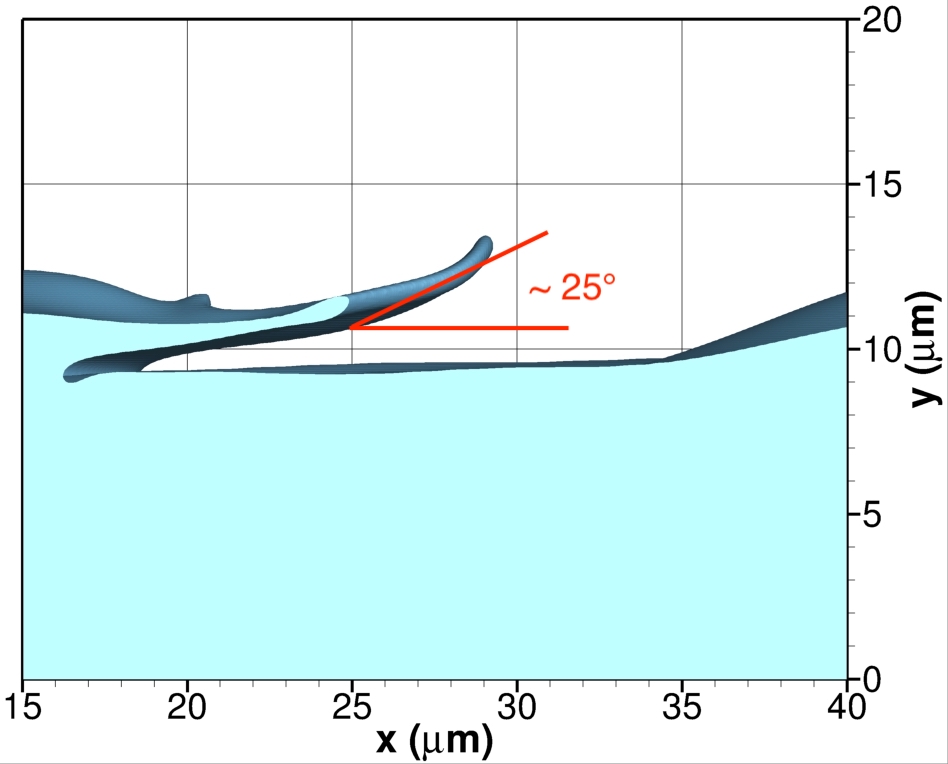}
  \label{subfig:100_50A_1p6mus_lobe_B}
\end{subfigure}%
\begin{subfigure}{0.33\textwidth}
  \centering
  \includegraphics[width=1.0\linewidth]{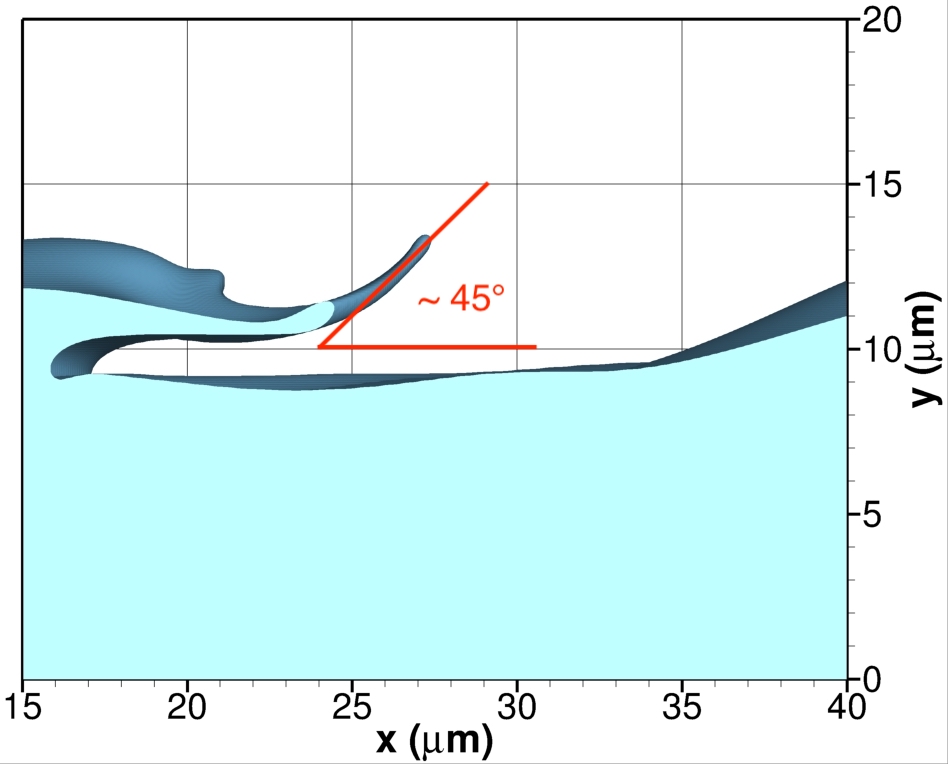}
  \label{subfig:150_30A_2p7mus_lobe_B2}
\end{subfigure}%
\\[-3ex]
\begin{subfigure}{0.33\textwidth}
  \centering
  \includegraphics[width=1.0\linewidth]{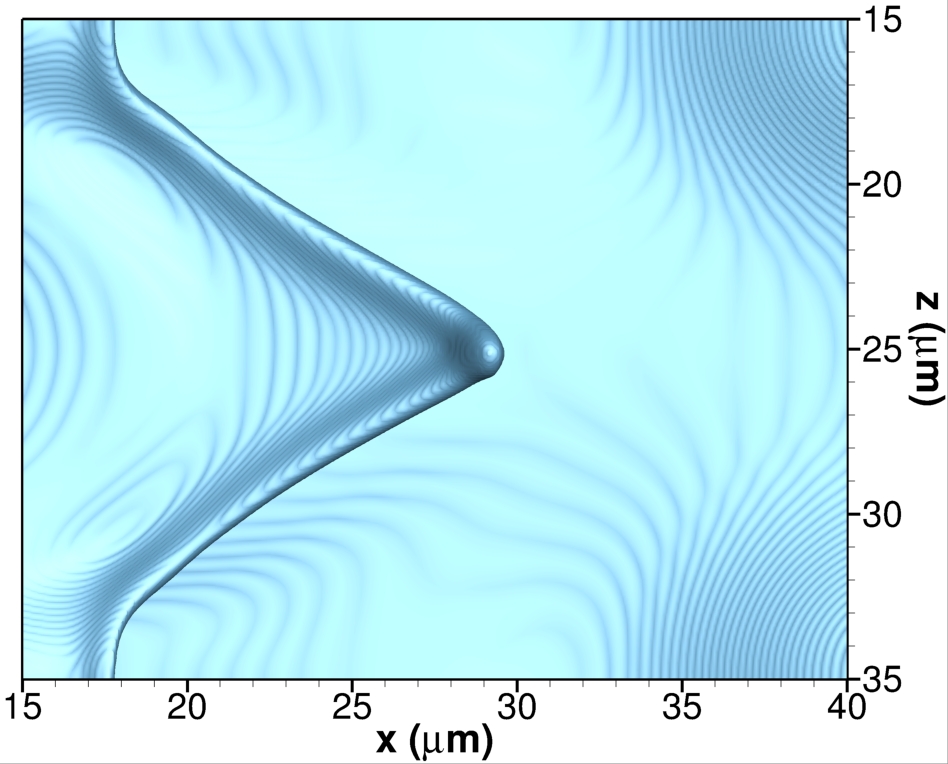}
  \caption{50 bar and \(u_G=70\) m/s (A2) at \(t^*=4.025\)} 
  \label{subfig:50_70A_1p1mus_lobe_D}
\end{subfigure}%
\begin{subfigure}{0.33\textwidth}
  \centering
  \includegraphics[width=1.0\linewidth]{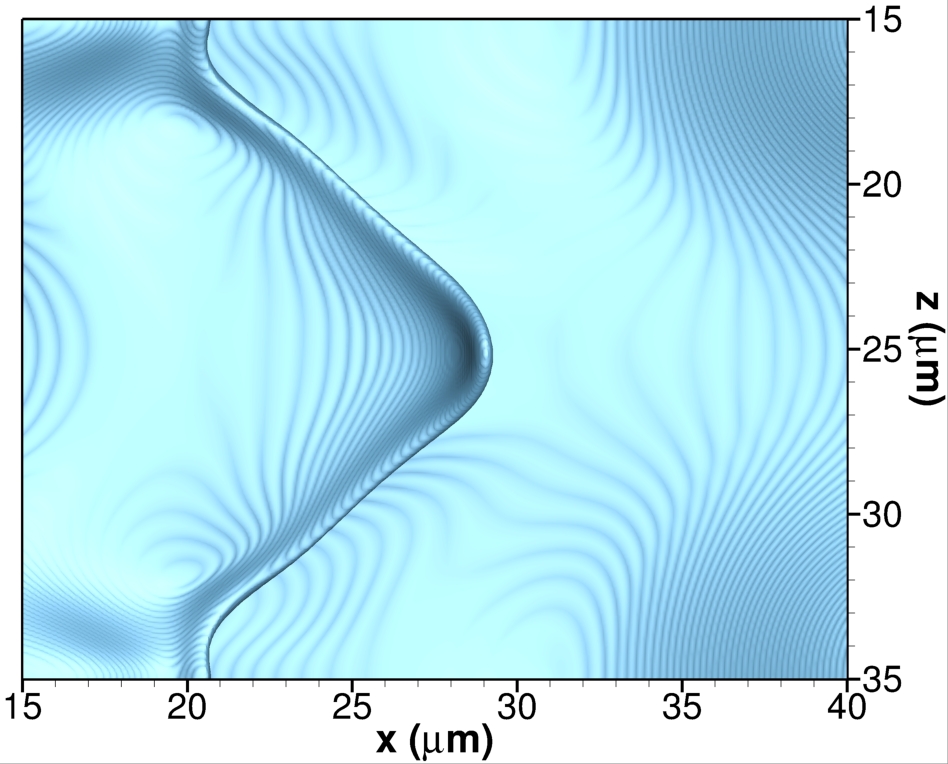}
  \caption{100 bar and \(u_G=50\) m/s (B1) at \(t^*=4.00\)}
  \label{subfig:100_50A_1p6mus_lobe_D}
\end{subfigure}%
\begin{subfigure}{0.33\textwidth}
  \centering
  \includegraphics[width=1.0\linewidth]{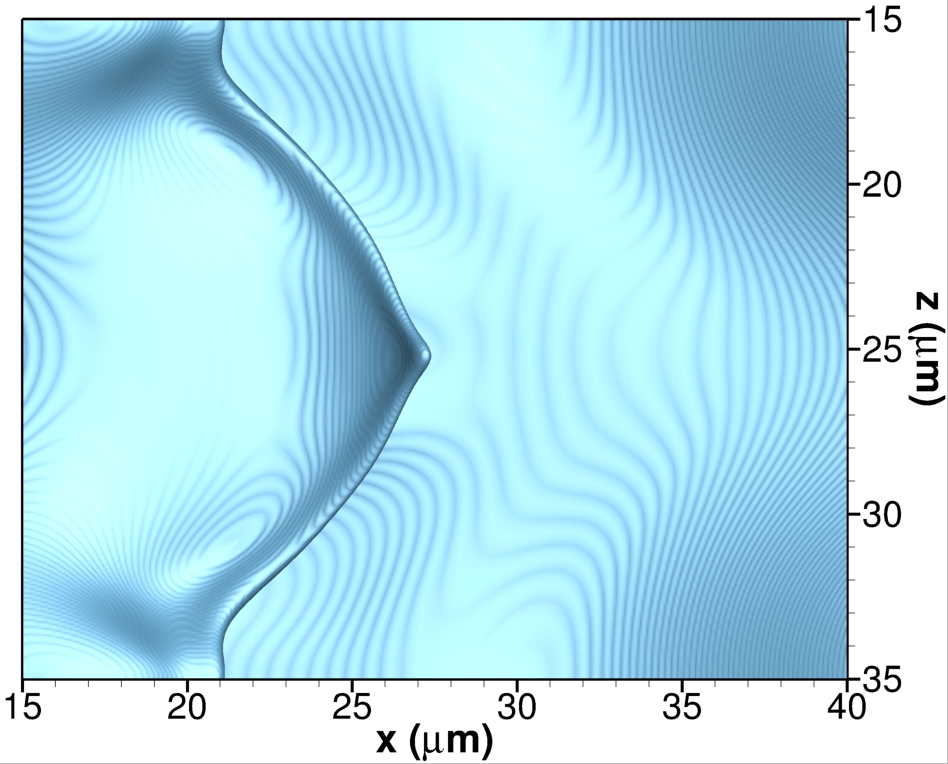}
  \caption{150 bar and \(u_G=30\) m/s (C1) at \(t^*=4.05\)}
  \label{subfig:150_30A_2p7mus_lobe_D2}
\end{subfigure}%
\caption{Pressure effects on the early lobe extension comparing cases A2, B1 and C1 with similar \(We_G\). The top figures show the side view from an \(xy\) plane located at \(z=40\) \(\mu\)m and the bottom figures show the top view from an \(xz\) plane located above the liquid surface. The interface location is identified as the isosurface with \(C=0.5\). A non-dimensional time is obtained as \(t^*=t/t_c=t\frac{u_G}{H}\). (a) case A2 at \(t^*=4.025\); (b) case B1 at \(t^*=4.00\); and (c) case C1 at \(t^*=4.05\).}
\label{fig:lobe_ext_pressure}
\end{figure}

Initially, all cases develop a lobe on the liquid surface due to the initial shear layer and surface shape, which eventually evolves following different patterns. Cases A2, B1 and C1 in Figures~\ref{fig:lobe_ext_time},~\ref{fig:lobe_bending_den_vis} and~\ref{fig:lobe_ext_pressure} present the extension of a lobe whose tip bends upward (i.e., rotates in the positive \(z\) direction) before high-pressure effects are observed. All three cases present a similar \(We_G\) between 476 and 679, but substantially different \(Re_L\) ranging from 1285 to 3777. In contrast, the initial lobe in case A1 with \(We_G=243\) reconnects with the liquid core due to higher surface tension. Also, cases B2, C2 and C3 with \(We_G>1000\) show a different lobe deformation mechanism explained in Subsection~\ref{subsubsec:lobe_fold}. \par 

The sequence of the deformation mechanism consisting of lobe extension, bending and perforation is presented in Figure~\ref{fig:lobe_ext_time} for the case at 150 bar with \(u_G=30\) m/s (i.e., case C1). As the lobe extends and gets thinner, mixing reduces the local liquid density and viscosity (see Figure~\ref{fig:lobe_bending_den_vis}). Therefore, vortical motion in the denser gas downstream of the lobe is able to bend it upwards. Then, part of the lobe faces the gas stream with a steep angle before being perforated. The hole expands rapidly and a thin bridge is formed, with some small droplets being generated during the process that quickly vaporize and disappear. Note that hole formation in VOF methods always has some mesh dependence once the liquid structure size falls below the mesh size (i.e., \(\Delta=0.0667\) \(\mu\)m in this computation). Nevertheless, the physics support the formation of the observed hole, and mesh resolution might have a minor effect on the exact time when the perforation event is observed. \par 

This deformation mechanism is directly linked to a range of \(We_G\). The different fluid behavior caused by the change of pressure (also reflected in \(Re_L\)) causes variations in this deformation process as seen in Figure~\ref{fig:lobe_ext_pressure} that are crucial for the later development of the jet. At lower pressures, the higher surface-tension coefficient is reflected in the lobe thickness and the radius of curvature it presents along its edge. Case A2 at 50 bar (see Figure~\ref{subfig:50_70A_1p1mus_lobe_D}) presents a higher lobe thickness which subsequently reduces as pressure increases toward 100 and 150 bar. The change in the bending angle is also noticeable. The 150-bar case shown in Figure~\ref{fig:lobe_ext_time} presents a bending angle of approximately 45 degrees, but it decreases to about 25 degrees at 100 bar and around 10 degrees at 50 bar. \par 

This trend is a result of various factors. The mixing effects on both phases increase as pressure increases, reducing rapidly the differences between the liquid phase and the gas phase. Therefore, the liquid phase is more easily affected by the gas flow dynamics at 150 bar than 50 bar. Moreover, the analyzed cases have a higher \(Re_L\) at low pressures than at high pressures, mainly because of the higher gas freestream velocities required to achieve similar \(We_G\). In the overall picture, the higher \(Re_L\) translates into inertial terms dominating over viscous terms, which might promote the extension of the lobe and prevent the upward bending to some extent. \par 

The implications of these features on the perforation of the lobe are crucial. For cases B1 and C1 (i.e., 100 and 150 bar), the lobe is sufficiently thin and bends enough toward the oxidizer stream that a hole is generated at approximately the same non-dimensional time \(t^*\approx 4.3\). On the other hand, case A2 at 50 bar does not show hole formation on the lobe. Rather, the lobe's tip transitions to ligament stretching as explained in Subsection~\ref{subsubsec:lig_stretch}. Notice from Figure~\ref{fig:Weg_vs_Rel_overview} that all three cases are well below the LoHBrLiD atomization sub-domain identified by Zandian et al.~\cite{zandian2017planar} but still hole formation is observed early in the computation. The formation of a hole at high pressures, subsequent bridge thinning and the breakup into ligaments and droplets may help induce further surface perturbations early in the liquid deformation process. \par 

Further proof of the influence of the mixing effects in the deformation process is obtained by comparing cases A2 with A2i and C1 with C1i. In the incompressible limit, phase change is not considered and the fluid properties of each phase remain constant and equal to the freestream conditions. The surface-tension coefficient is also constant and equal to the average surface-tension coefficient observed in cases A2 and C1, respectively, during the early times of each simulation. \par 

\begin{figure}[h!]
\centering
\begin{subfigure}{0.5\textwidth}
  \centering
  \includegraphics[width=1.0\linewidth]{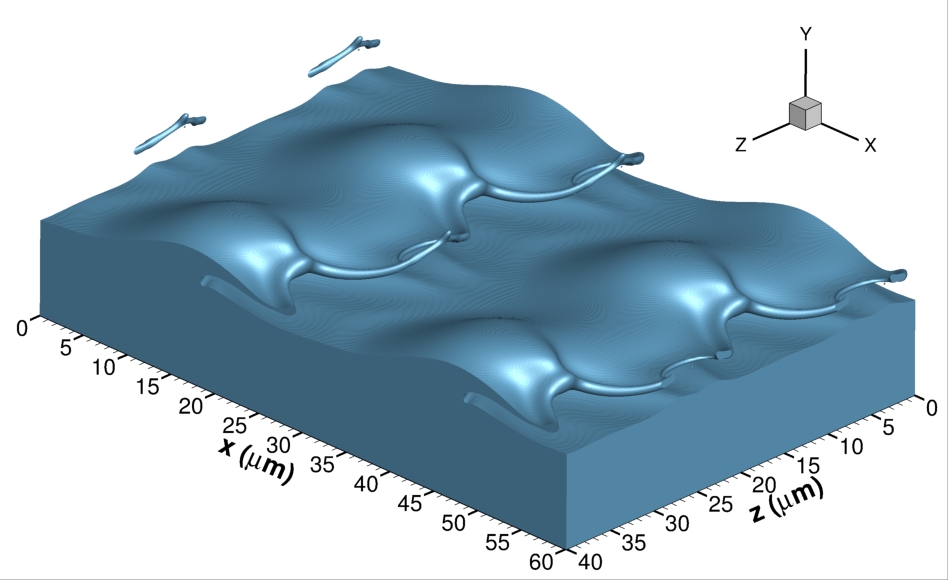}
  \caption{} 
  \label{subfig:150_30A_3mus_A}
\end{subfigure}%
\begin{subfigure}{0.5\textwidth}
  \centering
  \includegraphics[width=1.0\linewidth]{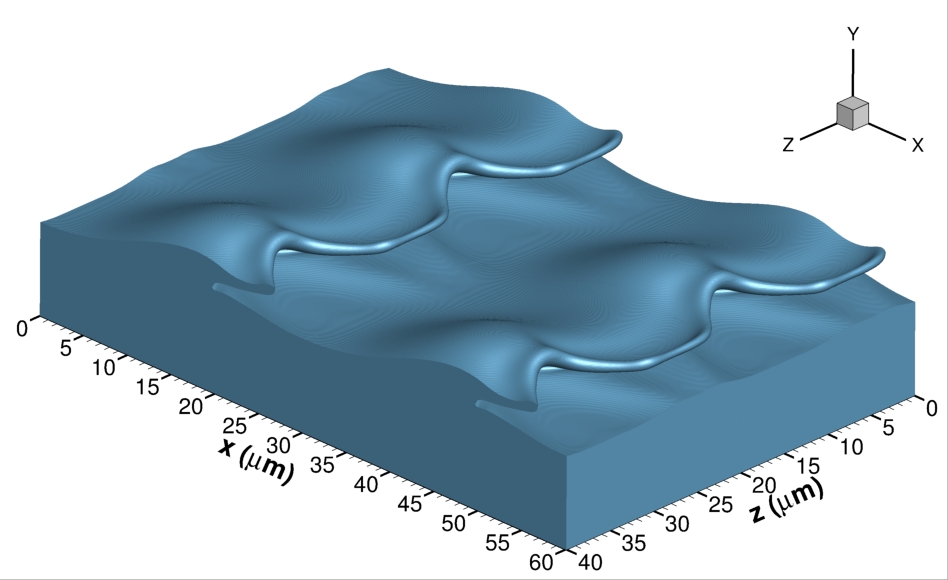}
  \caption{}
  \label{subfig:150_30A_incomp_3mus_A}
\end{subfigure}%
\caption{Lobe extension, bending and perforation at 150 bar with gas freestream velocity of \(u_G=30\) m/s (i.e., case C1) at the non-dimensional time \(t^*=t/t_c=t\frac{u_G}{H}=4.5\). The interface location is identified as the isosurface with \(C=0.5\). (a) compressible case C1; and (b) incompressible case C1i.}
\label{fig:lobe_bending_incomp}
\end{figure}

Initially, cases A2 and A2i behave very similarly because of the limited and slow mixing process. Only slight differences are observed during the lobe extension process. For example, the compressible case shows a slightly thinner lobe due to a higher interface temperature along the lobe's edge, which reduces the local surface-tension coefficient below the average value used in the incompressible simulation. On the other hand, the differences between cases C1 and C1i are striking. The lobe extension or stretching is reduced in the incompressible case as it has a much higher density and viscosity than the compressible case. Moreover, the higher surface-tension coefficient translates into a thicker lobe. As a result, there is no lobe bending nor hole formation for a prolonged time during the simulation (see Figure~\ref{fig:lobe_bending_incomp}), which is more in line with the sub-domain classification for incompressible flows identified by Zandian et al.~\cite{zandian2017planar}. \par 

\subsubsection{Lobe and crest corrugation}
\label{subsubsec:lobe_fold}

\begin{figure}[h!]
\centering
\begin{subfigure}{0.25\textwidth}
  \centering
  \includegraphics[width=1.0\linewidth]{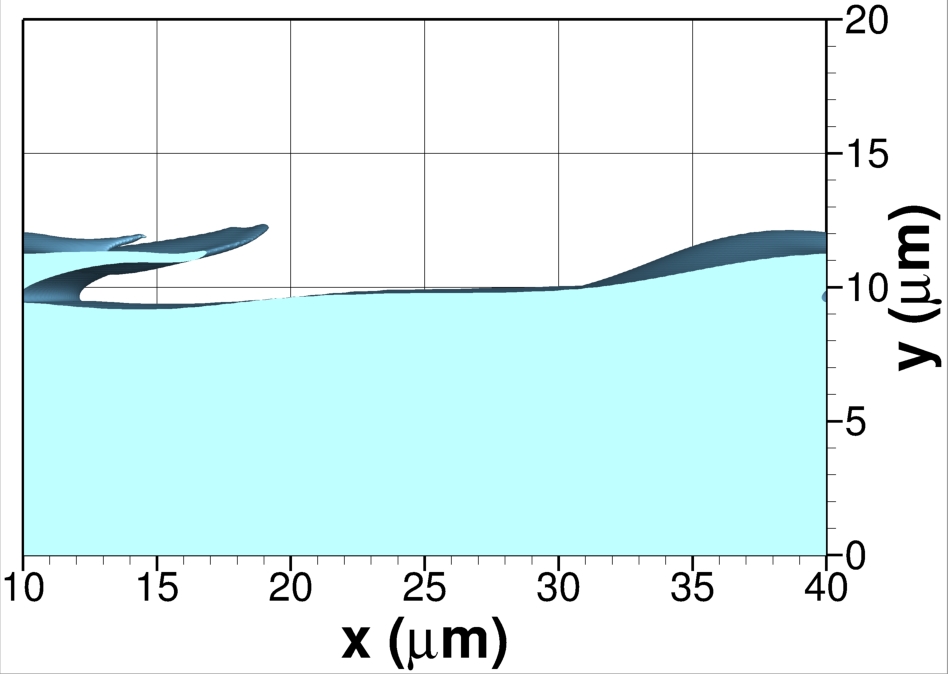}
  \label{subfig:150_50A_1p2mus_lobe_B}
\end{subfigure}%
\begin{subfigure}{0.25\textwidth}
  \centering
  \includegraphics[width=1.0\linewidth]{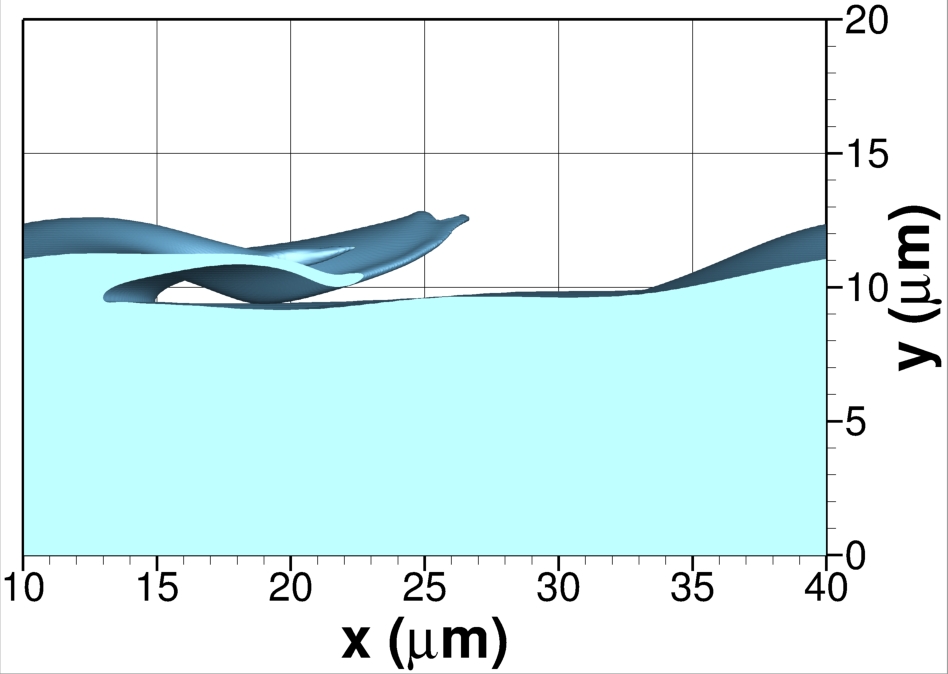}
  \label{subfig:150_50A_1p4mus_lobe_B}
\end{subfigure}%
\begin{subfigure}{0.25\textwidth}
  \centering
  \includegraphics[width=1.0\linewidth]{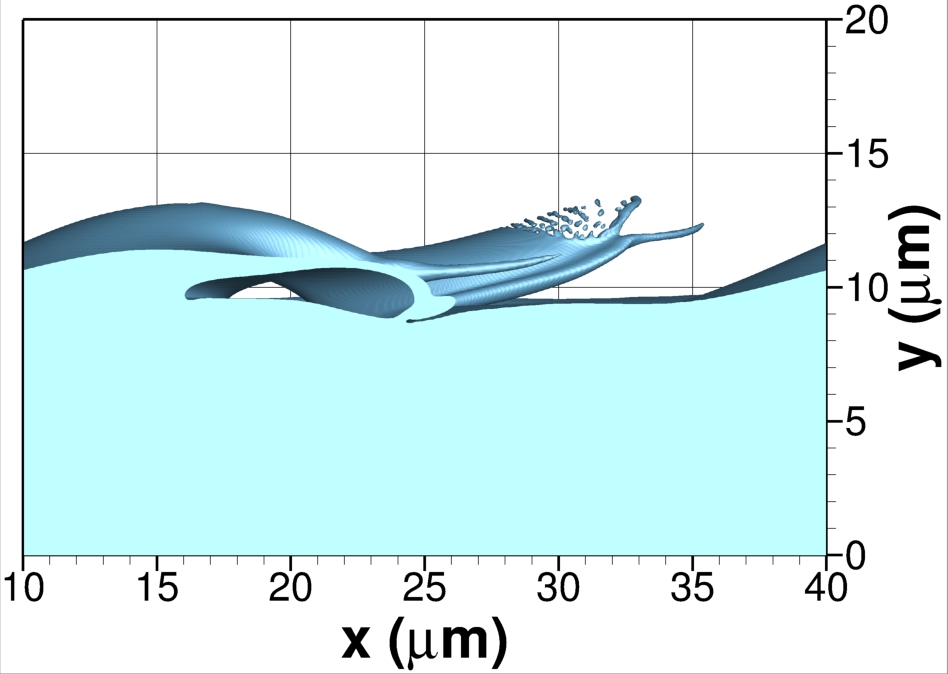}
  \label{subfig:150_50A_1p6mus_lobe_B}
\end{subfigure}%
\begin{subfigure}{0.25\textwidth}
  \centering
  \includegraphics[width=1.0\linewidth]{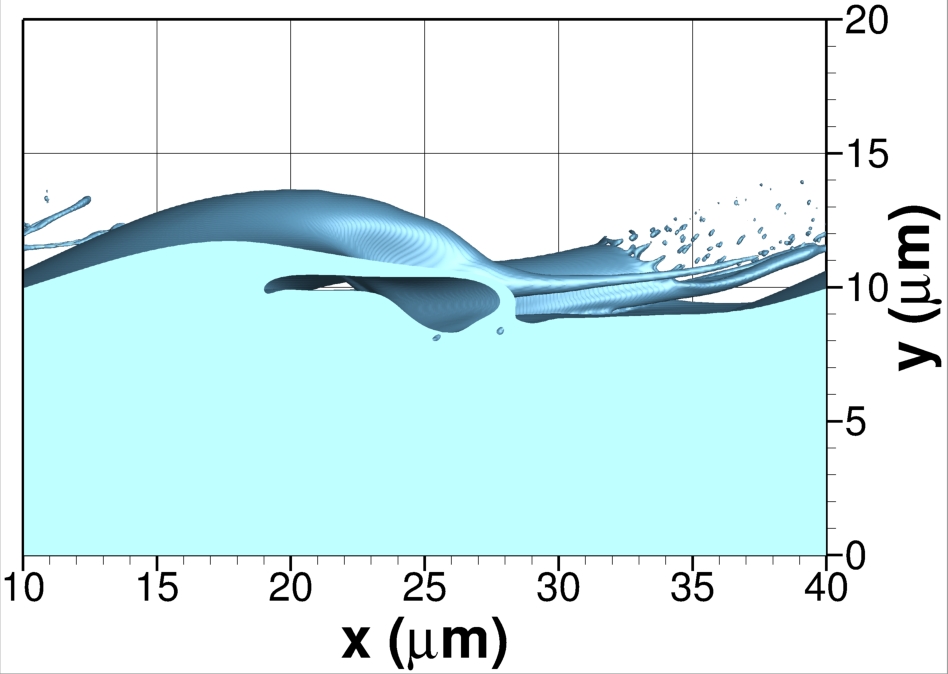}
  \label{subfig:150_50A_1p8mus_lobe_B}
\end{subfigure}%
\\[-3ex]
\begin{subfigure}{0.25\textwidth}
  \centering
  \includegraphics[width=1.0\linewidth]{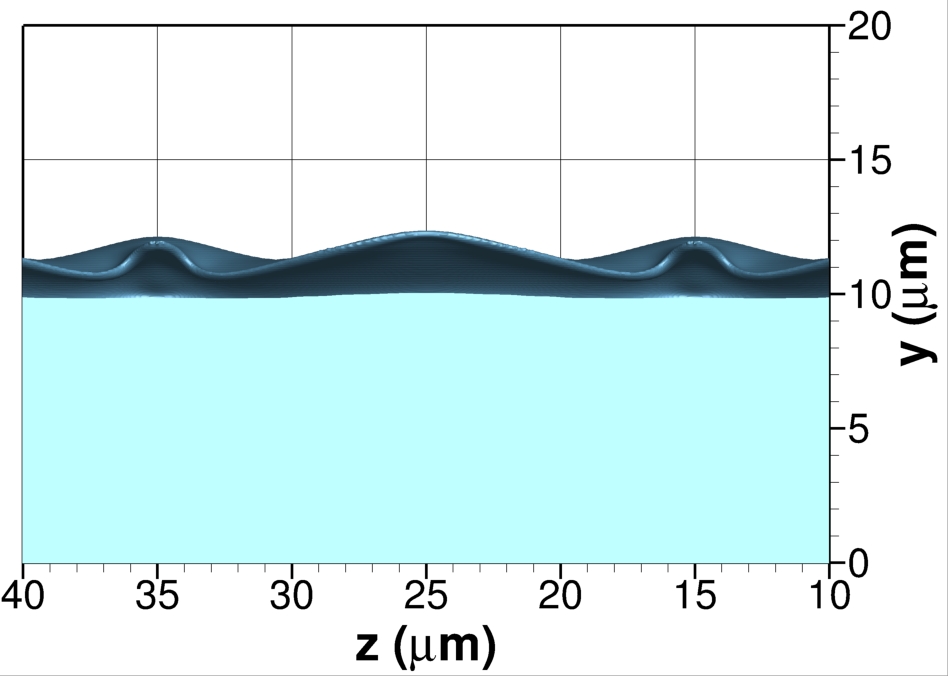}
  \label{subfig:150_50A_1p2mus_lobe_C}
\end{subfigure}%
\begin{subfigure}{0.25\textwidth}
  \centering
  \includegraphics[width=1.0\linewidth]{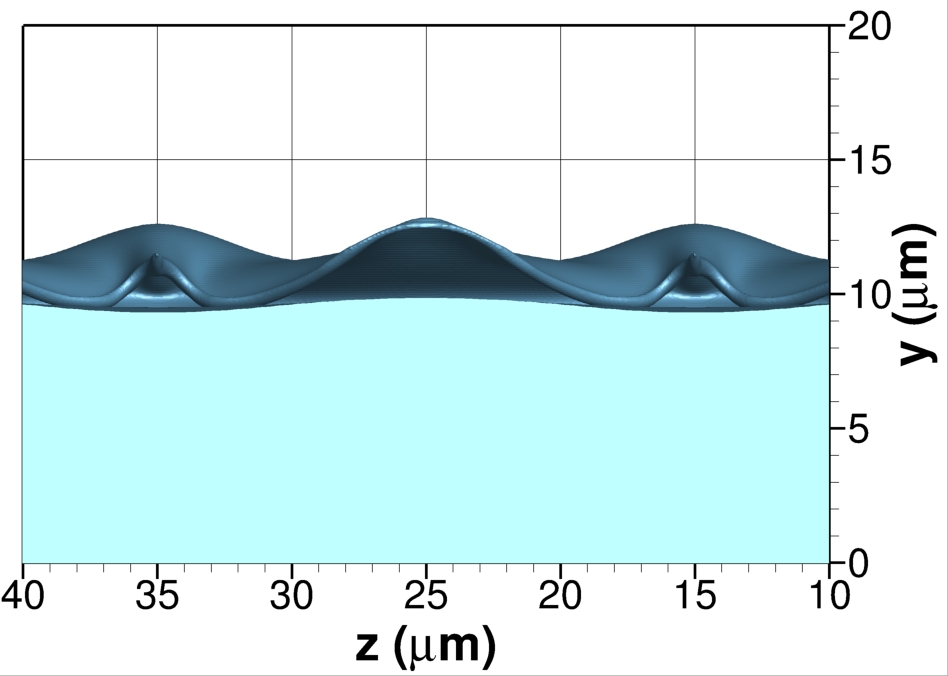}
  \label{subfig:150_50A_1p4mus_lobe_C}
\end{subfigure}%
\begin{subfigure}{0.25\textwidth}
  \centering
  \includegraphics[width=1.0\linewidth]{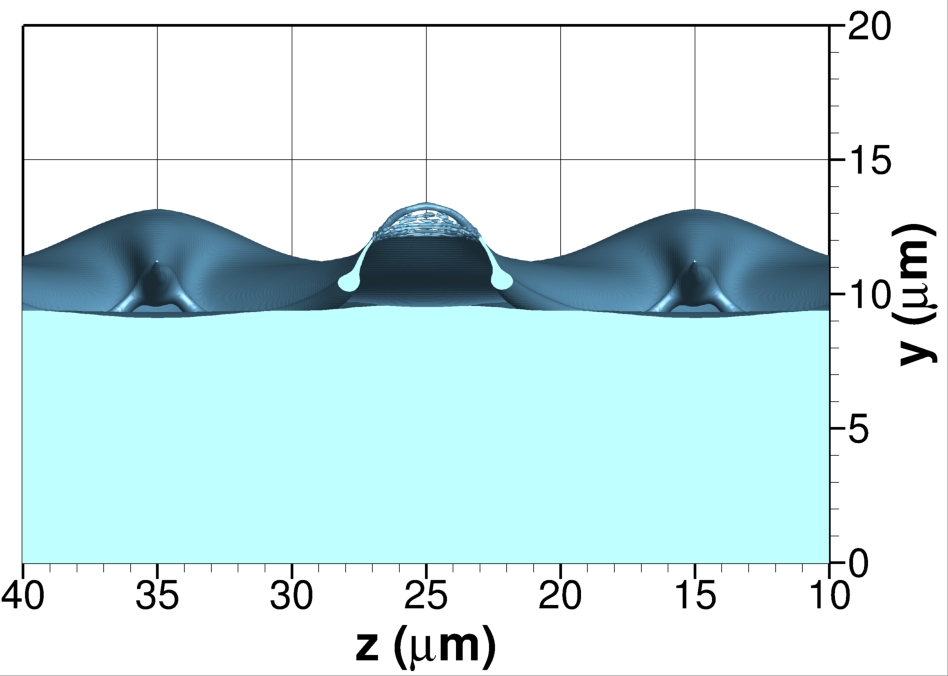}
  \label{subfig:150_50A_1p6mus_lobe_C}
\end{subfigure}%
\begin{subfigure}{0.25\textwidth}
  \centering
  \includegraphics[width=1.0\linewidth]{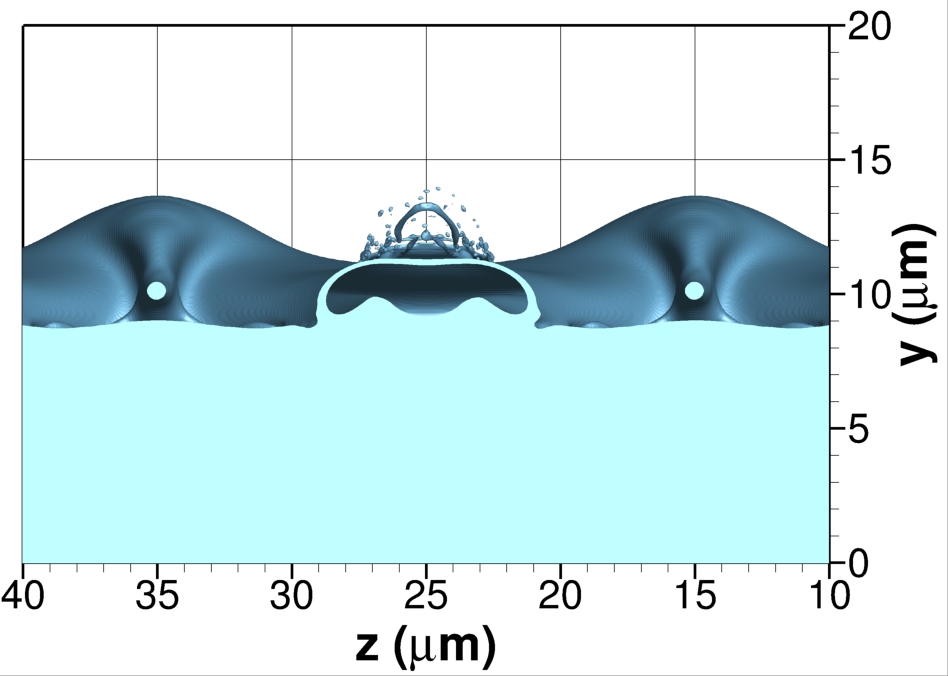}
  \label{subfig:150_50A_1p8mus_lobe_C}
\end{subfigure}%
\\[-3ex]
\begin{subfigure}{0.25\textwidth}
  \centering
  \includegraphics[width=1.0\linewidth]{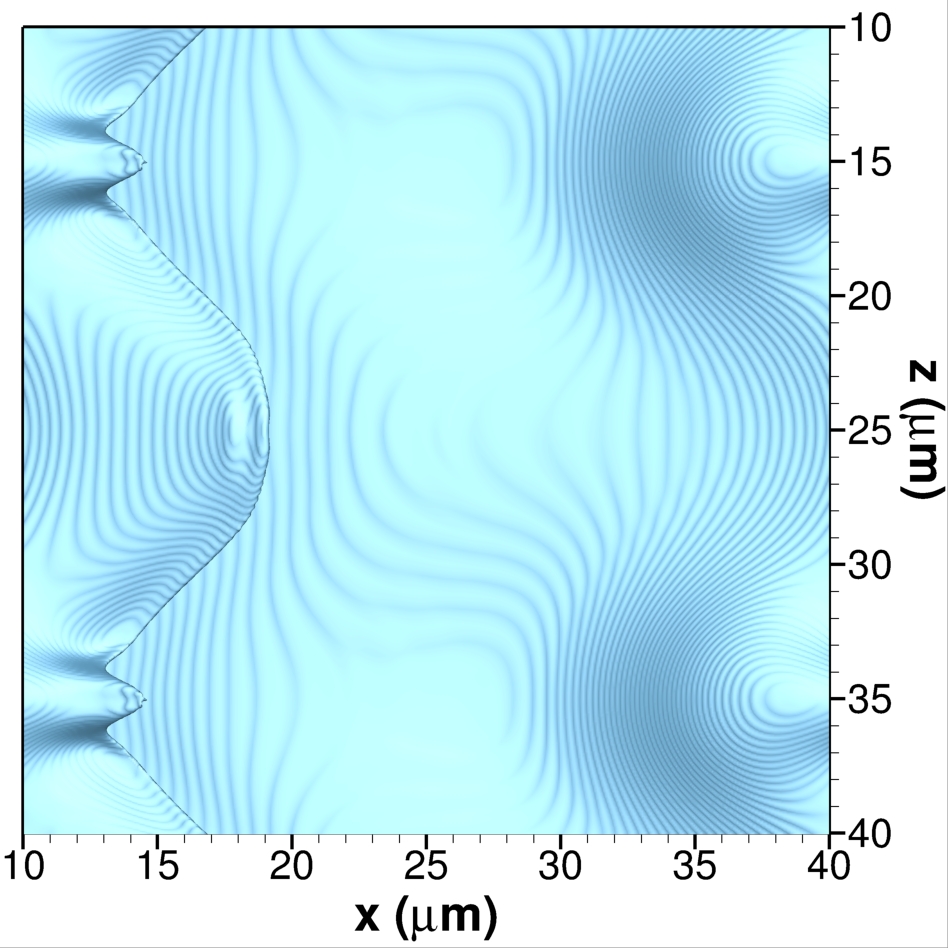}
  \caption{\(t^*=3.0\)} 
  \label{subfig:150_50A_1p2mus_lobe_D}
\end{subfigure}%
\begin{subfigure}{0.25\textwidth}
  \centering
  \includegraphics[width=1.0\linewidth]{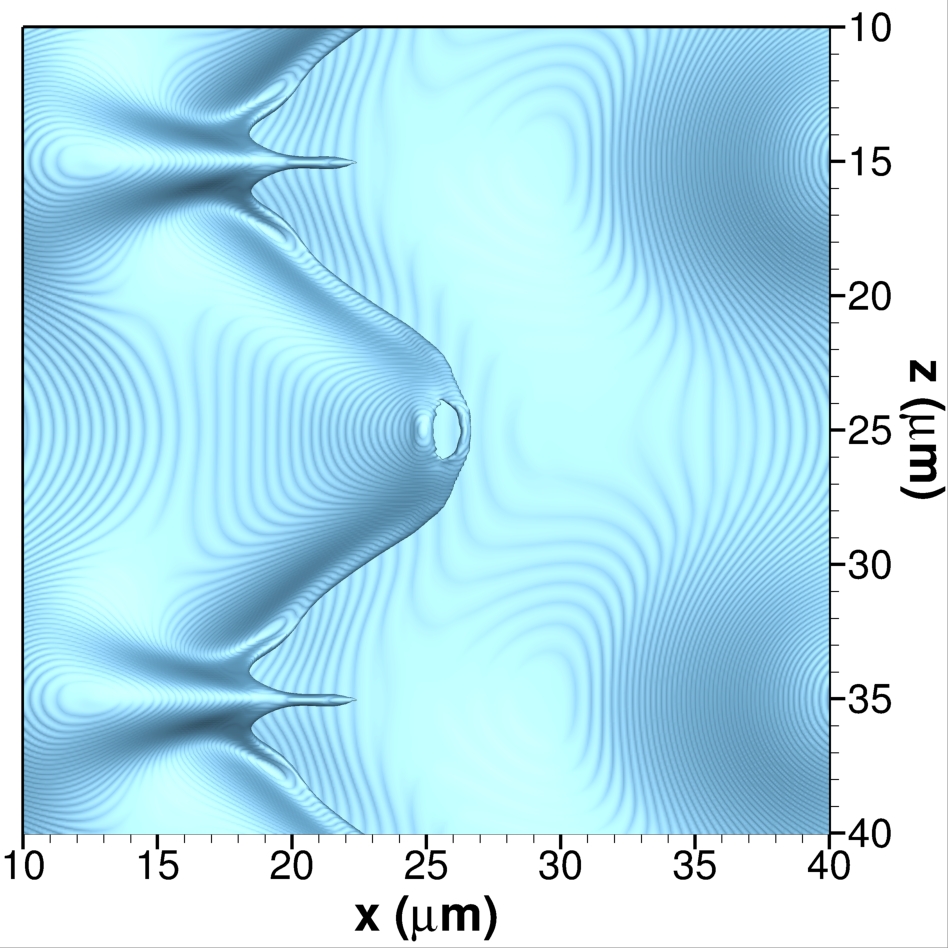}
  \caption{\(t^*=3.5\)}
  \label{subfig:150_50A_1p4mus_lobe_D}
\end{subfigure}%
\begin{subfigure}{0.25\textwidth}
  \centering
  \includegraphics[width=1.0\linewidth]{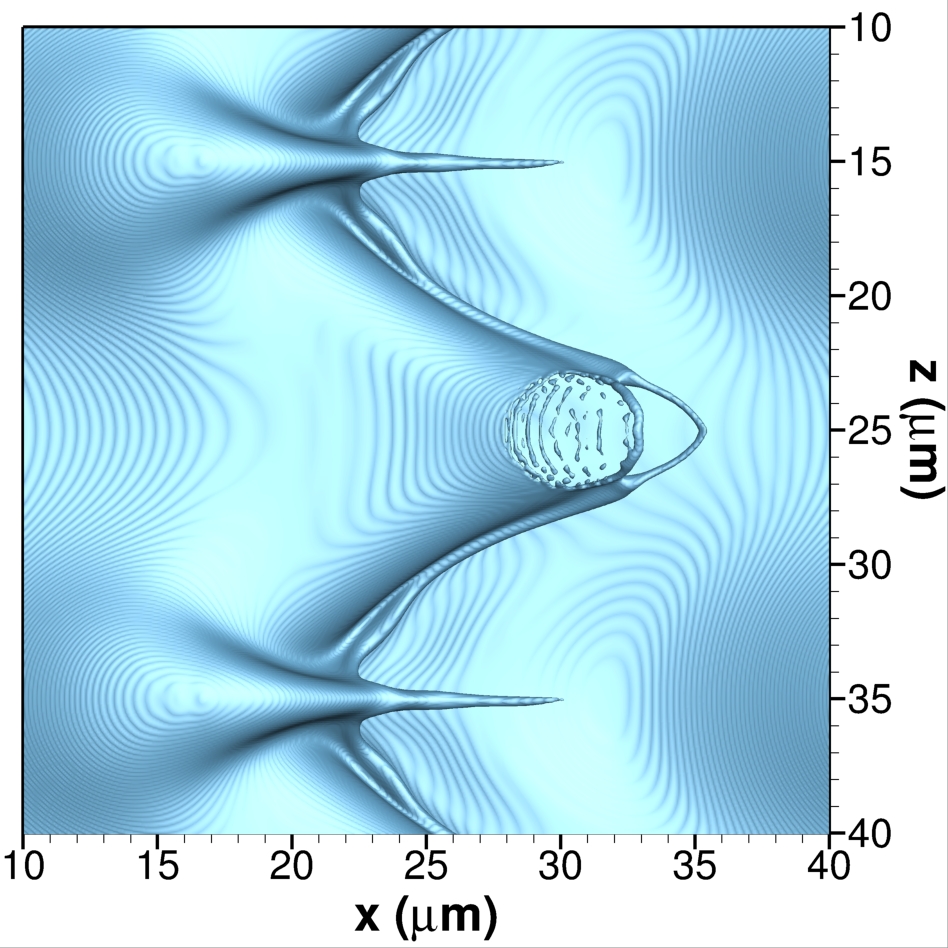}
  \caption{\(t^*=4.0\)}
  \label{subfig:150_50A_1p6mus_lobe_D}
\end{subfigure}%
\begin{subfigure}{0.25\textwidth}
  \centering
  \includegraphics[width=1.0\linewidth]{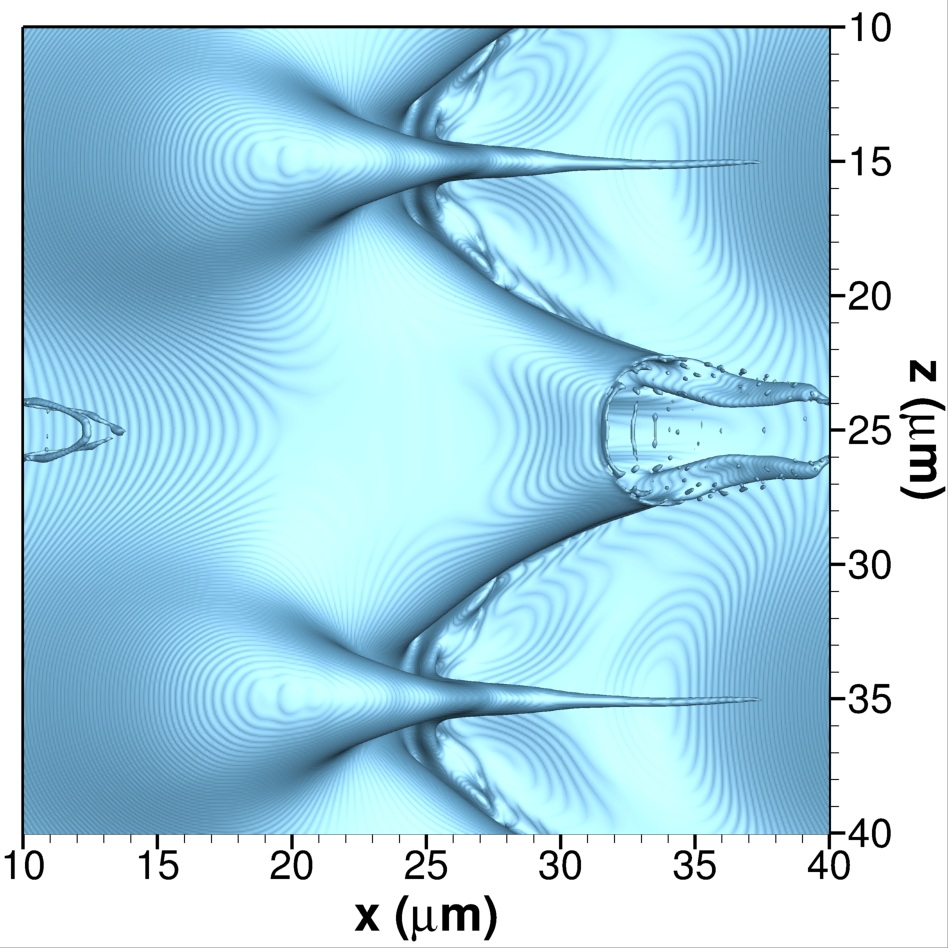}
  \caption{\(t^*=4.5\)}
  \label{subfig:150_50A_1p8mus_lobe_D}
\end{subfigure}%
\caption{Lobe and crest corrugation at 150 bar with gas freestream velocity of \(u_G=50\) m/s (i.e., case C2). The thin lobe bursts, similar to a bag breakup. The top figures show the side view from an \(xy\) plane located at \(z=40\) \(\mu\)m, the middle figures show the side view from an \(yz\) plane located at \(x=30\) \(\mu\)m and the bottom figures show the top view from an \(xz\) plane located above the liquid surface. The interface location is identified as the isosurface with \(C=0.5\). A non-dimensional time is obtained as \(t^*=t/t_c=t\frac{u_G}{H}\). (a) \(t^*=3.0\); (b) \(t^*=3.5\); (c) \(t^*=4.0\); and (d) \(t^*=4.5\).}
\label{fig:lobe_fold_time}
\end{figure}

A different lobe deformation mechanism is observed in cases B2, C2 and C3, which have a very high pressure of 100 bar and 150 bar with a reduced surface-tension coefficient compared to 50 bar. In all three cases, the gas Weber number based on freestream conditions is \(We_G>1000\). No substantial differences are observed as \(Re_L\) and \(We_G\) change across cases B2, C2 and C3. Figure~\ref{fig:lobe_fold_time} shows the evolution of the lobe for case C2 (i.e., 150 bar and \(u_G=50\) m/s). Initially, a thin lobe extends as described in Subsection~\ref{subsubsec:lobe_ext}. However, instead of the lobe's tip bending upward (i.e., rotating in the positive \(z\) direction), the lobe corrugates around the \(x\) direction engulfing the gas mixture before eventually bursting into droplets, similar to a bag breakup mechanism. After bursting, ligaments form that stretch into the oxidizer stream and the remaining of the lobe flattens while the main perturbation grows behind it. As discussed previously in Subsection~\ref{subsubsec:lobe_ext}, mesh resolution in VOF methods determines breakup events to some extent. Here, the bursting of the lobe might be delayed if better interface reconstruction methods were used, such as the two-plane reconstruction or R2P method~\cite{chiodi2018two,chiodi2020advancement,han2021liquid}. Nonetheless, the speed at which the lobe is corrugating at some location and thinning points to an eventual bursting event as observed with the current interface reconstruction methodology (i.e., PLIC). \par

The crest of the growing perturbation also corrugates, although the cause for this corrugating mechanism might be different from the one affecting the lobe. This issue will be investigated in future works where a deeper analysis will be done involving vorticity dynamics. As seen in Figure~\ref{fig:lobe_fold_time}, a nose-shaped formation appears as the perturbation crest grows (i.e., along the \(x\) direction at \(z=15\) \(\mu\)m or \(z=35\) \(\mu\)m). As the corrugation occurs, a ligament forms and is rapidly stretched into the oxidizer stream, similar to the ligaments forming after the bursting of the lobe or after the bridge breakup in the hole formation process explained in Subsection~\ref{subsubsec:lobe_ext}. The rapid stretching of ligaments at high pressures is explained in Subsection~\ref{subsubsec:lig_stretch}. \par 

Crest corrugation is not an inherent mechanism of high \(We_G\) cases only. It is also apparent in case B1 and to a lesser extent in cases A1, A2 and C1. Therefore, this behavior may be more related to the initial setup of the problem than any particular dynamical behavior. However, only the cases with a high gas Weber number show the simultaneous stretching of a ligament as the corrugation occurs. Interestingly, the corrugation of the liquid surface may induce surface instabilities in reduced surface-tension environments as explained in Subsection~\ref{subsec:instability}. Also, the burst of the lobe into droplets and the generation of thin ligaments observed in Figure~\ref{fig:lobe_fold_time} can be responsible for the generation of further surface instabilities later in time. \par

\subsubsection{Ligament stretching and shredding}
\label{subsubsec:lig_stretch}

\begin{figure}[h!]
\centering
\begin{subfigure}{0.33\textwidth}
  \centering
  \includegraphics[width=1.0\linewidth]{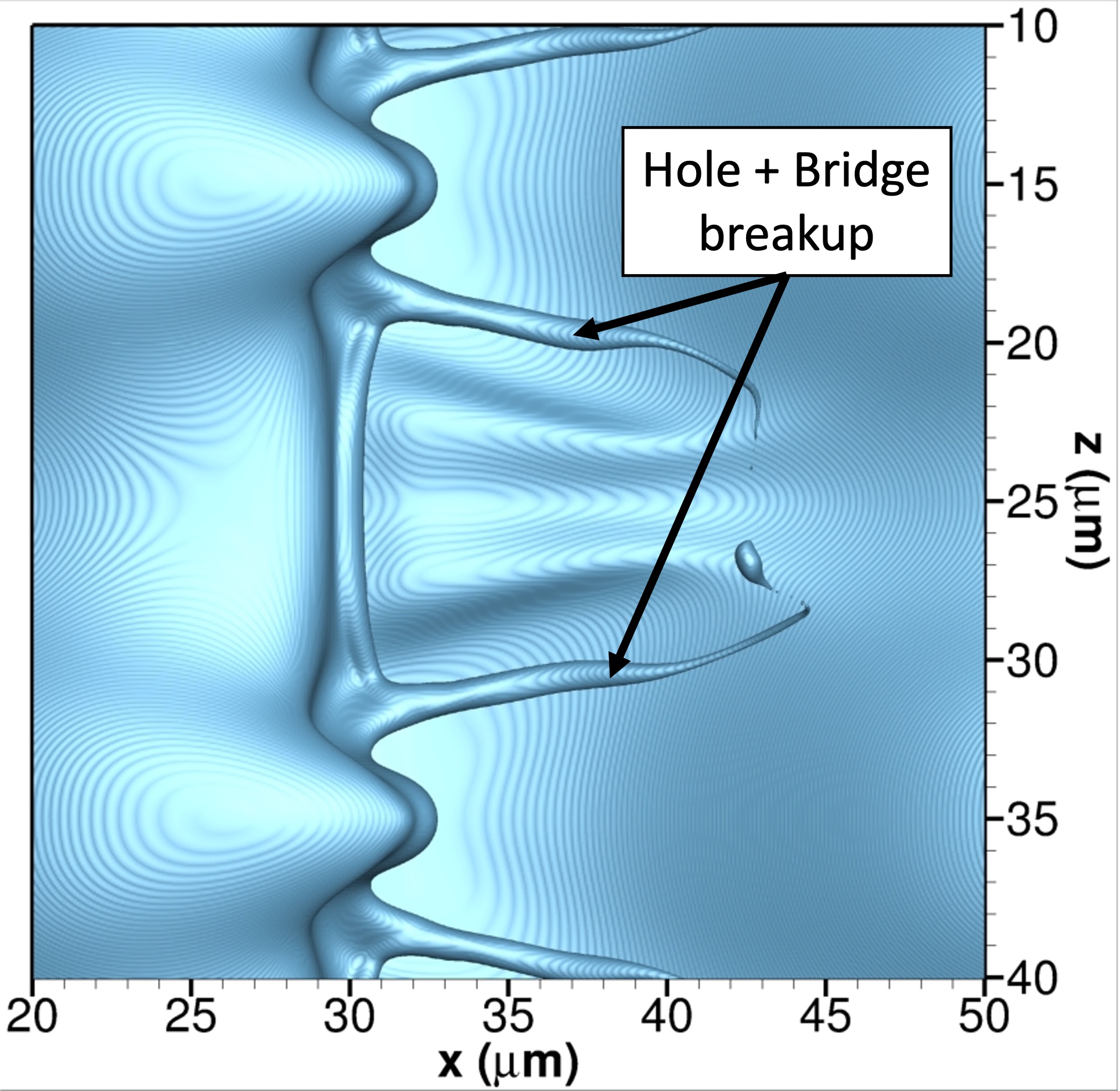}
  \caption{150 bar and \(u_G=30\) m/s (C1) at \(t^*=5.40\)} 
  \label{subfig:150_30A_3p6_ligament_stretching_2}
\end{subfigure}%
\begin{subfigure}{0.33\textwidth}
  \centering
  \includegraphics[width=1.0\linewidth]{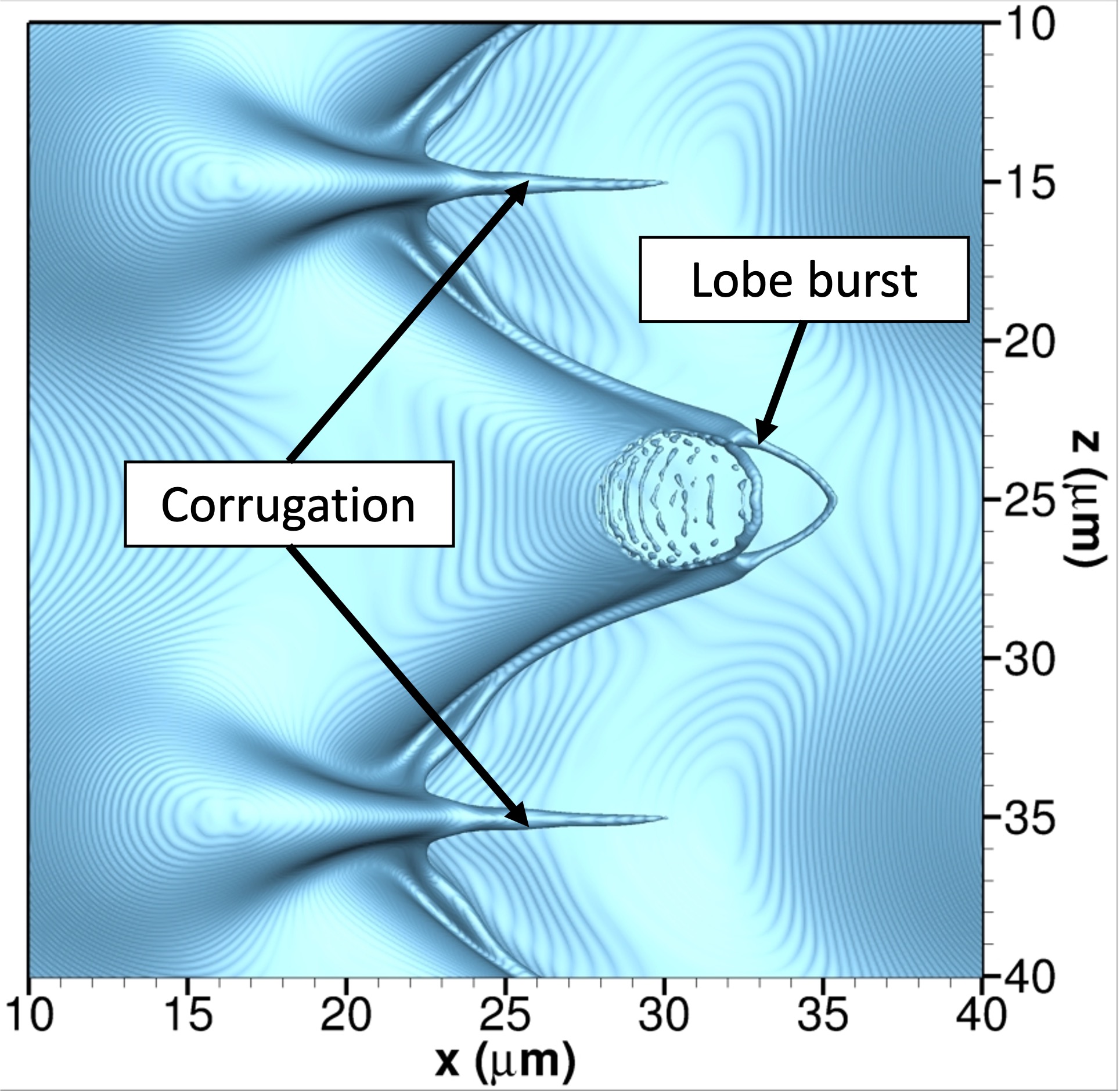}
  \caption{150 bar and \(u_G=50\) m/s (C2) at \(t^*=4.00\)}
  \label{subfig:150_50A_1p6_ligament_stretching_2}
\end{subfigure}%
\begin{subfigure}{0.33\textwidth}
  \centering
  \includegraphics[width=1.0\linewidth]{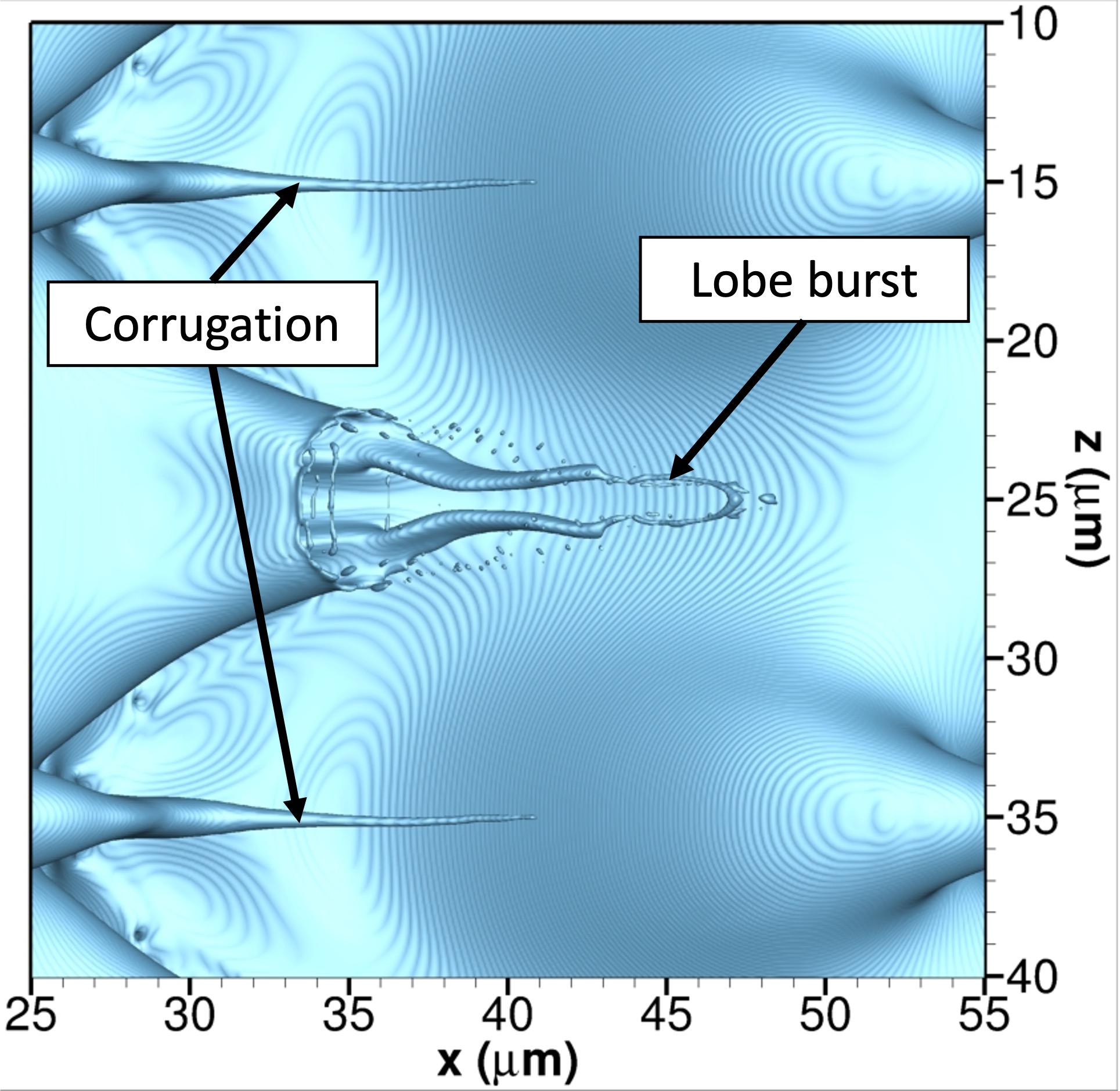}
  \caption{150 bar and \(u_G=50\) m/s (C2) at \(t^*=4.75\)}
  \label{subfig:150_50A_1p9_ligament_stretching_2}
\end{subfigure}%
\caption{Ligament formation and stretching from different precursor events: hole plus bridge breakup, crest corrugation and thin lobe burst. The top view is obtained from an \(xz\) plane located above the liquid surface. The interface location is identified as the isosurface with \(C=0.5\). A non-dimensional time is obtained as \(t^*=t/t_c=t\frac{u_G}{H}\). (a) case C1 at \(t^*=5.40\); (b) case C2 at \(t^*=4.00\); and (c) case C2 at \(t^*=4.75\).}
\label{fig:ligament_stretching}
\end{figure}

The formation of ligaments that stretch into the oxidizer stream before eventually breaking up into droplets or coalescing with the liquid is not unexpected. All analyzed cases fall inside the LoLiD and LoCLiD atomization sub-domains, as shown in Figure~\ref{fig:Weg_vs_Rel_overview}, which have this breakup mechanism as a characteristic feature. Figure~\ref{fig:ligament_stretching} summarizes different precursor events that exist before ligaments form, such as crest corrugation, hole formation with bridge breakup, and lobe burst or bag breakup. The problem configurations where these events exist have been identified in previous subsections. \par 

A higher tendency to form thin and elongated ligaments exists at supercritical pressures. As smaller liquid structures form, either from small protuberances on the liquid surface or along the edge of lobes (e.g., corrugations), they heat faster and the dissolution of oxygen increases. As previously explained, the surface-tension coefficient is reduced, the liquid density drops and the liquid viscosity presents gas-like values. These temperature and mixing effects are also observed in the ligaments forming from the hole formation on lobes or the bursting of the lobes. Similar to the lobe bending explained in Subsection~\ref{subsubsec:lobe_ext}, these liquid regions are more easily affected by the surrounding gas motion. That is, the local Reynolds number and Weber number increase. Thus, ligaments are formed and pulled away from the liquid surface if immersed in a faster-moving gas stream. This scenario is observed more frequently at 100 bar and 150 bar, especially at the higher gas freestream velocities of 50 m/s and 70 m/s. \par

Compared to the ligament behavior at low pressures, ligaments can become very thin at high pressures. Surface-tension forces become negligible and the capillary instabilities that promote neck formation and the ligament breakup into droplets are only important once ligaments reach a very small scale. Many ligaments experience a numerical breakup as they stretch and vaporize when their thickness falls below the mesh size before capillary instabilities can take place. As a result, frequent ligament shredding is observed in cases B2, C2 and C3. Thin ligaments are continuously being created, with some of them breaking up into a few droplets or just stretching and vaporizing. A detailed analysis of ligament and droplet formation under real-engine conditions is provided in Subsection~\ref{subsec:droplet}, which also discusses the influence of mesh resolution. \par 

\begin{figure}[h!]
\centering
\begin{subfigure}{0.5\textwidth}
  \centering
  \includegraphics[width=0.73\linewidth]{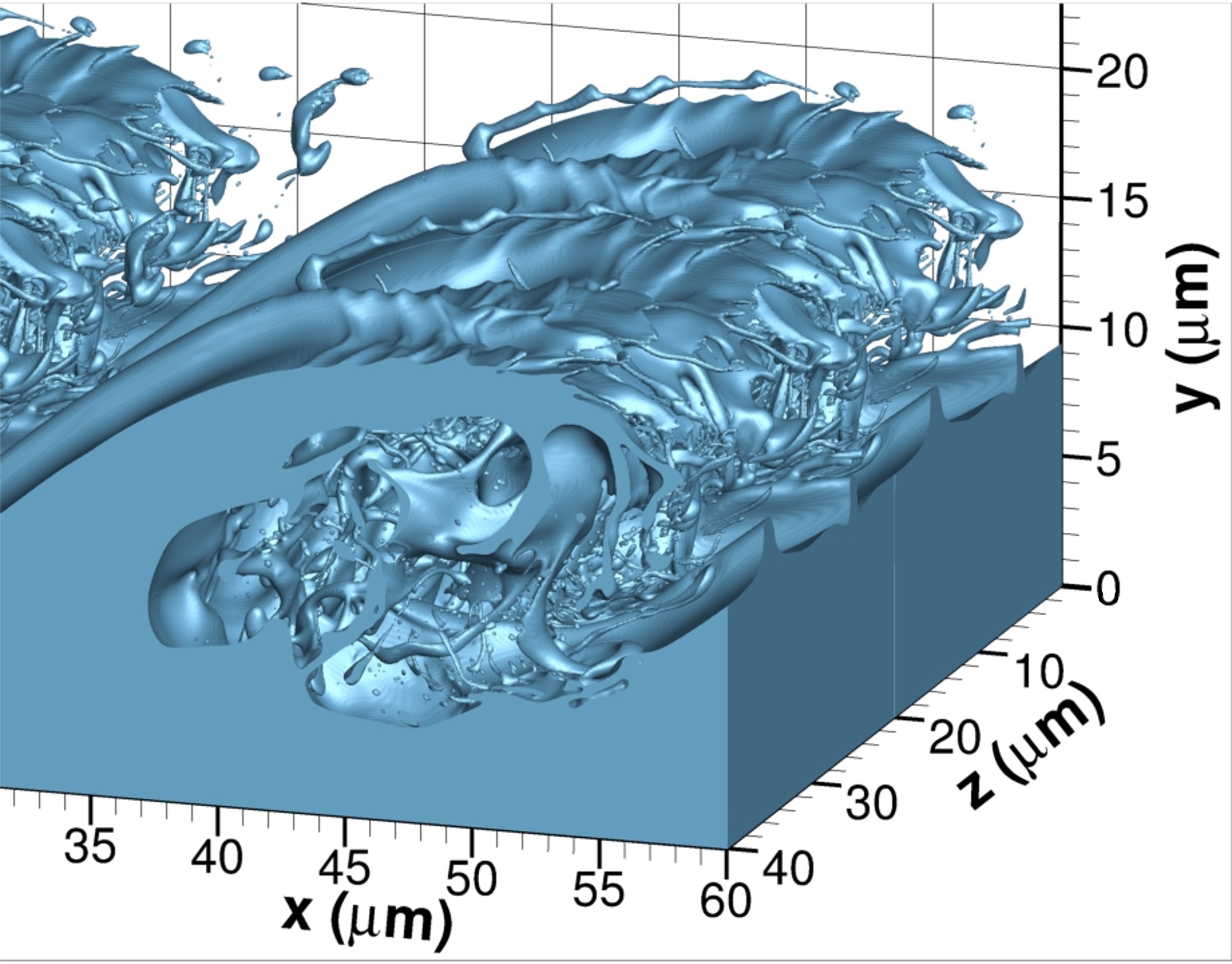}
  \caption{\(t^*=9.625\)} 
  \label{subfig:100_70A_2p75_shredding_2}
\end{subfigure}%
\begin{subfigure}{0.5\textwidth}
  \centering
  \includegraphics[width=1.0\linewidth]{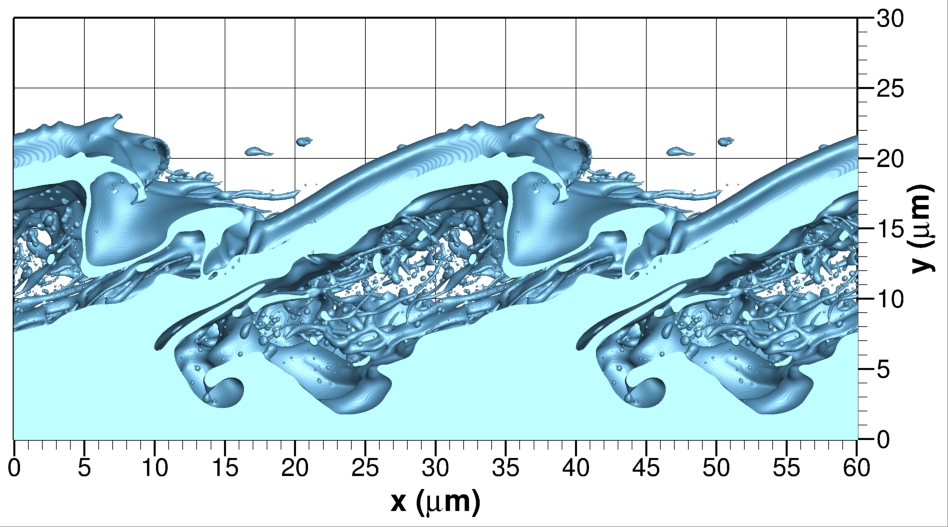}
  \caption{\(t^*=11.55\)}
  \label{subfig:100_70A_3p30mus_shredding}
\end{subfigure}%
\caption{Ligament shredding at 100 bar with gas freestream velocity of \(u_G=70\) m/s (i.e., case B2). The side view in the right figure is obtained from an \(xy\) plane located at \(z=40\) \(\mu\)m. The interface location is identified as the isosurface with \(C=0.5\). A non-dimensional time is obtained as \(t^*=t/t_c=t\frac{u_G}{H}\). (a) \(t^*=9.625\); and (b) \(t^*=11.55\).}
\label{fig:ligament_shredding}
\end{figure}

Similar findings are reported in Lagarza-Cort\'{e}s et al.~\cite{lagarza2019large}, where transcritical jet computations are presented with a diffuse-interface solver. A sharp liquid-gas interface is not identified. Nonetheless, the jet deformation is characterized by the interaction of hairpin vortices and the formation of finger-like elongated ligaments. \par

Figure~\ref{fig:ligament_shredding} shows the ligament shredding occurring in case B2 at 100 bar with a gas freestream velocity of \(u_G=70\) m/s. As observed in Figure~\ref{subfig:100_70A_2p75_shredding_2}, surface instabilities grow near the wave's crest, which promotes ligament shredding near the edge. As noted in Subsection~\ref{subsec:instability}, hotter interface regions, such as those near the wave crest, may grow surface instabilities more easily as the local surface-tension coefficient and liquid viscosity decrease. This surface behavior enhances mixing and the breakup of liquid structures and, overall, the liquid phase exhibits fluid properties more similar to the gas properties sooner. Moreover, Figure~\ref{subfig:100_70A_3p30mus_shredding} shows how there exists a tendency to trap the shredded ligaments and droplets underneath the growing perturbation. This feature can be described by the formation of liquid layers or sheets that eventually start overlapping each other. The layering mechanism is presented in detail in Subsection~\ref{subsubsec:layering}. \par

\subsubsection{Layering and liquid sheet tearing}
\label{subsubsec:layering}

The formation of liquid layers is observed at very high pressures of 100 bar and above. These layers or liquid sheets can be more or less perturbed depending on the \(We_G\) and \(Re_L\) of each particular case, but are seen as a common feature with important implications in the atomization process. \par

\begin{figure}[h!]
\centering
\begin{subfigure}{0.5\textwidth}
  \centering
  \includegraphics[width=1.0\linewidth]{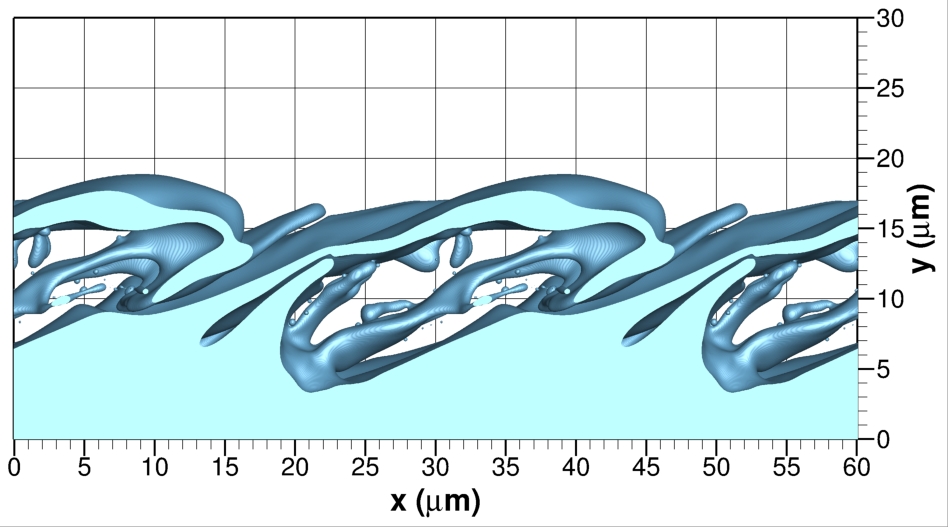}
  \caption{\(t^*=10.5\)} 
  \label{subfig:150_30A_7mus_B}
\end{subfigure}%
\begin{subfigure}{0.5\textwidth}
  \centering
  \includegraphics[width=1.0\linewidth]{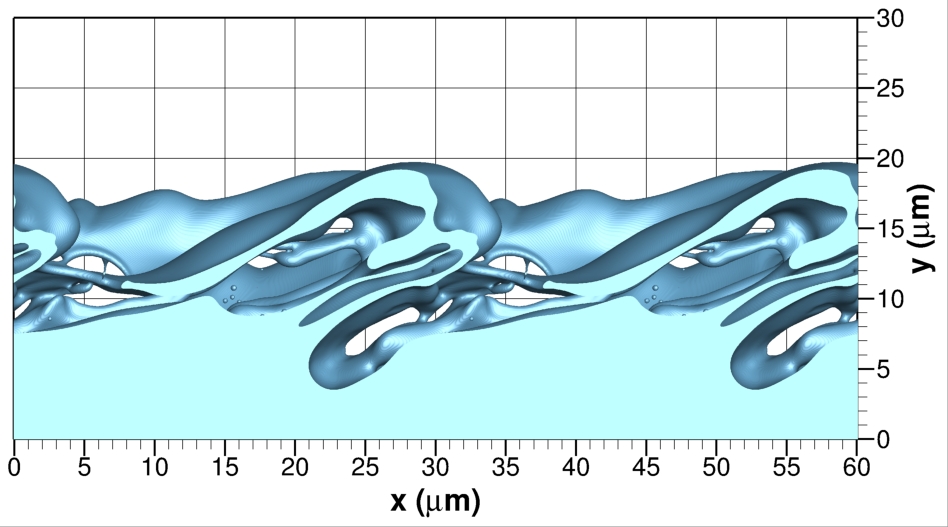}
  \caption{\(t^*=12.0\)}
  \label{subfig:150_30A_8mus_B}
\end{subfigure}%
\\
\begin{subfigure}{0.5\textwidth}
  \centering
  \includegraphics[width=1.0\linewidth]{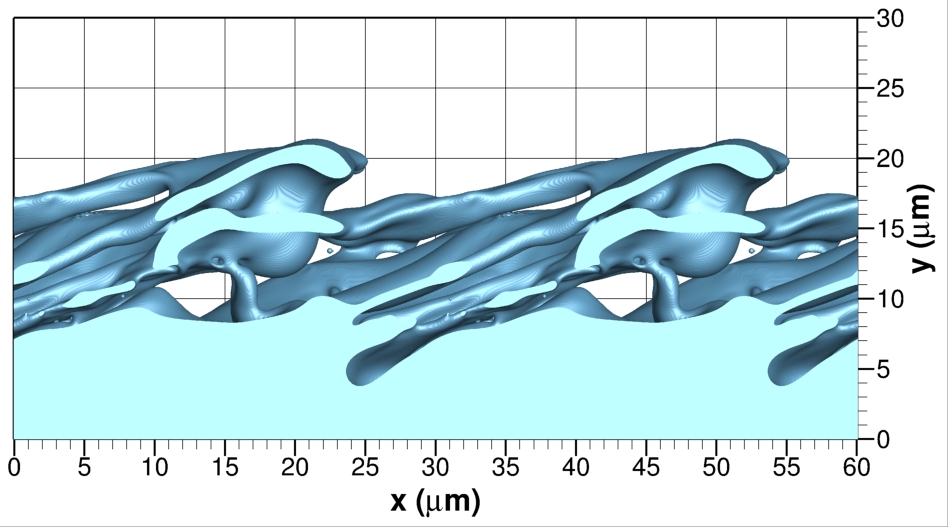}
  \caption{\(t^*=13.5\)} 
  \label{subfig:150_30A_9mus_B}
\end{subfigure}%
\begin{subfigure}{0.5\textwidth}
  \centering
  \includegraphics[width=1.0\linewidth]{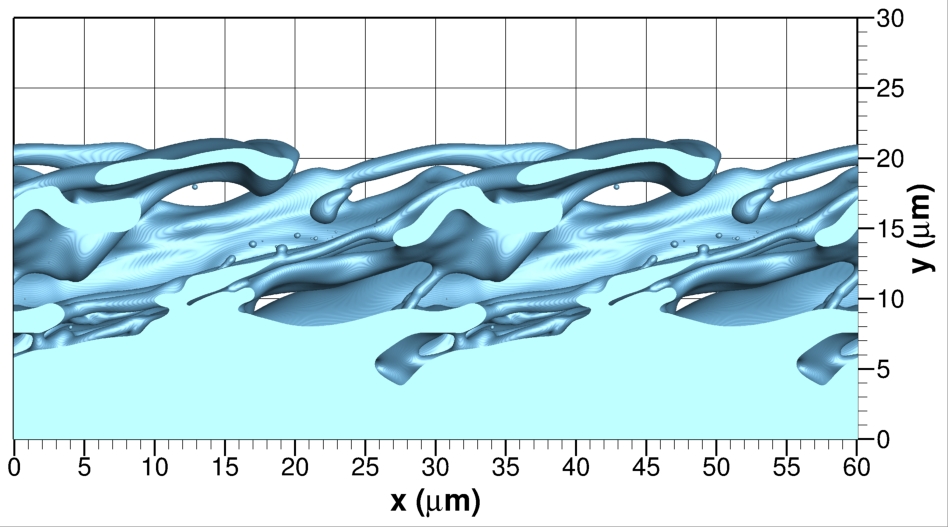}
  \caption{\(t^*=15.0\)}
  \label{subfig:150_30A_10mus_B}
\end{subfigure}%
\caption{Layering or overlap of liquid sheets at 150 bar with gas freestream velocity of \(u_G=30\) m/s (i.e., case C1). The side view is obtained from an \(xy\) plane located at \(z=40\) \(\mu\)m. The interface location is identified as the isosurface with \(C=0.5\). A non-dimensional time is obtained as \(t^*=t/t_c=t\frac{u_G}{H}\). (a) \(t^*=10.5\); (b) \(t^*=12.0\); (c) \(t^*=13.5\); and (d) \(t^*=15.0\).}
\label{fig:layering_xy}
\end{figure}

Initially, the perturbations growing in cases B1, B2, C1, C2 and C3 can easily roll over due to vortical motion. At lower pressures (i.e., 50 bar), wave crest rolling is a less common feature under the conditions analyzed in this work. After some time, the roller vortex weakens and, as seen in Figure~\ref{fig:layering_xy}, liquid sheets develop and readily stretch along the streamwise direction. Here, the results from case C1 are presented as they have a negligible amount of shredding and droplet formation, allowing for a clear observation of the layer formation. \par 

Liquid sheets quickly form and move past previously formed liquid sheets as they reach faster-moving gas streams. For instance, case C1 shows four layers overlapping each other at the end of the computation. During this process, liquid sheets are compressed and the transverse development of the two-phase mixture is limited. That is, layering can limit gas entrainment into the liquid jet, which helps push liquid structures away from the liquid core. \par

We name this deformation mechanism ``layering", although previous works showing similar mixing patterns refer to it as ``folding". Buch and Dahm~\cite{buch1996experimental,buch1998experimental} analyze the mixing structures of conserved scalars in turbulent flows and show a mixing process whereby structures stretch, fold and form layers in a cake-like pattern. \par

\begin{figure}[h!]
\centering
\includegraphics[width=1.0\linewidth]{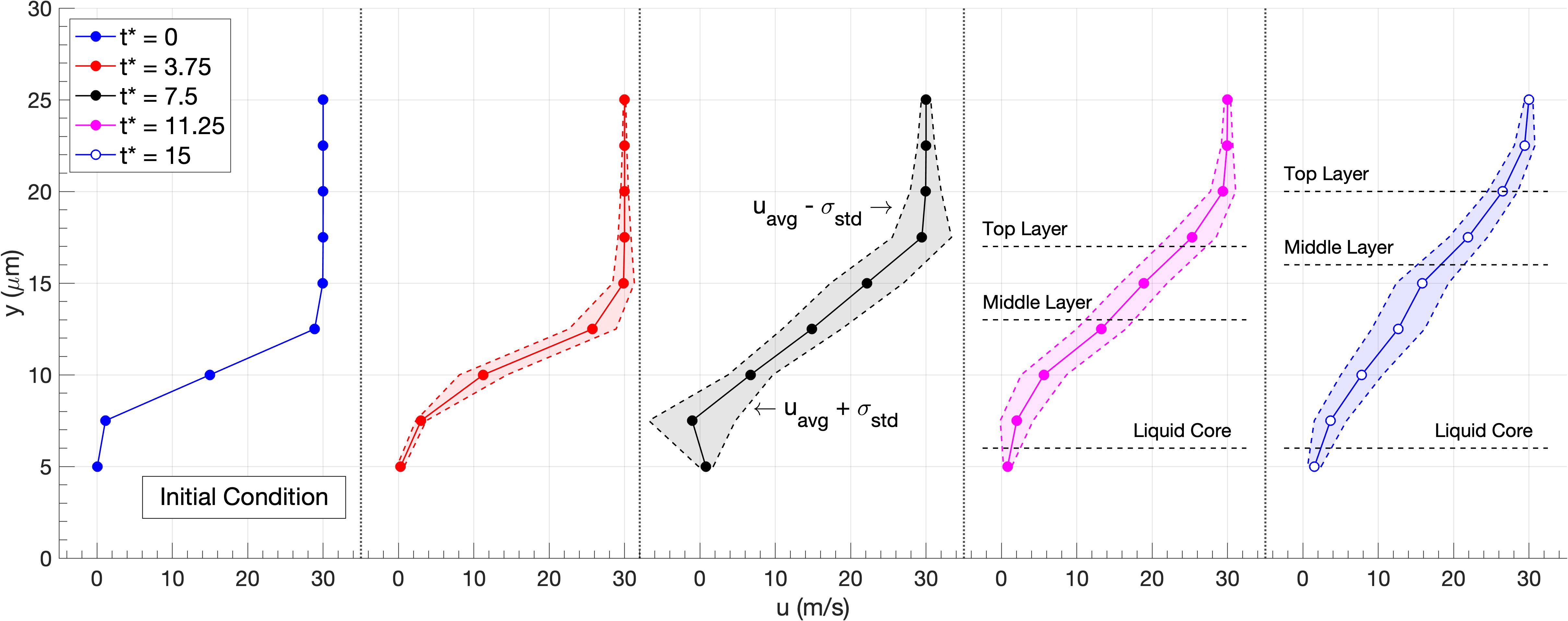}
\caption{Analysis of the streamwise velocity component at 150 bar with gas freestream velocity of \(u_G=30\) m/s (i.e., case C1). Various times are analyzed and the average streamwise velocity component, \(u_\text{avg}\), as well as the dispersion around one standard deviation, \(\sigma_\text{std}\), is shown. Once the layering deformation mechanism becomes dominant (i.e., \(t^*>10\)), different locations within the layers are presented.}
\label{fig:layering_velocity}
\end{figure}

To quantify the rate of layer formation in case C1, the streamwise velocity component, \(u\), is analyzed on various \(xz\) planes at different transverse locations. Figure~\ref{fig:layering_velocity} presents the average streamwise velocity component, \(u_\text{avg}\), at the non-dimensional times of \(t^*=0\), \(t^*=3.75\), \(t^*=7.5\), \(t^*=11.25\) and \(t^*=15\) where data have been analyzed for each \(xz\) plane located at \(y=5\) \(\mu\)m, \(y=7.5\) \(\mu\)m, \(y=10\) \(\mu\)m, \(y=12.5\) \(\mu\)m, \(y=15\) \(\mu\)m, \(y=17.5\) \(\mu\)m, \(y=20\) \(\mu\)m, \(y=22.5\) \(\mu\)m and \(y=25\) \(\mu\)m. Moreover, the standard deviation, \(\sigma_\text{std}\), of \(u\) is used to plot the dispersion of the streamwise velocity component on each \(xz\) plane one standard deviation around the mean value. Note the initial condition is an exact two-dimensional distribution, which is represented with a smooth distribution in Figure~\ref{subfig:initial_vel}. \par

The thickness of the momentum mixing layer grows over time as the sharp initial velocity profile diffuses. The first half of the computation for case C1 is dominated by the growth of the surface perturbation and strong vortical motion. For example, gas entrainment under the rolling wave is strong at \(t^*=7.5\), where the streamwise component of the velocity field becomes negative just above the liquid surface between 5 \(\mu\)m and 10 \(\mu\)m. Apart from the skewed velocity distribution toward negative streamwise values, the more significant standard deviation at \(t^*=7.5\) also suggests a more chaotic velocity field. Once the liquid-layer formation becomes the primary deformation mechanism, the streamwise velocity profile is distributed more evenly and smoothly. \par 

Figure~\ref{fig:layering_velocity} tracks the main liquid layer observed in Figure~\ref{fig:layering_xy} as it stretches and deforms. A reference value for the location of the layer's top is provided, as well as a midpoint location for the same layer and the surface of the liquid jet core. The transverse development of the layering mechanism follows the growth of the momentum mixing layer thickness. Streamwise acceleration is observed, mainly affecting the layering process within the two-phase mixture. The top region of the layers approximately moves at a constant velocity of around 25 m/s, while the middle layer that has been tracked shows an acceleration from around 14 m/s to 18 m/s. In contrast, the surface of the main liquid core is displaced at a velocity between 2 m/s and 3 m/s, very slowly compared to the upper regions of the two-phase mixture. This behavior suggests that longer times may be required to atomize the liquid jet as its core is little perturbed during the analyzed time frame. Once layering formation is dominant, the upper layers may overlap the jet core approximately every 1.4 \(\mu\)s. \par 

\begin{figure}[h!]
\centering
\begin{subfigure}{0.5\textwidth}
  \centering
  \includegraphics[width=1.0\linewidth]{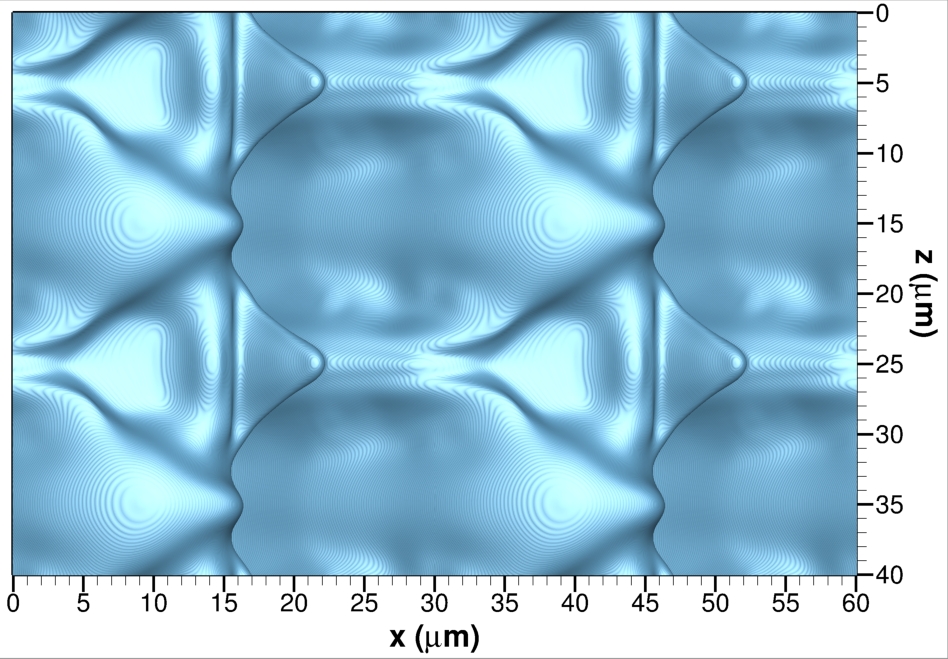}
  \caption{\(t^*=10.5\)} 
  \label{subfig:150_30A_7mus_D}
\end{subfigure}%
\begin{subfigure}{0.5\textwidth}
  \centering
  \includegraphics[width=1.0\linewidth]{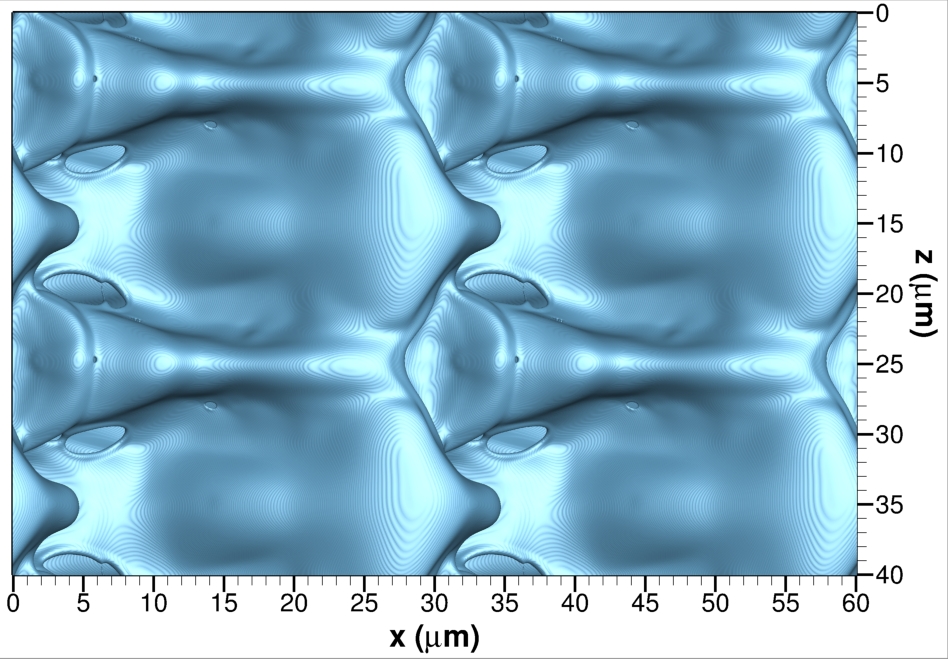}
  \caption{\(t^*=12.0\)}
  \label{subfig:150_30A_8mus_D}
\end{subfigure}%
\\
\begin{subfigure}{0.5\textwidth}
  \centering
  \includegraphics[width=1.0\linewidth]{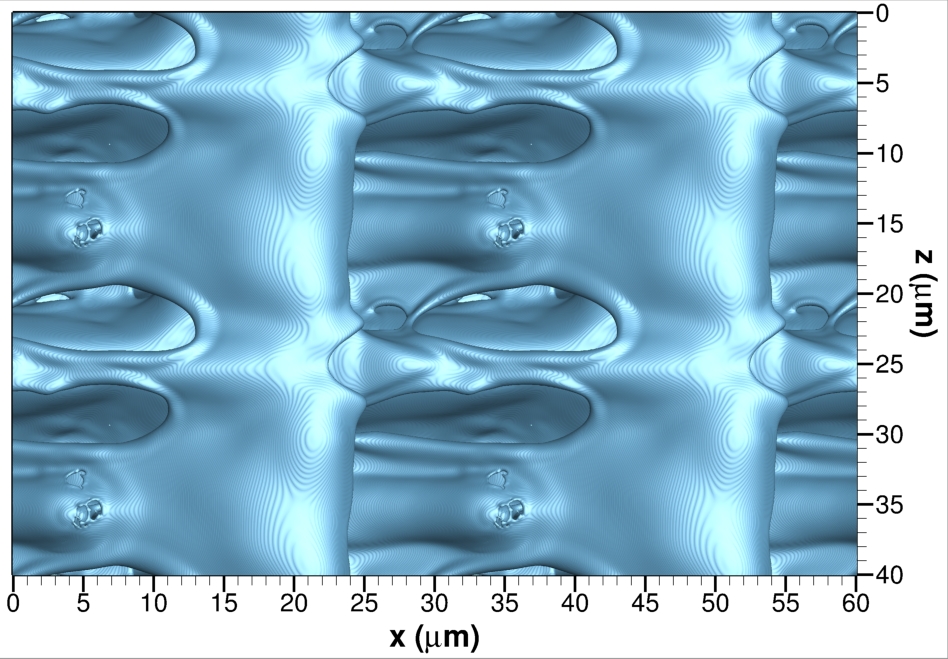}
  \caption{\(t^*=13.5\)} 
  \label{subfig:150_30A_9mus_D}
\end{subfigure}%
\begin{subfigure}{0.5\textwidth}
  \centering
  \includegraphics[width=1.0\linewidth]{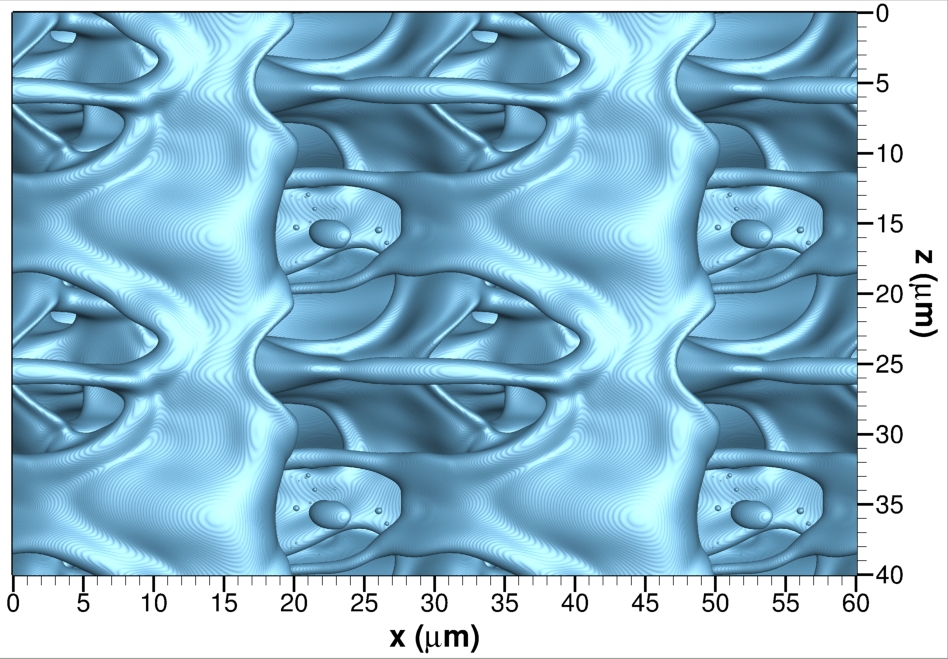}
  \caption{\(t^*=15.0\)}
  \label{subfig:150_30A_10mus_D}
\end{subfigure}%
\caption{Layering or overlap of liquid sheets at 150 bar with gas freestream velocity of \(u_G=30\) m/s (i.e., case C1). The top view is obtained from an \(xz\) plane located above the liquid surface. The interface location is identified as the isosurface with \(C=0.5\). A non-dimensional time is obtained as \(t^*=t/t_c=t\frac{u_G}{H}\). (a) \(t^*=10.5\); (b) \(t^*=12.0\); (c) \(t^*=13.5\); and (d) \(t^*=15.0\).}
\label{fig:layering_xz}
\end{figure}

As the liquid sheets stretch and become very thin, frequent formation of holes is observed (see Figure~\ref{fig:layering_xz}). In other words, continuous stretching and thinning of the liquid layers cause the sheet to tear under perforation events caused by the gas phase or other liquid structures. As previously discussed, mesh resolution also influences the exact time of the breakup event in this case. Nevertheless, it becomes important to understand why liquid sheets are formed so easily, how they are able to stretch and form various liquid layers at such high pressures and what impact they have on the atomization of the liquid jet. Localized sheet tearing can also be observed at lower pressures (e.g., cases A1 and A2 at 50 bar), but without layering being a main deformation feature. \par 

Layering is a direct result of the reduced surface-tension force coupled with the relatively small thickness of the initial shear layer between both fluids. In traditional subcritical atomization, surface tension becomes relevant at bigger length scales. Therefore, liquid sheets can rarely be stretched so easily by the oxidizer stream before surface tension and capillary instabilities deform and break up the liquid. High-pressure mixing influences the layering process, but does not determine whether layering occurs or not. For instance, the incompressible case C1i also shows a clear formation of overlapping liquid sheets. \par 

\begin{figure}[h!]
\centering
\begin{subfigure}{0.5\textwidth}
  \centering
  \includegraphics[width=1.0\linewidth]{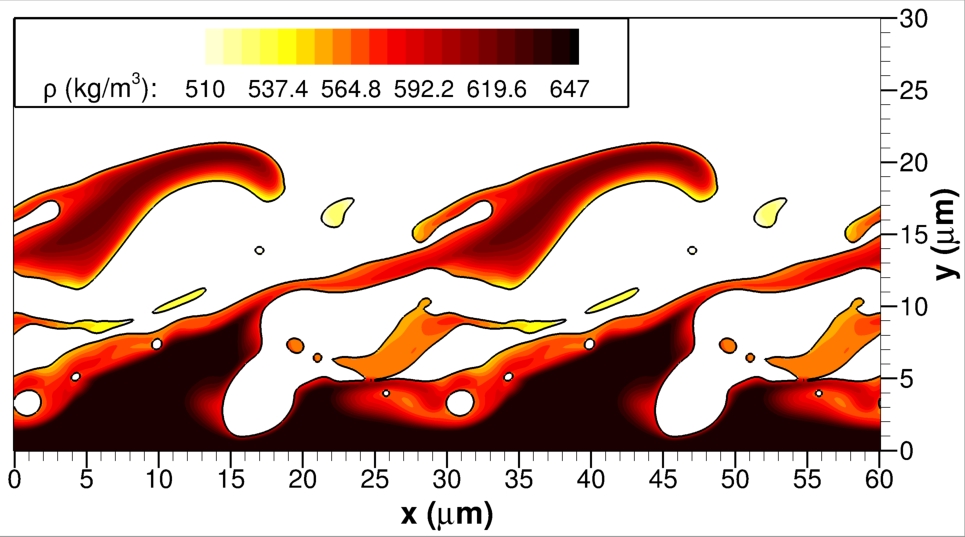}
  \caption{} 
  \label{subfig:150_30A_10mus_den_slice_z15}
\end{subfigure}%
\begin{subfigure}{0.5\textwidth}
  \centering
  \includegraphics[width=1.0\linewidth]{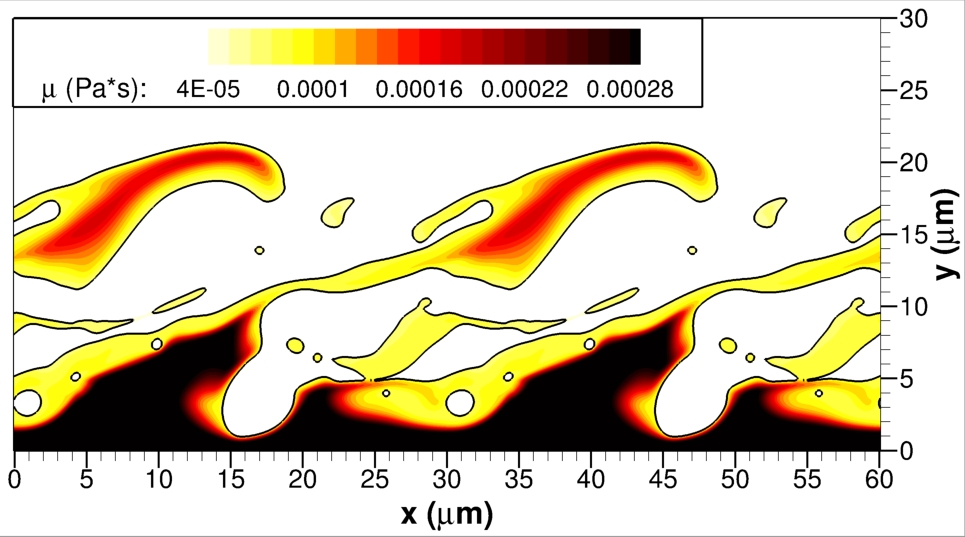}
  \caption{}
  \label{subfig:150_30A_10mus_vis_slice_z15}
\end{subfigure}%
\caption{Layering or overlap of liquid sheets at 150 bar with gas freestream velocity of \(u_G=30\) m/s (i.e., case C1) at the non-dimensional time \(t^*=t/t_c=t\frac{u_G}{H}=15\). A slice through the three-dimensional domain is obtained from an \(xy\) plane located at \(z=15\) \(\mu\)m. The interface location is identified as the isosurface with \(C=0.5\). (a) liquid density; and (b) viscosity of the two-phase mixture.}
\label{fig:layering_den_vis}
\end{figure}

Similar to the lobe extension process described in Subsection~\ref{subsubsec:lobe_ext}, mixing in the liquid phase allows for the development of thinner liquid structures as they are more easily stretched due to the reduced density and viscosity. This situation favors the formation of holes on the liquid sheets (or sheet tearing) in the compressible case compared to an incompressible fluid behavior (i.e., case C1 vs. case C1i). Figure~\ref{fig:layering_den_vis} shows the liquid density and the viscosity of both liquid and gas phases in a slice of the three-dimensional domain identified as the \(xy\) plane with \(z=15\) \(\mu\)m for case C1 at 150 bar. The non-dimensional time is \(t^*=15\) (i.e., it is a slice of Figures~\ref{subfig:150_30A_10mus_B} and~\ref{subfig:150_30A_10mus_D}). The internal structure of the liquid phase shows that pure liquid-decane properties only exist near the jet center. The liquid sheets and layers that have formed present a lower liquid density and a sharp reduction in liquid viscosity due to the dissolution of oxygen and heating. \par 

As explained in Subsection~\ref{subsec:mixing_class} (see Figure~\ref{fig:avg_den_vis}), the fluid properties of both phases change considerably over time as mixing occurs. Initially, only localized regions in the liquid phase close to the liquid-gas interface present higher concentrations of oxygen and higher temperatures and are easily affected by the motion of the dense gas. Examples of this behavior are the lobe deformation patterns described in Subsections~\ref{subsubsec:lobe_ext} and~\ref{subsubsec:lobe_fold} or the ligament stretching and shredding discussed in Subsection~\ref{subsubsec:lig_stretch}. All these features are local phenomena caused by the high-pressure environment, while the bulk of the liquid phase still presents the fluid properties of the injected fuel. After some time, mixing regions in the liquid have grown enough in size so that lower densities and lower viscosities are more representative of the outer regions of the liquid phase. This transition point where liquid sheets start to form enhances mixing even more as the surface area grows and the liquid sheet thickness reduces due to stretching. Therefore, further formation of liquid sheets and layering occurs. For configurations with higher velocities, local ligament and droplets formation are observed alongside the formation and stretching of liquid layers. \par 

\subsection{Surface instability triggers}
\label{subsec:instability}

This subsection discusses the various events that can trigger the growth of surface instabilities at high pressures. As a reminder to the reader, Subsection~\ref{subsec:description} presents the importance of analyzing the influence of spurious currents generated by the numerical approach. The sharp-interface approach, as well as the pressure solver based on the FFT method, may cause the growth of numerical instabilities near the interface if spurious currents are not bounded by the mesh resolution and the problem configuration. The examples presented in this subsection have a physical explanation behind them and, to the authors' knowledge, the influence of the numerical method is well contained. Spurious currents become detrimental, especially at low density ratios \(\rho_G/\rho_L\). Nevertheless, no sign of numerical instability is observed at 50 bar where the lowest density ratio exists. Testing shows that grid resolution has a limited impact on the development of the instabilities presented in this subsection, and the mesh size provides at least a resolution of about 20 to 30 cells per wavelength of the observed growing small perturbations. \par 

\begin{figure}[h!]
\centering
\begin{subfigure}{0.33\textwidth}
  \centering
  \includegraphics[width=1.0\linewidth]{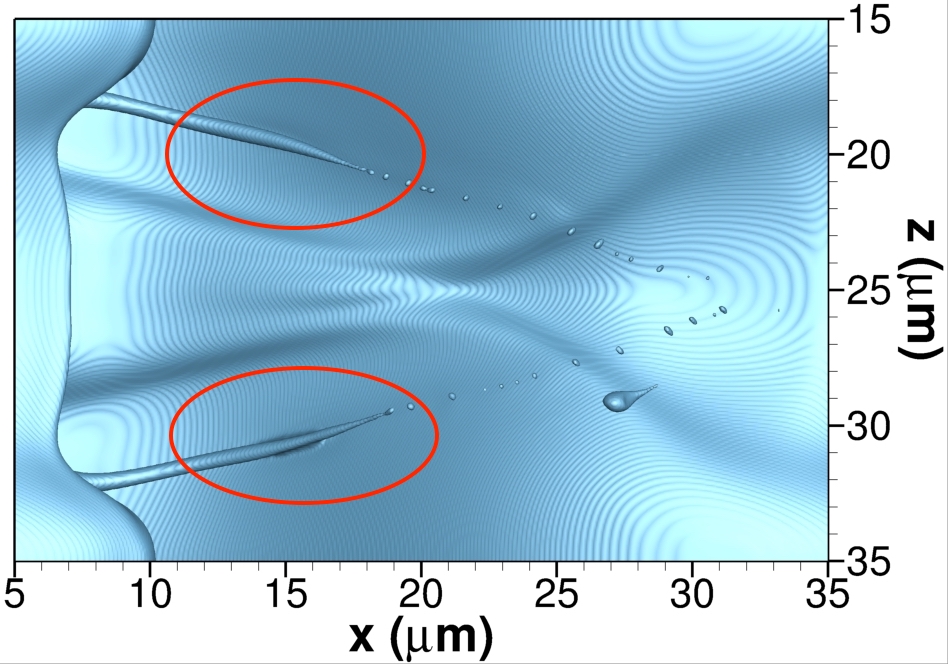}
  \caption{\(t^*=6.3\)} 
  \label{subfig:150_30A_4p2mus_impact_D}
\end{subfigure}%
\begin{subfigure}{0.33\textwidth}
  \centering
  \includegraphics[width=1.0\linewidth]{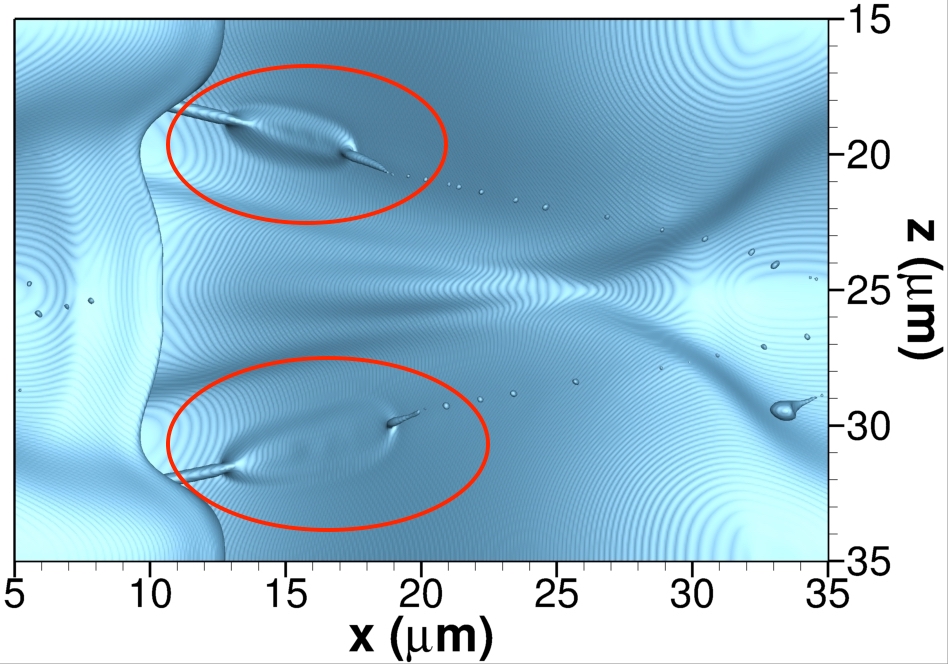}
  \caption{\(t^*=6.6\)}
  \label{subfig:150_30A_4p4mus_impact_D}
\end{subfigure}%
\begin{subfigure}{0.33\textwidth}
  \centering
  \includegraphics[width=1.0\linewidth]{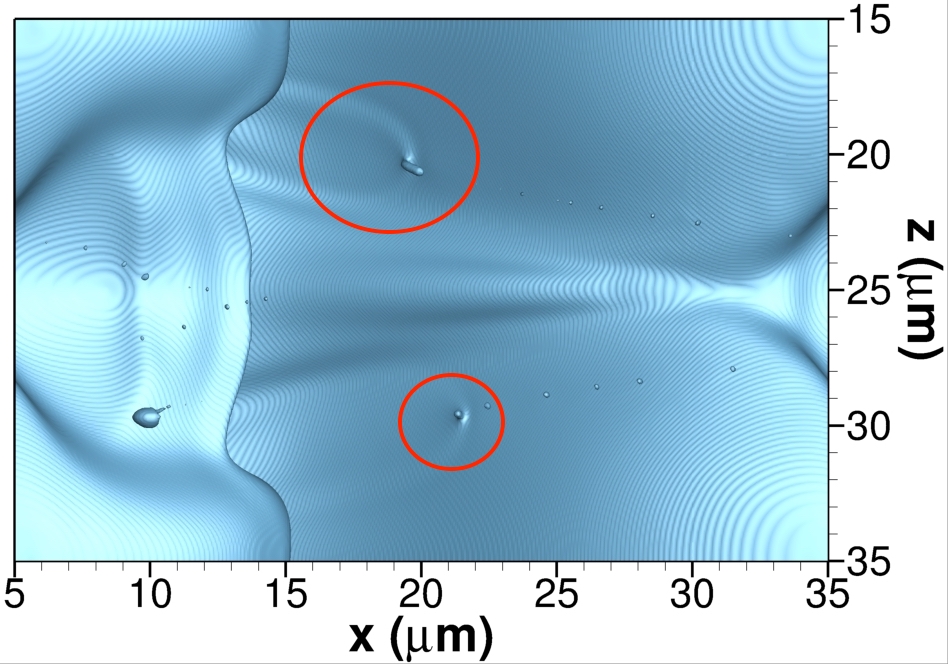}
  \caption{\(t^*=6.9\)}
  \label{subfig:150_30A_4p6mus_impact_D}
\end{subfigure}%
\\
\begin{subfigure}{0.33\textwidth}
  \centering
  \includegraphics[width=1.0\linewidth]{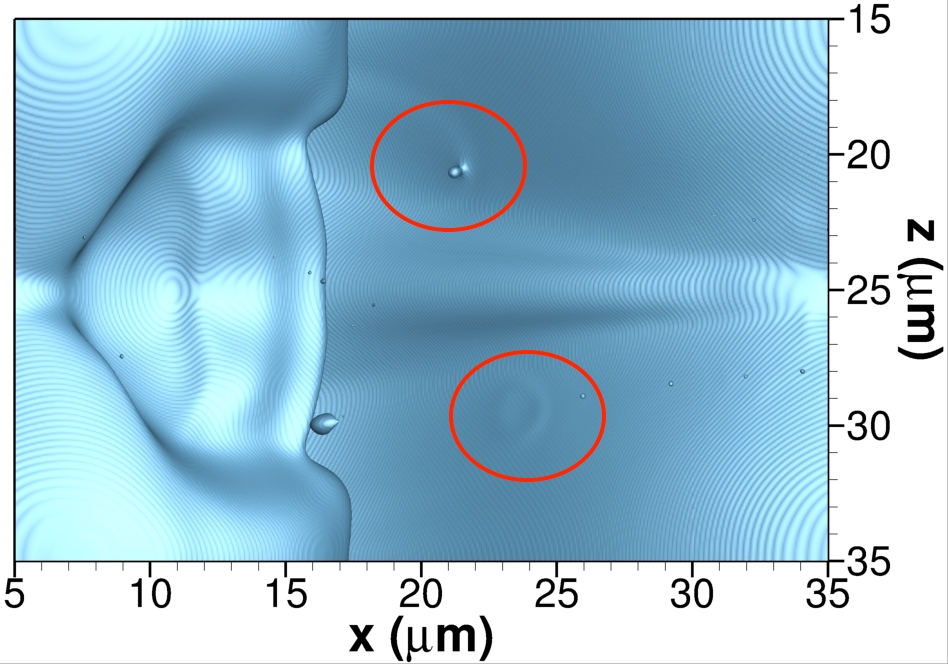}
  \caption{\(t^*=7.2\)} 
  \label{subfig:150_30A_4p8mus_impact_D}
\end{subfigure}%
\begin{subfigure}{0.33\textwidth}
  \centering
  \includegraphics[width=1.0\linewidth]{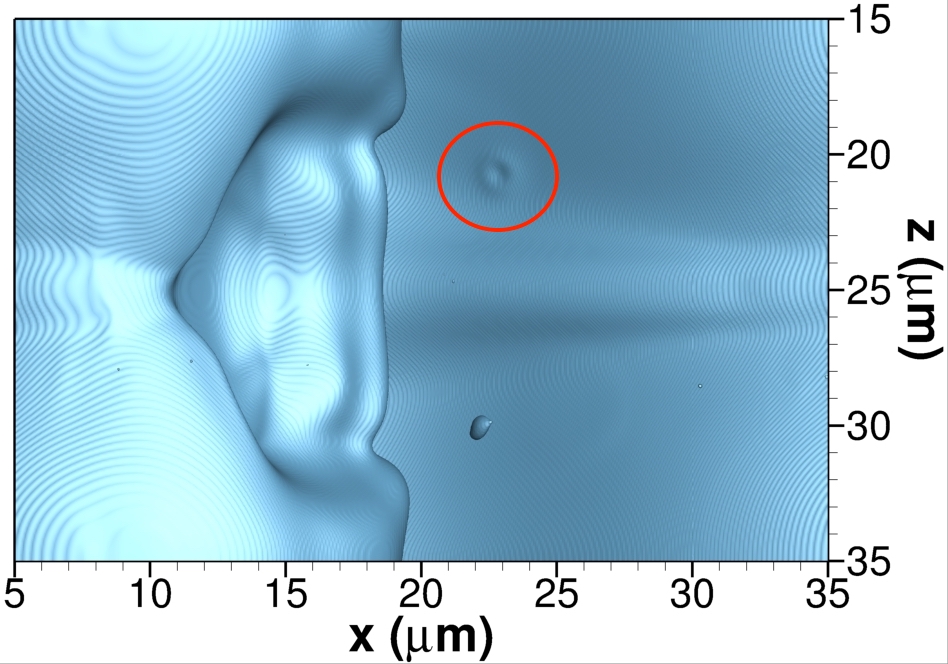}
  \caption{\(t^*=7.5\)}
  \label{subfig:150_30A_5mus_impact_D}
\end{subfigure}%
\begin{subfigure}{0.33\textwidth}
  \centering
  \includegraphics[width=1.0\linewidth]{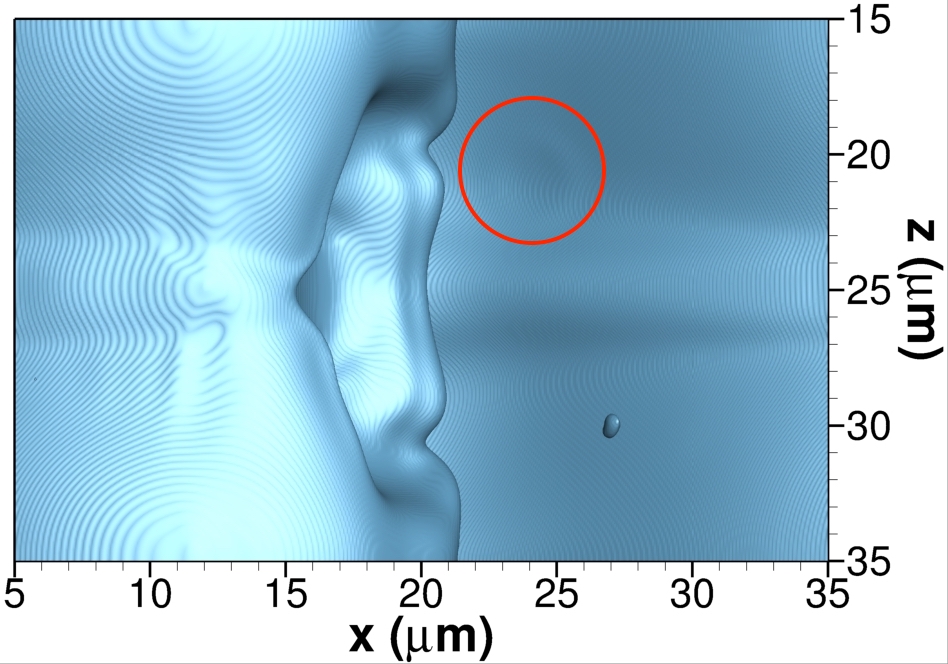}
  \caption{\(t^*=7.8\)}
  \label{subfig:150_30A_5p2mus_impact_D}
\end{subfigure}%
\caption{Stable impact or coalescence events at 150 bar with gas freestream velocity of \(u_G=30\) m/s (i.e., case C1). Impact regions are circled in red. The top view is obtained from an \(xz\) plane located above the liquid surface. The interface location is identified as the isosurface with \(C=0.5\). A non-dimensional time is obtained as \(t^*=t/t_c=t\frac{u_G}{H}\). (a) \(t^*=6.3\); (b) \(t^*=6.6\); (c) \(t^*=6.9\); (d) \(t^*=7.2\); (e) \(t^*=7.5\); and (f) \(t^*=7.8\).}
\label{fig:impact_stable}
\end{figure}

The reduced surface-tension coefficient in transcritical two-phase flows facilitates the growth of surface instabilities from perturbations caused by impact events (i.e., droplet collision and coalescence of liquid structures) or flow disturbances near the liquid-gas interface. These examples are not unique to high-pressure environments, but appear more frequently than in subcritical conditions. Stable events are also observed. For example, Figure~\ref{fig:impact_stable} shows the coalescence of ligaments and droplets for case C1 at 150 bar and with a gas freestream velocity of \(u_G=30\) m/s. The ligaments and some droplets that form after the bridge created during the hole formation process presented in Figure~\ref{fig:lobe_ext_time} breaks collide with the liquid surface near the trough of the perturbation. The combination of low velocities together with the higher surface-tension coefficient and higher liquid viscosity around the trough is stabilizing. The impact energy is absorbed and the amplitude of the perturbation decays. For reference, Figures~\ref{fig:150_30A_inter_5mus} and~\ref{fig:150_30A_slices_5mus} show the interface temperature, the surface-tension coefficient and the liquid fluid properties for case C1 at \(t=5\) \(\mu\)s, which corresponds to the non-dimensional time \(t^*=7.5\) shown in Figure~\ref{subfig:150_30A_5mus_impact_D}. Other stable impact events occur in case C1 between \(t^*=8.4\) and \(t^*=9.75\), where a big droplet collides with the liquid surface near the perturbation crest, and in case B1 from \(t^*=7.5\) to \(t^*=9.25\) with droplets colliding with the liquid surface near the perturbation trough. \par

\begin{figure}[h!]
\centering
\begin{subfigure}{0.33\textwidth}
  \centering
  \includegraphics[width=1.0\linewidth]{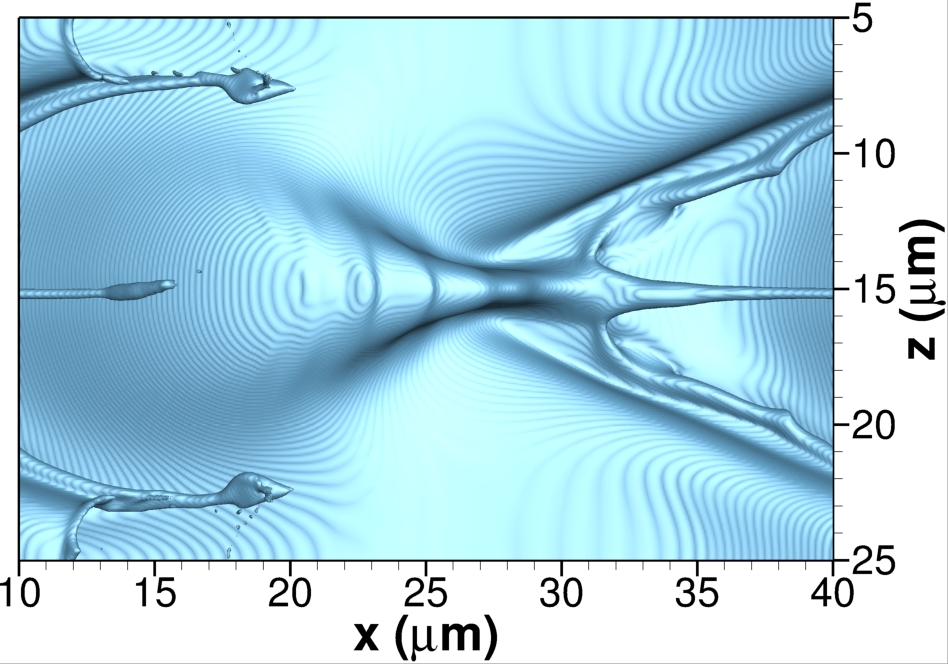}
  \caption{\(t^*=5.075\)} 
  \label{subfig:100_70A_1p45mus_impact_D}
\end{subfigure}%
\begin{subfigure}{0.33\textwidth}
  \centering
  \includegraphics[width=1.0\linewidth]{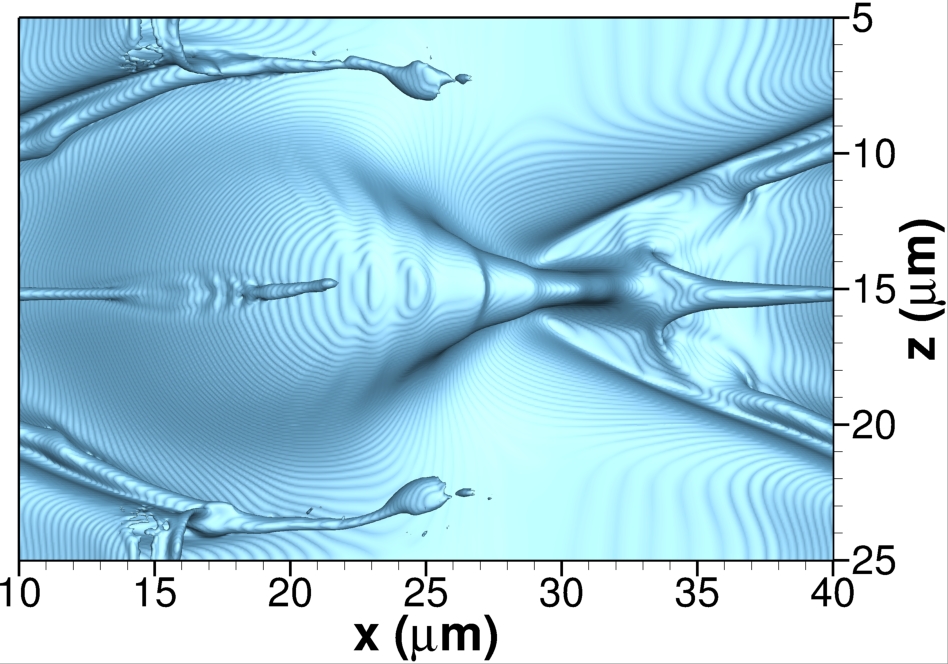}
  \caption{\(t^*=5.425\)}
  \label{subfig:100_70A_1p55mus_impact_D}
\end{subfigure}%
\begin{subfigure}{0.33\textwidth}
  \centering
  \includegraphics[width=1.0\linewidth]{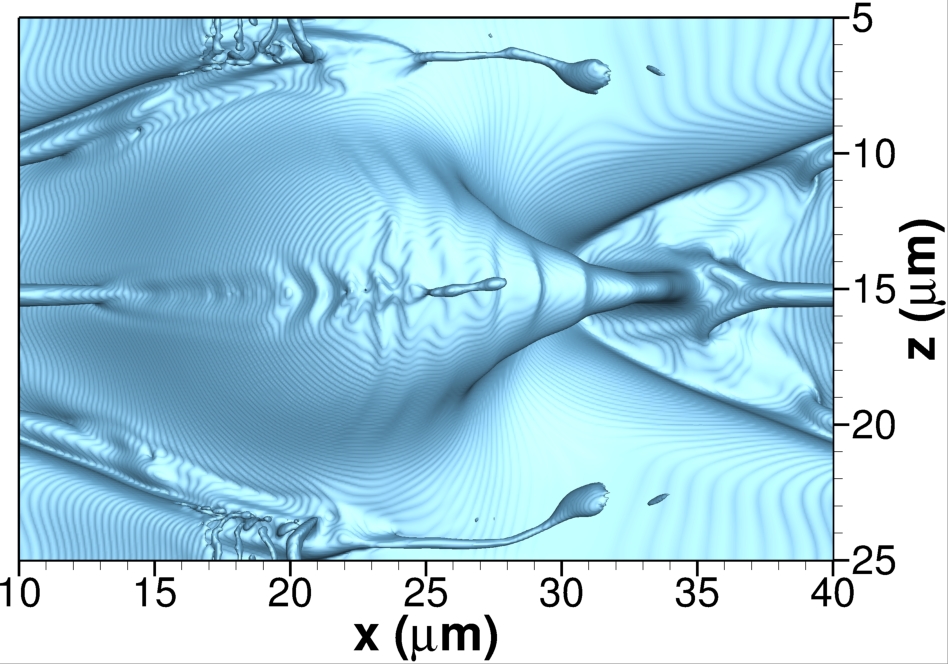}
  \caption{\(t^*=5.775\)}
  \label{subfig:100_70A_1p65mus_impact_D}
\end{subfigure}%
\\
\begin{subfigure}{0.33\textwidth}
  \centering
  \includegraphics[width=1.0\linewidth]{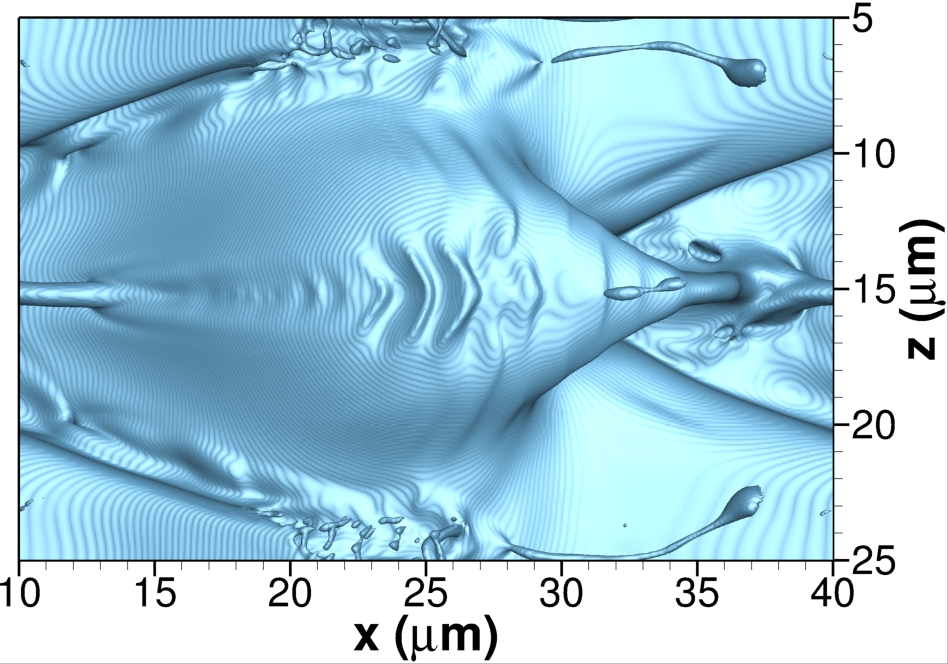}
  \caption{\(t^*=6.125\)} 
  \label{subfig:100_70A_1p75mus_impact_D}
\end{subfigure}%
\begin{subfigure}{0.33\textwidth}
  \centering
  \includegraphics[width=1.0\linewidth]{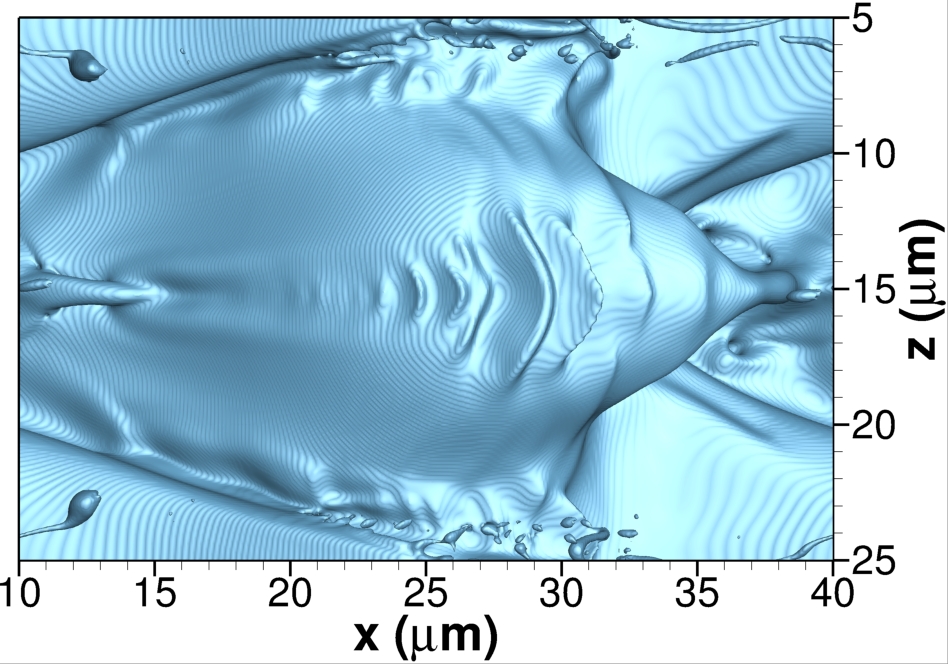}
  \caption{\(t^*=6.475\)}
  \label{subfig:100_70A_1p85mus_impact_D}
\end{subfigure}%
\begin{subfigure}{0.33\textwidth}
  \centering
  \includegraphics[width=1.0\linewidth]{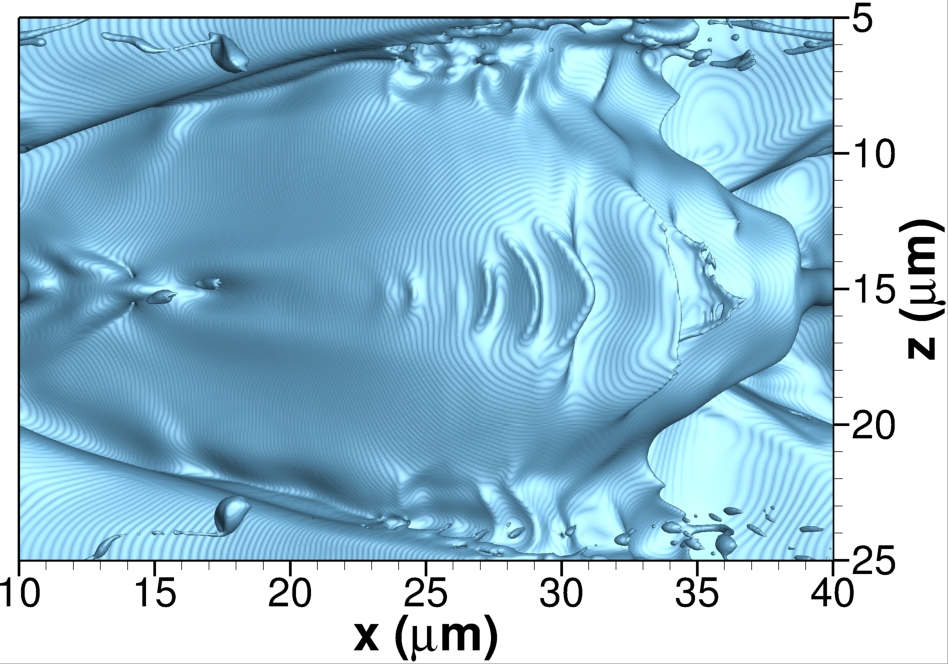}
  \caption{\(t^*=6.825\)}
  \label{subfig:100_70A_1p95mus_impact_D}
\end{subfigure}%
\caption{Unstable impact or coalescence events at 100 bar with gas freestream velocity of \(u_G=70\) m/s (i.e., case B2). The top view is obtained from an \(xz\) plane located above the liquid surface. The interface location is identified as the isosurface with \(C=0.5\). A non-dimensional time is obtained as \(t^*=t/t_c=t\frac{u_G}{H}\). (a) \(t^*=5.075\); (b) \(t^*=5.425\); (c) \(t^*=5.775\); (d) \(t^*=6.125\); (e) \(t^*=6.475\); and (f) \(t^*=6.825\).}
\label{fig:impact_unstable}
\end{figure}

There are many examples of unstable impact or coalescence events, especially at higher gas freestream velocities. Figure~\ref{fig:impact_unstable} presents the growth of surface instabilities at 100 bar and \(u_G=70\) m/s (i.e., case B2) following the collision of a ligament near the perturbation crest around \(z=15\) \(\mu\)m, which is also observed, for instance, in case C2 at 150 bar and \(u_G=50\) m/s. This event triggers a series of waves on the liquid surface that grow over time and define the formation of small lobes near the edge of the main wave. These lobes extend into the gas phase and precede the formation of small thin ligaments and droplets (e.g., shredding). Interestingly, the growth of these perturbations induces the formation of new surface waves upstream, which does not occur in lower pressure cases such as case A1 at 50 bar with \(u_G=50\) m/s. In case A1, a surface wave propagates from \(t^*=6.25\) to \(t^*=10\), forming a lobe at the edge of the perturbation without inducing upstream perturbations.  \par 

\begin{figure}[h!]
\centering
\begin{subfigure}{0.33\textwidth}
  \centering
  \includegraphics[width=1.0\linewidth]{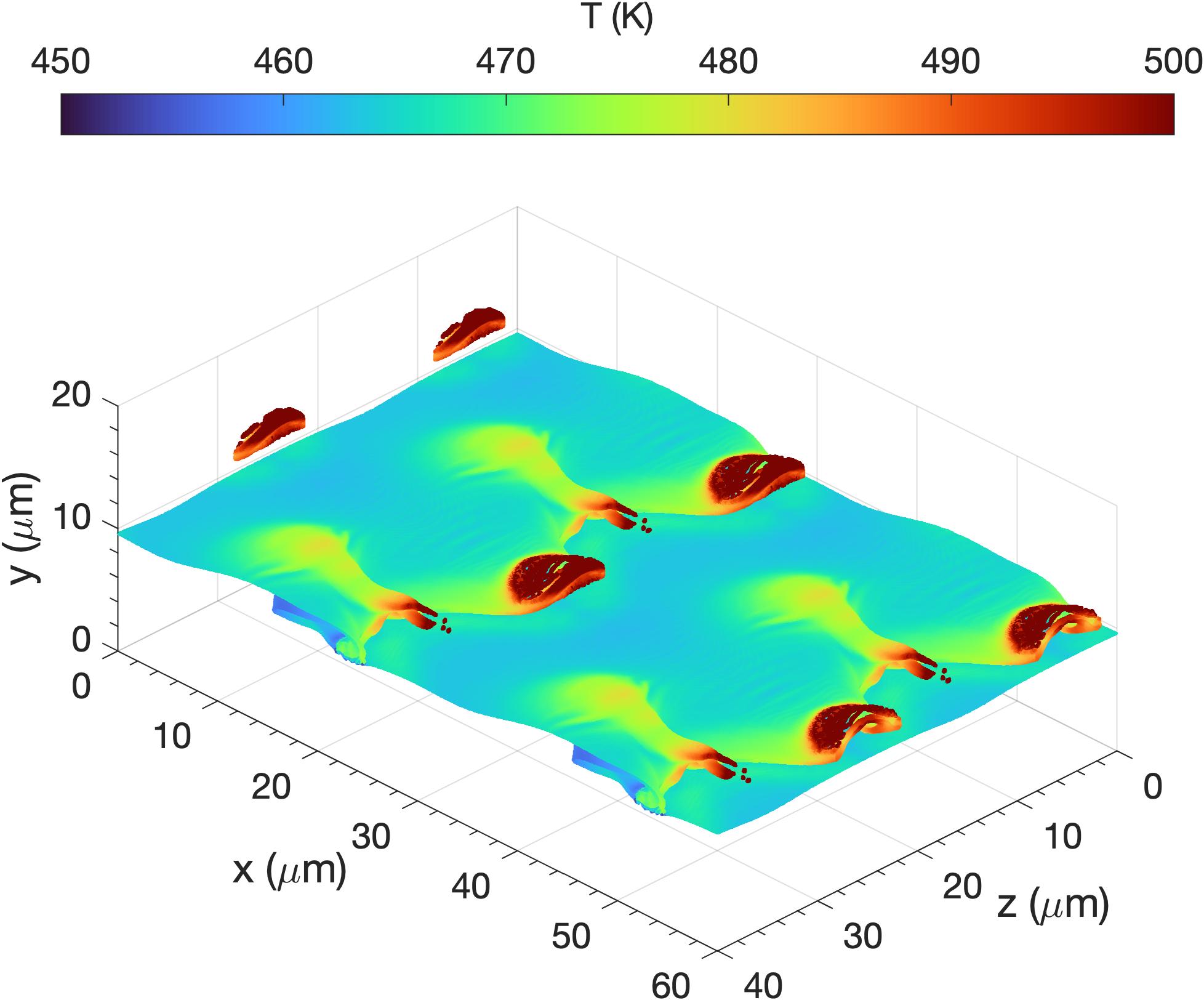}
  \caption{\(t^*=3.85\)}
  \label{subfig:150_70A_int_T_1p10mus}
\end{subfigure}%
\begin{subfigure}{0.33\textwidth}
  \centering
  \includegraphics[width=1.0\linewidth]{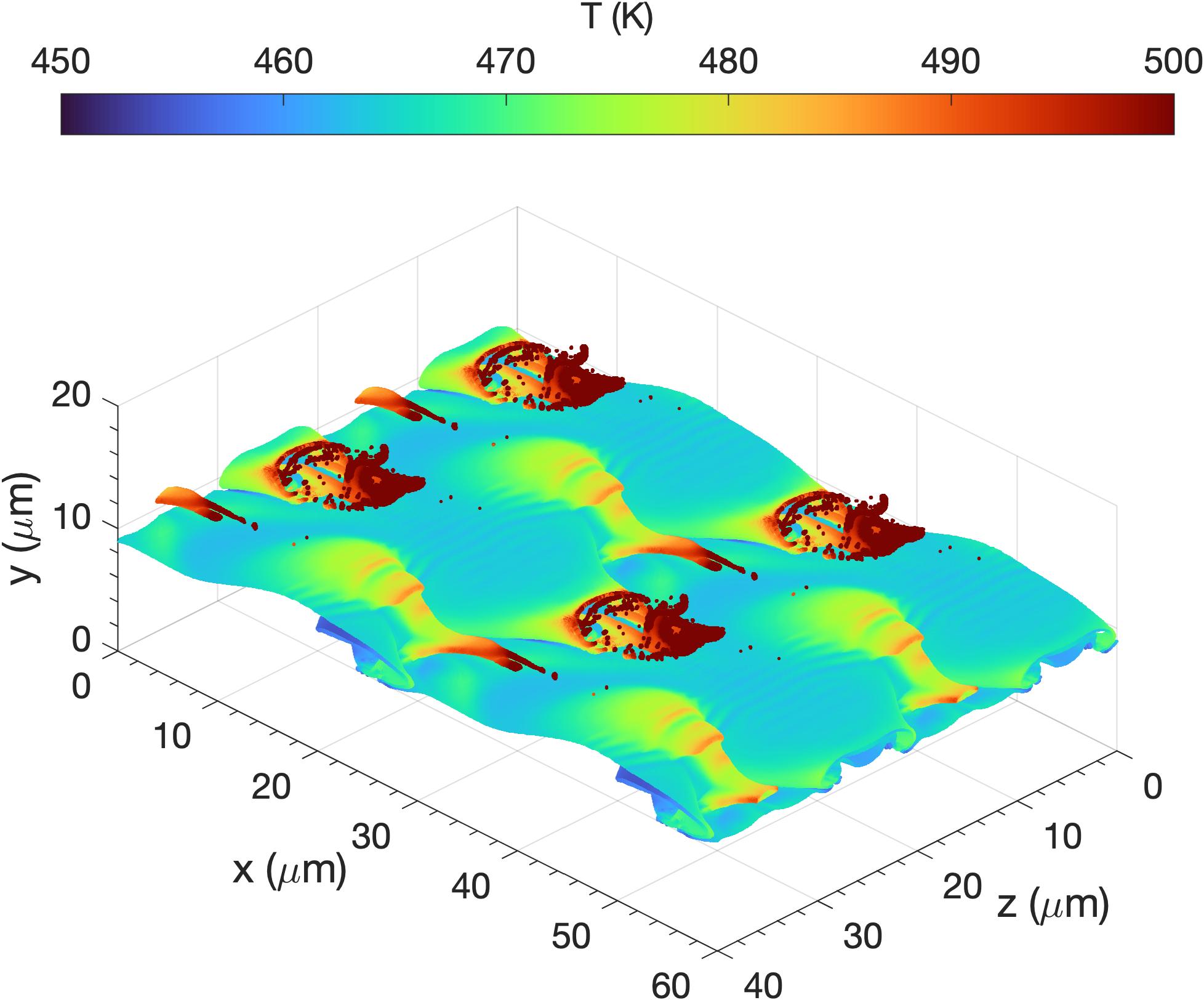}
  \caption{\(t^*=4.55\)} 
  \label{subfig:150_70A_int_T_1p30mus}
\end{subfigure}%
\begin{subfigure}{0.33\textwidth}
  \centering
  \includegraphics[width=1.0\linewidth]{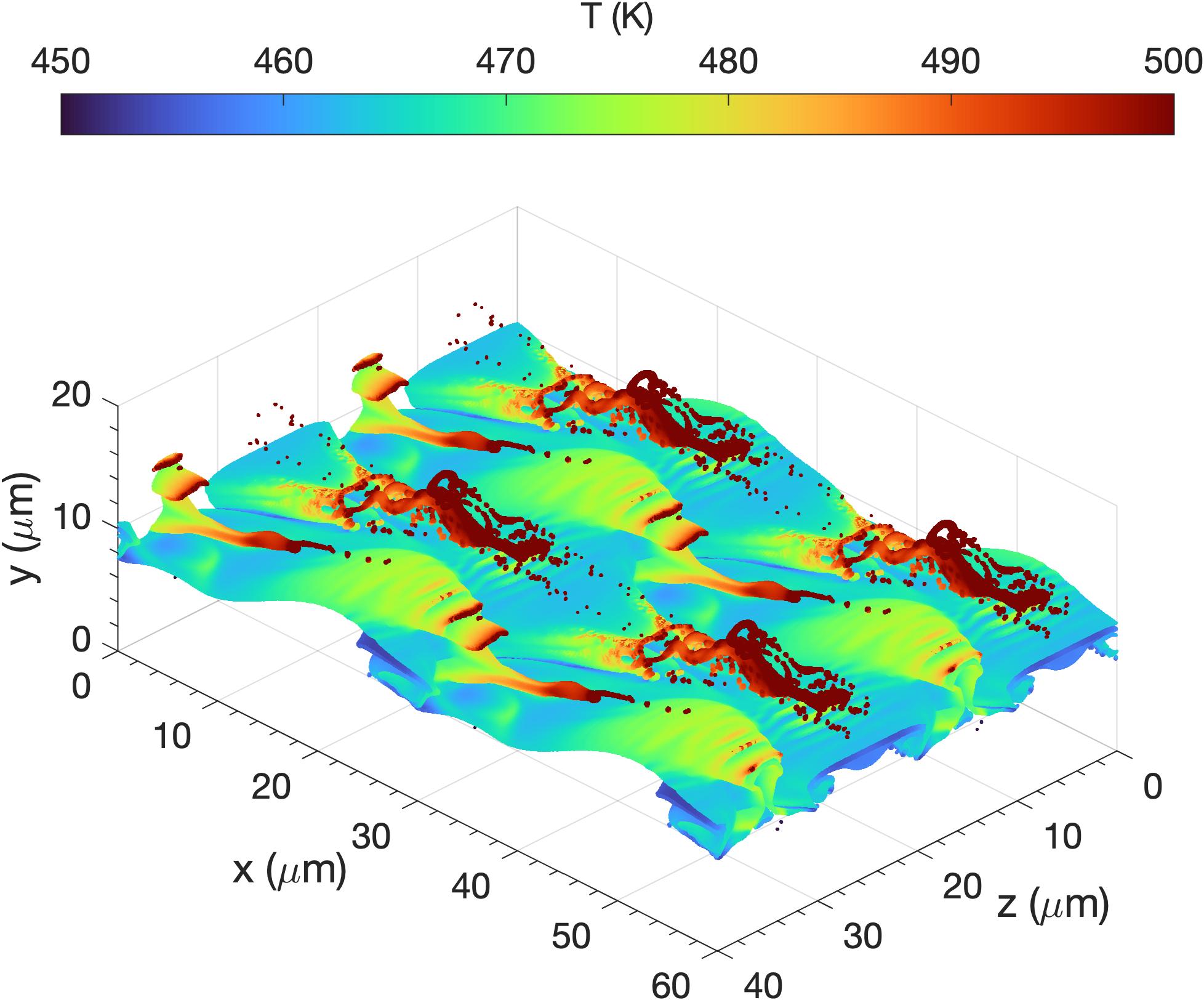}
  \caption{\(t^*=5.25\)}
  \label{subfig:150_70A_int_T_1p50mus}
\end{subfigure}%
\caption{Interface temperature at 150 bar with gas freestream velocity of \(u_G=70\) m/s (i.e., case C3). The interface location is identified at the centroid of each local interface plane defining the interface orientation at a cell level. A non-dimensional time is obtained as \(t^*=t/t_c=t\frac{u_G}{H}\). (a) \(t^*=3.85\); (b) \(t^*=4.55\); and (c) \(t^*=5.25\).}
\label{fig:surface_temperature_waves}
\end{figure}

Other events that trigger the growth of surface instabilities are flow disturbances caused by liquid structures flowing very close to the liquid surface. These perturbations are seen, for example, in cases C2 and C3 at 150 bar. When the initial lobe bursts into droplets and ligaments, as shown in Subsection~\ref{subsubsec:lobe_fold}, a train of droplets and other liquid structures is carried by the faster gas and distorts the flow field around it enough to affect the surface below. Surface waves also appear at high pressures and high velocities near regions where the liquid surface corrugates (see the crest corrugation mechanism explained in Subsection~\ref{subsubsec:lobe_fold}). A weak wave formation is observed around the corrugating crest in case C2 at 150 bar, and a mild formation is apparent in case B2 at 100 bar. At 150 bar and \(u_G=70\) m/s (i.e., case C3), the waves that form as the perturbation crest corrugates from \(t^*=3.15\) to \(t^*=5.95\) become unstable and define the later formation of ligaments and droplets (see Figure~\ref{fig:surface_temperature_waves}). \par

The coupling of mixing effects with the interface dynamics and its local thermodynamic state may facilitate the growth of surface instabilities beyond the observed behavior in incompressible scenarios. For instance, cases A2i and C1i show limited growth of short-wavelength surface instabilities. Only the collision of two surface waves in case A2i after \(t^*>10.15\) along \(z=15\) \(\mu\)m generates surface instabilities near the perturbation crest where a faster streamwise velocity component exists. In the real-fluid scenario, surface waves traveling near the perturbation crest are usually subject to not only faster velocities and stronger shear, but also to higher temperatures as they are closer to the hotter oxidizer stream. At higher temperatures, the interface equilibrium solution involves a smaller surface-tension coefficient and liquid viscosity near it. Thus, those regions will be more susceptible to the unstable growth of surface waves. \par 

The described features are presented in Figure~\ref{fig:surface_temperature_waves}, which shows the interface temperature for case C3. It can be observed that the formation of waves on the hotter surface generates hotter wave crests and colder wave troughs. Thus, the wave crest presents a smaller surface-tension coefficient and liquid viscosity than the wave trough. Moreover, small structures such as ligaments, lobes and droplets present higher surface temperatures as the enclosed liquid volume heats faster. \par 

The variable fluid properties along the wave, together with the faster surrounding gas stream, may generate an unstable mechanism with rapid growth. Impact or coalescence events may accelerate this behavior. As observed in Figure~\ref{fig:surface_temperature_waves}, once the amplitude of the wave has increased sufficiently, the crest is rapidly stretched into the oxidizer stream forming small lobes that precede the eventual shredding of small ligaments and droplet formation. Further investigation of this coupled mechanism involving variable fluid properties is required in future works. Nevertheless, the described features highlight the importance of strong variations in fluid properties in the atomization process of real liquids at supercritical pressures. \par 

Lastly, linear theory can provide a rough estimate of the growth rates and most unstable wavelengths of surface instabilities. Assuming the Kelvin-Helmholtz instability as the trigger for the observed short-wavelength perturbations, we use the linear analysis of small-amplitude interface perturbations from Joseph et al.~\cite{joseph2007potential}, which includes the effect of viscous normal stress, to estimate the perturbation growth rate as

\begin{equation}
\label{eqn:KHeq}
\begin{split}
\epsilon = & -i\frac{k(\rho_{g} u_{\infty_{g}}+\rho_{l} u_{\infty_{l}})}{\rho_{g}+\rho_{l}}-k^2\frac{\mu_{g}+\mu_{l}}{\rho_{g}+\rho_{l}}  \pm \bigg[\frac{\rho_{g} \rho_{l} k^2 (u_{\infty_{g}}-u_{\infty_{l}})^2}{(\rho_{g} + \rho_{l})^2} - \\
& -\frac{\sigma k^3}{\rho_{g} + \rho_{l}} + \frac{k^4(\mu_{g}+\mu_{l})^2}{(\rho_{g}+\rho_{l})^2} + 2ik^3 \frac{(\rho_{g}\mu_{l}-\rho_{l}\mu_{g})(u_{\infty_{g}}-u_{\infty_{l}})}{(\rho_{g}+\rho_{l})^2}\bigg]^{1/2}
\end{split}
\end{equation}

\noindent
where \(\epsilon=\epsilon_R + i\epsilon_I\) is the complex growth rate of a perturbation whose amplitude evolves as \(\Omega(x,t)=\hat{\Omega}\exp(\epsilon t)\exp(ikx)\) with \(\hat{\Omega}\) being the initial perturbation amplitude. Thus, the real part of the growth rate, \(\epsilon_R\), determines whether an oscillation is stable or not (i.e., stable for \(\epsilon_R<0\)). \(k\) represents the wave number and, for the following analysis, the velocities of the gas and liquid streams are defined as \(u_{\infty_{g}}=u_G\) and \(u_{\infty_{l}}=0\). Moreover, the effect of gravity is neglected in Eq.~(\ref{eqn:KHeq}). As a limiting case, \(\rho_{g}\), \(\rho_{l}\), \(\mu_{g}\) and \(\mu_{l}\) are chosen as the interface density and viscosity of each phase obtained from phase equilibrium at a given temperature. \par 

\begin{figure}[h!]
\centering
\includegraphics[width=0.5\linewidth]{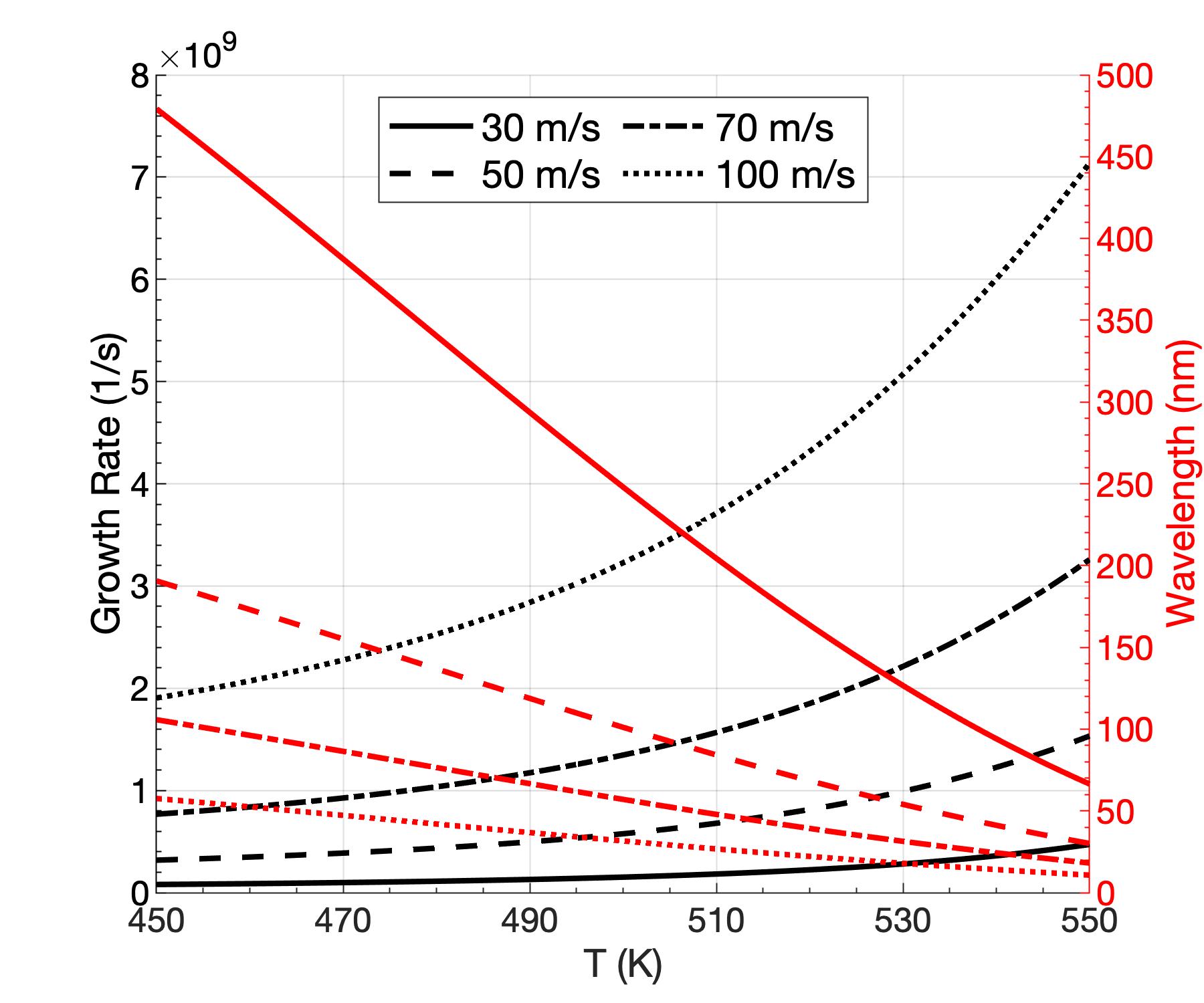}
\caption{Analysis of the wavelength of the most unstable wave and its amplitude growth rate using the linear theory, Eq.~(\ref{eqn:KHeq}), for an \textit{n}-decane/oxygen interface at 150 bar. Fluid properties are assumed to be the interface fluid properties obtained as a function of temperature from phase-equilibrium data.}
\label{fig:KH_study}
\end{figure}

Figure~\ref{fig:KH_study} presents the wavelength of the most unstable wave and its amplitude growth rate as a function of the interface temperature and \(u_G\), obtained from Eq.~(\ref{eqn:KHeq}) for an \textit{n}-decane/oxygen interface at 150 bar. The magnitudes of the most unstable wavelengths range from a few nanometers to hundreds of nanometers, depending on the configuration. The predicted wavelengths from the linear theory are much smaller than the short-wavelength perturbations observed on the liquid surface (i.e., around 3 \(\mu\)m). This result is expected since this theory is based on a viscous potential flow with only normal stresses and, thereby, the stabilizing shear stress is ignored. Also, the interface fluid properties are considered here, which differ somewhat with respect to the liquid and gas fluid properties away from the interface. Nevertheless, the linear theory results offer some guidance on how much more unstable the interface can become as the interface temperature increases. \par 

From an interface temperature of 450 K to 500 K, the most unstable wavelengths at a given velocity difference are approximately reduced by a factor of 2. Similarly, their growth rates increase by a factor of approximately 1.5. Therefore, it is expected that hotter interface regions subject to faster velocities may develop unstable surface waves that do not appear in colder regions, especially if a trigger even exists (e.g., collision). Note that the most unstable wavelength does not scale with the interface temperature alone in the analyzed cases. Since colder interface regions are usually near the bottom of the two-phase mixing layer, they are subject to much smaller velocities and, as observed in Figure~\ref{fig:KH_study}, the most unstable wavelength rapidly grows an order of magnitude as the velocity difference is reduced. \par 

Further analysis of the configurations that develop short-wavelength surface instabilities shows that a trend exists whereby lower pressures and lower velocities develop slightly longer waves, as expected. For instance, case C2 (i.e., 150 bar and \(u_G=50\) m/s) develops waves with wavelengths between 3 \(\mu\)m and 4 \(\mu\)m, whereas case C3 (i.e., 150 bar and \(u_G=70\) m/s) develops surface instabilities with wavelengths between 2 \(\mu\)m and 3 \(\mu\)m. \par

\subsection{Ligament and droplet formation}
\label{subsec:droplet}

This subsection describes the formation of detached ligaments and droplets in each analyzed configuration. Moreover, trapped bubbles and gas pockets are also identified. Compared to previous Subsections~\ref{subsec:deformation} and~\ref{subsec:instability}, which mainly showed individual features, the focus now is on the breakup process of the liquid jet as a whole. Structures connected to the bulk liquid, such as connected ligaments, are not considered due to difficulties in implementing an efficient identification algorithm to capture such cases. Nevertheless, ligaments eventually separate from the bulk liquid and are identified. \par 

Detached structures from the main liquid jet, whether liquid or gas, are identified from an interface data set which includes the location of interface cells, the relevant geometrical information and information on the thermodynamic state of the interface (e.g., temperature, mass flux per unit area). This data set is used to plot, for instance, the interface temperature in Figures~\ref{fig:150_30A_inter_5mus} and~\ref{fig:surface_temperature_waves}. With this data set, more information is readily available to analyze than other techniques that identify structures directly from the volume fraction field. \par 

\begin{figure}[h!]
\centering
\begin{subfigure}{0.33\textwidth}
  \centering
  \includegraphics[width=1.0\linewidth]{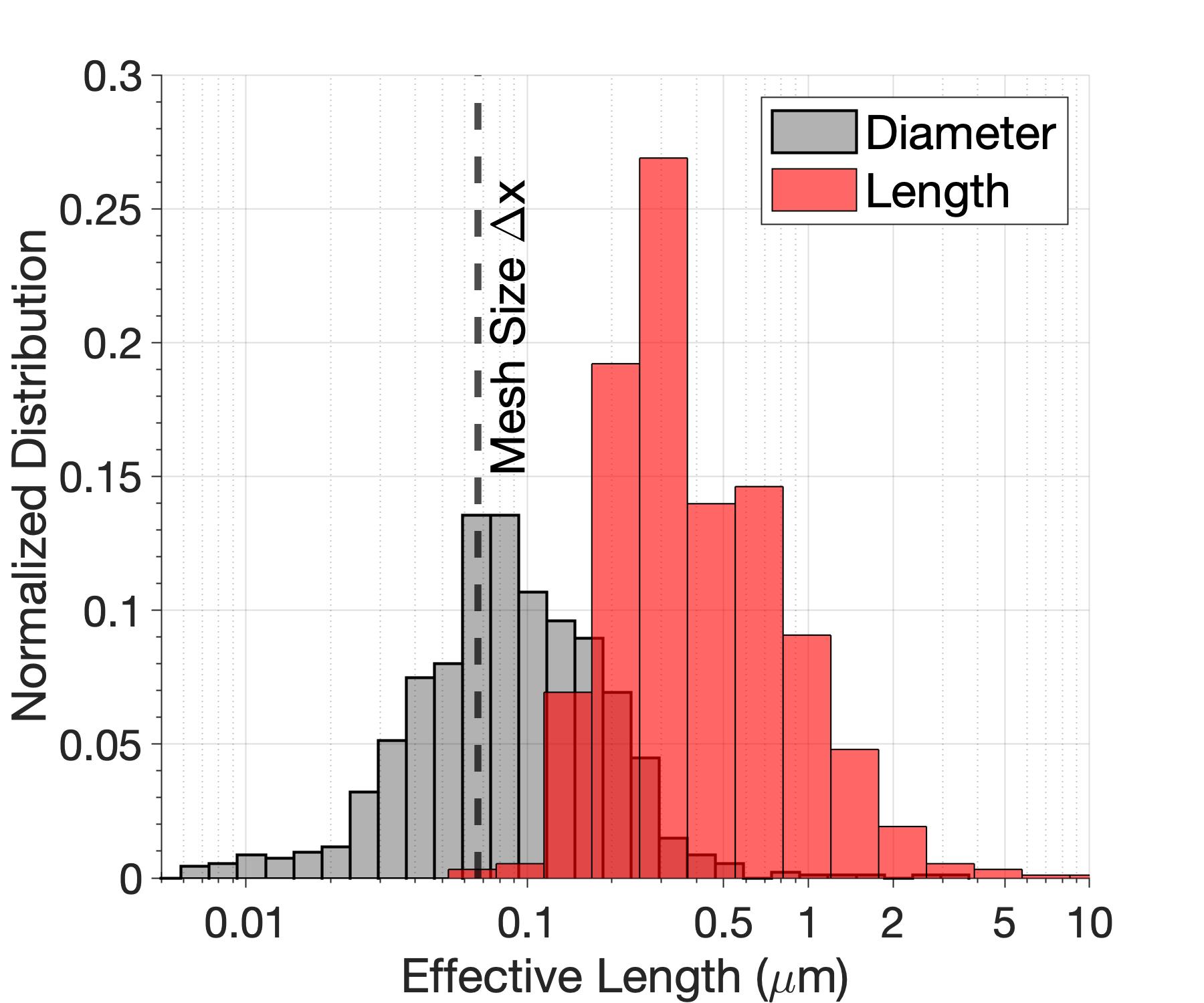}
  \caption{} 
  \label{subfig:lig_distr_150_70A}
\end{subfigure}%
\begin{subfigure}{0.33\textwidth}
  \centering
  \includegraphics[width=1.0\linewidth]{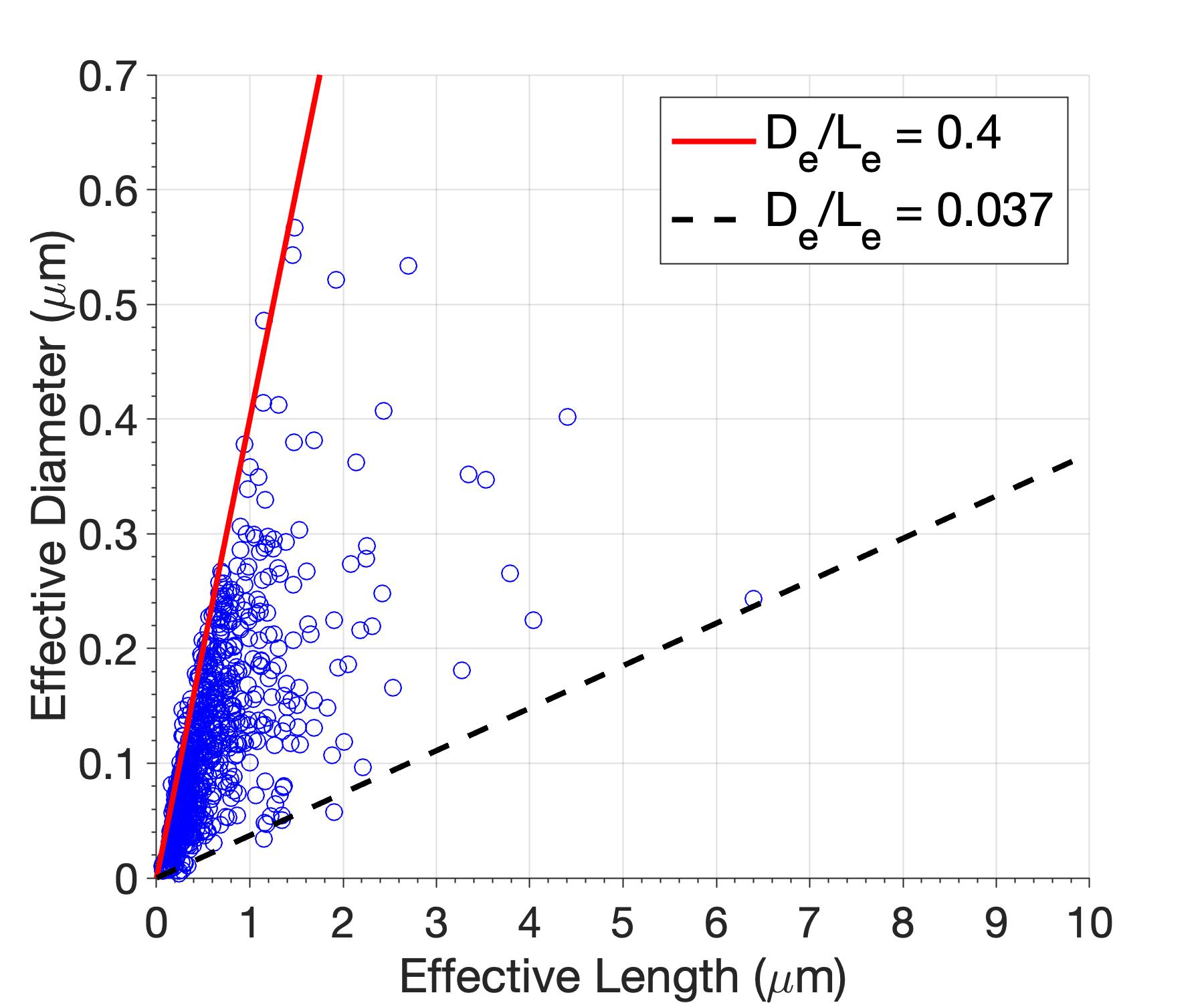}
  \caption{}
  \label{subfig:lig_corr_150_70A}
\end{subfigure}%
\begin{subfigure}{0.33\textwidth}
  \centering
  \includegraphics[width=1.0\linewidth]{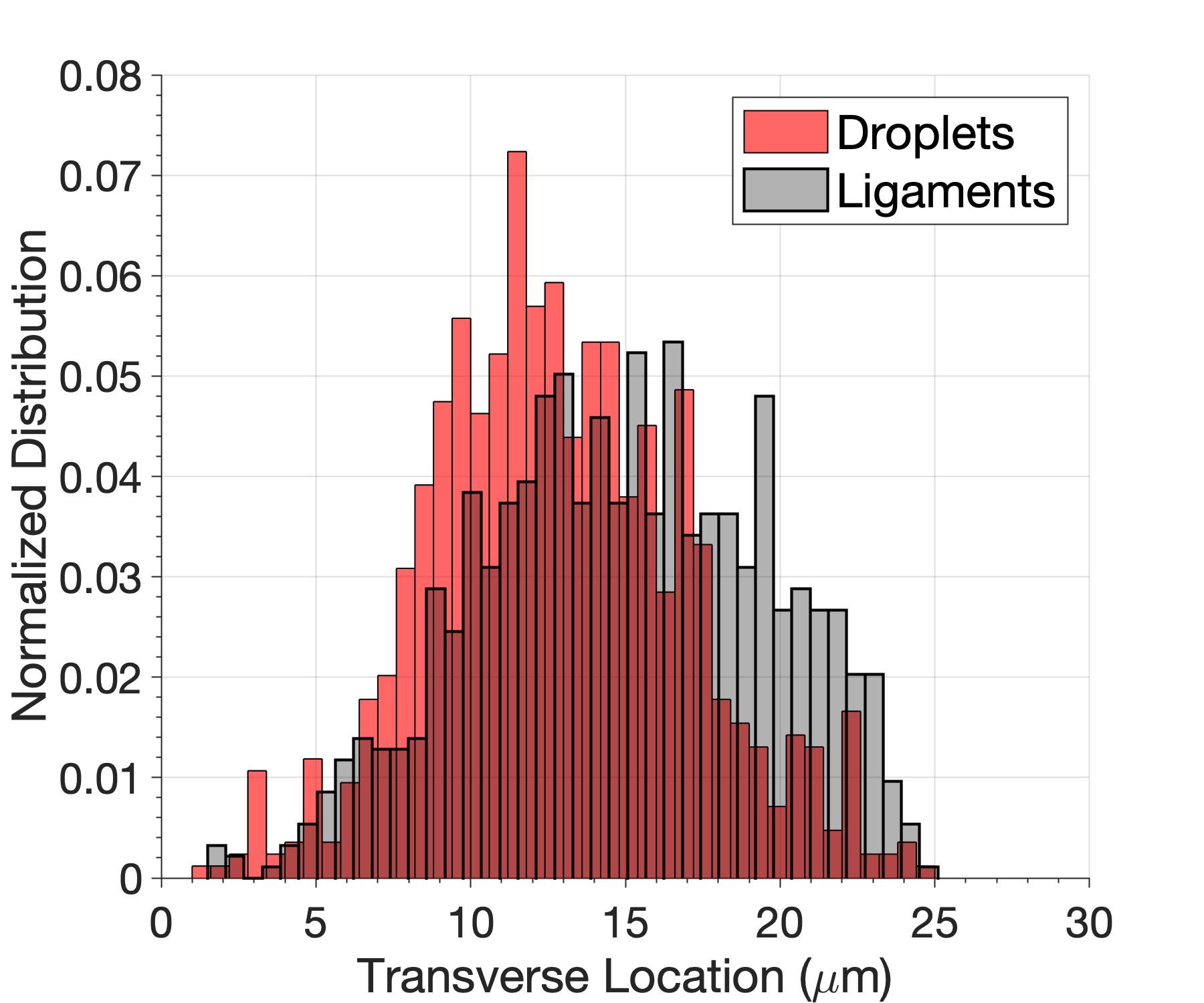}
  \caption{}
  \label{subfig:loc_distr_150_70A}
\end{subfigure}%
\caption{Effective ligament length and diameter and spatial distribution of ligaments and droplets at 150 bar and \(u_G=70\) m/s (i.e., case C3) for the non-dimensional time of \(t^*=14.7\). (a) effective diameter and effective length distribution. Notice the logarithmic scale used in the horizontal axis; (b) correlation between effective diameter and effective length; and (c) spatial distribution of ligaments and droplets.}
\label{fig:ligaments}
\end{figure}

A structure involves a set of interface cells adjacent to each other. In a three-dimensional domain, two interface cells are considered adjacent to each other if they fall inside a cube of three cells per side centered around either interface cell. Once the structure's shape is known (i.e., distribution of the interface cells defining the structure), its surface area is computed by adding together the areas of each local interface plane obtained from PLIC. If needed, the enclosed volume can be obtained from the volume fraction field contained by the structure and the interface cells defining its surface. However, the volume fraction field belongs to a different data set in our computations, which cannot be post-processed efficiently to determine the volume of detached structures. Thus, this traditional path is only considered when volume information is necessary for the analysis (e.g., the ligament size study presented in Figure~\ref{fig:ligaments}). \par

\begin{figure}[h!]
\centering
\begin{subfigure}{0.4\textwidth}
  \centering
  \includegraphics[width=1.0\linewidth]{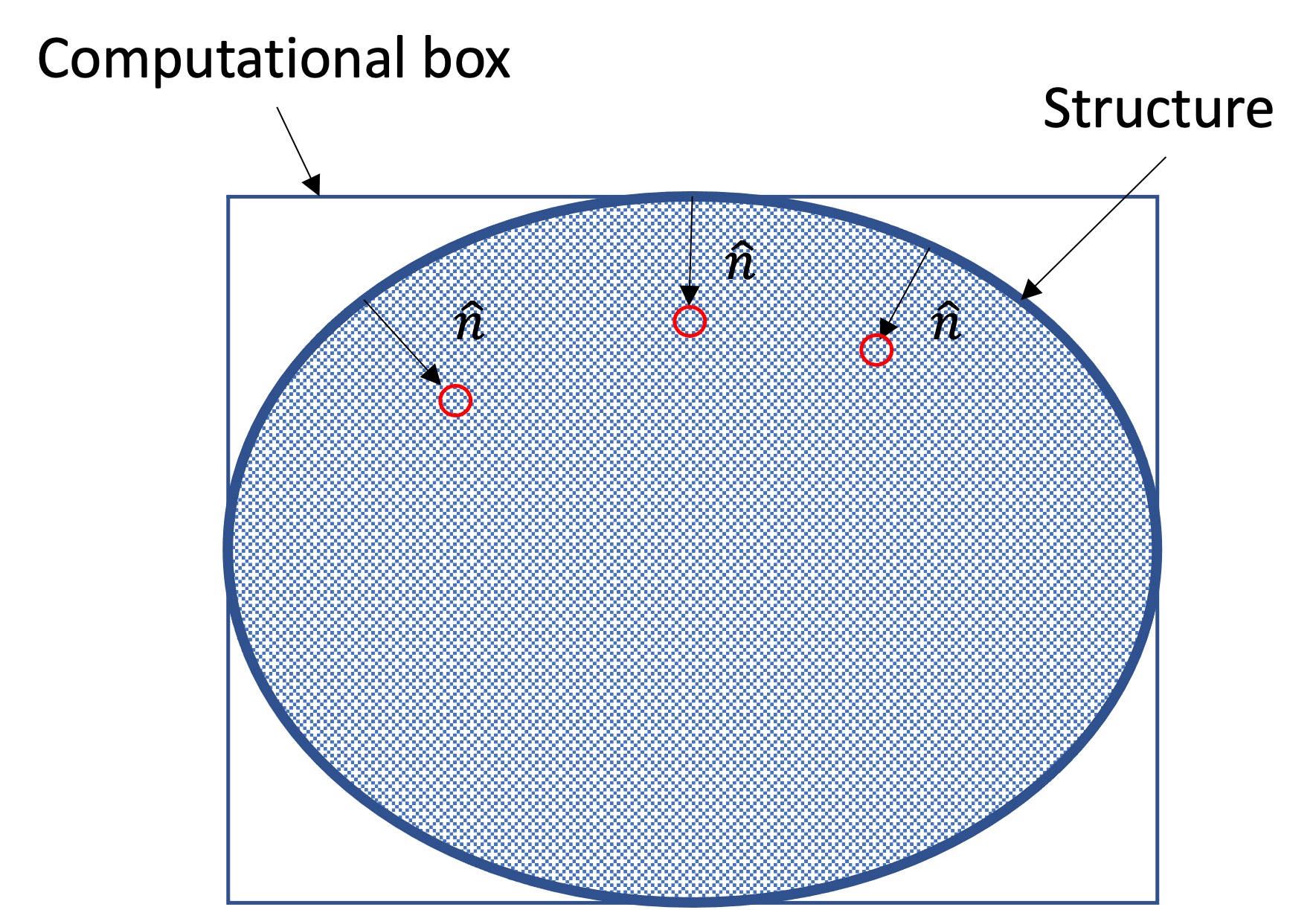}
  \caption{Liquid structure} 
  \label{subfig:liquid_structure}
\end{subfigure}%
\begin{subfigure}{0.4\textwidth}
  \centering
  \includegraphics[width=1.0\linewidth]{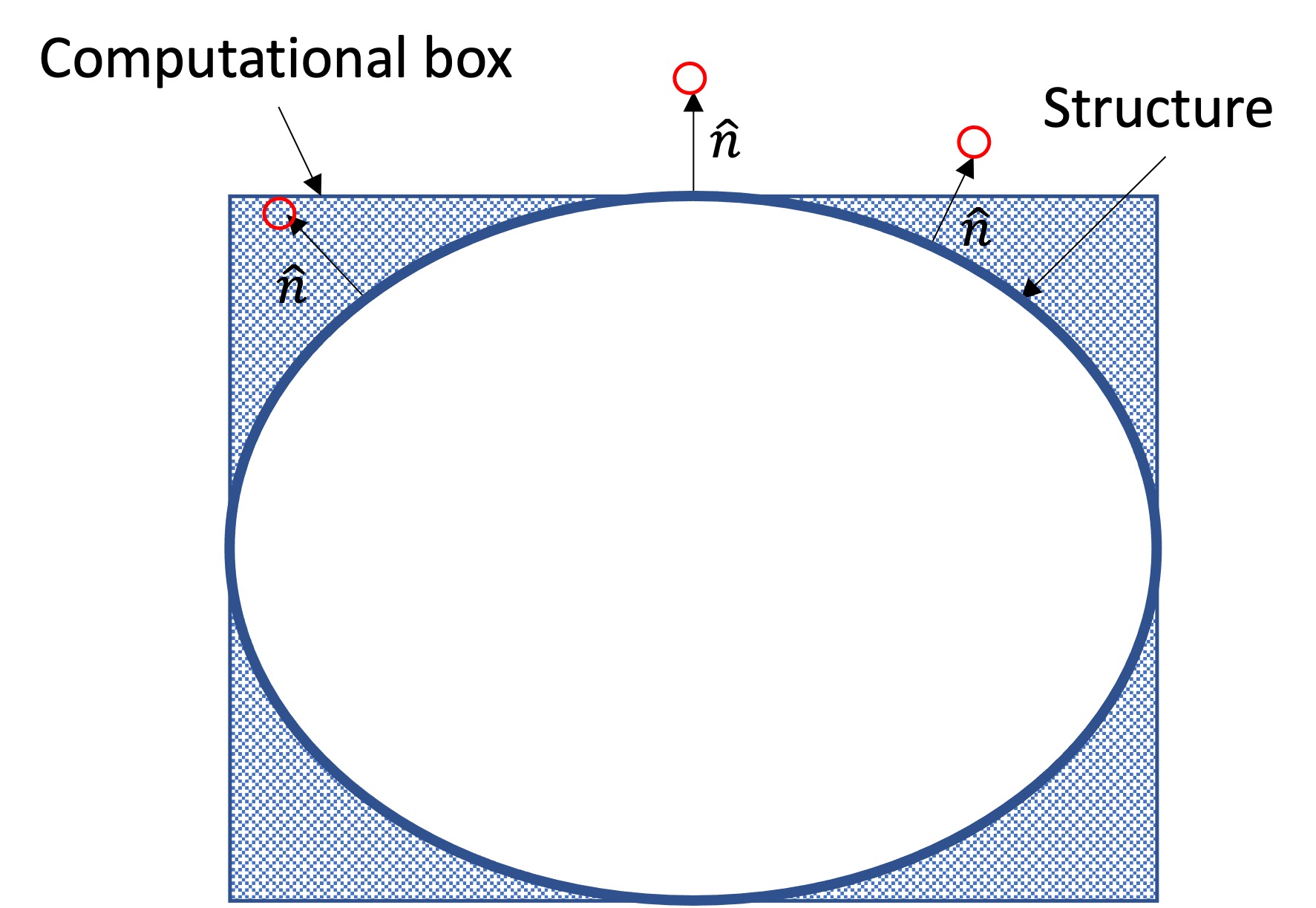}
  \caption{Gas structure}
  \label{subfig:gas_structure}
\end{subfigure}%
\caption{Two-dimensional sketch showing the geometry constraint based on the interface normal unit vector, \(\hat{n}\), to determine whether a detached structure (e.g., droplet, ligament, bubble or gas pocket) encloses liquid phase or gas phase. The red circles represent the test points obtained by extending a point from the local interface plane centroid a distance of \(\Delta x\) following \(\hat{n}\). (a) liquid structure; and (b) gas structure.}
\label{fig:structure_determination}
\end{figure}

Because the post-processed data set only includes information of the interface cells, the volume fraction, \(C\), of the enclosed cells is not available to determine whether the structure defines a liquid body (e.g., droplet or ligament) or a gaseous body (e.g., bubble or gas pocket). Instead, the interface normal unit vector is used. Under the VOF framework implemented in this work, the normal unit vector points toward the liquid phase~\cite{baraldi2014mass}. Thus, the fluid phase of the enclosed volume is determined from a geometry constraint. Note that \(\hat{n}=-\Vec{n}\) as defined in Section~\ref{sec:governing}. Figure~\ref{fig:structure_determination} presents a two-dimensional sketch of the methodology, which can easily be extended to three-dimensional configurations. Each detached structure is contained in a computational box, and it may touch the sides of such a box at many different locations. Figure~\ref{fig:structure_determination} shows the simple example of a structure resembling an ellipsoid. Knowing the direction of the interface normal unit vector, \(\hat{n}\), a test point can be obtained for each interface cell defining the structure, which extends from the centroid of the local interface plane a distance of \(\Delta x\) following \(\hat{n}\). All test points fall inside the computational box for a liquid structure (i.e., droplet or ligament). On the other hand, the enclosed volume is gas if at least one test point falls outside the computational box. The method is robust since the computational box and the structure will usually touch tangentially. Only very under-resolved configurations might fail. Nevertheless, this possibility has a negligible impact on the study of droplets and other structures. \par

Once the structure phase has been identified, the shape of the structure is used to classify among discrete droplets, ligaments, bubbles and gas pockets. To do so, the shape of the computational box containing the structure is considered first. Three characteristic lengths are considered, one per axes: \(\delta x = x_\text{max}-x_\text{min}\), \(\delta y = y_\text{max}-y_\text{min}\) and \(\delta z = z_\text{max}-z_\text{min}\). The subscripts ``max" and ``min" denote the maximum and minimum locations along the coordinate axes of the interface cells defining the structure. Also, the computational box volume is evaluated as \(\delta V = \delta x \delta y \delta z\). Normalizing by the streamwise length (i.e., \(\delta \hat{x} = 1\), \(\delta \hat{y} = \delta y/\delta x\) and \(\delta \hat{z} = \delta z/\delta x\)), the conditions \(0.8 < \delta \hat{y} < 1.2\) and \(0.8 < \delta \hat{z} < 1.2\) are checked. If satisfied, the computational box is approximately cubic and may contain a spherical shape or similar. Otherwise, the computational box is at least larger in one direction compared to the other two, which may indicate that it contains an elongated structure. Here, a deviation of up to 20\% in the length of \(\delta \hat{y}\) and \(\delta \hat{z}\) with respect to \(\delta \hat{x}\) is allowed. Normalizing by streamwise length may appear arbitrary, yet it is done because ligaments are predicted to stretch primarily in the streamwise direction. Other normalization criteria, such as normalizing by the intermediate length, may be investigated in future works (i.e., if \(\delta y<\delta z<\delta x\), normalize the lengths dividing by \(\delta z\)). \par

For approximately cubic computational boxes, another criterion is used to determine whether they contain a spherical shape or not. Using the ratio between the structure's surface and the volume of the computational box, \(S/\delta V\), and an effective cube length, \(\delta x_c = (\delta V) ^{1/3}\), a sphere has an exact ratio \(S/\delta V = \pi/\delta x_c\). Here, the threshold \(S/\delta V = 2.5/\delta x_c\) is implemented. For \(S/\delta V > 2.5/\delta x_c\), the structure is classified as a droplet or bubble depending on whether the enclosed volume is liquid or gas, even if it is slightly deformed. For \(S/\delta V \leq 2.5/\delta x_c\), the structure is identified as a detached ligament or gas pocket. If the computational box is not cubic, the structures are directly classified as discrete ligaments or gas pockets. \par 

Before proceeding with the analysis, note that the studied configurations do not let us prove statistical convergence of results such as droplet distributions. Only one case per set of initial conditions has been simulated using the same mesh size. However, we assess whether under-resolution affects or not the presented data. \par

\begin{figure}[h!]
\centering
\begin{subfigure}{0.5\textwidth}
  \centering
  \includegraphics[width=1.0\linewidth]{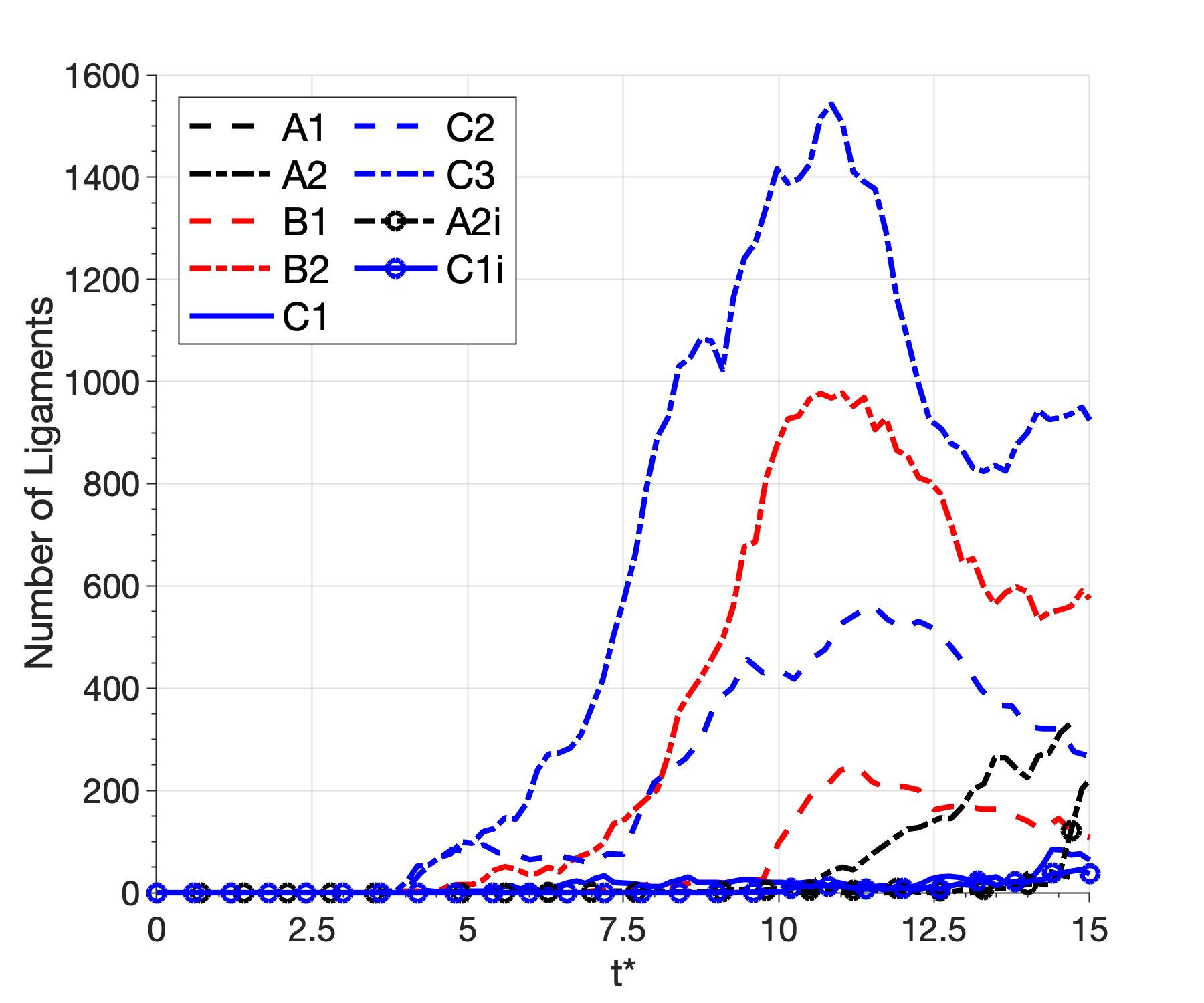}
  \caption{} 
  \label{subfig:num_ligaments}
\end{subfigure}%
\begin{subfigure}{0.5\textwidth}
  \centering
  \includegraphics[width=1.0\linewidth]{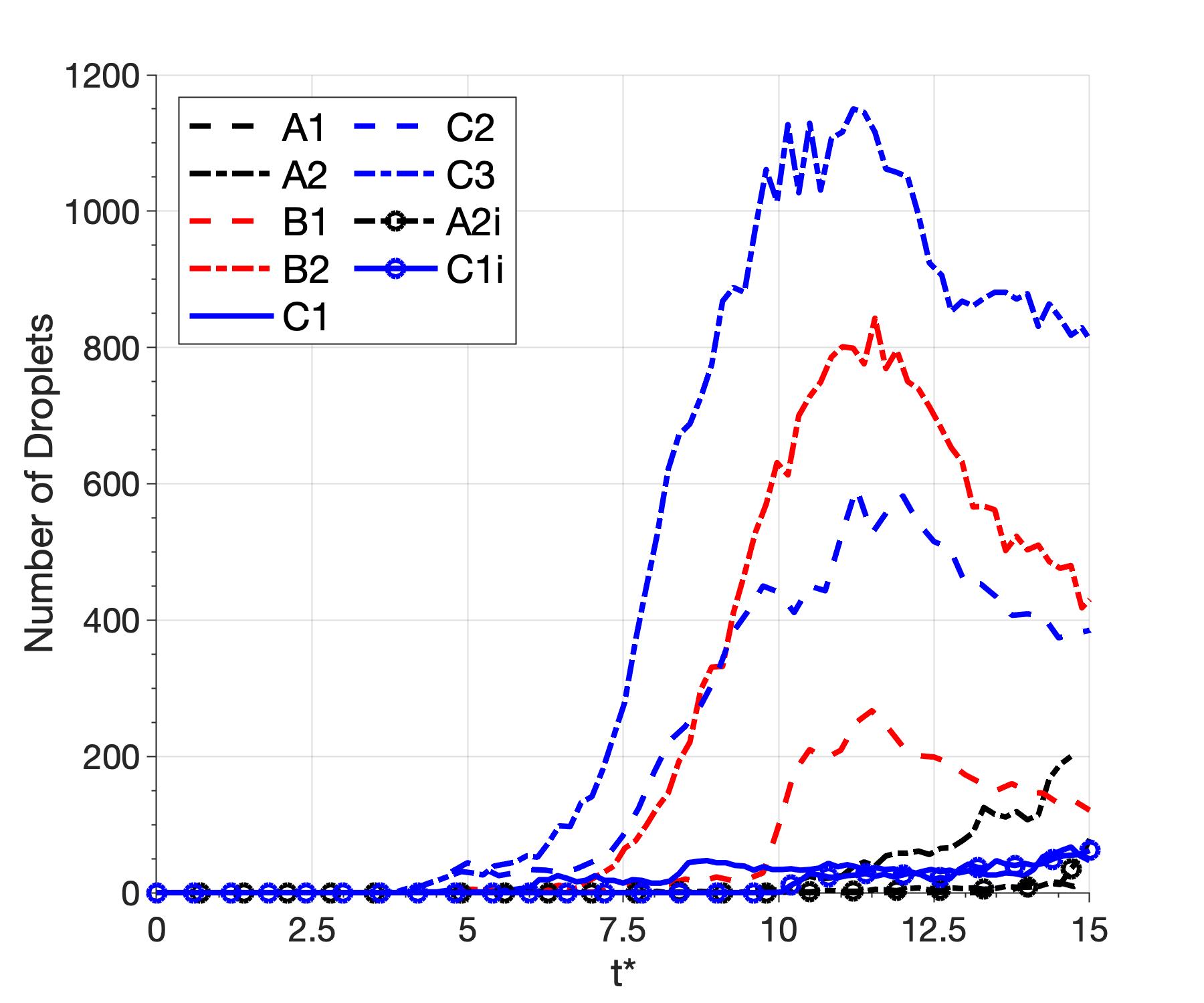}
  \caption{}
  \label{subfig:num_droplets}
\end{subfigure}%
\\
\begin{subfigure}{0.5\textwidth}
  \centering
  \includegraphics[width=1.0\linewidth]{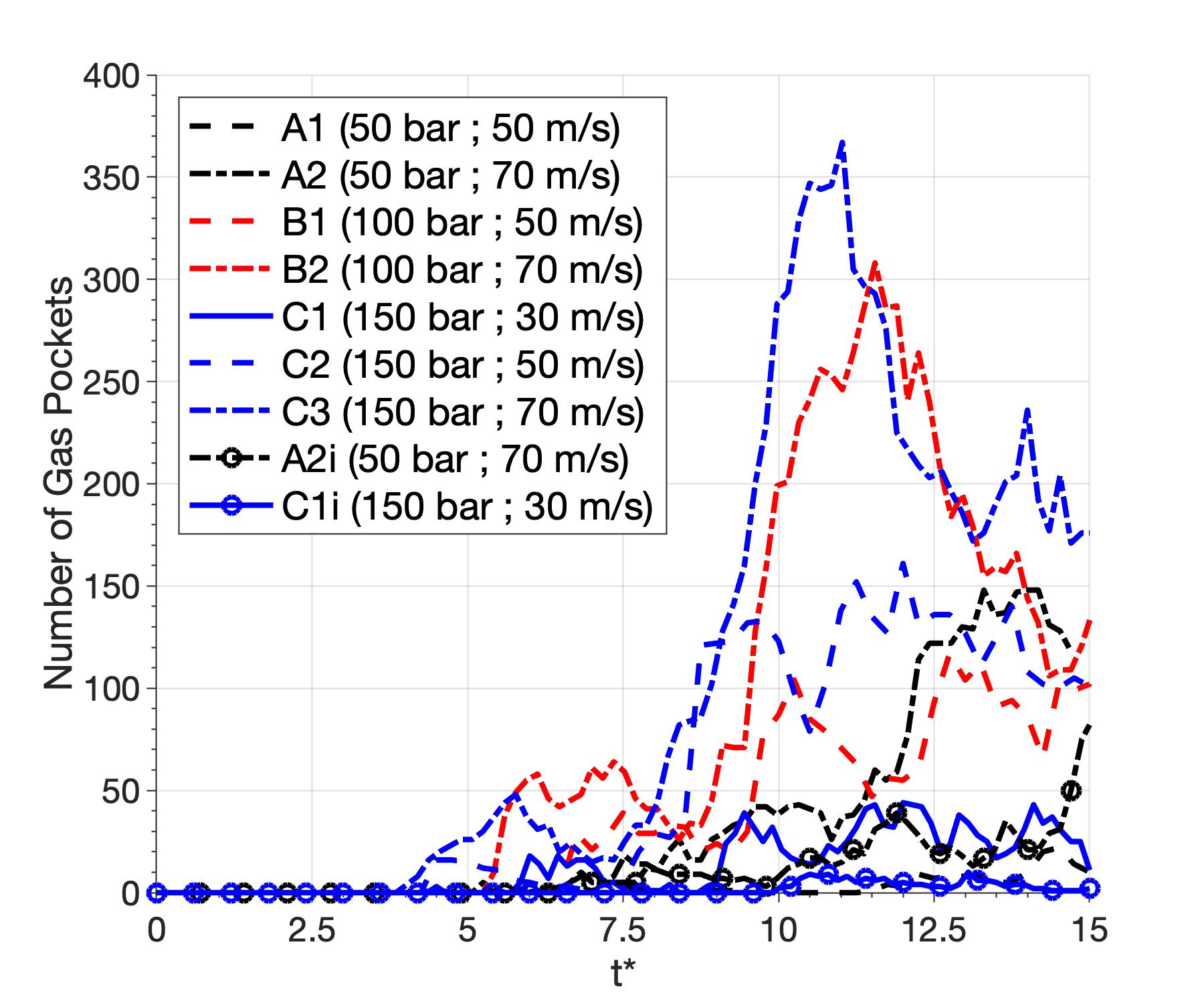}
  \caption{} 
  \label{subfig:num_gaspockets}
\end{subfigure}%
\begin{subfigure}{0.5\textwidth}
  \centering
  \includegraphics[width=1.0\linewidth]{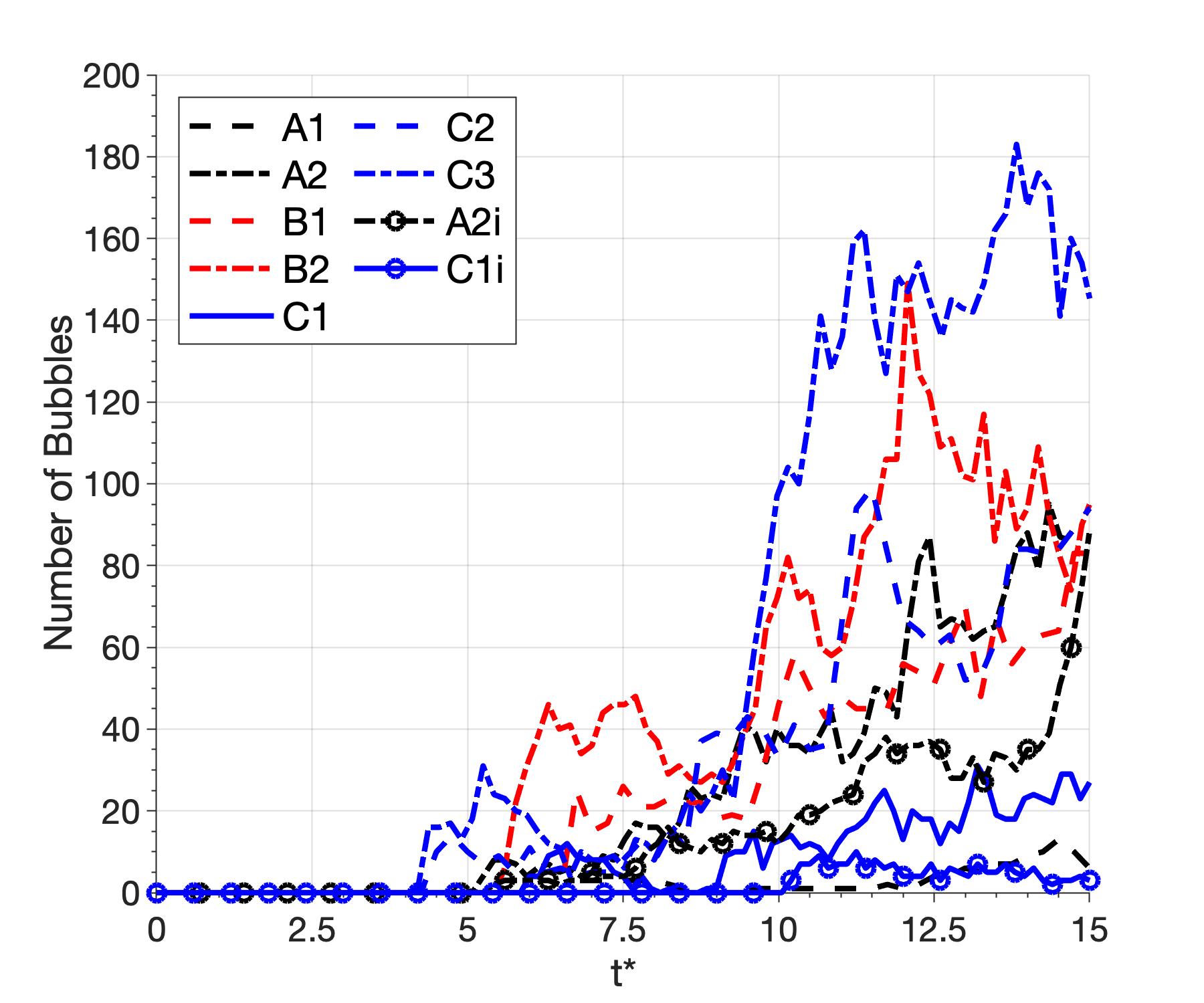}
  \caption{}
  \label{subfig:num_bubbles}
\end{subfigure}%
\caption{Number of detached structures over time contained inside the computational domain defined in Figure~\ref{subfig:initial_domain}. A non-dimensional time is obtained as \(t^*=t/t_c=t\frac{u_G}{H}\). (a) number of ligaments; (b) number of droplets; (c) number of gas pockets; and (d) number of bubbles.}
\label{fig:num_structures}
\end{figure}

Figure~\ref{fig:num_structures} presents the number of ligaments, droplets, gas pockets and bubbles generated over time for each analyzed case. The detached structures are contained in the computational domain described in Figure~\ref{subfig:initial_domain}. That is, the bigger domain presented in other figures which took advantage of periodicity to represent the surface evolution and generation of liquid structures better is not considered. The expected trend is observed whereby cases with higher \(We_G\) and \(Re_L\) present the generation of additional detached structures as the flow becomes more chaotic. The generation of trapped gas pockets and bubbles partially depends on the initial lobe development and how it reconnects with the liquid jet. Additionally, the enhanced generation of ligaments and droplets in cases where inertia forces dominate and mixing effects are stronger also promotes more liquid coalescence events that may trap the gas. \par

The effect of \(Re_L\) becomes clear when analyzing the generation of detached liquid structures. As shown in Table~\ref{tab:cases_2}, cases A2, B1 and C1 have a similar \(We_G\). However, \(Re_L\) decreases with pressure. This trend translates into a higher presence of ligaments and droplets over time at lower pressures. Viscous forces decrease relative to inertial terms, facilitating the breakup of liquid structures from the main liquid jet. Also, some mixing effects are observed. For instance, case B1 shows the presence of more droplets than case A2 until \(t^*\approx 14\) and case C1 shows more droplets than case A2 until \(t^*\approx 11.5\). A similar trend is observed with the number of ligaments. This behavior results from local regions of the liquid jet being more easily deformed due to a lower density and viscosity, as discussed in previous subsections. Comparing cases C1 and B1, the same scenario exists where the higher pressure develops droplets sooner. For cases with higher \(We_G\) (i.e., cases B2 and C2), the differences caused by the change in \(Re_L\) with a similar \(We_G\) are bigger. Also, note that the incompressible cases A2i and C1i present limited generation of ligaments and droplets when compared to their compressible counterparts. This behavior is expected since mixing in the compressible case enhances the generation of liquid structures, as discussed throughout this paper.  \par 

Figure~\ref{fig:num_structures} also provides details on the effect of layering on the early atomization process. For all cases where layering exists (i.e., high pressures of 100 bar and 150 bar), layering starts developing approximately after \(t^*=10\). Around the same non-dimensional time, a peak in the formation of droplets and ligaments is observed for cases B1, B2, C2 and C3. These results are consistent with the layering features described in Subsection~\ref{subsubsec:layering}. The liquid-sheet formation fractures vortices 
and limits the gas entrainment. Moreover, the liquid layers press toward each other as they overlap. These features limit the generation of ligaments and droplets and promote coalescence events. \par 

The slides provided in the Supplemental Material may give the false impression that hundreds of ligaments and droplets do not exist in some configurations, as shown in Figure~\ref{fig:num_structures}. This is a result of the size of such structures and the limited viewpoints provided. Some ligaments and droplets may be hidden underneath bigger liquid structures and also their size may be small enough that the plotting software cannot reconstruct an isosurface with \(C=0.5\). The structure identification and classification algorithm used in this work has been tested and validated. For instance, global parameters such as total surface area and total liquid volume are conserved, and each interface cell is assigned uniquely to a single structure. \par 

\begin{figure}[h!]
\centering
\begin{subfigure}{0.33\textwidth}
  \centering
  \includegraphics[width=1.0\linewidth]{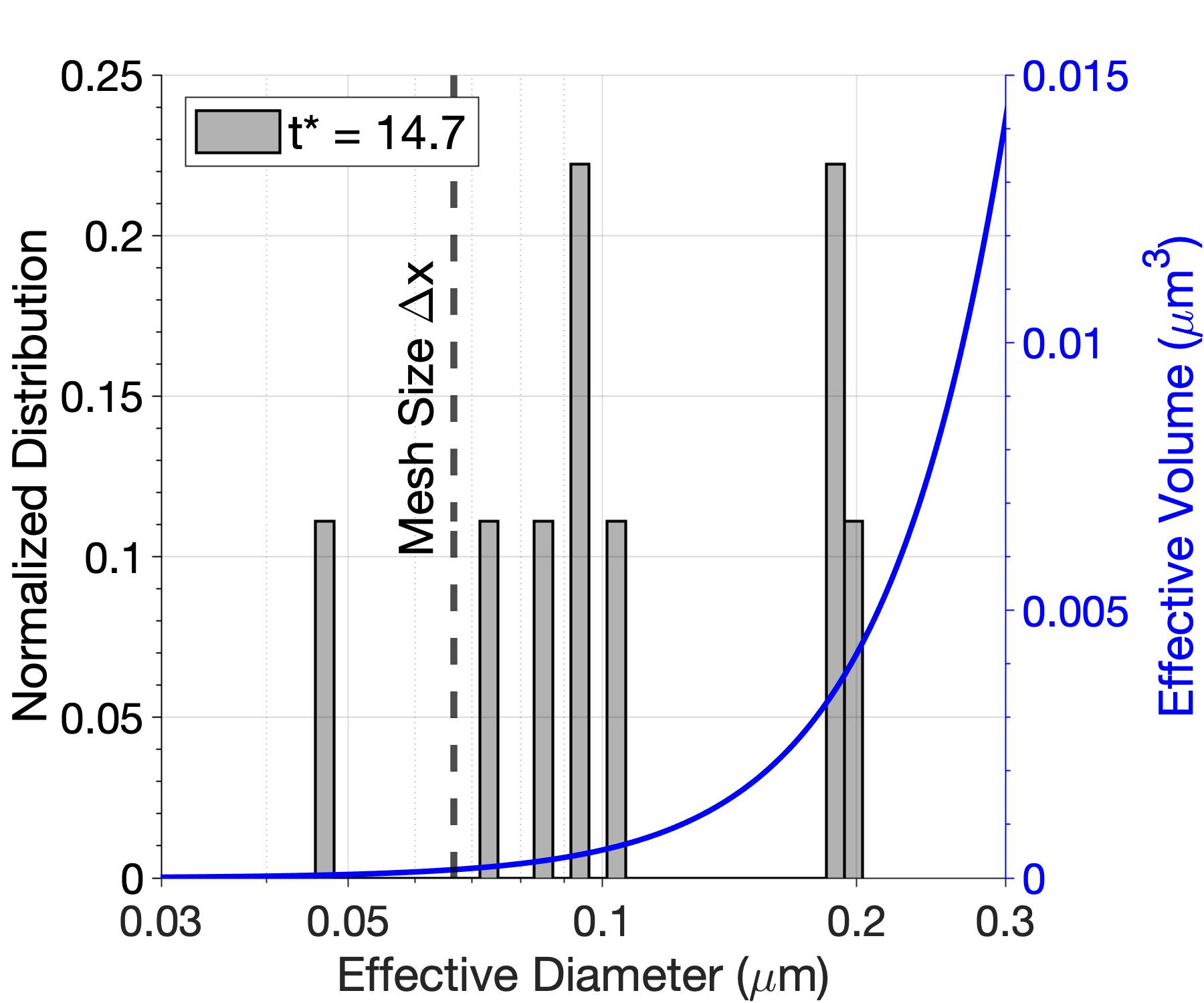}
  \caption{50 bar and \(u_G=50\) m/s (A1)} 
  \label{subfig:drop_distr_50_50A}
\end{subfigure}%
\begin{subfigure}{0.33\textwidth}
  \centering
  \includegraphics[width=1.0\linewidth]{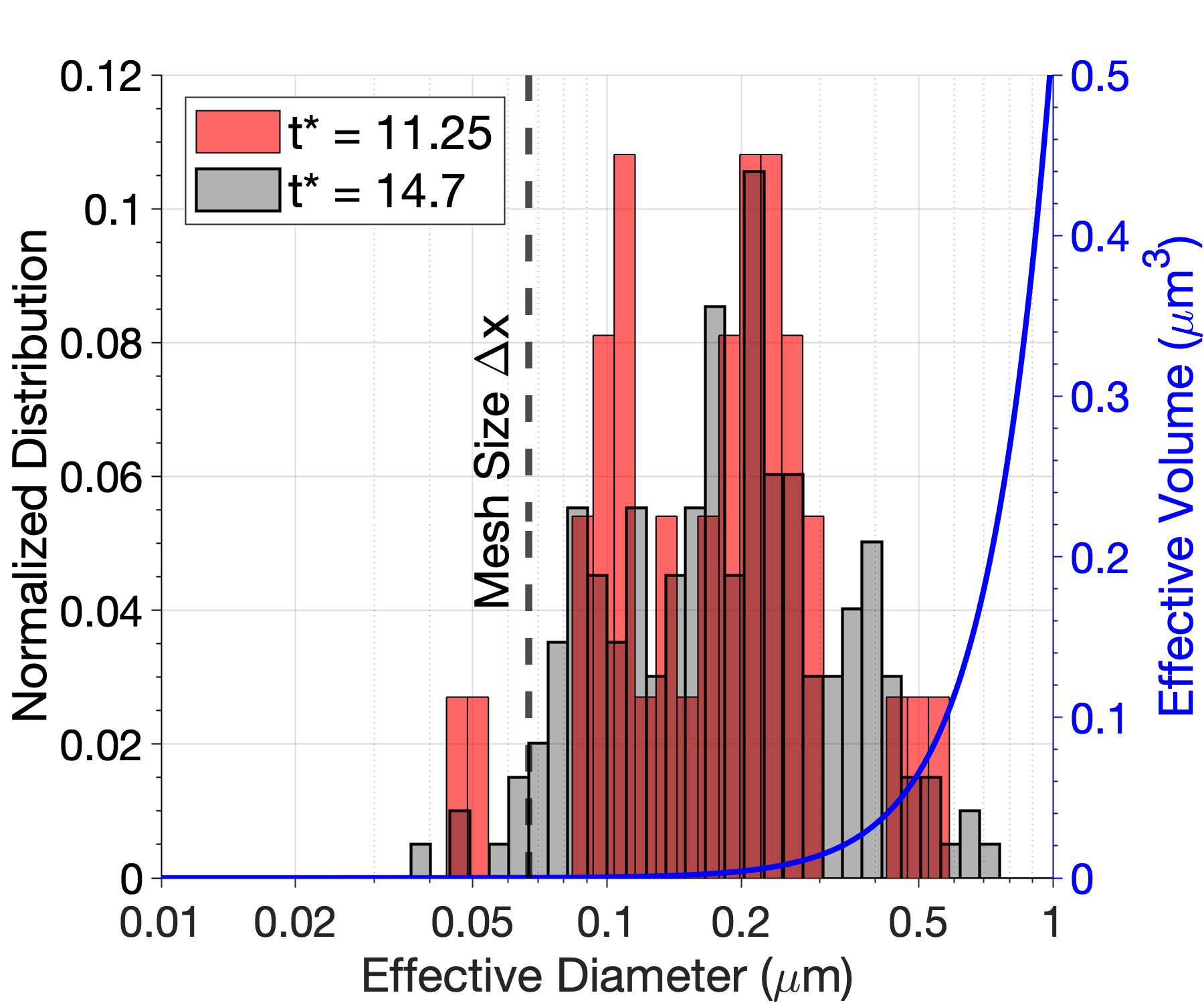}
  \caption{50 bar and \(u_G=70\) m/s (A2)}
  \label{subfig:drop_distr_50_70A}
\end{subfigure}%
\begin{subfigure}{0.33\textwidth}
  \centering
  \includegraphics[width=1.0\linewidth]{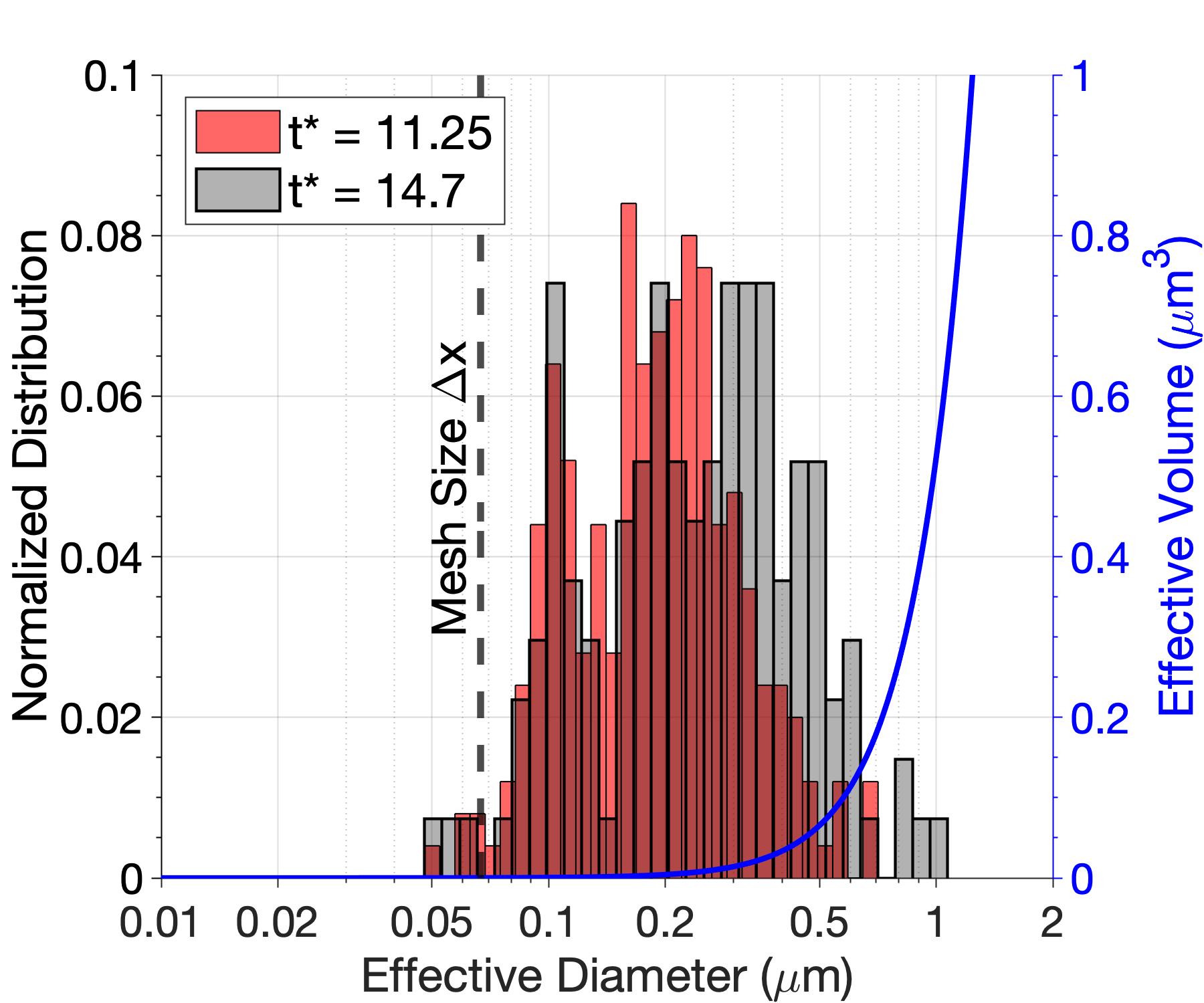}
  \caption{100 bar and \(u_G=50\) m/s (B1)}
  \label{subfig:drop_distr_100_50A}
\end{subfigure}%
\\
\begin{subfigure}{0.33\textwidth}
  \centering
  \includegraphics[width=1.0\linewidth]{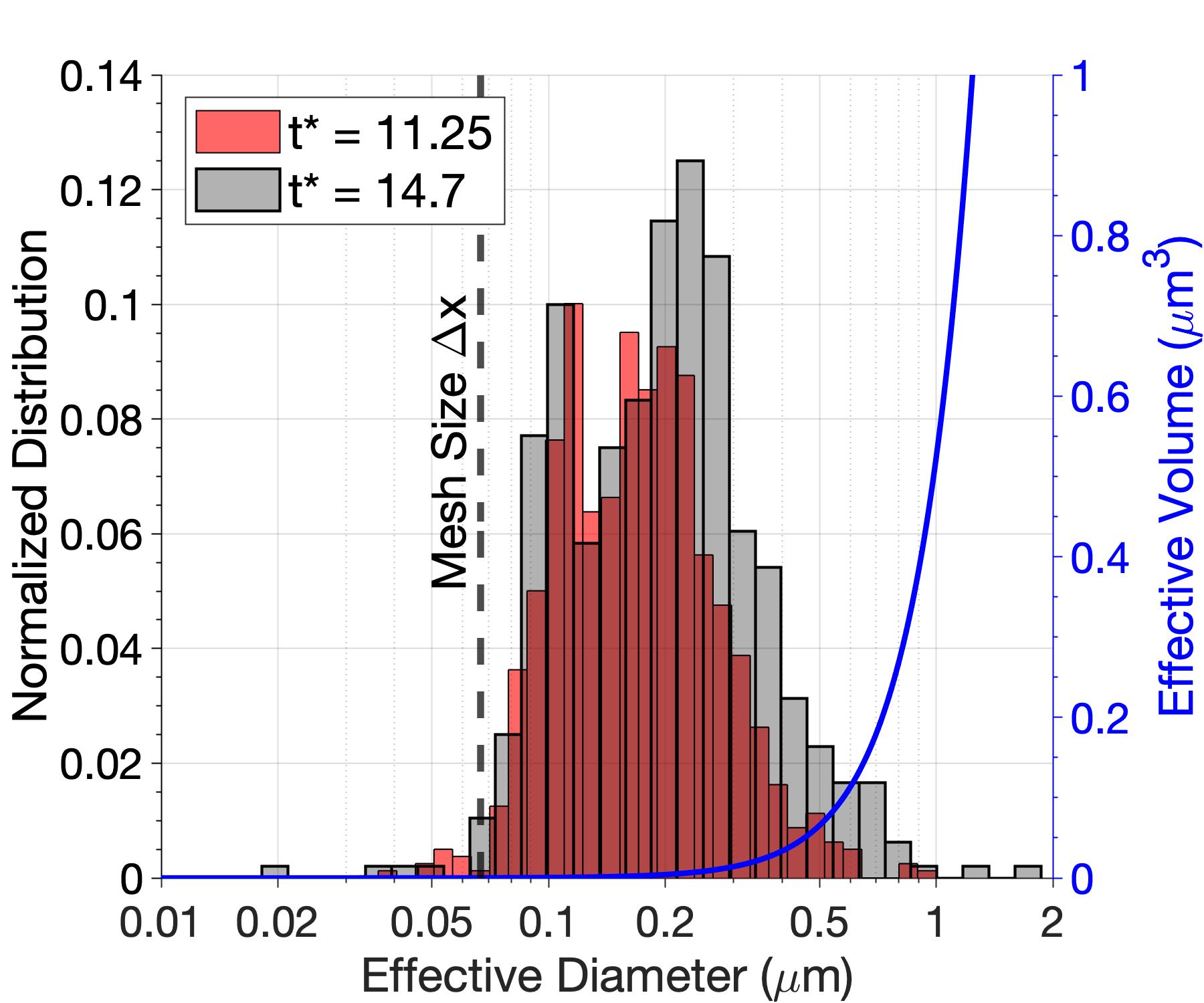}
  \caption{100 bar and \(u_G=70\) m/s (B2)} 
  \label{subfig:drop_distr_100_70A}
\end{subfigure}%
\begin{subfigure}{0.33\textwidth}
  \centering
  \includegraphics[width=1.0\linewidth]{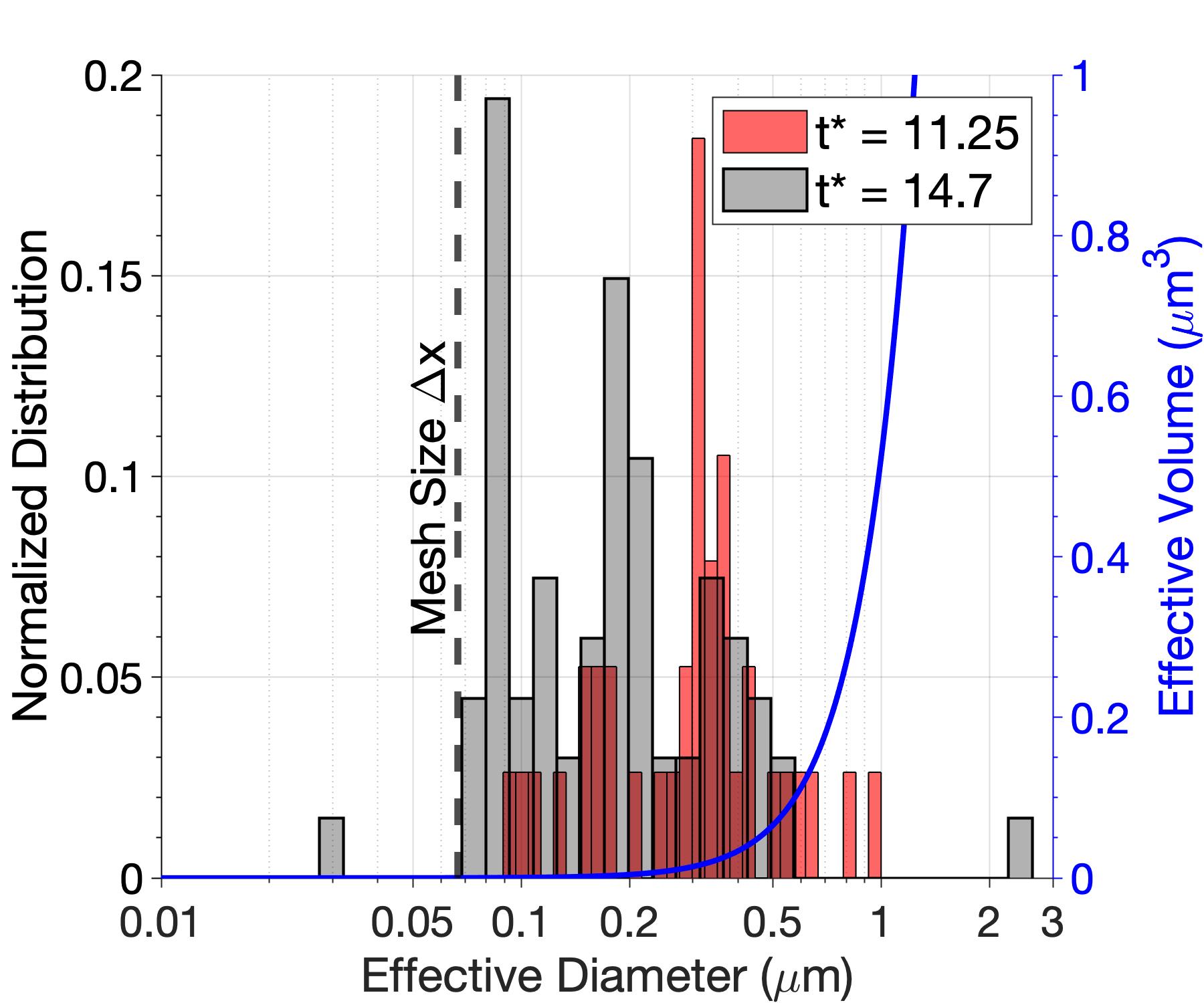}
  \caption{150 bar and \(u_G=30\) m/s (C1)}
  \label{subfig:drop_distr_150_30A}
\end{subfigure}%
\begin{subfigure}{0.33\textwidth}
  \centering
  \includegraphics[width=1.0\linewidth]{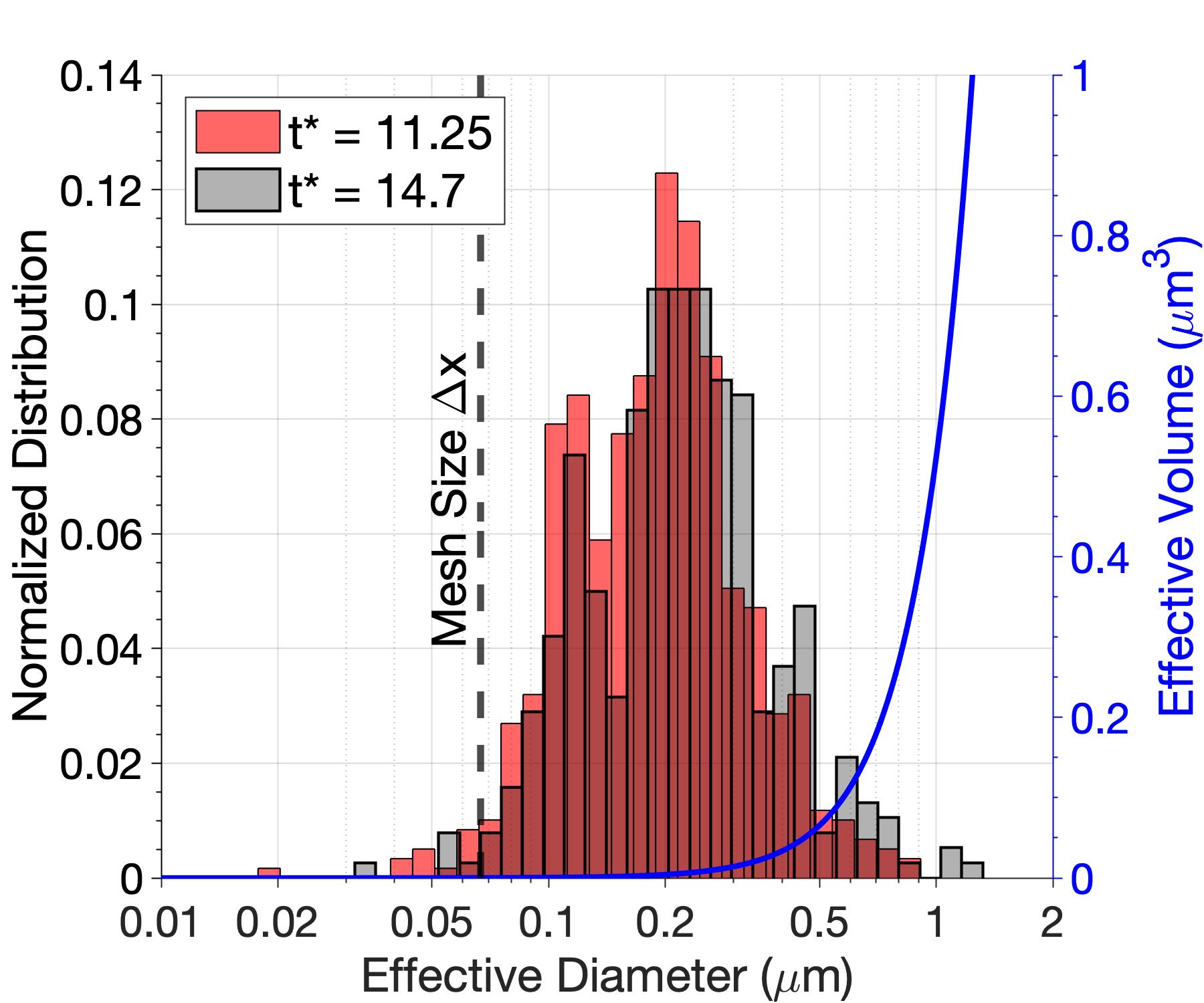}
  \caption{150 bar and \(u_G=50\) m/s (C2)}
  \label{subfig:drop_distr_150_50A}
\end{subfigure}%
\\
\begin{subfigure}{0.33\textwidth}
  \centering
  \includegraphics[width=1.0\linewidth]{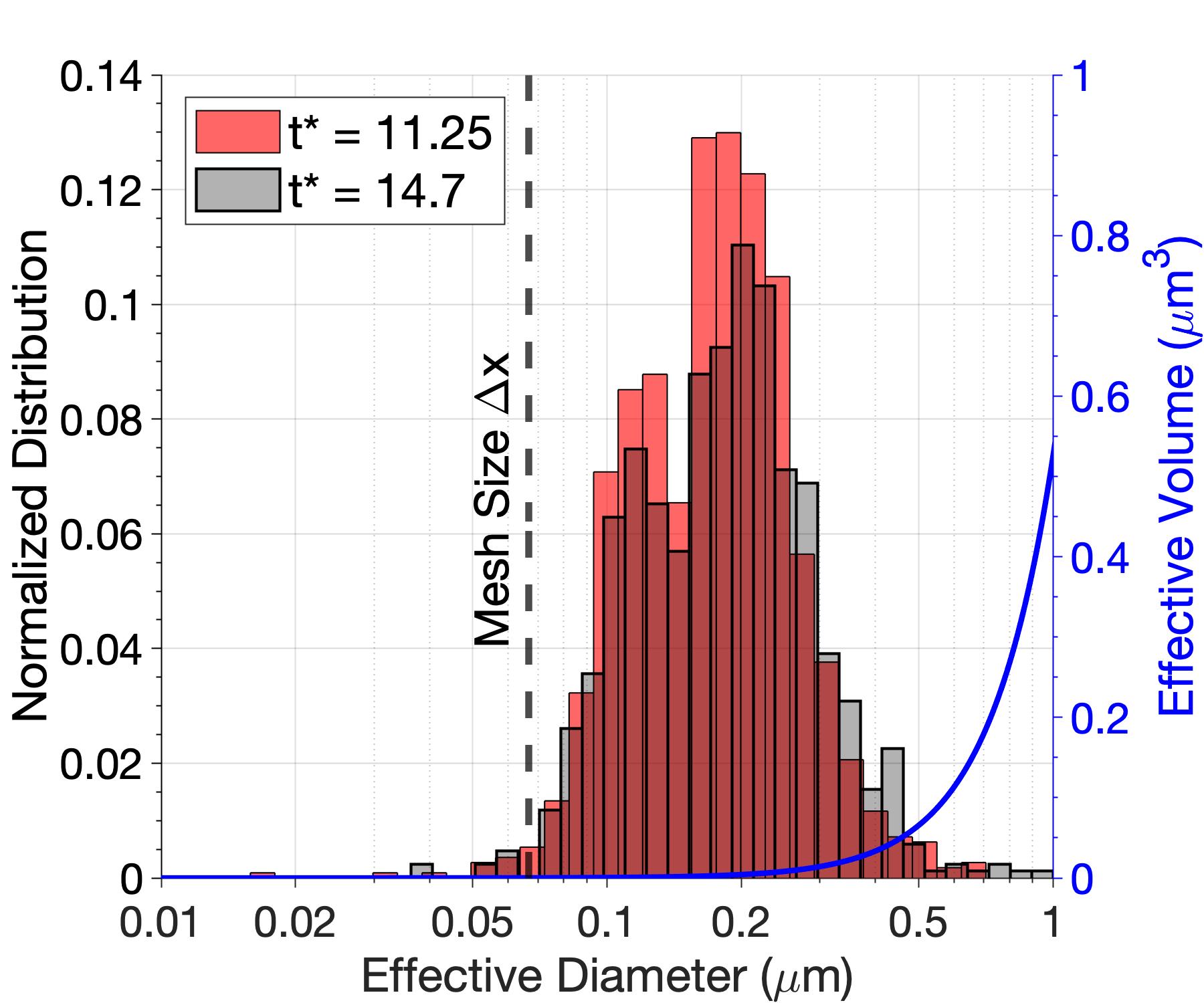}
  \caption{150 bar and \(u_G=70\) m/s (C3)} 
  \label{subfig:drop_distr_150_70A}
\end{subfigure}%
\begin{subfigure}{0.33\textwidth}
  \centering
  \includegraphics[width=1.0\linewidth]{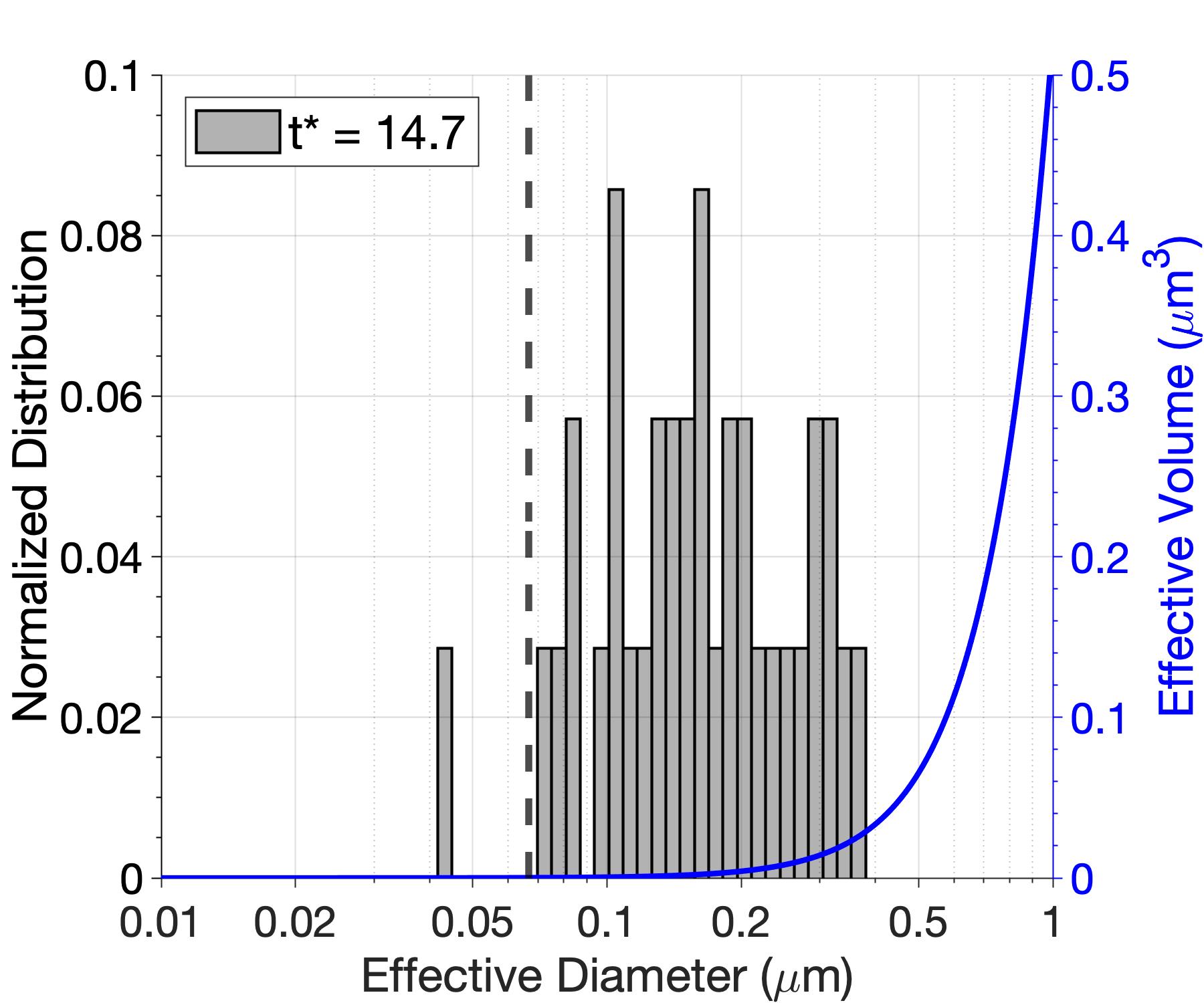}
  \caption{50 bar and \(u_G=70\) m/s (A2i)} 
  \label{subfig:drop_distr_50_70A_incomp}
\end{subfigure}%
\begin{subfigure}{0.33\textwidth}
  \centering
  \includegraphics[width=1.0\linewidth]{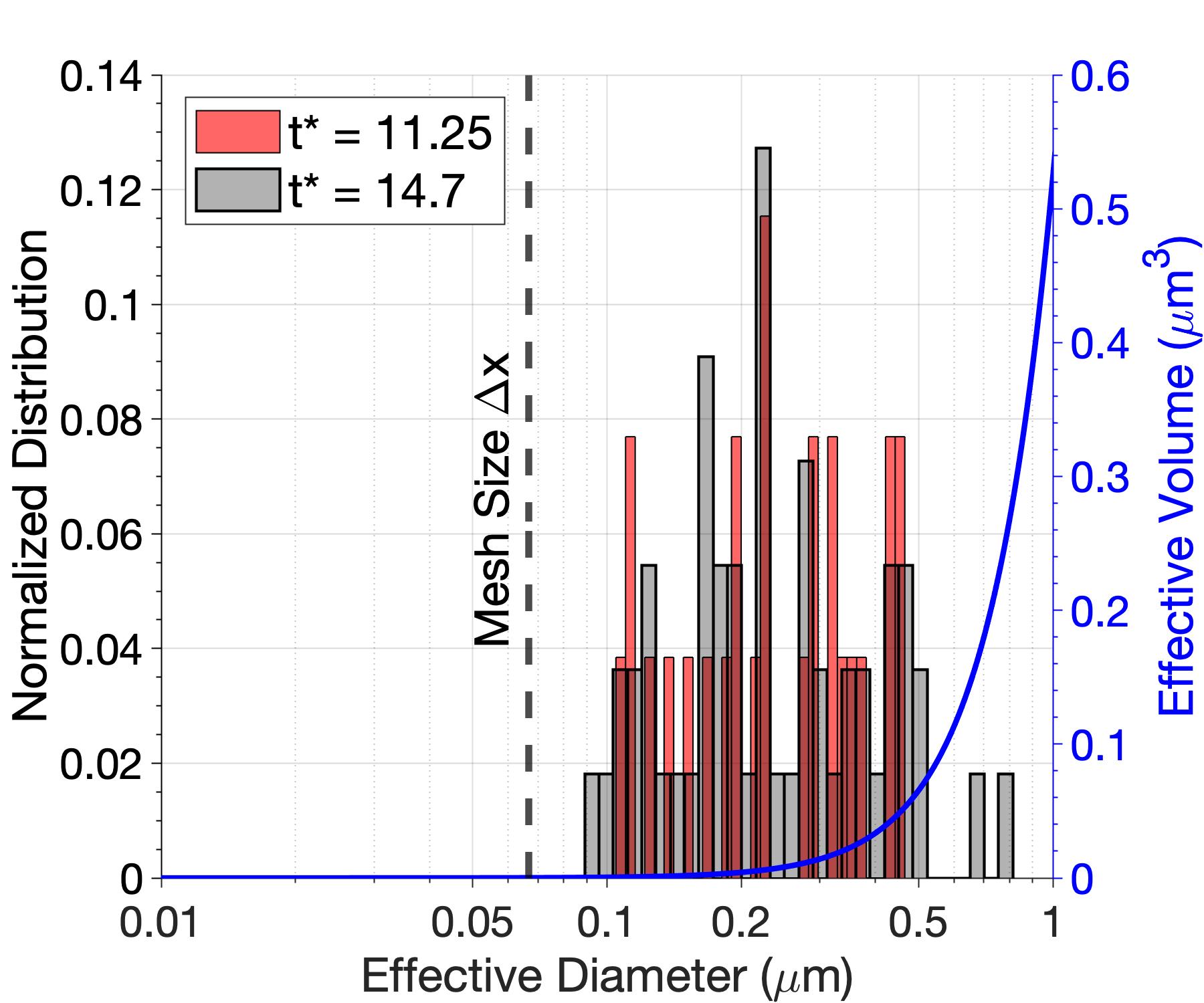}
  \caption{150 bar and \(u_G=30\) m/s (C1i)} 
  \label{subfig:drop_distr_150_30A_incomp}
\end{subfigure}%
\caption{Droplet effective diameter distribution at different instants of time for all analyzed cases. A non-dimensional time is obtained as \(t^*=t/t_c=t\frac{u_G}{H}\). The associated droplet effective volume is represented by the blue curve and the mesh resolution limit is represented by the vertical dashed line. Notice the logarithmic scale used in the horizontal axis. (a) case A1; (b) case A2; (c) case B1; (d) case B2; (e) case C1; (f) case C2; (g) case C3; (h) case A2i; and (i) case C1i.}
\label{fig:drop_distribution}
\end{figure}

Figure~\ref{fig:drop_distribution} presents the normalized distribution of the droplet size for all analyzed cases at two different instants of time \(t^*=11.25\) and \(t^*=14.7\), when available. Once a liquid structure has been identified as a droplet, an effective diameter, \(D_e\), is obtained by assuming a spherical shape and using the information on the surface area of the liquid structure such that \(S=\pi D_{e}^{2}\). The normalization of the droplet-size distribution is performed by using the total number of droplets at each analyzed time and for each case. \par

The droplet effective diameter distribution shows that almost all droplets are sub-micron, and only a few of them have effective diameters in the micron scale. This result is consistent with the reduced surface tension at high pressures. Droplet formation from capillary instabilities is minimized and only important at very small scales. For instance, assuming that droplets can maintain a spherical shape for very small local Weber numbers (i.e., in the range of 1 to 10), expected droplet diameters based on the range of fluid properties observed during the computations can be estimated. Under an extreme case with a very low surface-tension coefficient and a very fast relative velocity of 70 m/s, the droplet diameter estimate is in the order of a few nanometers. This estimate is physically questionable, both from the continuum approach and because a droplet that size would almost instantly vaporize. With more conservative estimates and lower velocities between 5 m/s to 10 m/s, expected droplet diameters are more in line with those obtained in the computations, between 0.01 \(\mu\)m and 1 \(\mu\)m. As seen in Figure~\ref{subfig:loc_distr_150_70A}, the droplets are mostly found in slower-moving flow zones, and there are essentially none in the faster-moving gaseous stream. \par

Some spherical droplets are analyzed to provide examples of local droplet Weber number, defined as \(We_\text{d}=\frac{\rho_\text{d}\Delta U^2 D_\text{d}}{\sigma_\text{d}}\). \(\rho_\text{d}\) is an approximate droplet density, \(\Delta U\) is an approximate relative velocity between the droplet and the surrounding gas, \(D_\text{d}\) is an approximate droplet diameter and \(\sigma_\text{d}\) is an average surface-tension coefficient based on the droplet's surface composition and temperature. Table~\ref{tab:sphericaldrops} presents three different spherical droplets at different pressures centered at (\(x_\text{d}\), \(y_\text{d}\), \(z_\text{d}\)): a droplet from case A2 at \(t^*=14.7\) located at (11.4, 17.3, 16.4) \(\mu\)m, a droplet from case B2 at \(t^*=13.75\) located at (22.9, 10.0, 2.1) \(\mu\)m and a droplet from case C1 at \(t^*=15\) located at (17.0, 13.9, 15.0) \(\mu\)m. The diameter of these droplets ranges from 0.47 \(\mu\)m to 0.67 \(\mu\)m and the local Weber numbers are well below unity (i.e., \(We_\text{d}<1\)). Therefore, the droplet can maintain a spherical shape. \par

\begin{table}[h!]
\begin{center}
\setlength{\tabcolsep}{4pt}
\begin{tabular}{|r|r|r|r|r|r|r|r|r|} 
\multicolumn{1}{c|}{Case} & 
\multicolumn{1}{c|}{\(x_\text{d}\) (\(\mu\)m)} & 
\multicolumn{1}{c|}{\(y_\text{d}\) (\(\mu\)m)} & 
\multicolumn{1}{c|}{\(z_\text{d}\) (\(\mu\)m)} & 
\multicolumn{1}{c|}{\(\rho_\text{d}\) (kg/m\(^3\))} &
\multicolumn{1}{c|}{\(D_\text{d}\) (\(\mu\)m)} &
\multicolumn{1}{c|}{\(\Delta U\) (m/s)} &
\multicolumn{1}{c|}{\(\sigma_\text{d}\) (mN/m)} &
\multicolumn{1}{c}{\(We_\text{d}\)}\\    
\hline
\hline
\multicolumn{1}{c|}{A2} & 
\multicolumn{1}{c|}{11.4} & 
\multicolumn{1}{c|}{17.3} &
\multicolumn{1}{c|}{16.4} & 
\multicolumn{1}{c|}{564} &
\multicolumn{1}{c|}{0.67} &
\multicolumn{1}{c|}{3.0} &
\multicolumn{1}{c|}{6.3} &
\multicolumn{1}{c}{0.54} \\
\multicolumn{1}{c|}{B2} & 
\multicolumn{1}{c|}{22.9} & 
\multicolumn{1}{c|}{10.0} & 
\multicolumn{1}{c|}{2.1} & 
\multicolumn{1}{c|}{557} &
\multicolumn{1}{c|}{0.60} &
\multicolumn{1}{c|}{1.25} &
\multicolumn{1}{c|}{4.4} &
\multicolumn{1}{c}{0.12} \\
\multicolumn{1}{c|}{C1} & 
\multicolumn{1}{c|}{17.0} & 
\multicolumn{1}{c|}{13.9} & 
\multicolumn{1}{c|}{15.0} & 
\multicolumn{1}{c|}{510} &
\multicolumn{1}{c|}{0.47} &
\multicolumn{1}{c|}{0.50} &
\multicolumn{1}{c|}{1.8} &
\multicolumn{1}{c}{0.033}
\end{tabular}
\end{center}
\caption{Spherical droplets at three different thermodynamic pressures: case A2 at 50 bar and \(t^*=14.7\), case B2 at 100 bar and \(t^*=13.75\), and case C1 at 150 bar and \(t^*=15\). Their location, approximate size and properties are shown, as well as their local Weber number.}
\label{tab:sphericaldrops}
\end{table}

The mesh resolution seems to be small enough to capture the majority of the droplets. Only a few droplets fall into a sub-grid scale, but it is noticeable that droplets are almost non-existent below a certain size. It might be possible for this size boundary to be mesh-dependent. However, it can hardly be argued that the lifetime of such droplets is relevant for the flow development. As discussed in previous lines, such small droplets will vaporize very quickly and may not be accurately represented by the discrete numerical methodology. Nevertheless, resolving such small droplets should not be a realistic expectation for a two-phase solver that considers phase change. \par 

Similarly, many droplets present only two or three cells across their diameter. This issue introduces some degree of under-resolution to the results, but the influence on the overall development of the liquid jet is expected to be minimal. As shown in Figure~\ref{fig:drop_distribution}, the effective volume of the droplets (i.e., \(V_e = \pi D_{e}^{3}/6\)) is skewed toward the larger diameters. Thus, almost all the volume contained in the droplets is represented by better-resolved droplets with five cells per diameter or more. \par 

The small droplet size and the relatively similar size distributions among all cases raise the question of whether numerical breakup has an impact on the droplet formation. For instance, it has been discussed in previous subsections that the mesh resolution can partially influence hole formation or hole burst. Nevertheless, there are substantially different interface deformation patterns across all cases to support the fact that droplet size distributions at late times are related to the two-phase flow dynamics rather than the particular numerical configuration. In any case, these results highlight the important role that ligaments may play on two-phase mixing at high pressures since droplet formation is reduced to the very small scales. \par 

An analysis on the ligament characteristic size is performed. Here, the assumption that the ligament is an exact cylinder is made. This assumption simplifies the study due to the difficulty of including other geometric parameters that may define the ligament's shape. The real picture differs somewhat, since many ligaments may be curled or curved by the flow dynamics, especially under the large stretching observed at high pressures. Given the surface area and volume of a ligament, its diameter and length can be obtained under the cylinder assumption. However, this approach does not have the desired robustness. When a ligament's shape differs considerably from an exact cylinder, predicted lengths might become nonphysical (e.g., negative or imaginary). Thus, a different path is sought to obtain effective lengths to define the ligament's size. \par 

The volume-to-surface ratio of a cylinder is \(V/S=DL/(2D+4L)\) where \(D\) is the cylinder's diameter and \(L\) its length. Under the assumption that \(L>>D\), the approximation \(2D+4L \approx 4L\) is made; thus, \(V/S\approx D_e/4\) defines an effective diameter, \(D_e\), for the ligament. Then, the effective length, \(L_e\), is obtained from the volume of a cylinder \(V=\pi D_{e}^{2} L_e /4\) instead of using the surface area of a cylinder. This way, a robust method is used where \(L_e\) is always positive. Only a few ligaments show inconsistent effective lengths (e.g., too small or too large). \par

Figure~\ref{subfig:lig_distr_150_70A} presents the normalized distribution of \(L_e\) and \(D_e\) for case C3 at \(t^*=14.7\), similar to how the effective diameter of the droplets has been presented in Figure~\ref{fig:drop_distribution}. The effective ligament length can be an order of magnitude greater than the ligament's effective diameter, suggesting the formation of very elongated ligaments. On the other hand, the effective diameter of ligaments can easily become under-resolved by the mesh as ligaments stretch as filaments and vaporization rates accelerate. This behavior is expected in regions where inertial terms dominate considerably (i.e., regions with higher velocities). Similar to the under-resolved droplets, it is not realistic to demand a finer mesh to capture the ligament thickness better as it stretches. The reduced surface tension limits the possibility of droplets forming from ligament necking, and vaporization will naturally and quickly reduce the liquid volume at such small scales, so that the lifetime of the thin region of the ligament will be short. When comparing among all cases, similar effective lengths and diameters are obtained, whereas each individual distribution differs somewhat. \par

An analysis of the correlation between the effective length and the effective diameter for case C3 at \(t^*=14.7\) shows a general trend whereby longer ligaments present larger diameters (see Figure~\ref{subfig:lig_corr_150_70A}). A boundary defined by \(D_e/L_e=0.4\) exists for all cases and almost all ligaments present a similar \(D_e/L_e\) ratio. However, some deviations from this trend exist and each individual case has some ligaments with larger length-to-diameter ratios. For case C3, a trend line defined by \(D_e/L_e=0.037\) represents the upper limit defining elongated ligaments. The boundary given by \(D_e/L_e=0.4\) appears due to the classification criteria between ligament and droplet used in this work. Any structure described under the cylindrical ligament assumption is actually classified as a droplet for approximately \(D_e/L_e>0.4\). \par 

Lastly, a distribution of the transverse location of each liquid structure is shown. An average transverse location obtained from all interface cells defining the structure is used. Such distribution is plotted in Figure~\ref{subfig:loc_distr_150_70A} for case C3 at \(t^*=14.7\). Note that both ligaments and droplets are present within the two-phase mixture region. Ligaments are more concentrated near the layering region where they are generated, while droplets concentrate closer to the center plane of the jet. The breakup of the liquid phase occurs in a well-bounded region and no structures are found above the top layers or liquid sheets that form at 150 bar. This observation can be a consequence of two factors. First, layering limits the entrainment of the gas phase as explained in Subsection~\ref{subsubsec:layering}. Therefore, the transverse development of the two-phase mixture is limited. Then, it is conceivable that any small droplets or ligaments that can be ejected into the hotter gas stream evaporate fast enough that they cannot reach too far above the liquid layers. As shown in Subsection~\ref{subsec:surfarea}, lower pressure cases (i.e., 50 bar) show a faster transverse development later in time with the presence of some droplets above the main liquid jet. This situation occurs because layering is not a dominant deformation mechanism at 50 bar. \par

\subsection{Surface-area growth and transverse development of the jet}
\label{subsec:surfarea}

\begin{figure}[h!]
\centering
\begin{subfigure}{0.33\textwidth}
  \centering
  \includegraphics[width=1.0\linewidth]{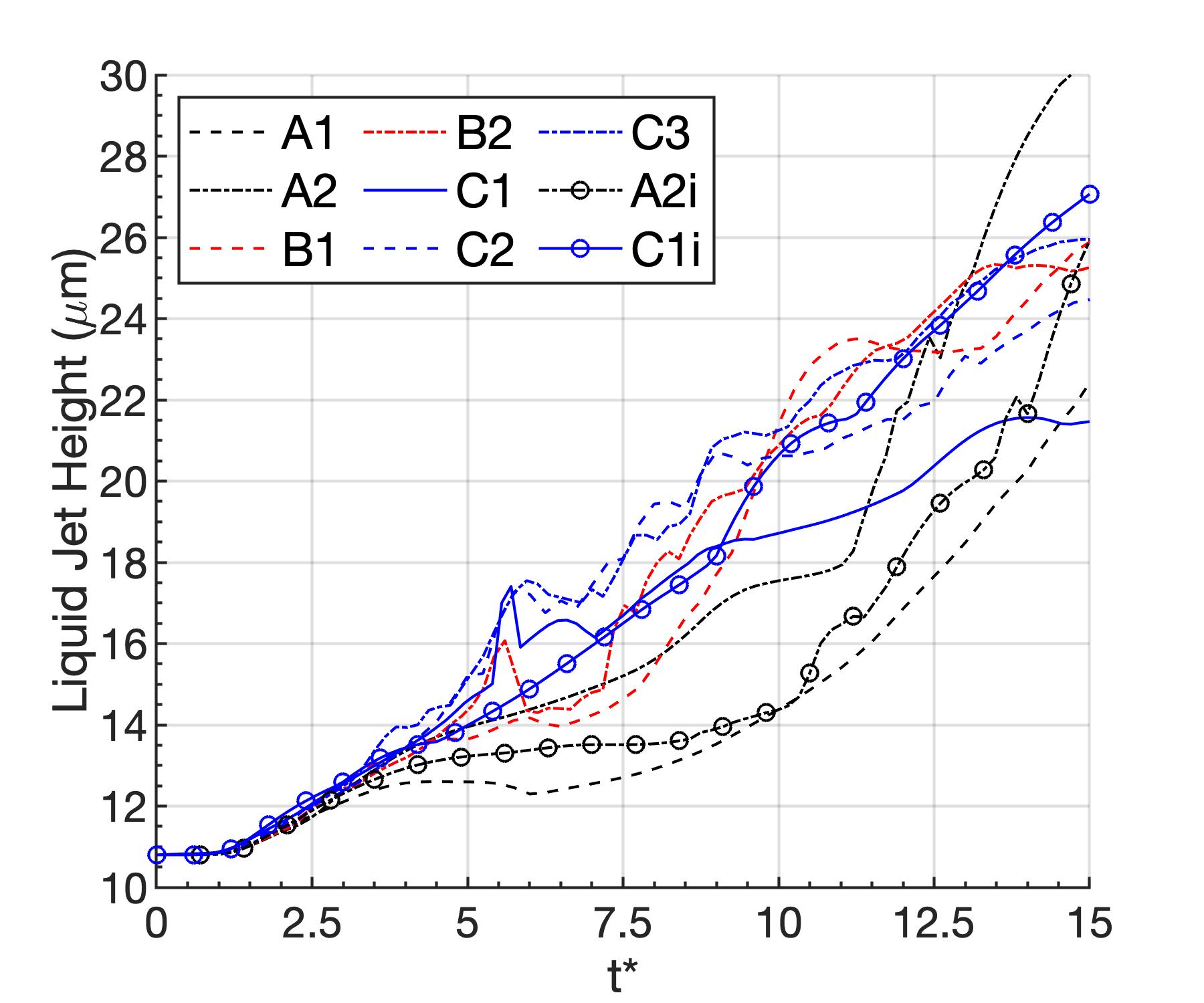}
  \caption{} 
  \label{subfig:max_transverse_location}
\end{subfigure}%
\begin{subfigure}{0.33\textwidth}
  \centering
  \includegraphics[width=1.0\linewidth]{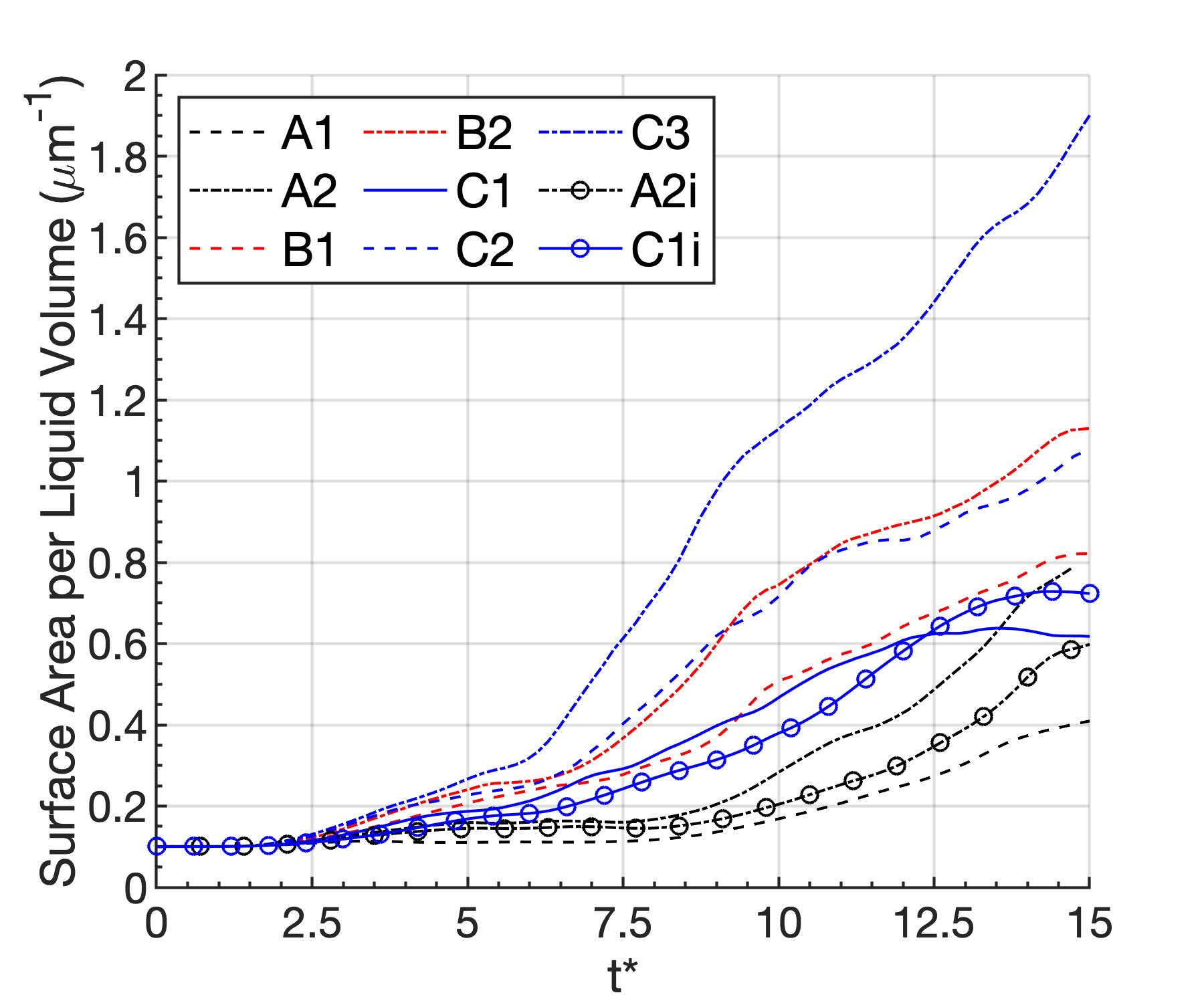}
  \caption{}
  \label{subfig:surftovoljet}
\end{subfigure}%
\begin{subfigure}{0.33\textwidth}
  \centering
  \includegraphics[width=1.0\linewidth]{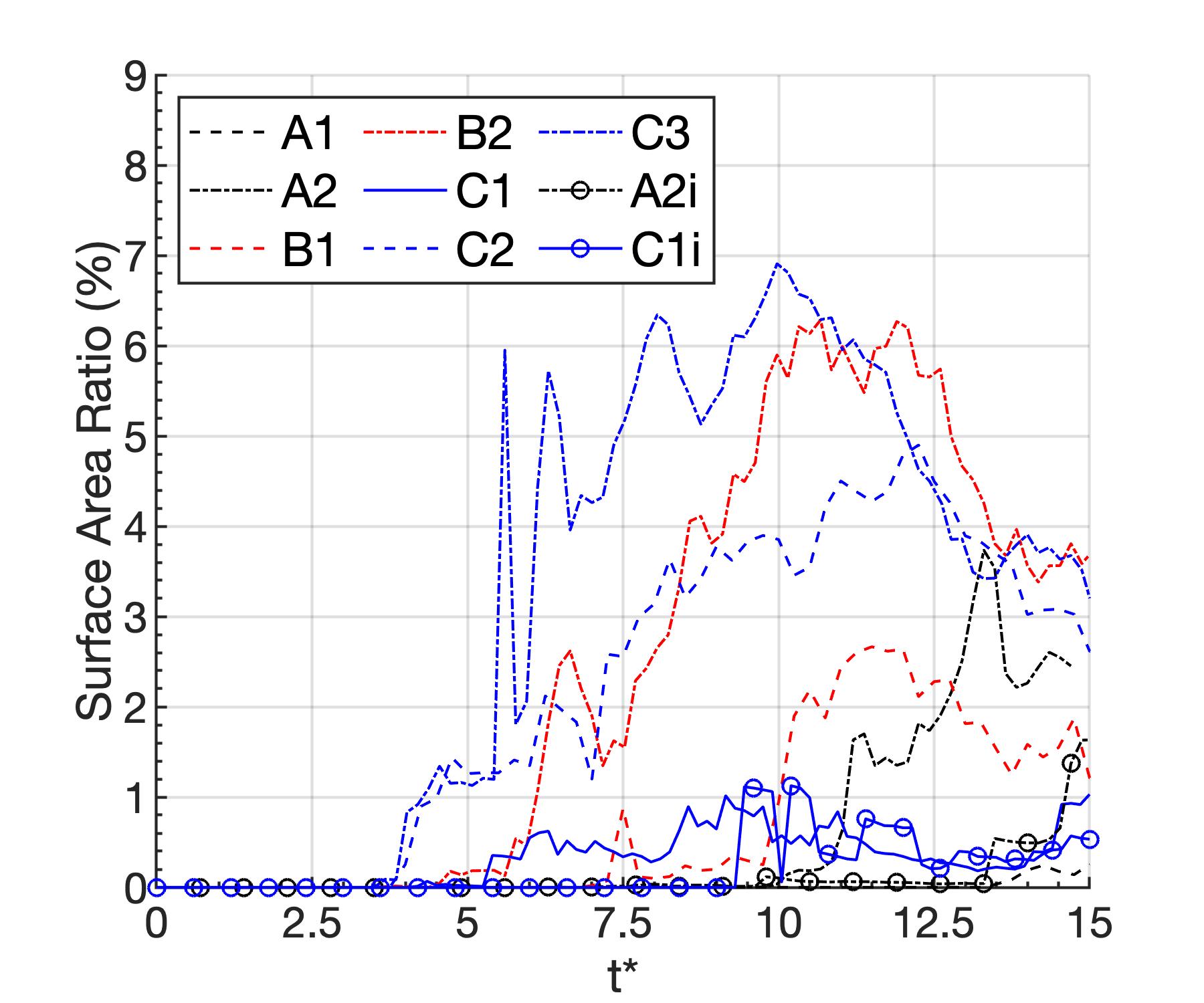}
  \caption{}
  \label{subfig:surfacearearatio}
\end{subfigure}%
\caption{Temporal evolution of the transverse development or height of the liquid jet, the ratio between the surface area and the volume of the liquid jet and the surface area ratio between detached liquid structures (i.e., ligaments and droplets) and the total liquid-phase surface area. A non-dimensional time is obtained as \(t^*=t/t_c=t\frac{u_G}{H}\). (a) surface area per liquid volume; (b) liquid jet height in the transverse direction; and (c) surface area ratio of detached liquid structures.}
\label{fig:surfacearea}
\end{figure}

This subsection discusses some general deformation features of transcritical liquid jets. Figure~\ref{fig:surfacearea} presents the transverse development of the liquid jet, the surface-area growth of the jet divided by the liquid volume and the surface area ratio between detached liquid structures and the total liquid-phase surface area. The first parameter provides a measure of how fast the jet expands away from the centerline, while the second sub-figure not only provides information on the surface-area growth, but it also provides a characteristic length of the jet structure as it deforms. The last sub-figure highlights the two-phase mechanism that drives surface-area growth. \par

The analysis of the liquid jet height evolution (see Figure~\ref{subfig:max_transverse_location}) shows that for very high pressures of 100 bar and above, the transverse development of the jet is similar. Surface-tension forces are very small and only affect the smaller scales. Thus, the initial perturbation growth, rolling and later formation of liquid sheets becomes a dominant feature that defines the transverse evolution of the jet. As discussed in Subsection~\ref{subsubsec:layering}, layer formation limits the entrainment of the gas phase, which would enhance the transverse expansion of the two-phase mixture. At 50 bar, the initial surface perturbation does not grow as fast as at higher pressures. However, the limited layering promotes the transverse development of the jet later in time and, eventually, case A1 shows a similar jet height as case C1 and case A2 is able to expand faster than higher-pressure cases. That is, some ligaments and droplets are easily propelled away from the main jet. \par 

Although the incompressible case C1i shows a faster transverse development than its compressible counterpart (i.e., case C1), both cases present a similar surface deformation over time. Differences appear because the incompressible case is not affected by mixing and, therefore, it does not deform as much as the compressible case. That is, the edges of the liquid layers do not roll as easily and holes are formed less frequently. These differences are enough to cause the upper liquid layer to extend more rapidly in the transverse direction. On the other hand, the faster transverse development of the compressible case A2 compared to the incompressible case A2i is entirely related to the generation of ligaments and droplets in the compressible configuration. \par 

The surface-area growth divided by the liquid volume (see Figure~\ref{subfig:surftovoljet}) is also representative of the absolute surface-area growth. As seen in Figure~\ref{subfig:liqvolume}, the volume of the liquid phase does not change much during the time interval covered in this work. Another advantage of representing the surface area per unit volume is that it eliminates the dependence on the size of the domain that is being covered. Moreover, all cases start with a surface area per unit volume of approximately 0.1 \(\mu\)m\(^{-1}\), which identifies the characteristic length of the jet, \(L_\text{jet}=V_\text{jet}/S_\text{jet}\), to be \(L_\text{jet}=10\) \(\mu\)m or the jet half-thickness. Over time, this characteristic length is reduced as lobes, ligaments and droplets form. At \(t^*=15\), the 50-bar cases have \(L_\text{jet}=1.27-2.44\) \(\mu\)m, the 100-bar cases have \(L_\text{jet}=0.89-1.22\) \(\mu\)m and the 150-bar cases have \(L_\text{jet}=0.53-1.62\) \(\mu\)m. As expected, the surface-area growth scales with \(We_G\). The relative strength of the inertial forces as the surface-tension coefficient decreases with pressure and the gas freestream velocity increases generates smaller structures, thereby increasing the surface area. \par

Nevertheless, an analysis of the individual impact of detached liquid structures on the total surface area suggests that the main mechanism for surface-area growth may not be the classical formation of ligaments and droplets as a result of the atomization of the liquid jet. Figure~\ref{subfig:surfacearearatio} presents the surface area ratio between the surface area of detached ligaments and droplets and the total surface area of the liquid phase. Even in cases with a strong formation of ligaments and droplets, the surface area contained in detached liquid structures is never more than 7\% of the total surface area. Moreover, it can be inferred that the liquid volume contained in detached structures is much less than 7\% of the total liquid volume, as these small detached structures will tend to have a higher surface area per unit volume compared to the rest of the liquid jet. Note that surface tension tends to minimize surface area per unit volume, but the deformation and breakup of a liquid jet into smaller structures increases the available surface area per unit volume. \par 

For cases with pressures of 100 bar and above, Figure~\ref{fig:num_structures} shows that the number of ligaments and droplets is reduced when the layering mechanism becomes dominant. Some recombination of liquid structures exists, yet the surface area of the liquid phase keeps growing. This observation is further evidence that the increase in surface area during the early times is primarily related to the surface deformation of the liquid jet (e.g., lobe and layer formation). This conclusion is not surprising given the reduced surface-tension coefficient. Capillary instabilities driving the deformation of the liquid phase and the breakup of ligaments into droplets can only take place at the very small scales. Combined with the fluid mixing effects at very high pressures, the liquid-phase evolution predominantly shows features resembling a gas-like behavior instead of the classical subcritical behavior of liquid injection and spray formation. \par

\subsection{Mass exchange across the interface}
\label{subsec:massexchange}

Phase change is also an important feature to analyze for liquid injection at supercritical pressures. When liquid fuel is injected into a hotter gas environment, the available energy drives the vaporization of the fuel and the reduction of the liquid volume. However, similar situations at very high pressures (i.e., well above the critical pressure of the fuel) can show a thermodynamic reversal whereby net condensation occurs while the fuel species still vaporizes and mixes with the oxidizer~\cite{poblador2018transient,poblador2019axisymmetric,davis2019development,poblador2021selfsimilar,poblador2021liquidjet}. At low pressures, the change across the interface in specific enthalpy, \(h\), and specific internal energy, \(e\), from liquid to gas phase follow the same direction. That is, for a hotter gas \(\Delta h >0\) and \(\Delta e >0\), thus net vaporization occurs. It can be shown that, for a constant-pressure process, \(\Delta e = \Delta h - p\Delta v\), where \(v\) is the specific volume. Usually, phase change is analyzed at low-pressure conditions where the term \(p\Delta v\) is very small (i.e., \(p\Delta v < \Delta h\)). Therefore, \(\Delta h\) and \(\Delta e\) have the same sign. However, for sufficiently high pressures, it is possible for \(p\Delta v>\Delta h\). In this scenario, a configuration with a hotter gas and \(\Delta h>0\) may show net condensation as \(\Delta e<0\). \par 

\begin{figure}[h!]
\centering
\begin{subfigure}{0.5\textwidth}
  \centering
  \includegraphics[width=1.0\linewidth]{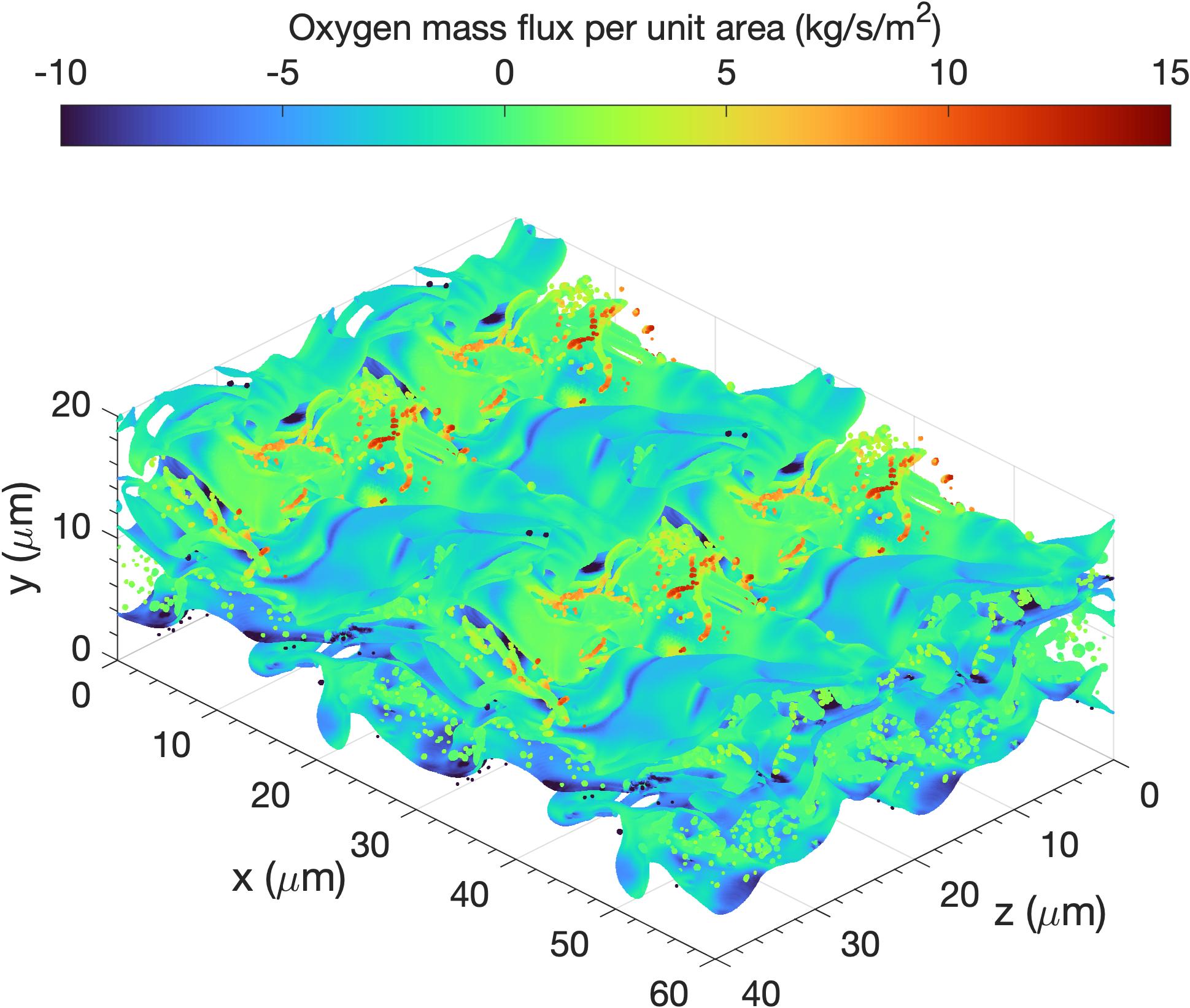}
  \caption{} 
  \label{subfig:150_50A_int_mflux_Oliq_5mus}
\end{subfigure}%
\begin{subfigure}{0.5\textwidth}
  \centering
  \includegraphics[width=1.0\linewidth]{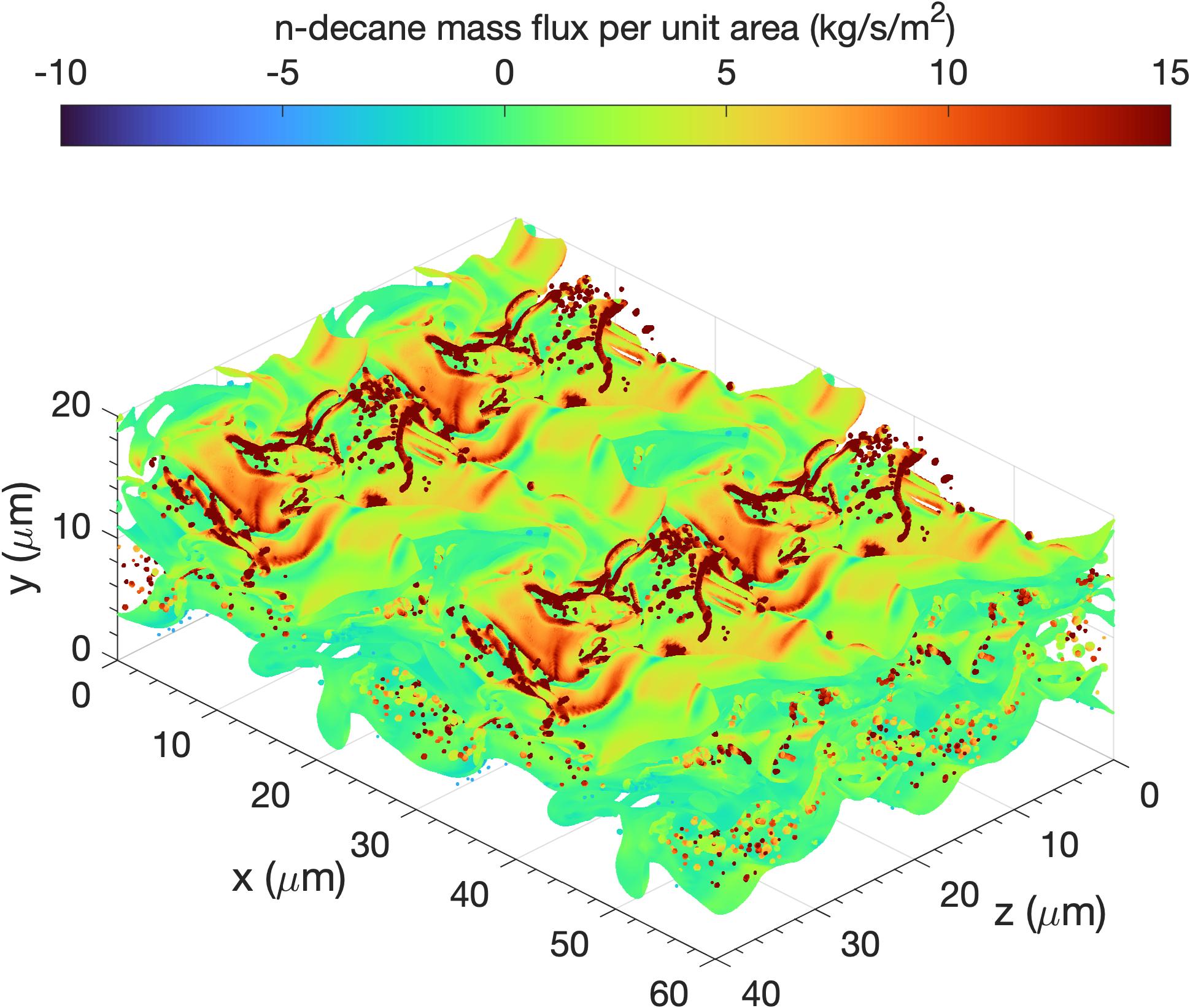}
  \caption{} 
  \label{subfig:150_50A_int_mflux_Fliq_5mus}
\end{subfigure}%
\\
\begin{subfigure}{0.5\textwidth}
  \centering
  \includegraphics[width=1.0\linewidth]{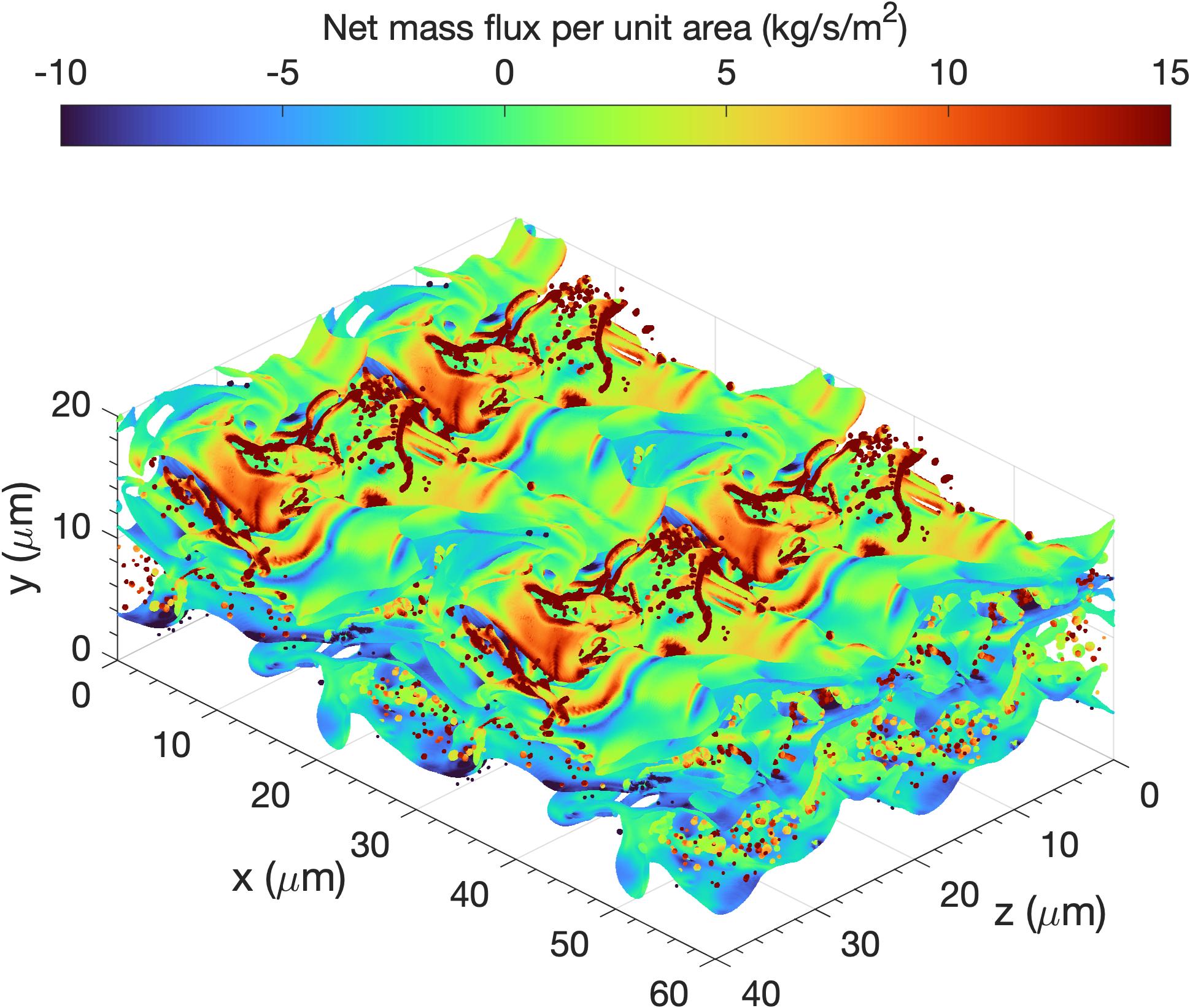}
  \caption{} 
  \label{subfig:150_50A_int_mflux_5mus}
\end{subfigure}%
\begin{subfigure}{0.5\textwidth}
  \centering
  \includegraphics[width=1.0\linewidth]{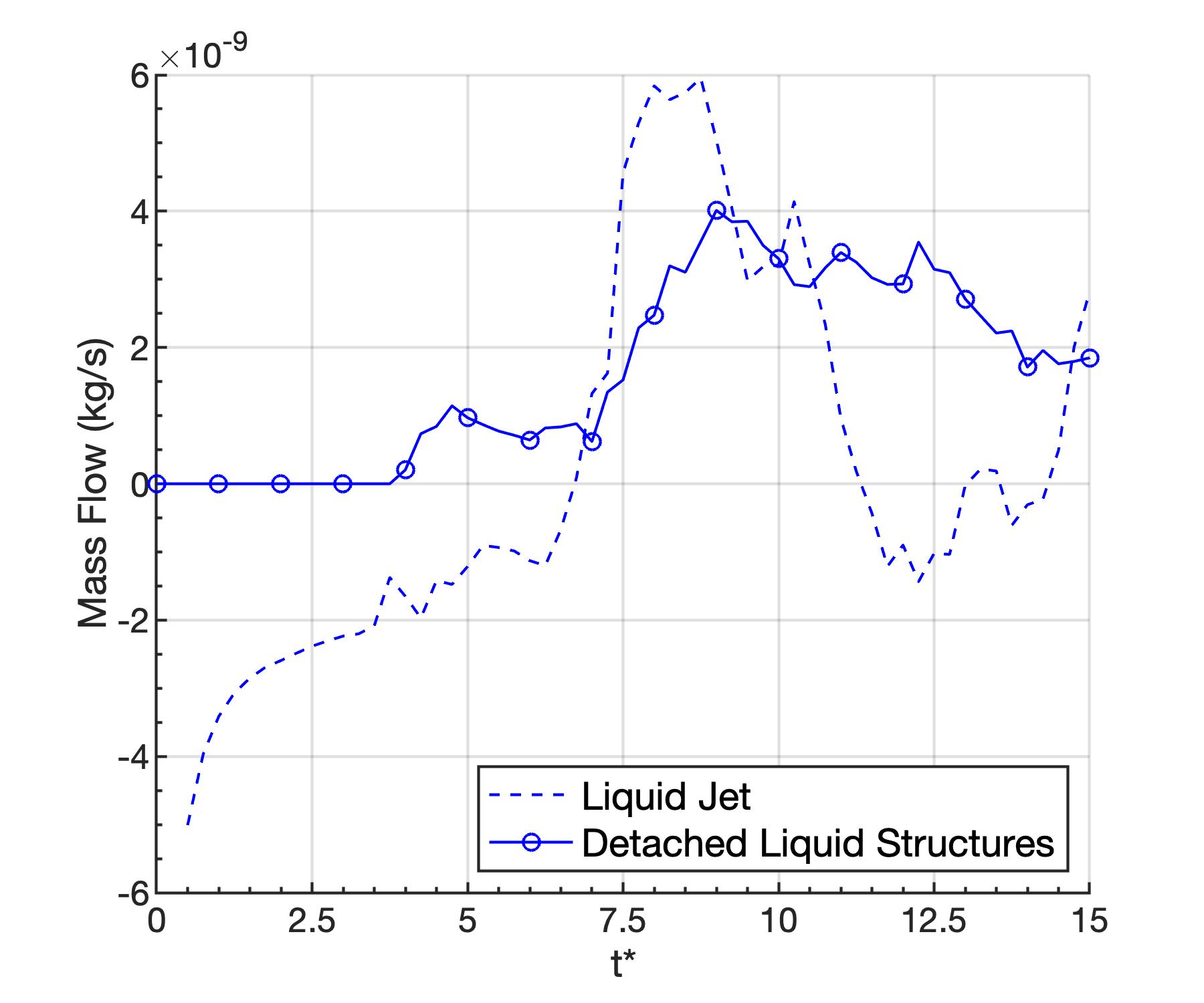}
  \caption{} 
  \label{subfig:massflow_150_50A}
\end{subfigure}%
\caption{Local mass flux per unit area across the interface at \(t^*=12.5\) and temporal evolution of the mass flow across the liquid surface at 150 bar and \(u_G=50\) m/s (i.e., case C2). A comparison between the mass flow across the main liquid jet and the detached liquid structures is provided. A non-dimensional time is obtained as \(t^*=t/t_c=t\frac{u_G}{H}\). (a) oxygen mass flux per unit area; (b) \textit{n}-decane mass flux per unit area; (c) net mass flux per unit area; and (d) mass flow across liquid structures.}
\label{fig:massflow_150_50A}
\end{figure}

This thermodynamic reversal directly results from the energy balance across the interface, which is affected by the local flow conditions near the interface (e.g., mixing layer compression and expansion, thermal and concentration gradients). One-dimensional configurations with the binary mixture of \textit{n}-decane and oxygen with the same initial temperatures as in this work show that net condensation may occur for pressures above 50 bar~\cite{poblador2018transient,davis2019development,poblador2021selfsimilar}. For liquid atomization at very high pressures, a complex picture exists where net vaporization and net condensation can happen simultaneously in different interface regions, as seen in Figure~\ref{subfig:150_50A_int_mflux_5mus}, which presents the local net mass flux per unit area across the interface for case C2 at \(t^*=12.5\), with vaporization rate as positive and condensation rate as negative. \par 

Net condensation appears in liquid regions that present enough liquid volume subject to the colder liquid jet conditions. That is, regions where the liquid can easily absorb the heat flux coming from the hotter gas. As smaller structures form and less liquid volume is available in a certain region, the liquid heats faster and, eventually, the energy input from the gas phase is enough to vaporize the liquid mixture. Therefore, net vaporization at very high pressures primarily occurs in thin lobes and liquid layers and detached liquid structures, such as ligaments and droplets. Figure~\ref{subfig:massflow_150_50A} shows the temporal evolution of the total mass flow across the contiguous liquid jet and the detached liquid structures for case C2. The net contribution of mass exchange across the interface produces net condensation for a long time because the liquid jet represents the majority of the interface surface area. However, vaporization in ligaments and droplets is very strong and may counteract the net condensation effect in the rest of the liquid phase. \par 

Figures~\ref{subfig:150_50A_int_mflux_Oliq_5mus} and~\ref{subfig:150_50A_int_mflux_Fliq_5mus} depict the dissolution rates of oxygen and the vaporization rates of \textit{n}-decane, respectively, for case C2 at \(t^*=12.5\). The fuel vaporizes everywhere, with strong vaporization rates across the small liquid structures. On the other hand, oxygen mainly dissolves into the liquid phase. Dissolution is strong near the liquid core or in cold liquid regions, overcoming the \textit{n}-decane vaporization rates and causing the net condensation shown in Figure~\ref{subfig:150_50A_int_mflux_5mus}. Nonetheless, the dissolved oxygen can vaporize again in smaller and hotter liquid structures. Similarly, it might be possible for \textit{n}-decane to condense again despite not being observed in Figure~\ref{subfig:150_50A_int_mflux_Fliq_5mus}. The strong dissolution of oxygen into the liquid phase resulting from LTE at high pressures is another reason why net condensation may occur. \par

\begin{figure}[h!]
\centering
\begin{subfigure}{0.33\textwidth}
  \centering
  \includegraphics[width=1.0\linewidth]{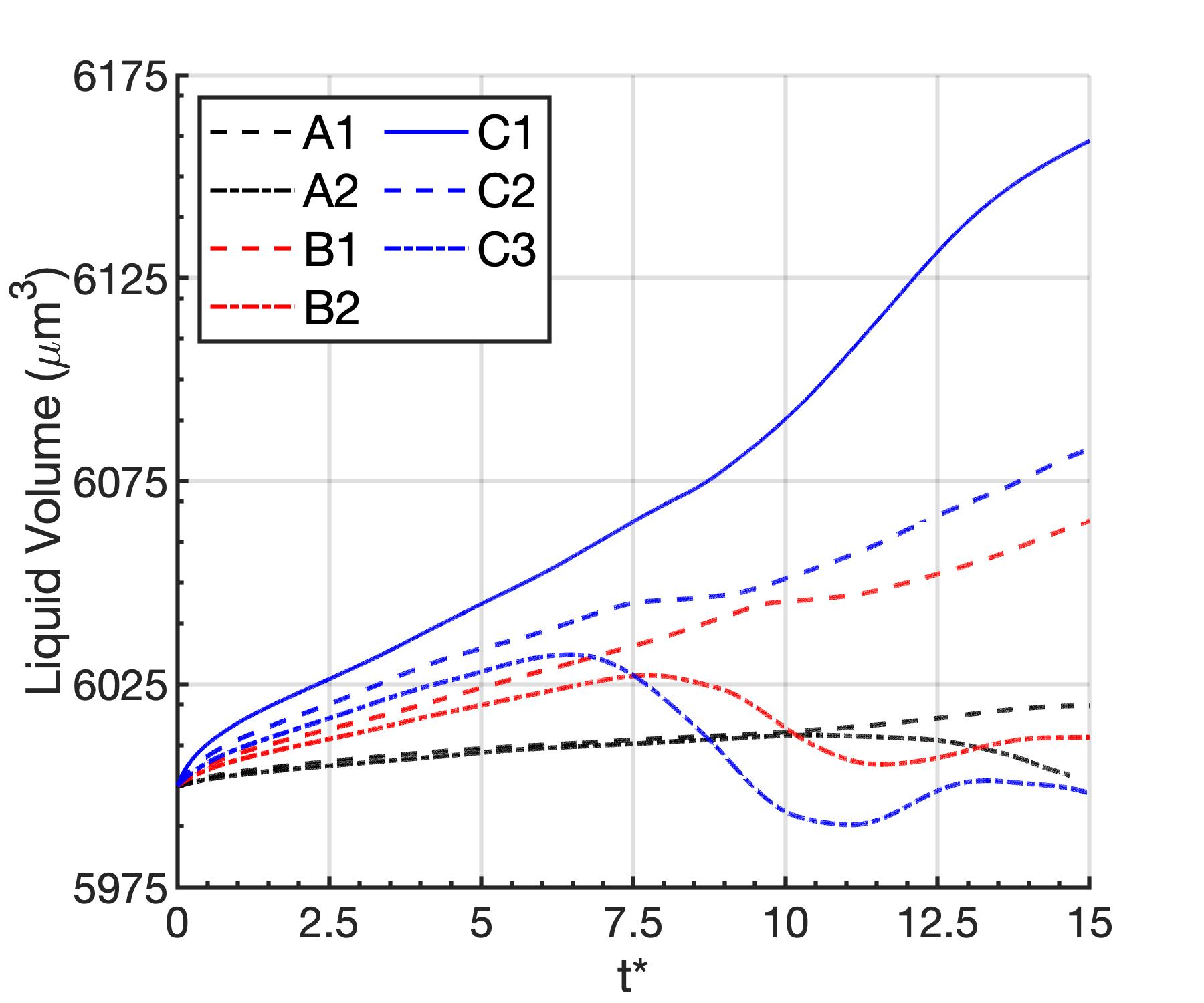}
  \caption{} 
  \label{subfig:liqvolume}
\end{subfigure}%
\begin{subfigure}{0.33\textwidth}
  \centering
  \includegraphics[width=1.0\linewidth]{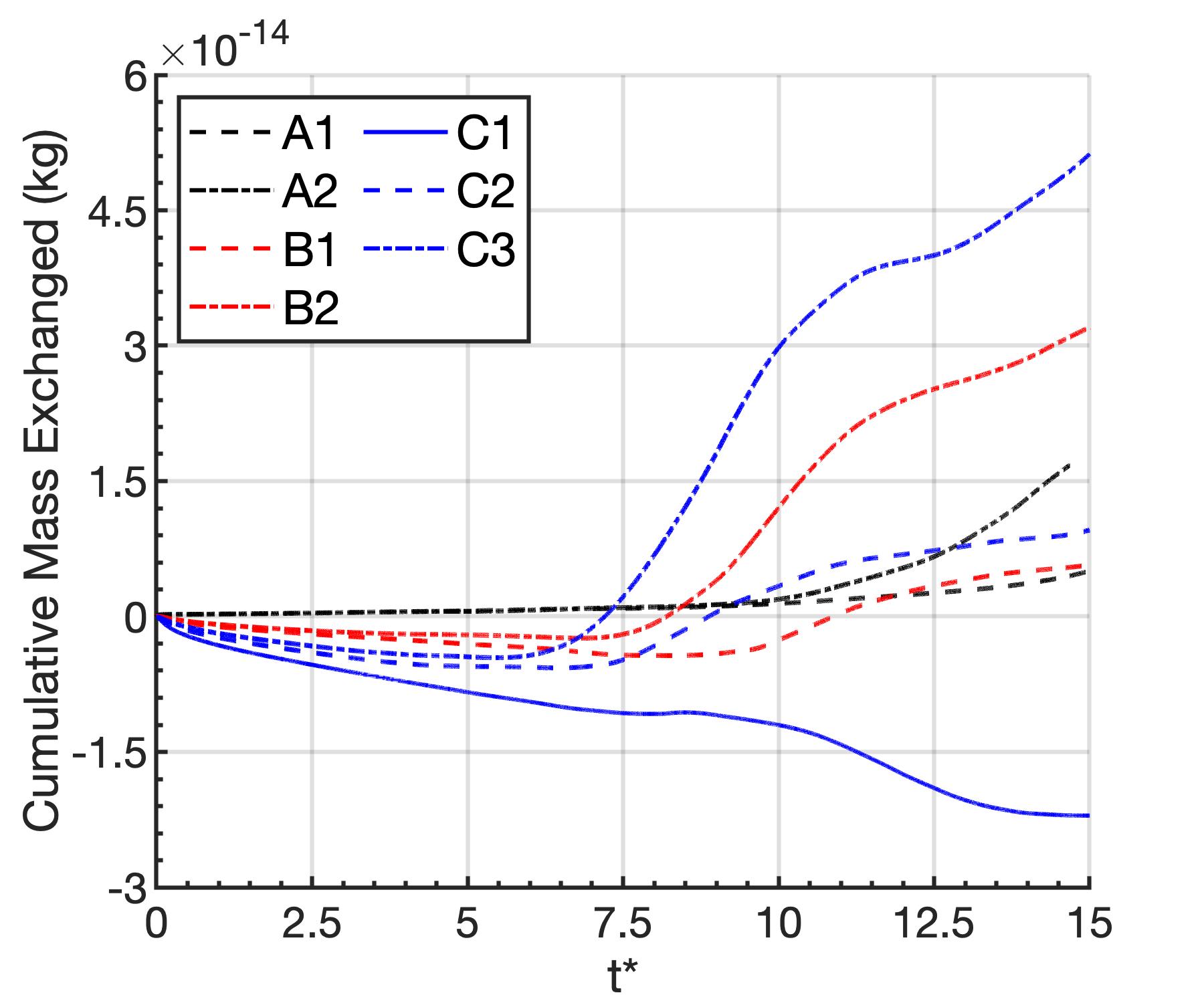}
  \caption{}
  \label{subfig:massexchanged}
\end{subfigure}%
\begin{subfigure}{0.33\textwidth}
  \centering
  \includegraphics[width=1.0\linewidth]{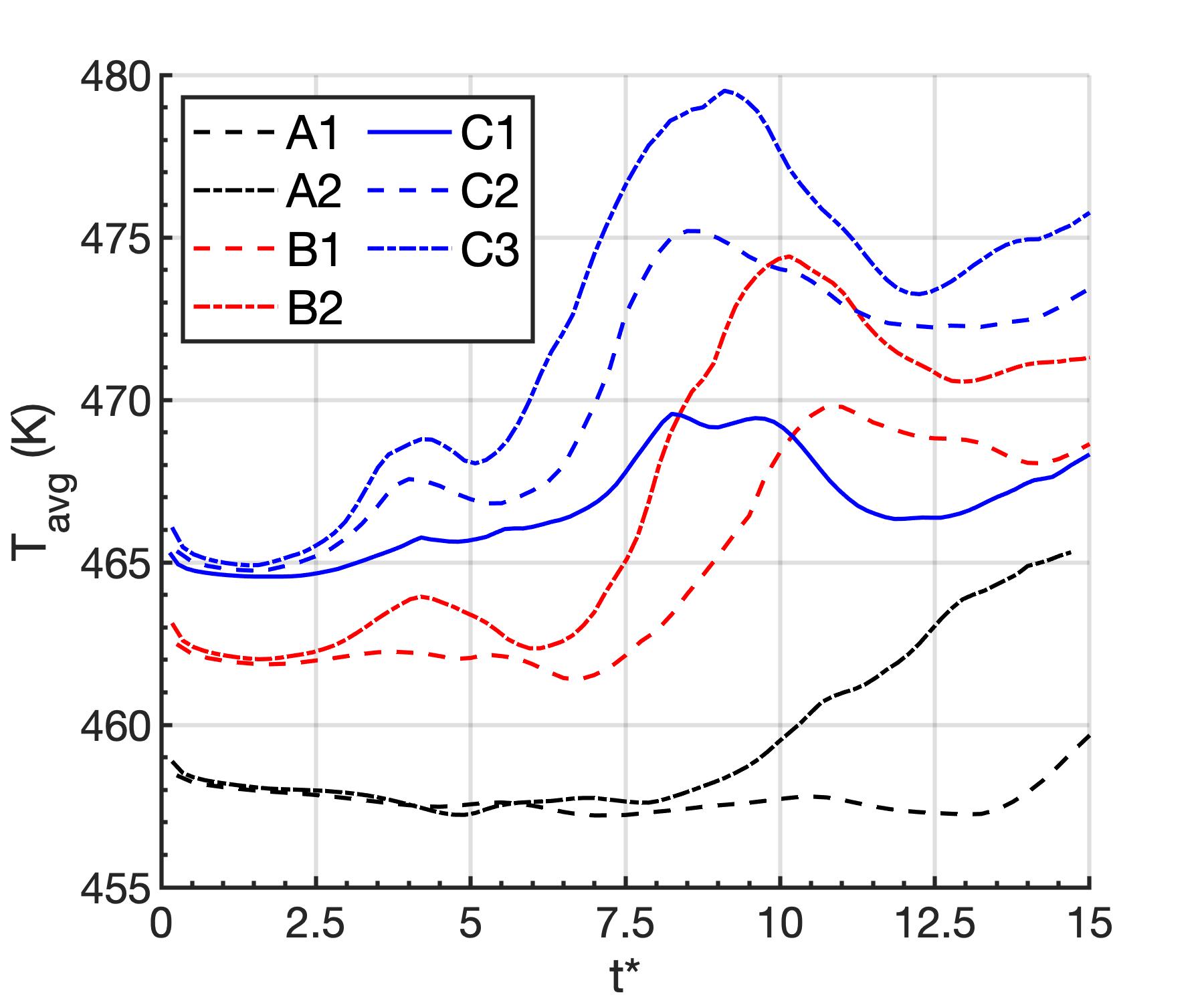}
  \caption{}
  \label{subfig:Tavg}
\end{subfigure}%
\caption{Temporal evolution of the volume of the liquid phase, the cumulative mass exchanged across the interface and the average interface temperature. A non-dimensional time is obtained as \(t^*=t/t_c=t\frac{u_G}{H}\). (a) liquid phase volume; (b) cumulative mass exchanged across the interface; and (c) average interface temperature.}
\label{fig:liqvolume_and_massexchanged}
\end{figure}

Although the net result of mass exchange tends to be the vaporization of the liquid phase, the liquid volume may not be reduced over time. Figure~\ref{subfig:liqvolume} shows the temporal evolution of the liquid-phase volume for all analyzed cases and, for example, even though case C2 shows net vaporization starting around \(t^*=6.5\), the liquid volume keeps increasing. That is, the heating of the liquid phase together with the dissolution of oxygen into the liquid, which is enhanced as pressure increases, can cause enough local liquid volume expansion such that it overcomes the volume reduction caused by a vaporizing interface. \par 

As seen in Figure~\ref{subfig:massexchanged}, the formation of ligaments and droplets, which is a predominant feature in cases A2, B2, C2 and C3, accelerates the vaporization of the liquid phase. In some situations, vaporization rates can be large enough that they can cause a reduction in the overall liquid volume (e.g., cases A2, B2 and C3). On the other hand, limited formation of such structures translates into a slowly vaporizing case (e.g., case A1) or into a continuously condensing case (e.g., case C1). The formation of smaller liquid structures tends to be accompanied by an increase in the average interface temperature (see Figure~\ref{subfig:Tavg}). The heat coming from the hotter gas can more easily raise the temperature of a small liquid structure, as well as cause its vaporization. \par 

Lastly, the effect of the layer formation at pressures above 100 bar is apparent and shows how this deformation mechanism can impact the mixing and vaporization of the liquid jet. As discussed in Subsection~\ref{subsec:droplet}, the onset of the layering process around \(t^*=10\) defines a peak in the formation of ligaments and droplets. Similarly, a peak exists in the average interface temperature as the detached structures are shielded from the hotter gas and their rate of formation slows down. With the full vaporization of some of the remaining ligaments and droplets, as well as the recombination of some liquid structures, the liquid vaporization rate also slows down and the total liquid volume may increase again under the influence of local liquid expansion. \par

\section{Summary and conclusions}
\label{sec:summary_and_conclusions}

A low-Mach-number, two-phase temporal analysis has been performed for the early atomization characteristics of a transcritical liquid \textit{n}-decane planar jet injected into oxygen under various configurations with different ambient thermodynamic pressures and gas freestream velocities, as summarized in Table~\ref{tab:cases}. The physical and numerical model handles non-reactive two-phase flows in the transcritical region where the pressure is supercritical for the injected fuel, but the temperature in the vicinity of the interface is still below the mixture critical temperature. Thus, downstream phenomena during the injection process into combustion chambers that may be affected by a hotter environment (e.g., energy release, flame front) are not captured. However, all physics involved during the early times of the injection process can be analyzed to determine what features describe the problem and how they affect the transition to full atomization and spray formation of the liquid fuel jet. \par

The deformation of the liquid jet at supercritical pressures using an initial configuration relevant to transcritical injection occurs rather rapidly, especially at the very high pressures of 100 bar and 150 bar. All the results presented in this work correspond to less than 10 \(\mu\)s in physical time. The reduced surface tension, as well as the mixing effects on the variation of fluid properties in both the liquid and gas phases explains this behavior. At high pressures, the surface-tension coefficient drops substantially as the oxygen dissolution into the liquid phase is enhanced. Moreover, the gas becomes denser and the mixing of a lighter species into the liquid phase causes a considerable reduction in the liquid density and a sharp decrease in liquid viscosity to gas-like values. Therefore, perturbations with a shorter wavelength than traditional subcritical or incompressible studies emerge. \par

\begin{figure}[h!]
\centering
\includegraphics[width=0.5\linewidth]{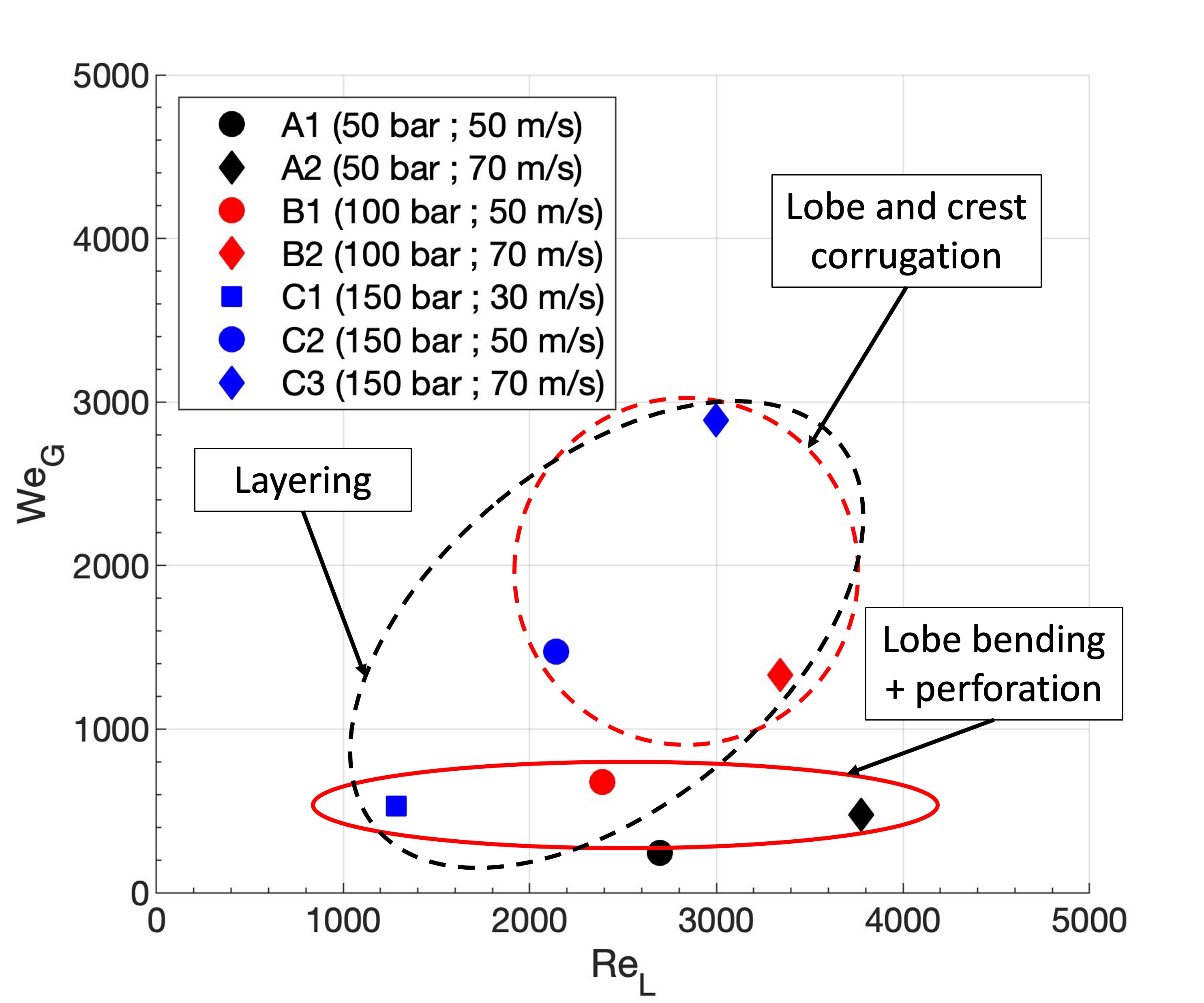}
\caption{Classification of the analyzed cases based on \(We_G\) and \(Re_L\) using freestream properties. The early deformation mechanisms are identified in the plot. The red solid curve identifies the lobe bending and perforation mechanism, the red dashed curve identifies the lobe and crest corrugation mechanism and the black dashed curve identifies the layering mechanism.}
\label{fig:Weg_Rel_deformation}
\end{figure}

It has been shown how mixing plays an important role in the local deformation of the jet, especially at 100 bar and 150 bar. Localized mixing generates thin liquid structures that can easily deform under the influence of the vortical motion in the gas phase. Eventually, such structures show hole formation and the generation of ligaments and droplets. Various predominant deformation features have been identified (e.g., lobe corrugation) according to the configuration parameters of each case. They are summarized in Table~\ref{tab:features} and presented in Figure~\ref{fig:Weg_Rel_deformation}. Additionally, the varying interface surface properties (e.g., temperature and surface-tension coefficient) are responsible for an unstable surface behavior. Hotter interface regions submerged into the faster oxidizer easily develop short-wavelength perturbations from trigger events such as surface impacts. These small perturbations grow small lobes on the liquid surface that eventually break up into small and elongated ligaments. Then, ligament shredding is observed.  \par 

For 100 bar and 150 bar, these local features overlap with the main deformation mechanism defined by the continuous layering of liquid sheets. This mechanism appears as a combined result of the low surface-tension force and the relatively thin initial shear layer that has been imposed. Compared to subcritical configurations where surface-tension forces are dominant at bigger length scales, here the main mechanism for surface-area growth is the surface deformation and layer formation rather than the atomization into ligaments and droplets. The analyzed cases show that the surface area represented by detached ligaments and droplets is never more than 7\% of the total surface area for the analyzed times. \par 

The formation of these layers has a crucial impact on the local formation of ligaments and droplets and the mass exchange across the interface. Layering causes a shielding effect that isolates part of the two-phase mixing region from the oxidizer stream. For example, it limits the entrainment of hotter gas. As a result, the formation of ligaments and droplets is reduced, the average interface temperature drops, and vaporization rates drop. This observation is important since net condensation may occur at very high pressures (i.e., above 50 bar for the analyzed binary configuration), limiting the volume reduction of the liquid phase. Only detached liquid structures can heat sufficiently such that the heat coming from the gas phase results in net vaporization. Nevertheless, the fuel species is vaporizing and mixing with the gas phase at all times. \par

The described features, which mainly impact very high-pressure scenarios, point to the mixing between fuel and oxidizer being driven by diffusion and convective transport in each phase rather than classical two-phase mixing or spray formation. Also, they help explain why experiments of transcritical liquid injection using traditional visualization techniques (e.g., shadowgraphy or ballistic imaging) have trouble capturing any two-phase behavior. Varying fluid properties across mixing regions coupled to the formation of small structures or fine droplets can be a source of scattering and refraction. Therefore, any evidence of two-phase flows may be hidden behind a cloudy structure resembling a turbulent gas jet. \par 

Some comments about necessary future work in the field and practical application of these results must be provided. First, the configurations presented in this work only represent a small data set in the big picture. Only a specific range in the thermodynamic space and injection velocities have been analyzed. Also, other multi-component configurations might show similar behaviors at different pressures and temperatures, or completely different behaviors. However, to the authors' best knowledge, this is the first work to show in detail the physical processes involved in transcritical injection and their effect on the early deformation characteristics of the liquid jet. \par

The time scales analyzed in this work are very short. The liquid phase is expected to vaporize completely for longer times, and the resulting gas mixture would become a dilute supercritical mixture of oxygen and \textit{n}-decane, unless other physics are involved (e.g., combustion reaction). During this vaporizing process, the liquid phase will stretch and break up in some localized regions, each mechanism being influenced by the ambient pressure. Moreover, length scales are small, and many of the liquid structures that form will have short lifespans. Nonetheless, understanding how high-pressure physics defines the early deformation mechanisms of liquid injection at transcritical conditions is necessary to improve the design and operation of combustion chambers. \par

Experimenters with practical continuous combustors (e.g., the gas-turbine-engine combustor) have found that atomization at moderate pressures has resulted in a non-monotonic spatial variation of mixture ratio prior to combustion whether the mixture is gas-phase solely or two-phase. These local composition variations are critical in the development of combustion instabilities~\cite{o2015transverse}. The evidence here at higher pressures from these early deformation studies is that a similar qualitative situation can arise. The non-uniformity of the vorticity field due to vortex stretching and roll-up and the variation of the mixture composition due to local condensation or vaporization contribute to this finding. \par

Regarding future studies, it is still unclear what mixing patterns between the fuel and the oxidizer develop. Because mixing is mainly driven by diffusion and convective transport in each phase, it is expected that vortex dynamics will explain the causes for the deformation of the liquid surface and phase-wise mixing patterns. At the same time, the varying fluid properties may impact the development of vortical structures. Simulation capabilities should be improved to include the modeling of the transition from a sharp two-phase interface to a diffuse single-phase mixing between the compressible liquid and the dense gas to capture scenarios where phase equilibrium is no longer valid; that is, near and beyond the mixture critical point. Additionally, a revision of the low-Mach-number formulation is necessary to analyze injection at faster velocities. However, the computational efficiency introduced by the FFT pressure solver would be lost. Therefore, the level of resolution offered in this work might not be a realistic possibility for years to come for full-scale simulations of spatially-developing jets. Nevertheless, the insight offered by this type of study may be crucial to developing reduced-order models or sub-grid models to be used in full-scale analyses of injection of transcritical liquid fuels into combustion chambers. For instance, models might be developed that quantify the generation of droplets, the increase in surface area per unit volume, and vaporization rates below the mesh size. \par

\section*{Conflict of interest}

The authors declared that there is no conflict of interest.

\section*{Acknowledgments}

The authors are grateful for the support of the NSF grant with Award Number 1803833 and Dr. Ron Joslin as Scientific Officer. The authors are also grateful for the helpful discussions in developing this work with Prof. Antonino Ferrante and his student Pablo Trefftz-Posada, from the University of Washington. The incompressible VOF subroutines from his group shared with us are also appreciated.  \par

This work utilized the infrastructure for high-performance and high-throughput computing, research data storage and analysis, and scientific software tool integration built, operated, and updated by the Research Cyberinfrastructure Center (RCIC) at the University of California, Irvine (UCI). The RCIC provides cluster-based systems, application software, and scalable storage to directly support the UCI research community. https://rcic.uci.edu \par

Also, this work used the Extreme Science and Engineering Discovery Environment (XSEDE), which is supported by the NSF grant number ACI-1548562. The computational resources used through the startup allocation MCH210008 were \textit{Stampede2} at the University of Texas at Austin, \textit{Bridges2} at the University of Pittsburgh and \textit{Expanse} at the University of California San Diego. \par 



\newpage
\appendix

\section{Volume-corrected Soave-Redlich-Kwong equation of state}
\label{apn:srk}

The volume-corrected SRK equation of state is expressed in terms of the compressibility factor, \(Z\), as

\begin{equation}
\label{eqn:SRKEoS}
Z^3+(3B_{*}-1)Z^2+\big[B_{*}(3B_{*}-2)+A-B-B^2\big]Z+B_{*}(B_{*}^{2}-B_{*}+A-B-B^2)-AB=0
\end{equation}

\noindent
with

\begin{equation}
Z = \frac{p}{\rho RT} \quad ; \quad A = \frac{a(T)p}{R_{u}^{2}T^2} \quad ; \quad B = \frac{bp}{R_uT} \quad ; \quad B_{*} = \frac{c(T)p}{R_uT}
\end{equation}

Some details regarding the parameters of the equation of state and the mixing rules are the following. \(a(T)\) is a temperature-dependent cohesive energy parameter; \(b\) represents a volumetric parameter related to the space occupied by the molecules; and \(c(T)\) is a temperature-dependent volume correction. \(R\) and \(R_u\) are the specific gas constant and the universal gas constant, respectively. Quadratic mixing rules are used to determine \(a(T)\), \(b\) and \(c(T)\)~\cite{soave1972equilibrium,lin2006volumetric}, which are shown to work well for non-polar fluids. Binary interaction parameters, \(k_{ij}\), are obtained from vapor-liquid equilibrium experimental data, when available. Models can be developed for each equation of state, such as the correlations presented by Soave et al.~\cite{soave2010srk}. However, data availability is scarce, in which case approximations have to be made, such as neglecting the influence of the binary interaction parameters by setting \(k_{ij}=0\). This work deals with an oxygen-alkane mixture, for which the authors have not been able to find useful data to estimate \(k_{ij}\) for the SRK equation of state. Nevertheless, \(k_{ij}\approx 0\) for nitrogen-alkane mixtures~\cite{soave2010srk} and, under the assumption that nitrogen and oxygen are similar components, the binary interaction coefficient is neglected and set to zero. \par

\section{Split pressure-gradient method for low-Mach-number flows}
\label{apn:pressuremethod}

A predictor-projection method is used to solve the pressure-velocity coupling in the momentum equation. With a first-order temporal integration, the predicted one-fluid velocity, \(\vec{u}^p\), is obtained from

\begin{equation}
\label{eqn:mom_pred}
\vec{u}^p = \frac{\rho^n\vec{u}^n}{\rho^{n+1}} + \frac{\Delta t}{\rho^{n+1}}\bigg[ -\nabla\cdot \big(\rho\vec{u}\vec{u}\big)^n + \nabla\cdot\bar{\bar{\tau}}^n + \frac{\rho^{n+1}}{\langle\rho\rangle}\bigg(\sigma^{n+1}\kappa^{n+1}\nabla C^{n+1} + \nabla_s\sigma^{n+1}|\nabla C^{n+1}|\bigg)\bigg]
\end{equation}

\noindent
where the surface-tension force has been included as a localized source term under the CSF framework~\cite{brackbill1992continuum,kothe1996volume,seric2018direct}. \(\langle\rho\rangle=\frac{1}{2}(\rho_G+\rho_L)\), where \(\rho_G\) and \(\rho_L\) are the initial pure gas and pure liquid densities, respectively. \par

The pressure field is obtained from a pressure Poisson equation (PPE) using the split pressure-gradient technique~\cite{dodd2014fast,pobladoribanez2021volumeoffluid} as

\begin{equation}
\label{eqn:ppe2}
\nabla^2 p^{n+1} = \nabla \cdot \bigg[ \bigg( 1-\frac{\rho_0}{\rho^{n+1}}\bigg)\nabla \hat{p} \bigg] + \frac{\rho_0}{\Delta t}\bigg(\nabla\cdot\vec{u}^p-\nabla\cdot\vec{u}^{n+1}\bigg)
\end{equation}

\noindent
where \(\rho_0 = \text{min}(\rho)\) and \(\hat{p}=2p^n-p^{n-1}\) is an explicit linear extrapolation in time of pressure. The PPE is a valid equation to obtain the pressure field under the low-Mach-number modeling. Eq. (\ref{eqn:ppe2}) is a constant-coefficient PPE for \(p^{n+1}\) which can be solved with a computationally efficient Fast Fourier Transform (FFT) method~\cite{dodd2014fast,costa2018fft}. The computational savings introduced by the split pressure-gradient method are significant since adding a thermodynamic model and a resolved interface adds extra computational cost that scales up with the surface-area growth in atomization problems. Once the pressure field is obtained, the predicted velocity is corrected to get the one-fluid velocity field as

\begin{equation}
\label{eqn:mom_proj2}
\vec{u}^{n+1} = \vec{u}^p - \Delta t \bigg[ \frac{1}{\rho_0}\nabla p^{n+1} + \bigg(\frac{1}{\rho^{n+1}}-\frac{1}{\rho_0}\bigg)\nabla \hat{p} \bigg]
\end{equation}

The one-fluid continuity constraint, \(\nabla\cdot\vec{u}^{n+1}\), is obtained as presented in Duret et al.~\cite{duret2018pressure} (see Eq. (\ref{eqn:divnewtime})). For the sake of a cleaner notation, the \(n\)+1 superscript has been dropped from the terms on the right hand side of the equation. \par 

\begin{equation}
\label{eqn:divnewtime}
\nabla\cdot\vec{u}^{n+1}=-(1-C)\frac{1}{\rho_g}\frac{D\rho_g}{Dt}-C\frac{1}{\rho_l}\frac{D\rho_l}{Dt} + \dot{m}\bigg(\frac{1}{\rho_g}-\frac{1}{\rho_l}\bigg)
\end{equation}

In Eq.~(\ref{eqn:divnewtime}), \(\dot{m} = \dot{m}'\delta_\Gamma\) is the mass flux per unit volume, which is related to the mass flux per unit area using the concept of interfacial surface area density, \(\delta_\Gamma\)~\cite{palmore2019volume}. This term activates phase-change effects only at interface cells. Then, fluid compressibilities for the binary mixture are obtained from the thermodynamic model and the solution of the governing equations for species mass fraction and enthalpy as 

\begin{equation}
\label{eqn:rhochange2}
-\frac{1}{\rho}\frac{D\rho}{Dt}  = \frac{1}{c_p\bar{v}}\frac{\partial \bar{v}}{\partial T}\bigg|_{Y_i}\frac{Dh}{Dt} + \Bigg(\frac{\rho}{W_O}\frac{\partial \bar{v}}{\partial X_O}\bigg|_{T,X_{j\neq i}} -\frac{\rho}{W_F}\frac{\partial \bar{v}}{\partial X_F}\bigg|_{T,X_{j\neq i}} - \frac{h_O-h_F}{c_p\bar{v}}\frac{\partial \bar{v}}{\partial T}\bigg|_{Y_i}\Bigg)\frac{DY_O}{Dt}
\end{equation} 

\noindent
where the thermodynamic derivatives are evaluated at constant pressure and at time \(n\)+1. \(\bar{v}\), \(W_O\) and \(W_F\) are the mixture molar volume, the molecular weight of the oxidizer species and the molecular weight of the fuel species, respectively. \par

Eq.~(\ref{eqn:rhochange2}) can be used to obtain the compressibility of each fluid in pure cells (i.e., \(C=0\) or \(C=1\)). However, to obtain each phase's compressibility at interface cells and across the interface into a narrow band of cells of about 2\(\Delta x\), the values for \(-\frac{1}{\rho}\frac{D\rho}{Dt}\) are extrapolated using the PDE technique proposed by Aslam~\cite{aslam2004partial} adapted to VOF methods. Instead of the iterative approach, the fast marching method (FMM) proposed by McCaslin et al.~\cite{mccaslin2014fast} is used. The same extrapolation PDEs are solved directly at steady state. The method becomes computationally efficient since no iterations on a pseudo-time are required. For stability purposes, a constant extrapolation is desired. \par

The extrapolation of the fluid compressibility is also a necessary step to define phase-wise velocities, \(\vec{u}_g\) and \(\vec{u}_l\). These velocities are obtained with an extrapolation technique that ensures they satisfy the previously extrapolated fluid compressibility (i.e., \(\nabla\cdot\vec{u}_l=-\frac{1}{\rho_l}\frac{D\rho_l}{Dt}\))~\cite{pobladoribanez2021volumeoffluid}. Phase-wise velocities are necessary to obtain consistent numerical stencils when discretizing the governing equations and advecting the liquid phase. \par

\newpage
\typeout{}
\bibliography{journal_bib}

\end{document}